\PassOptionsToPackage{table}{xcolor}
\documentclass[iicol,pdflatex,sn-mathphys-ay]{sn-jnl}

\usepackage{graphicx}
\usepackage{amsmath,amssymb,amsfonts,amsthm}
\usepackage{booktabs}
\usepackage{array}
\usepackage{float}
\floatstyle{ruled}
\newfloat{algorithm}{tp}{loa}
\floatname{algorithm}{Algorithm}
\floatstyle{plain}          
\usepackage{enumitem}
\setlist{nosep,leftmargin=1.4em}
\graphicspath{{figures/}}

\newlength{\snfullfig}
\newlength{\snhalffig}
\newcommand{\fullfig}[1]{\makebox[\textwidth][c]{\includegraphics[width=\snfullfig]{#1}}}

\theoremstyle{plain}
\newtheorem{theorem}{Theorem}
\newtheorem{lemma}{Lemma}
\newtheorem{proposition}{Proposition}
\newtheorem{corollary}{Corollary}

\newcommand{\E}{\mathbb{E}}
\newcommand{\Prob}{\mathbb{P}}

\newcommand{\tox}{\tilde{x}}
\newcommand{\Sinf}{S_{\infty}}

\newcommand{\tableliu}{%
projection & 0.047 & covariate-space & 0.594 & 0.698 & 0.911 & 0.734 \\
\textbf{DeepGOF-1} & \textbf{0.051} & partition (learned) & \textbf{0.465} & \textbf{0.612} & \textbf{0.834} & \textbf{0.637} \\
Stukel & 0.062 & link & 0.436 & 0.556 & 0.838 & 0.610 \\
BAGofT & 0.048 & partition (adaptive) & 0.373 & 0.590 & 0.838 & 0.600 \\
GiViTI & 0.060 & calibration curve & 0.412 & 0.541 & 0.842 & 0.598 \\
Stute-Zhu & 0.059 & covariate-space & 0.392 & 0.516 & 0.832 & 0.580 \\
HL & 0.044 & partition & 0.332 & 0.466 & 0.772 & 0.523 \\
Pigeon-Heyse & 0.028 & partition & 0.332 & 0.466 & 0.772 & 0.523 \\
mHL & 0.065 & partition & 0.280 & 0.441 & 0.813 & 0.511 \\
HL-equalwidth & 0.043 & partition & 0.300 & 0.410 & 0.752 & 0.488 \\
\midrule
Osius-Rojek (inflated) & 0.260 & global & 0.377 & 0.449 & 0.766 & 0.531 \\
}

\newcommand{\narmsvalid}{%
10}

\newcommand{\narmsword}{%
eleven}

\newcommand{\narmsvalidword}{%
ten}

\newcommand{\rankdeepvalidword}{%
second}

\newcommand{\rankdeepnfiftyword}{%
second}

\newcommand{\rankdeepnfivehundredword}{%
fifth}

\newcommand{\powdeep}{%
0.637}

\newcommand{\narms}{%
11}

\newcommand{\sizecap}{%
0.075}

\newcommand{\powstukeldeclared}{%
0.643}

\newcommand{\rankstukeldeclaredword}{%
second}

\newcommand{\powproj}{%
0.734}

\newcommand{\powstukel}{%
0.610}

\newcommand{\sizeproj}{%
0.047}

\newcommand{\sizestukel}{%
0.062}

\newcommand{\sizegiviti}{%
0.060}

\newcommand{\sizemhl}{%
0.065}

\newcommand{\sizemhlmax}{%
0.222}

\newcommand{\sizemhlsfour}{%
0.134}

\newcommand{\sizegivitimax}{%
0.084}

\newcommand{\margHL}{%
0.13/0.15/0.06}

\newcommand{\margHLW}{%
0.16/0.20/0.08}

\newcommand{\gaptoproj}{%
0.13/0.09/0.08}

\newcommand{\diffstukel}{%
-0.027}

\newcommand{\sediffstukel}{%
0.005}

\newcommand{\sdsizedeep}{%
0.004}

\newcommand{\sdsizeproj}{%
0.012}

\newcommand{\sdsizestukel}{%
0.014}

\newcommand{\tstatweak}{%
-2.9}

\newcommand{\tstatstrong}{%
-12.0}

\newcommand{\riseproj}{%
+0.95}

\newcommand{\risestutezhu}{%
+0.29}

\newcommand{\risehl}{%
+0.21}

\newcommand{\risestukel}{%
+0.30}

\newcommand{\risedeep}{%
+0.51}

\newcommand{\risebagoft}{%
+0.90}

\newcommand{\osiussizebadthree}{%
0.199}

\newcommand{\osiussizebadfour}{%
0.356}

\newcommand{\sizedeep}{%
0.0506}

\newcommand{\sizeosius}{%
0.260}

\newcommand{\bagpow}{%
0.600}

\newcommand{\bagsize}{%
0.048}

\newcommand{\bagdiscord}{%
389:581}

\newcommand{\bagmcnemar}{%
$7.6\times10^{-10}$}

\newcommand{\bagmcnemarcl}{%
$4.6\times10^{-8}$}

\newcommand{\bagseboot}{%
0.007}

\newcommand{\gridnullstotal}{%
2,000}

\newcommand{\gridnullsheld}{%
1,000}

\newcommand{\bagsecslo}{%
64}

\newcommand{\bagsecshi}{%
179}

\newcommand{\tablecert}{%
step / threshold & $2.200$ & $1.075$ \\
curvature stripe (omitted $x^{2}$) & $1.729$ & $1.644$ \\
local bump & $1.506$ & $1.493$ \\
diffuse calibration drift & $1.464$ & $1.463$ \\
interaction saddle & $1.188$ & $1.141$ \\
\midrule
most negative direction found (checkerboard) & \multicolumn{2}{c}{$-0.281$} \\
}

\newcommand{\sizegrid}{%
0.0502}

\newcommand{\tableniche}{%
interaction & 0.3 & 200 & \textbf{.079} & .058 & .072 & .073 & .058 & .052 & .070 \\
interaction & 0.5 & 200 & \textbf{.165} & .060 & .127 & .136 & .088 & .082 & .050 \\
interaction & 0.8 & 200 & \textbf{.351}$^{*}$ & .114 & .291 & .291 & .152 & .116 & .140 \\
interaction & 0.3 & 500 & \textbf{.152} & .076 & .186 & .183 & .088 & .099 & -- \\
interaction & 0.5 & 500 & \textbf{.375} & .114 & .375 & .366 & .161 & .163 & -- \\
interaction & 0.8 & 500 & \textbf{.826}$^{*}$ & .296 & .734 & .724 & .392 & .360 & -- \\
quadratic & 0.3 & 200 & \textbf{.142} & .071 & .111 & .116 & .073 & .063 & .070 \\
quadratic & 0.5 & 200 & \textbf{.397}$^{*}$ & .170 & .234 & .241 & .146 & .110 & .230 \\
quadratic & 0.8 & 200 & \textbf{.782}$^{*}$ & .449 & .418 & .423 & .214 & .142 & .470 \\
quadratic & 0.3 & 500 & \textbf{.357}$^{*}$ & .145 & .255 & .256 & .119 & .118 & -- \\
quadratic & 0.5 & 500 & \textbf{.810}$^{*}$ & .475 & .621 & .610 & .311 & .273 & -- \\
quadratic & 0.8 & 500 & \textbf{.999}$^{*}$ & .929 & .886 & .883 & .593 & .518 & -- \\
\midrule
null size & 0 & 200 & \textbf{.047} & .056 & .054 & .075 & .051 & .130 & .030 \\
null size & 0 & 500 & \textbf{.050} & .056 & .049 & .068 & .045 & .388 & -- \\
}

\newcommand{\nichenlead}{%
7}

\newcommand{\nicheleadproj}{%
11}

\newcommand{\nichedminproj}{%
.021}

\newcommand{\nichedmaxproj}{%
.530}

\newcommand{\nichesizestukel}{%
.075/.068}

\newcommand{\nichesizemhl}{%
.130/.388}

\newcommand{\nichesizese}{%
.005}

\newcommand{\nichebagsemax}{%
.050}

\newcommand{\nichebagnlead}{%
4}

\newcommand{\nichenhighest}{%
10}

\newcommand{\nichenquadhighest}{%
6}

\newcommand{\nichentests}{%
seven}

\newcommand{\abLiuNet}{.535}
\newcommand{\abLiuSS}{.510}
\newcommand{\abLiuMax}{.365}

\newcommand{\abLiuSSAhead}{5}

\newcommand{\abLiuSSLevel}{26}

\newcommand{\abLiuCells}{36}
\newcommand{\abNicheNet}{.452}
\newcommand{\abNicheSS}{.359}
\newcommand{\abNicheMax}{.220}

\newcommand{\abNicheSSAhead}{6}
\newcommand{\abNicheMaxAhead}{10}

\newcommand{\locInterRej}{.86}
\newcommand{\locInterRight}{97\%}
\newcommand{\locInterLocated}{73\%}
\newcommand{\locInterOtherRej}{.07 and .13}

\newcommand{\locNullFalseAll}{.012}
\newcommand{\locNullFalseAxes}{.006}
\newcommand{\locUSplineLocated}{95\%}
\newcommand{\locUAllLocated}{14\%}
\newcommand{\locThreshSpline}{83\%}
\newcommand{\locThreshAll}{2\%}

\newcommand{\extThreshDG}{.80}
\newcommand{\extThreshGiv}{.36}
\newcommand{\extThreshBest}{.36}

\newcommand{\extUDG}{.99}
\newcommand{\extUGiv}{.34}
\newcommand{\extUBest}{.42}

\newcommand{\extUweakDG}{.40}
\newcommand{\extUweakGiv}{.06}
\newcommand{\extUweakBest}{.10}

\newcommand{\extInterDG}{.51}
\newcommand{\extInterGiv}{.24}
\newcommand{\extInterBest}{.43}

\newcommand{\extInterWeakDG}{.15}
\newcommand{\extInterWeakBest}{.16}
\newcommand{\extLevelDGlo}{.038}
\newcommand{\extLevelDGhi}{.063}
\newcommand{\extLevelOthHi}{.076}
\newcommand{\plFullDGlo}{.020}
\newcommand{\plFullDGhi}{.040}
\newcommand{\plFullReps}{100}
\newcommand{\plDGlo}{.030}
\newcommand{\plDGhi}{.060}
\newcommand{\plProjPowK}{.57}
\newcommand{\plDGPowK}{.25}
\newcommand{\plClassPowK}{.04}
\newcommand{\tmProjHours}{9~h~53~min}

\newcommand{\tmProjGB}{4.5}
\newcommand{\tmDGaxes}{13}
\newcommand{\tmDGall}{71}
\newcommand{\tmRatio}{2{,}700}
\newcommand{\tmRatioAll}{500}
\newcommand{\tmProjP}{.005}
\newcommand{\xsSplits}{20}
\newcommand{\xsNval}{4{,}437}
\newcommand{\xsMzeroDG}{20}
\newcommand{\xsMzeroGiv}{2}
\newcommand{\xsMzeroHL}{11}
\newcommand{\xsMzeroStuk}{7}
\newcommand{\xsMzeroDGmed}{.007}
\newcommand{\xsMzeroGivmed}{.34}
\newcommand{\xsMzeroSlope}{.99}
\newcommand{\xsMzeroCitl}{.01}

\newcommand{\xsMzeroResp}{75\%}
\newcommand{\coDGten}{.088}
\newcommand{\coStukten}{.627}
\newcommand{\coGivten}{.397}
\newcommand{\coHLten}{.058}
\newcommand{\coDGfive}{.055}
\newcommand{\coStukfive}{.398}
\newcommand{\coGivfive}{.251}
\newcommand{\coDGclean}{.051}
\newcommand{\coDGxeight}{.16}
\newcommand{\coProjxeight}{.22}
\newcommand{\coStukxeight}{.49}

\newcommand{\coBagclean}{.030}
\newcommand{\coProjlo}{.040}
\newcommand{\coProjhi}{.100}
\newcommand{\coProjN}{200}
\newcommand{\coBagN}{100}
\newcommand{\coDiscMax}{12}
\newcommand{\tablecorruptmain}{1{,}000 & none & 1{,}000 & .051 & .040 & .044 & .042 & .055 (200) & .030 (100) \\
1{,}000 & $\times4$, 1 record & 1{,}000 & .047 & .051 & .121 & .089 & .080 (200) & -- \\
1{,}000 & $\times4$, 2 records & 1{,}000 & .050 & .044 & .217 & .146 & .040 (200) & -- \\
1{,}000 & $\times4$, 5 records & 1{,}000 & .055 & .042 & .398 & .251 & .100 (200) & -- \\
1{,}000 & $\times4$, 10 records & 1{,}000 & .088 & .058 & .627 & .397 & .085 (200) & .030 (100) \\}

\begin{document}

\makeatletter
\if@filesw \immediate\write\@auxout{\string\citation{snsettings}}\fi
\makeatother

\title[Where does a logistic risk model fail?]{Where Does a Logistic Risk Model Fail? An Audited Neural Goodness-of-Fit Test for Model Development and External Validation}

\author*[1]{\fnm{Ebrahim Khaled} \sur{Ebrahim}}\email{ebrahimkhaled32@gmail.com}

\author[1]{\fnm{Osama Abd El-Aziz} \sur{Hussein}}

\author[1]{\fnm{Ahmed} \sur{El-Kotory}}\email{ahmed.elkatory@alexu.edu.eg}

\affil[1]{\orgdiv{Department of Applied Statistics, Faculty of Business},
\orgname{Alexandria University}, \orgaddress{\city{Alexandria}, \postcode{5424074}, \country{Egypt}}}

\abstract{Goodness-of-fit tests for logistic regression are routine in clinical risk
modelling, yet the classical tests say whether a model misfits, not where. A pretrained network is not a test
until its level, power and blind spots are established. We audit DeepGOF-1, which renders the
residuals of a fitted logistic model as a map over the ranks of two covariates, scores the map with a
frozen convolutional network, and calibrates the score by the analyst's parametric bootstrap. The
level is first-order valid under standard bootstrap regularity and two conditions, consistency
against a named alternative is decided by one forward pass, and local power has an explicit limit.
In simulation the level holds from 20 patients to 8,873;
the network adds power over a chi-square statistic of the same map when misfit is sparse among many
covariates; the map finds a missed interaction that one-covariate diagnostics cannot see; and, applied
to a published risk model on new patients, it is an exactly valid test of calibration within patient
subgroups, where the calibration belt has little power. In the SUPPORT study of 8,873 inpatients,
DeepGOF-1 rejects a linear in-hospital mortality model, its map shows the U-shaped risks of blood
pressure and respiratory rate that a calibration curve hides, and it guides the repairs. It ran in 13
seconds; the projection test, the strongest omnibus rival, took 9 hours 53 minutes and 4.5 GB on the
same data, too slow for routine use. The test is
\texttt{deepgof1()} in the R package \texttt{ebrahim.gof}.}



\maketitle

\section{Introduction}\label{sec:intro}

Prognostic systems for critically ill adults, such as APACHE III \citep{knaus1991apache} and the
SUPPORT prognostic model \citep{knaus1995support}, turn a handful of physiological measurements into
a predicted risk of death, and clinicians read those predictions most closely at the extremes of
physiology, where decisions are hardest. APACHE III scores each physiological variable so that
abnormally low and abnormally high values both add risk, so a logistic model that enters such a
variable linearly is wrong exactly where its predictions matter most. Violating the assumptions of a
clinical prediction model degrades its calibration \citep{austin2025ph}, and calibration is assessed
far less often than discrimination \citep{krikella2025}. A goodness-of-fit test should detect such
misfit, and the modeller also needs to know where the model fails in order to repair it.

The standard instruments are partition tests: the Hosmer--Lemeshow test \citep{hosmerlemeshow1980},
its equal-width variant and refinements such as that of \citet{pigeonheyse1999} group observations,
compare observed with expected counts and refer the discrepancy to a $\chi^{2}$ law; for generalized linear models the Pearson statistic coincides with the Rao score statistic \citep{lovison2005}. Their documented
weakness is one of level \citep{hosmer1997}: at $n$ between fifty and a few hundred with modest event
counts, the rejection rate on a correctly specified model departs from $\alpha$ in both directions,
and Pearson's statistic performs poorly on sparse binary data \citep{xu2026}. A benchmark of more
than twenty such tests under sparse data finds several of them liberal and others with little power
\citep{ebrahim2026benchmark}, and the test of \citet{osius1992} rejects a true model several times
more often than its stated level on half of the designs of \citet{liu2024}. Recent work avoids
grouping altogether, through non-parametric inference on the calibration of predicted risks
\citep{sadatsafavi2024}, or gains power from the structure of case-control sampling \citep{wang2024casecontrol}. Goodness-of-fit testing has also been extended to discrete-response regressions, the binomial model included, calibrated by the parametric bootstrap \citep{meintanis2025}.
Other checks point to where a model fails: tests that partition the covariate space rather than the
predicted risk \citep{tsiatis1980}, and cumulative sums of residuals against each covariate,
calibrated by simulation \citep{suwei1991,linweiying2002}.

Machine learning now trains networks once, on large or simulated data, and applies them frozen, as
TabPFN does for small tabular datasets \citep{hollmann2025}. Such a network returns an answer but no
guarantee on how often the answer is wrong for the data at hand, and that error rate is what a test
must control. The mechanisms that give a learned statistic finite-sample control each solve a
different problem: permutation and conformal methods for classifiers in the two-sample problem
\citep{kim2021c2st}; co-sufficient sampling, which \citet{barberjanson2022} show to be powerless for
logistic regression, and their approximate version, whose guarantee holds only under
maximum-likelihood asymptotics; a finite-sample test around a black-box classifier for a fixed, not
an estimated, model \citep{javanmard2024}; and an adaptively learned partition calibrated by sample
splitting \citep{zhang2023}. Pretrained neural tests of normality exist \citep{simic2021,kim2025};
their authors note that a learned statistic does not by itself control the false-rejection rate
\citep{simic2021} and that size worsens when a nuisance parameter is misspecified \citep{kim2025}.

For a composite regression null with estimated nuisance parameters the standard route is the
parametric bootstrap, which gives first-order validity for regular statistics of the fitted residual
process \citep{stute1998boot,genest2008,baillo2026}. For a frozen learned statistic what must be
added is a check of its conditions; for a piecewise-affine network the only condition that depends on
the network is the continuity of the null limit law of the score. We take that route and audit one
instance, DeepGOF-1 (Section~\ref{sec:test}). It renders the misfit of the fitted model as a
\emph{residual map}, a $6\times6$ grid of standardized residual sums over the ranks of the two
strongest covariates; a small convolutional network, trained once on simulated departures and shipped
as $18{,}273$ frozen numbers, scores the map; and the analyst's own parametric bootstrap calibrates the
score through the rank p-value of \citet{besagclifford1989}. The network was trained against
bootstrap replicates, so it is trained against the null it is calibrated against.

The result is a test that answers questions the classical tests cannot: where a model fails,
including in a missed interaction of two covariates that tools examining one covariate at a time do
not see (Section~\ref{sec:location}); whether a published risk model is calibrated within groups of
new patients, by an exact test of frozen predictions (Section~\ref{sec:external}); and both at the size
of a clinical registry, in seconds (Section~\ref{sec:support}). Before it is trusted with these
questions, the test is audited, in four parts.

\emph{Validity} (Section~\ref{sec:validity}). Under standard bootstrap regularity, a unique pair of
strongest covariates and a continuous null limit law of the score (checked numerically), the test is
first-order valid. Its finite-sample level is measured: on a sixty-cell grid
(Section~\ref{sec:size}), networks trained by the shipped recipe hold the nominal band in $58$ of $60$
cells (mean $\sizegrid$), and the shipped network itself, on the same grid's held-out null datasets,
in all $20$ of its cells (mean $.050$). A size bound with one measured unknown also holds at every
sample size, though it is loose.

\emph{What the test can see} (Section~\ref{sec:consistency}). The network is piecewise affine, so
along every direction its large-signal behaviour is linear, with a slope given by the same network
with its offsets removed; once the map direction of a named alternative has been computed under an
assumed covariate law, one forward pass on the frozen weights decides consistency against it. The
directions where it fails, the \emph{blind cone}, are computed, and
misfit that integrates to zero within every cell defeats any statistic of the map.

\emph{How much power} (Section~\ref{sec:localpower}). The statistic is Lipschitz, which gives an
explicit local power limit at the deployed number of bootstrap draws, an expectation over a Gaussian
law that can be computed before data are collected, so the sample sizes that the test and its rivals
need can be compared in advance. With both covariates carrying the
signal the projection test of \citet{liu2024} needs fewer observations; the two are level at about two
added covariates that are only weakly predictive and carry no misfit, and with four or more the
projection test needs more.

\emph{Where it breaks} (Section~\ref{sec:corrupt}). At $n\le50$ a single corrupted covariate value
can decide a goodness-of-fit test. One study shows which tests break, and a second, on
fresh seeds, confirms a repair that restores DeepGOF-1 at no measurable cost in power. At $n=1{,}000$
DeepGOF-1 keeps its level with up to five corrupted records, where Stukel's test and GiViTI do not.

On the benchmark of \citet{liu2024}, extended to $\narms$ methods, the shipped network, run
as an analyst runs it, keeps its level and is third in power of the nine tests run on the same
datasets, behind the projection test and Stukel's test; read over all covariate pairs it is second
(Section~\ref{sec:liubench}). Networks of the same architecture trained on each setting's design rank
\rankdeepvalidword\ of the \narmsvalidword\ tests that hold their level, behind only the projection
test, with the steadiest level; counting an undefined p-value as a non-rejection moves Stukel's test
ahead of them. On that benchmark a plain $\chi^{2}$ statistic of the same map has about the same
power as the network, so the power there comes from the map and its calibration. With two active
covariates among ten (Section~\ref{sec:niche}) the network earns its place: DeepGOF-1 has the highest
matched power in \nichenhighest\ of twelve settings, and the network is ahead of the $\chi^{2}$ of
its own map. The all-pairs map locates a missed interaction of two covariates, which tools that
examine one covariate at a time cannot see (Section~\ref{sec:location}). Applied to a published risk
model on new patients, with nothing refitted, DeepGOF-1 is an exact test of calibration within
patient subgroups, and it has the most power where the calibration belt and the other tests along
the predicted risk are nearly blind (Section~\ref{sec:external}). In the SUPPORT study
(Section~\ref{sec:support}) the classical tests detect the misfit; the map places it in one test, as a
covariate-by-covariate spline screen also does, and guides its repair; on the study's own design
DeepGOF-1 keeps its level up to all $8{,}873$ patients, and in a split-sample validation it rejects the
linear model in every split while the calibration belt passes it in nearly all. Proofs and supporting studies are in Online Resource 1.

\section{The test}\label{sec:test}

\subsection{The residual map}\label{sec:map}

\begin{figure*}[!tbp]
\centering
\fullfig{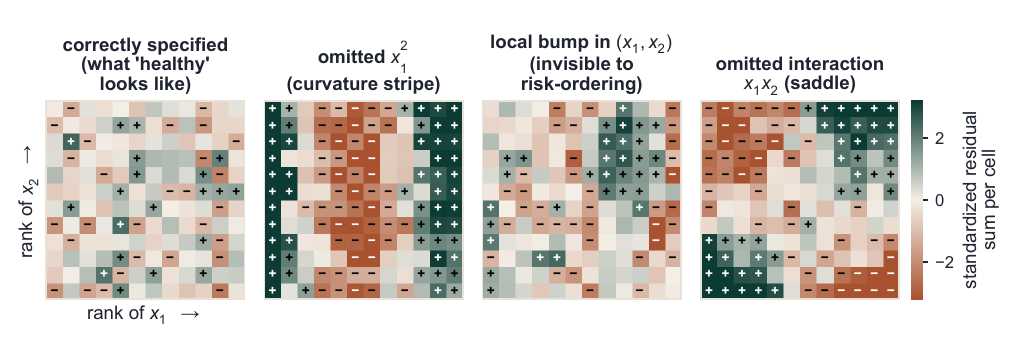}
\caption{Residual maps render misfit as pictures, so a departure has a \emph{location}.
Under a correct model (left) the map is noise; an omitted quadratic, a local bump and an
omitted interaction each leave a characteristic spatial pattern. Each cell is approximately
$\mathcal{N}(0,1)$ under a correct model, and the grid is over covariate \emph{ranks}, so the map
is scale-free. A ``$+$'' or ``$-$'' marks a cell whose standardized residual sum exceeds $1$ in
absolute value. The maps are high-$n$ renderings on a $12\times12$ grid for visibility; the test
uses a $6\times6$ grid}
\label{fig:maps}
\end{figure*}

Let $(x_{i},y_{i})$, $i=1,\dots,n$, be i.i.d.\ with $y\mid x\sim
\mathrm{Bern}(\pi^{*}(x))$, and let $\hat\beta$ be the maximum likelihood fit of the
logistic model $\pi_{\beta}(x)=\Lambda(\tox^{\top}\beta)$,
$\Lambda(u)=(1+e^{-u})^{-1}$, $\tox=(1,x^{\top})^{\top}$. Among the covariates, the two
with the largest $|\hat\beta_{j}|\,\hat\sigma_{j}$ (coefficient times sample standard
deviation) are selected as axes. The selection runs over the columns of the design matrix, so the
analyst should read which axes were selected. Ties within a selected covariate are broken by one
random permutation of the observations, held fixed for the observed map and every bootstrap map, so
that the row order cannot carry the outcome. With $K=6$, observation $i$ falls in cell $C_{k\ell}$
when the empirical rank of its first selected covariate lies in the $k$-th sixth and
that of its second in the $\ell$-th sixth. The \emph{residual map} is the
$K\times K$ array
\[
m_{k\ell}\;=\;\frac{\sum_{i\in C_{k\ell}}\bigl(y_{i}-\hat\pi_{i}\bigr)}
{\sqrt{\sum_{i\in C_{k\ell}}\hat\pi_{i}(1-\hat\pi_{i})}},
\qquad \hat\pi_{i}=\pi_{\hat\beta}(x_{i}),
\]
each cell approximately standard normal under a correct model. Misfit acquires a
\emph{location}: an omitted quadratic paints a stripe, an omitted interaction a saddle,
a local departure a bump (Figure~\ref{fig:maps}). Ranks make the map invariant to
monotone rescaling of the covariates, so the same frozen network applies to any dataset.

The grid resolution $K=6$ was chosen from $K\in\{4,6,8,12\}$ in pilot experiments: a $12\times12$
grid leaves too few observations per cell at the sample sizes this test targets, and $K=4$ discards
spatial detail. A covariate beyond the selected two lies off the grid, a cost that
Section~\ref{sec:power} measures and Corollary~\ref{cor:representation} explains. The package also
offers an \emph{all-pairs reading}: the largest score over the maps of every pair of covariates, taken
again in every bootstrap replicate, so the choice of pair needs no correction and does not depend on
the linear effects. Its level is first-order valid without condition A3 (Online Resource 1,
remark after Proposition~S.3). The axis rule is the test this paper studies;
Sections~\ref{sec:liubench}, \ref{sec:niche} and~\ref{sec:support} compare the two readings.

\subsection{The pretrained score and the deployed test}\label{sec:score}

\begin{figure*}[!tbp]
\centering
\fullfig{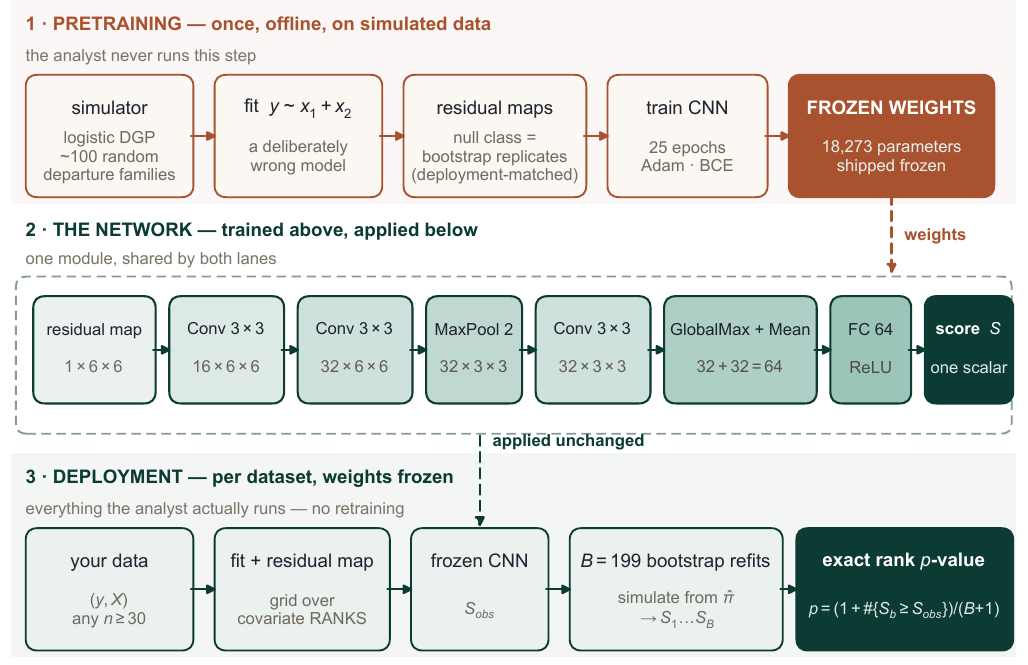}
\caption{The test is trained once and then only applied. \emph{Top:} pretraining, run once on
simulated data, with bootstrap replicates as the null class. \emph{Middle:} the frozen network;
global max pooling finds a localized bump and global mean pooling a diffuse trend. \emph{Bottom:}
what the analyst runs: fit, map, score, and the parametric bootstrap that gives the rank p-value of
\eqref{eq:pval}, so the level is a property of the calibration rather than of the learning}
\label{fig:arch}
\end{figure*}

The score is $T=S\bigl((m-a)/s\bigr)$, where $(a,s)$ is a fixed standardization and $S$
a small convolutional network: three $3\times3$ convolution layers, one max-pooling
stage, a global max and mean readout, a $64$-unit fully connected rectifier layer and a linear head,
$18{,}273$ parameters in all, trained once by us and shipped frozen (Figure~\ref{fig:arch}, which
writes $S$ for the score). Training data are
simulated: null maps and alternative maps at $n=200$, the alternatives drawn from
roughly one hundred random departure families (polynomial, interaction, oscillatory,
threshold, local, and index-type terms with random amplitudes). The \emph{null class is
bootstrap replicates}: a null training map is produced by fitting a correct model, simulating from
the fit, and refitting, which is exactly the object the deployed test compares against; training on
plain correct-model maps would reward the score for separating data nulls from bootstrap nulls, the
difference the calibration must not see (Section~\ref{sec:validity}). The departure prior is
deposited with the weights, because the power credited to a learned test depends on it.
Algorithm~\ref{alg:deepgof} is the whole of the deployed test, and its p-value is

\begin{algorithm*}[!tbp]
\caption{DeepGOF-1, as deployed. Inputs: a fitted \texttt{glm} with binomial family and
$d\ge2$ covariates; the shipped weights $(S,a,s)$; $B=199$; $K=6$.}
\label{alg:deepgof}
\begin{enumerate}\itemsep2pt
\item[1.] \textbf{Axes.} Choose the two covariates with the largest
$|\hat\beta_{j}|\hat\sigma_{j}$, ties broken by model order.
\item[2.] \textbf{Map.} Bin observations by the empirical ranks of those two covariates
into a $K\times K$ grid, breaking ties within a covariate by one random permutation of the
observations, drawn once per call and held fixed for the observed map and every bootstrap map; set
$m_{k\ell}=\sum_{C_{k\ell}}(y_{i}-\hat\pi_{i})/\{\sum_{C_{k\ell}}\hat\pi_{i}(1-\hat\pi_{i})\}^{1/2}$,
with an empty cell set to $0$ and the denominator floored at $10^{-8}$.
\item[3.] \textbf{Score.} $T_{\mathrm{obs}}=S((m-a)/s)$.
\item[4.] \textbf{Calibrate.} For $b=1,\dots,B$: draw
$y^{*b}_{i}\sim\mathrm{Bern}(\hat\pi_{i})$ at the observed covariates, refit the
\emph{same} formula, and repeat steps 1--3 to get $T^{*b}$. A refit that fails is
scored $+\infty$, so it counts against rejection.
\item[5.] \textbf{Report.}
$p=\bigl(1+\#\{b: T^{*b}\ge T_{\mathrm{obs}}\}\bigr)/(B+1)$.
\end{enumerate}
\end{algorithm*}

\begin{equation}\label{eq:pval}
p\;=\;\frac{1+\#\{b:\ T^{*b}\ge T_{\mathrm{obs}}\}}{B+1}.
\end{equation}

The work is $B+1$ forward passes and $B$ logistic refits: a complete test of a model with $n=150$
takes $1.6$ seconds on one desktop core, and about $15$ seconds at $n=8{,}873$
(Section~\ref{sec:support}). The implementation is pure base R. The shipped weights compute every
DeepGOF-1 number of this paper except the classifier diagnostic of Section~\ref{sec:validity} and the
main runs of Sections~\ref{sec:size} and~\ref{sec:power}, which used networks of the same architecture
and training loop trained per simulation block; the shipped network was then run on the same
datasets. The other tests are computed with
\texttt{ebrahim.gof} and, for GiViTI and BAGofT, with their authors' R packages. The first author used a large language model to improve the language and readability of this manuscript and to assist with the implementation of the simulation code; the authors reviewed and edited all content and take full responsibility for it.

\section{Theory}\label{sec:theory}

Proofs are in Online Resource 1, which numbers its results independently:
Lemmas~\ref{lem:exact} and~\ref{lem:recession} are its Lemmas~S.2 and~S.1,
Propositions~\ref{prop:valid} and~\ref{prop:fsbound} its Propositions~S.3 and~S.7,
Theorem~\ref{thm:consistency} its Theorem~S.4, Corollaries~\ref{cor:cone}--\ref{cor:representation}
its Corollaries~S.5--S.6, and Theorem~\ref{thm:local} its Theorem~S.9. Proposition~\ref{prop:valid}
and Theorem~\ref{thm:local} apply standard bootstrap and contiguity arguments to the network. What is new is the consistency certificate and the computed blind cone
(Section~\ref{sec:consistency}), and the contrast with the projection test when added covariates
carry no misfit (Section~\ref{sec:irrelevantmain}).

The conditions invoked below are stated in full in Online Resource 1; in words they are
that the data are i.i.d.\ with $0<\pi^{*}<1$ and $\E\lVert\tox\rVert^{2}<\infty$, with
the two selected covariates having continuous marginals near their population quantiles
(so a binary or otherwise atomic selected covariate is not covered) and every population cell
carrying positive mass (A1); that the quasi-maximum-likelihood
estimator is asymptotically linear about a unique Kullback--Leibler projection
$\beta^{*}$, as under the conditions of \citet{white1982} (A2); that the population axis
criterion has a unique top-two set (A3); that the misfit leaves a nonzero footprint on
the map, $\mu\ne0$ (A4, consistency only); that the certificate is positive,
$\Sinf(\mu/s)>0$ (A5, consistency only); and that the null limit law of the score is
continuous (A6). A1--A3 are the standard regularity of a grouped test with estimated parameters,
and A6 is the only condition that depends on the shipped weights. Throughout, $\theta_{0}$ denotes
the true parameter under $H_{0}$.

\subsection{Validity of the level}\label{sec:validity}

\begin{lemma}[Exactness in the pivotal idealization]\label{lem:exact}
Fix the covariate configuration and suppose the law of the score is the same under
every null parameter. Then, for every $k$,
$\Prob\bigl(p\le k/(B+1)\bigr)\le k/(B+1)$, with equality when that law is atomless: the
test that rejects when $p\le\alpha$ has size at most
$\lfloor\alpha(B+1)\rfloor/(B+1)\le\alpha$ at every sample size.
\end{lemma}

Pivotality cannot hold exactly for a fixed network, because the score is computed from fitted
probabilities; the lemma is the benchmark that training on bootstrap replicates aims at. A classifier
trained to separate scores on true-null data from scores on bootstrap replicates attains area under
the curve $.513$ on average over the sixty cells of Section~\ref{sec:size} (range $.500$--$.550$),
against the chance value $.5$. This check is necessary, not sufficient, and the validity of the
deployed test does not rest on it:

\begin{proposition}[First-order validity of the deployed test]\label{prop:valid}
Under $H_{0}$ and regularity conditions (A1--A3, A6 of Online Resource 1; A2 is White
regularity for the fitted logistic model), the conditional law of the bootstrap score
converges uniformly to the null limit law of the observed score. Consequently, at the deployed fixed $B$, $\Prob(p\le k/(B+1))\to k/(B+1)$ for
every $k$ --- the test is asymptotically exact at the $B+1$ attainable levels, of which
the nominal $.05$ is one --- and if additionally $B\to\infty$,
$p\rightsquigarrow\mathrm{Unif}(0,1)$.
\end{proposition}

A2 constrains the estimator and A3 the design; the network enters only through A6, the continuity
of the null limit law of the score. A6 is assumed, not proved: a piecewise-affine network can have an
atom where all its final units are inactive. It was checked on the shipped weights: $200{,}000$
draws from the Gaussian null limit of the map gave one repeated score, and a fresh $20{,}000$ standard
Gaussian maps gave none, so an atom, if there is one, carries mass of order $10^{-4}$ at most and can
move the level by no more than that. Unlike a grouped
statistic referred to a fixed $\chi^{2}$ law, the test needs only the gap between two laws at the
same sample size, the null law of the observed score and the conditional law of the bootstrap score,
to vanish. That gap is measured in Sections~\ref{sec:size} and~\ref{sec:power} and bounded at every
sample size by the next result.

\begin{proposition}[Finite-sample size bound with a measurable calibration
gap]\label{prop:fsbound}
Assume only that $H_{0}$ holds. Fix the covariate configuration and let
$F_{\hat\theta}$ and $F_{\theta_{0}}$ be the distribution functions of the score when
the responses are drawn at the fitted parameter and at the truth, so that $F_{\hat\theta}$
is the law the bootstrap samples from. Let
$\Delta^{+}=\sup_{t}\bigl(F_{\hat\theta}(t)-F_{\theta_{0}}(t)\bigr)_{+}$ be the
one-sided calibration gap, take any constant $\varepsilon\ge0$, and put
$\zeta=\Prob(\Delta^{+}>\varepsilon)$. Then for every sample size $n$, every $B$ and
every $k\in\{1,\dots,B+1\}$,
\[
\Prob\Bigl(p\le\tfrac{k}{B+1}\Bigr)\ \le\ \frac{k}{B+1}+\varepsilon+\zeta,
\]
and hence
\[
\Prob(p\le\alpha)\ \le\ \alpha+\varepsilon+\zeta
\]
for every $\alpha\in(0,1)$. The constant multiplying $\varepsilon$ is one, it does not
depend on $B$, and no smaller constant is available.
\end{proposition}

The bound needs nothing about the estimator, a limit law or the network, and Lemma~\ref{lem:exact}
is its corner $\varepsilon=\zeta=0$. On one design of Section~\ref{sec:size} the ninetieth percentile
of the tail-localized gap (Online Resource 1, Proposition~S.8) is $.030$ at $n=50$ and $.007$ at
$n=200$; the measured part of the resulting bound is about $.19$ with the Monte Carlo margin, before
two terms that were not measured, against a measured size near $.05$ (Online Resource 1,
Section 6). The bound is therefore loose, and the evidence for the level is the measured sizes of
Sections~\ref{sec:size}--\ref{sec:corrupt}. The maximized Monte Carlo test of \citet{dufour2006} is
exact at every sample size, but on that design at $n=100$ its power is $.17$ against $.35$ for the
deployed test, at about ninety times the computation. Measured on $114{,}000$ null datasets across
three designs, the level error of the deployed test is $.0025$ or less at every $n\ge100$ (standard
error $.0014$; Online Resource 1, Section 8).

\subsection{Consistency and the blind cone}\label{sec:consistency}

The consistency theory rests on a structural fact about the network class itself.

\begin{lemma}[Recession function]\label{lem:recession}
For $S$ built from affine maps, rectifiers, and componentwise maxima, the limit
$\Sinf(u)=\lim_{t\to\infty}S(z_{0}+tu)/t$ exists for every direction $u$, does not
depend on $z_{0}$, and equals the same network with every offset set to zero. Moreover
$\sup_{z}|S(z)-\Sinf(z)|<\infty$.
\end{lemma}

Each ingredient is known. Networks of this class are exactly the finitely piecewise-affine
functions \citep{arora2018}; removing the offsets leaves a function positively homogeneous of
degree one, the standing assumption of the homogeneous-network literature \citep{lyuli2020}; and
the limit of $S(z_{0}+tu)/t$ along a ray is the recession function of convex analysis
\citep[\S8]{rockafellar1970}, here for a function that is piecewise affine rather than convex.
What is new is its use: other learned tests prove consistency under rate conditions \citep{zhang2023}
or leave power to simulation \citep{barberjanson2022,javanmard2024}, whereas the recession function
makes the consistency of a \emph{frozen} statistic decidable for a named alternative before any data
are seen. Under a fixed alternative the map grows like
$\sqrt n$ times a deterministic direction $\mu$, the standardized cell-wise integrals of the misfit
$\pi^{*}-\pi_{\beta^{*}}$, while the bootstrap scores stay bounded, so detection reduces to the sign
of one number that one forward pass returns.

\begin{theorem}[Conditional consistency]\label{thm:consistency}
Under conditions A1--A5 of Online Resource 1 --- in particular $\mu\ne0$ and the
\emph{certificate} $\Sinf(\mu/s)>0$ --- the power of the deployed test at any attainable
level tends to one.
\end{theorem}

\begin{corollary}[The blind cone]\label{cor:cone}
The exceptional set $\mathcal{B}=\{u:\Sinf(u/s)\le0\}$ is a closed cone computable from
the frozen weights. For DeepGOF-1: none of $250{,}000$ uniform random directions fell in
$\mathcal{B}$ (with $95\%$ confidence its spherical measure is below
$1.2\times10^{-5}$); multi-start minimization found a direction with slope $-0.281$, so
$\mathcal{B}\ne\{0\}$; and that direction is a checkerboard at exactly the cell scale
--- sub-resolution oscillation, the failure mode every binned statistic shares. The
canonical departure directions carry margins well above zero, from $1.08$ to $2.20$
(Table~\ref{tab:cert}).
\end{corollary}

\begin{table*}[!tbp]
\centering
\caption{The consistency certificate $\Sinf(\cdot/s)$ of the shipped network, by
departure direction (and its mirror image). One forward pass with the offsets removed
decides, for any hypothesized alternative, whether Theorem~\ref{thm:consistency}
guarantees eventual detection. All values reproduce from the released weights.}
\label{tab:cert}
\begin{tabular}{@{}lcc@{}}
\toprule
map direction & certificate & mirror \\
\midrule
\tablecert
\bottomrule
\end{tabular}
\end{table*}

\begin{corollary}[Representation limit]\label{cor:representation}
If the misfit integrates to zero within every population cell ($\mu=0$), the map is
uniformly tight: no fixed locally bounded statistic of it can diverge, so consistency
through this map is impossible for \emph{any} such test, and under nondegeneracy
conditions the deployed test's power converges to a limit strictly below one.
\end{corollary}

Corollary~\ref{cor:representation} is the formal version of the cost of a third covariate that
lies off the grid (Section~\ref{sec:power}) and of the near-blindness of every grouped statistic to
probit-type alternatives, whose $\mu$ is tiny.

\section{Local power and power certificates}\label{sec:localpower}

\subsection{Local asymptotic power}\label{sec:local}

Consider local alternatives $\pi_{n}(x)=\pi_{\beta_{0}}(x)+n^{-1/2}h(x)$ with $|h|\le c\,v$, where
$v=\pi_{\beta_{0}}(1-\pi_{\beta_{0}})$. This covers every logit-scale departure
$\eta_{0}+n^{-1/2}\tau g$ with bounded $g$, and it keeps $\pi_{n}$ inside $(0,1)$. For the population
cells $C_{c}$ of the grid write $d_{c}=\E[v\,1_{C_{c}}]$, $g_{c}=\E[v\,\tox\,1_{C_{c}}]$,
$\mathcal I=\E[v\,\tox\tox^{\top}]$, $a_{c}(h)=\E[h\,1_{C_{c}}]$ and $b(h)=\E[\tox h]$. Put
$D=\mathrm{diag}(d_{c})$, let $G$ have rows $g_{c}^{\top}$, and define
\begin{align*}
\delta(h)&=D^{-1/2}\bigl\{a(h)-G\mathcal I^{-1}b(h)\bigr\},\\
\Sigma&=I-D^{-1/2}G\mathcal I^{-1}G^{\top}D^{-1/2}.
\end{align*}
$\Sigma$ is the null covariance of the fitted map. Its eigenvalues lie in $[0,1]$, and generically
only the intercept direction is null. $\delta(h)$ is the part of the departure that the fit cannot
absorb, cell by cell. Let $F_{0}$ be the distribution function of $\tilde S(Z)$,
$Z\sim N(0,\Sigma)$, where $\tilde S(u)=S((u-a)/s)$.

\begin{theorem}[Local asymptotic power]\label{thm:local}
Let A1--A3 and A6 hold at $\beta_{0}$, let $\mathcal I$ be nonsingular and every population
cell mass $d_{c}$ positive, and let $|h|\le c\,v$. Then, for the selected-axes statistic of
Section~\ref{sec:test}, for every $B$ and every $k\in\{1,\dots,B+1\}$,
\begin{multline*}
\Prob_{n}\Bigl(p\le\tfrac{k}{B+1}\Bigr)\longrightarrow\\
\E\Bigl[\mathrm{Bin}\bigl(k-1;\,B,\;1-F_{0}(\tilde S(\delta(h)+Z))\bigr)\Bigr],
\end{multline*}
where $\mathrm{Bin}(\cdot;B,q)$ is the binomial distribution function. As $B\to\infty$ the limit
becomes $\Prob\{\tilde S(\delta(h)+Z)>F_{0}^{-1}(1-\alpha)\}$, and with $h=0$ it is $k/(B+1)$.
\end{theorem}

The theorem covers the statistic of Algorithm~\ref{alg:deepgof} on the two selected axes, as
\texttt{deepgof1()} computes it. No expansion of the network is needed: $\tilde S$ is piecewise
affine, hence Lipschitz, and the continuous mapping theorem carries the Gaussian limit of the map
through it. The other steps are standard: contiguity \citep[Lemma 6.4]{vaart1998}, a first-order
expansion of the fit, and a triangular-array central limit theorem for the map in the manner of
\citet{moorespruill1975}; at fixed $B$ the limit is the Monte Carlo power formula of
\citet{dufour2006} evaluated at the local limit law (Online Resource 1, Section 11). If
$\delta(h)=\lambda u$ with a positive certificate $\Sinf(u/s)>0$, the local power tends to one as
$\lambda$ grows, which joins the two parts of the theory.

Given the covariate law and the departure, $\delta(h)$ and $\Sigma$ are population quantities, so the
limit can be computed before any data are seen, for the test and for its rivals on the same drift.
Table~\ref{tab:are} gives the \emph{local sample-size ratio at power one half}, the ratio of the local
sample sizes two tests need to reach limit power one half; unlike a Pitman efficiency, it depends on
the power level and on the design.

\begin{table}[!tbp]
\centering
\caption{Local sample-size ratio of DeepGOF-1 at power one half,
$(\tau_{50}^{\mathrm{rival}}/\tau_{50}^{\mathrm{DeepGOF}})^{2}$, where $\tau_{50}$ is the local
amplitude giving limit power one half at $\alpha=.05$ and $B=199$ (DeepGOF-1 and the projection
test). Values above one mean DeepGOF-1 needs fewer observations; values below one mean the rival
does. Design: two standard normal covariates, $\eta_{0}=-0.5+x_{1}+0.6x_{2}$. HL: Hosmer--Lemeshow;
projection: the test of \citet{liu2024}, whose limit is evaluated at twelve reference samples of
$1{,}000$ covariate draws (Monte Carlo standard error in parentheses); the HL and Stukel columns are
exact limits. Stukel: Stukel's test in the SstBoth form of \citet{liu2024}.
From Theorem~\ref{thm:local} and the corresponding limits of the
rivals (Online Resource 1).}
\label{tab:are}
\small
\begin{tabular}{@{}lccc@{}}
\toprule
departure & HL & Stukel & projection \\
\midrule
quadratic, $x_{1}^{2}-1$ & 1.62 & 0.55 & 0.62 (.01) \\
interaction, $x_{1}x_{2}$ & 2.52 & 0.82 & 0.73 (.02) \\
cubic, $x_{1}^{3}-3x_{1}$ & 1.01 & 0.68 & 0.26 (.01) \\
local bump & 3.36 & 2.61 & 0.87 (.03) \\
threshold, $1\{x_{2}>1\}$ & 43.7 & 15.6 & 0.67 (.01) \\
\bottomrule
\end{tabular}
\end{table}

With both covariates carrying signal, the projection test needs $0.26$ (cubic) to $0.87$ (bump) times
the observations DeepGOF-1 needs. DeepGOF-1 needs fewer observations than the Hosmer--Lemeshow test
on four of the five departures (about $44$ times fewer on the threshold, tied on the cubic), and fewer
than Stukel's test on the bump and the threshold but up to $1.8$ times as many on the quadratic.
Stukel's two score statistics are correlated ($\rho=-.61$), so its limit size is $.064$ rather than
$.05$ and its column slightly favours it. Simulating the deployed test with $199$ bootstrap refits at
$n=32{,}000$, all six cells (five departures and the null) agree with the formula within $1.3$
standard errors.

\subsection{Covariates that carry no misfit}\label{sec:irrelevantmain}

Table~\ref{tab:are} has two covariates, and both carry the misfit. Added covariates that carry no
misfit affect DeepGOF-1 and the projection test in opposite ways (Online Resource 1,
Proposition~S.10a). When they are independent of the two active covariates and absent from the true
predictor, the axis rule
picks the right pair with probability tending to one, so the limit power of DeepGOF-1 does not change
with their number $q$; the same holds for the Hosmer--Lemeshow and Stukel tests, which are built from
the fitted linear predictor. The projection statistic instead averages the residual process over all
directions of the $d$-dimensional covariate space, and a random direction puts only a share $2/d$ of
its squared length on the two active covariates, so its local signal-to-noise ratio falls like $1/d$
and the sample size it needs, relative to DeepGOF-1, grows about linearly in $d$ (Corollary~S.10b).

\begin{figure*}[!tbp]
\centering
\fullfig{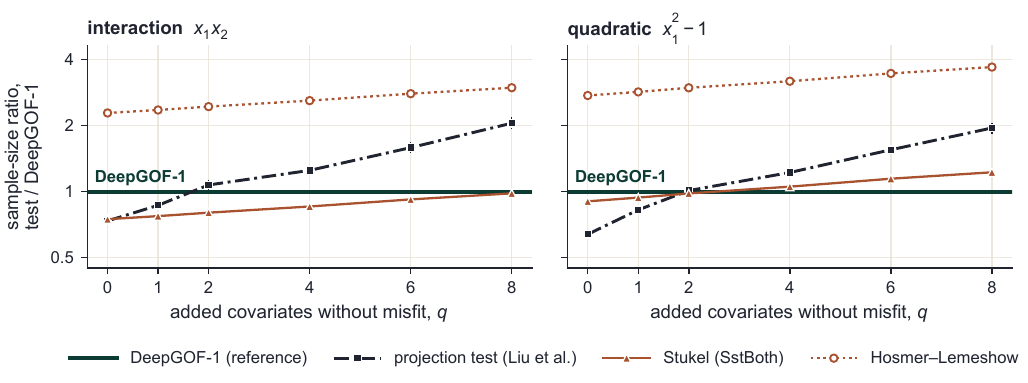}
\caption{Local sample-size ratio at limit power one half ($\alpha=.05$, $B=199$) against the
number $q$ of added covariates that are weakly predictive and carry no misfit, in the design of
Section~\ref{sec:niche}.
Each curve is the number of observations the test needs divided by the number DeepGOF-1 needs;
values above one mean DeepGOF-1 needs fewer. Left: interaction departure; right: quadratic. The
projection test of \citet{liu2024} reaches one at about $q=2$ and exceeds it from $q=4$;
Hosmer--Lemeshow stays above one at every $q$; Stukel's test (SstBoth) is below one for the
interaction at every $q$ and for the quadratic up to $q=2$. The projection curve's bars are $\pm1$
Monte Carlo standard error; the other curves are exact limits}
\label{fig:irrelevant}
\end{figure*}

Figure~\ref{fig:irrelevant} evaluates the limits in the design of Section~\ref{sec:niche}
($x\sim N(0,I_{2+q})$, $\eta_{0}=-0.5+x_{1}+0.8x_{2}+0.15\sum_{j\ge3}x_{j}$). The added covariates are
weakly predictive but carry no misfit, so the invariance is approximate: the amplitude at which
DeepGOF-1 reaches power one half moves by at most $1.4\%$ over $q$. The projection test needs
$0.74\pm0.03$ (interaction) and $0.64\pm0.02$ (quadratic) times the observations DeepGOF-1 needs at
$q=0$, about as many at $q=2$, $1.25$ and $1.22$ times as many at $q=4$, and about $2.0$ times as many
at $q=8$, the setting of Section~\ref{sec:niche}. At $q=8$ and the sample sizes of that simulation the
null spread of the projection statistic is still well above its limit, so the limit over-predicts the
projection test's power at the stronger departures (Online Resource 1, Section 12, Remark (f)).

A lower bound on power can also be computed from the weights for a Gaussian model of the map,
$m=t\,v+e$ with $e\sim N(0,I_{36})$. The deployed map is Gaussian only in the limit, has null
covariance $\Sigma$ and is calibrated by a bootstrap rank, so such certificates are statements about
the model, not finite-sample guarantees for the test. Linear relaxation of the network, the standard
verification tool, is far too loose ($19$ to $52$ times the sample size simulation needs); a bound by
Gaussian smoothing certifies power $.95$ at $1.02$ to $1.06$ times the true sample size for five
directions, all twenty certificates holding together with probability at least $.988$ (Online Resource 1, Proposition~S.12a and Section~13).

\section{Simulation I: the level on a sixty-cell grid}\label{sec:size}

\begin{figure*}[!tbp]
\centering
\fullfig{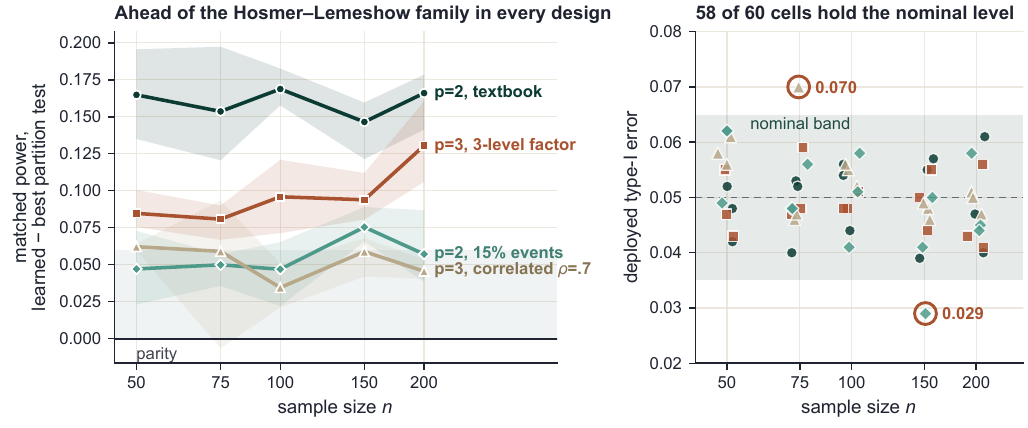}
\caption{The sixty-cell level study: 5 sample sizes $\times$ 4 designs
$\times$ 3 independent training corpora. \emph{Left:} matched-level power gain of DeepGOF-1 over the
strongest published partition test in each cell (Hosmer--Lemeshow, Pigeon--Heyse or the equal-width
variant, \emph{chosen per cell in hindsight}, so an exploratory comparison); the line marked
\emph{parity} is zero, and every test is randomized to exactly $\alpha=.05$ on held-out nulls before
power is read. \emph{Right:} type-I error of DeepGOF-1 in all sixty cells against the
band (58 of 60 inside; the two excursions circled and labelled). Shaded ribbons give the range across
the three corpora}
\label{fig:size}
\end{figure*}

The grid crosses five sample sizes ($n=50$ to $200$), four covariate
designs and three independent training corpora. The designs are the textbook one (two independent
continuous covariates, $50\%$ events), a three-covariate design with a three-level factor, a
correlated three-covariate design (AR(1), $\rho=0.7$) and a two-covariate design with $15\%$ events.
Each cell has $\gridnullstotal$ null replicates: the first $\gridnullsheld$ train that cell's
network, as bootstrap replicates, and the held-out $\gridnullsheld$ carry the deployed level, against
the band $[.035,.065]$. The networks, of the shipped architecture and training loop, were trained per
sample size and corpus (Section~\ref{sec:score}). Their level lies in the band in $58$ of $60$
cells, with mean $\sizegrid$ (Figure~\ref{fig:size}); the two excursions ($.070$ and $.029$) sit on
opposite sides and neither recurs across corpora, while the classical tests on the same designs drift
in both directions.

The deployed network was then run, as an analyst runs it (a plain \texttt{glm} fit and
\texttt{deepgof1()} with $B=199$), on the same held-out null datasets, regenerated from their seeds
and checked against the deposited p-values of the classical tests: $1{,}000$ per cell, one network,
so $20$ cells. Its size lies in the band in all $20$ cells, from $.039$ to $.059$ (standard error about
$.007$ per cell), with pooled size $.0495$ over the $20{,}000$ datasets (standard error $.0015$), and
every dataset returned a p-value.

\begin{figure}[!tbp]
\centering
\includegraphics[width=\snhalffig]{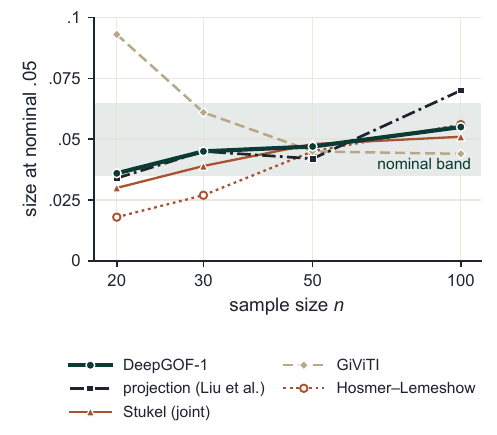}
\caption{Size at nominal $.05$ against sample size, two covariates, $2{,}000$ null datasets at $n=20$
and $30$ and $1{,}000$ at $n=50$ and $100$ (standard error at most $.007$). The shaded region is the
nominal band $[.035,.065]$. Only DeepGOF-1 stayed inside it at every $n$, but several other sizes are
within one or two standard errors of its edges; at $n=20$ DeepGOF-1 ($.036$) and the projection test
($.034$) cannot be told apart. Stukel: the joint score test}
\label{fig:smalln}
\end{figure}

A further study took the level below the grid, to $n=20$ and $30$, where the map holds fewer than one
patient per cell (two standard normal covariates, $\eta=0.25+1.5x_{1}+x_{2}$; $2{,}000$ null datasets
at $n=20$ and $30$, $1{,}000$ at $n=50$ and $100$; Figure~\ref{fig:smalln}). DeepGOF-1 is the only
test whose size stayed inside the band at every sample size ($.036$, $.045$, $.047$ and $.055$;
standard errors $.005$ to $.007$), although $7\%$ of the fits at $n=20$ are separated; several other
sizes were within one or two standard errors of the band, and at $n=20$ the projection test ($.034$)
and DeepGOF-1 cannot be told apart. The Hosmer--Lemeshow test falls to $.018$ and $.027$ at $n=20$ and
$30$, GiViTI rises to $.093$ at $n=20$ and returns no p-value on $6\%$ of those datasets, Stukel's test
is at $.030$ at $n=20$, and the projection test is at $.070$ at $n=100$. At matched level (Online Resource 1, Section 18) and $n\le50$, DeepGOF-1 has more power than the Hosmer--Lemeshow test in ten
of the twelve cells with an omitted quadratic or interaction, and less than the projection test in all
twelve and than Stukel's test in ten; against a wrong link only Stukel's test and GiViTI gain
appreciable power.

In an exploratory
comparison against the strongest published partition test in each cell, chosen in hindsight,
DeepGOF-1 is ahead in every design (Figure~\ref{fig:size}, left), by $+.09$ overall and $+.16$ on the
textbook design.

The level also transfers to real covariate geometry. On the design matrices of FDIC-insured banks in
six US states before the failures of 2009--2011 ($97$ to $655$ banks and $16$ to $69$ failures per
state, $2{,}000$ null outcome vectors each), a rare-event logistic risk model like a clinical score,
the size of DeepGOF-1 lies in the band in five of the six cells ($.044$ to $.068$), while the
Hosmer--Lemeshow and Pigeon--Heyse tests exceed $.065$ in the two cells with the fewest events for
their size (Online Resource 1, Section 20).

\section{Simulation II: a published benchmark, \narmsword\ methods}\label{sec:power}

\citet{liu2024} compare goodness-of-fit tests for logistic regression across four
settings (quadratic and off-index quadratic departures, interactions with binary and continuous
moderators) and conclude that on their Setting~2 the Hosmer--Lemeshow, Osius--Rojek and Stukel tests
fail while their projection test does not. We use their four data-generating processes at
$n\in\{50,100,500\}$ with two independent training corpora, and compare $\narms$ methods: their
tests --- Hosmer--Lemeshow \citep{hosmerlemeshow1980}, the modified Hosmer--Lemeshow test (mHL),
\citet{osius1992}, Stukel's test \citep{stukel1988} in their SstBoth form, \citet{stute2002} and their
projection test \citep{escanciano2006,liu2024}, on the raw covariates and calibrated by a parametric
bootstrap with $B=199$ refits --- together with \citet{pigeonheyse1999}, the equal-width grouping, the
GiViTI calibration test \citep{nattino2014,nattino2016}, the adaptive-partition test of
\citet{zhang2023}, and DeepGOF-1. Every test is placed at exactly level $.05$ on held-out nulls by a
randomized threshold, so power is compared at matched level, and each test is evaluated, for level
and power alike, on the datasets where it returns a p-value; the other convention, and why it inflates
matched power, is in Online Resource 1 (Section 26). Setting~1 has a single covariate, for
which \texttt{deepgof1()} uses a one-dimensional map of $36$ quantile cells; the shipped network,
trained on two-covariate maps only, holds its level there ($.055$) with about $.10$ less power. The
twenty-four blocks (four settings, three sample sizes, two corpora) use different dataset seeds and
are independent.

\subsection{The extended benchmark}\label{sec:liubench}

\begin{table*}[!tbp]
\centering
\caption{The extended \citet{liu2024} benchmark: empirical size at nominal $.05$ and
matched-level power by sample size, averaged over the four settings and departure
strengths. Tests that hold their level (mean size at most $\sizecap$) are ordered by power, and the
one marked \emph{(inflated)} is listed last, for completeness only. Each test is evaluated on the
datasets where it returns a p-value. mHL: modified Hosmer--Lemeshow, which holds its level on average
but not in every block (size $\sizemhlmax$ in Setting~1 and $\sizemhlsfour$ in Setting~4, both at
$n=500$); Stukel: Stukel's test in the
SstBoth form of \citet{liu2024}; projection: the statistic of \citet{liu2024} on the raw
covariates. Pigeon--Heyse orders the datasets almost exactly as Hosmer--Lemeshow does (rank
correlation at least $.9999$), so their matched-level power agrees although their sizes differ. The
Monte Carlo standard error of the overall column is about $.002$ for DeepGOF-1 and $\bagseboot$ for
BAGofT, which was run at a hundred replicates per power cell.}
\label{tab:liu}
\begin{tabular}{@{}lclcccc@{}}
\toprule
test & size & family & $n{=}50$ & $n{=}100$ & $n{=}500$ & overall \\
\midrule
\tableliu
\bottomrule
\end{tabular}
\end{table*}

BAGofT, because of its cost (\bagsecslo--\bagsecshi\ seconds per test), was run at a hundred
replicates per power cell and two hundred per null block, on one training corpus. Stukel's test
returns no p-value on $28.6\%$ of the null and $5.7\%$ of the alternative datasets, most often in
Setting~3, while no bootstrap refit failed for the projection test.

\emph{Level.} The useful statement is about stability rather than the mean. Averaged over the
twenty-four blocks DeepGOF-1 sits at $\sizedeep$, the projection test at $\sizeproj$ and Stukel's
test at $\sizestukel$, but across blocks the standard deviation of the realized size is
$\sdsizedeep$ for DeepGOF-1, against $\sdsizeproj$ and $\sdsizestukel$. Binomial noise alone would
give about $.005$, so DeepGOF-1's dispersion is no larger than sampling noise, and it is the smallest
of all the tests: a paired comparison of per-block distance from nominal favours it against every other test
that holds its level and was measured on all twenty-four blocks ($t$ from $\tstatstrong$ to
$\tstatweak$). The classical levels average near nominal because they drift in both directions as the
design changes; the learned level is near nominal within each design, which is what an analyst
holding one dataset needs.

The modified Hosmer--Lemeshow test holds its level on average ($\sizemhl$) but not in every block
(Table~\ref{tab:liu}), as \citet{liu2024} report, and GiViTI ($\sizegiviti$) is liberal at
$n=50$, up to $\sizegivitimax$ in one setting. The test of \citet{osius1992} does not hold its level
($\sizeosius$ on average, still $\osiussizebadthree$ and $\osiussizebadfour$ at $n=500$ in
Settings~3 and~4), because the variance of its normal reference collapses when the design absorbs
almost all of $(1-2\hat\pi)/\{\hat\pi(1-\hat\pi)\}$.

\emph{Family.} DeepGOF-1 reads a $6\times6$ grid, so it is a partition test, and it has higher power
than its own family: by $+\margHL$ over Hosmer--Lemeshow and $+\margHLW$ over the equal-width variant
at $n=50/100/500$, mainly a small-sample advantage. Against GiViTI and Stute--Zhu it is ahead at
$n\le100$ and level at $n=500$.

\emph{Ranking}, with each test evaluated on the datasets where it returns a p-value. Overall
DeepGOF-1 is \rankdeepvalidword\ of the $\narmsvalid$ tests that hold their level ($\powdeep$), and
\rankdeepvalidword\ of all $\narms$. One test is ahead of it: the projection test, well ahead
($\powproj$; $\gaptoproj$ ahead at $n=50/100/500$). Stukel's test follows at $\powstukel$; its
paired difference from DeepGOF-1 is $\diffstukel$ with a standard error of $\sediffstukel$.
Counting undefined p-values as non-rejections instead gives Stukel's test $\powstukeldeclared$ and \rankstukeldeclaredword\ place, because in
Setting~3 its p-value is undefined in most null datasets.
DeepGOF-1's rank falls from \rankdeepnfiftyword\ at $n\le100$ to \rankdeepnfivehundredword\ at
$n=500$, where GiViTI, Stukel's test and BAGofT also pass it, each by less than $.01$, consistent with
the information lost off the two-axis grid.

\emph{The shipped network.} Table~\ref{tab:liu} scores DeepGOF-1 with networks trained on each
setting's design. The deployed network was then run, as an analyst runs it, on Settings~2--4: the
$1{,}000$ held-out nulls and the first $500$ alternatives at each strength of one corpus, $27{,}000$
datasets, every one of which returned a p-value. Its level is unchanged ($.053$), and its matched power
is $.04$ lower ($.523$ against $.564$, paired standard error $.002$), mostly in Setting~3, which has a
binary covariate. On these settings it is third of the nine tests run on the same datasets, behind the
projection test ($.696$) and Stukel's test ($.534$), ahead of Stukel's test at $n\le100$ and behind it
at $n=500$, and ahead of GiViTI ($.506$), Stute--Zhu ($.496$) and every partition test ($.41$ to
$.42$). BAGofT was run on only a hundred replicates per cell and is not in this comparison; its raw
power there on Settings~2--4, about $.54$, suggests that the shipped network may be fourth once it is
counted. Its all-pairs reading (Section~\ref{sec:map}) is second: power $.608$ at level $.049$, ahead
of Stukel's test at every $n$ (paired standard errors at most $.004$). The gain is all in Setting~2,
where the misfit is a quadratic in a covariate with a small linear effect that the axis rule often
passes over ($.804$ against $.550$); Settings~3 and~4 have one pair, and the two readings coincide.
Over these nine blocks and the four sample sizes of the very-small-sample study
(Section~\ref{sec:size}), the shipped network is the only one of the five tests measured on both
whose size stayed inside the band every time; the Hosmer--Lemeshow, Stukel, GiViTI and projection
tests each left it at least once, some by no more than one or two standard errors.

\emph{The network or the map?} On Settings~2--4, with one set of refits per dataset, the shipped
network was compared with two simple statistics of the same axis-rule map, the sum of its squared cells
(a $\chi^{2}$ over a partition of two covariates) and its largest absolute cell, each at matched level
(Online Resource 1, Section~21). The network and the sum of squares have about the same power
($\abLiuNet$ and $\abLiuSS$; level in $\abLiuSSLevel$ of the $\abLiuCells$ cells, each ahead in
$\abLiuSSAhead$), and both are ahead of the largest cell ($\abLiuMax$). On this benchmark, then, the
power of DeepGOF-1 comes from the map and its bootstrap calibration, not from the learned score;
Section~\ref{sec:niche} shows where the score adds power.

\emph{Setting 2}, which \citet{liu2024} single out, confirms both halves of their conclusion: at the
strongest departure, less $\alpha$, Hosmer--Lemeshow rises by only $\risehl$, Stukel's test by
$\risestukel$ and Stute--Zhu by $\risestutezhu$, while their projection test rises by $\riseproj$. Of
the added tests only BAGofT also solves this setting ($\risebagoft$); DeepGOF-1 rises by $\risedeep$.

BAGofT \citep{zhang2023}, which learns a partition on the analyst's own data by sample splitting,
holds its level on average ($\bagsize$) and has power $\bagpow$, below DeepGOF-1. On identical
datasets the two tests disagree \bagdiscord\ times (only BAGofT rejects, only DeepGOF-1 rejects), an
exact McNemar p-value of \bagmcnemar\ (\bagmcnemarcl\ allowing for clustering); BAGofT is ahead only
on Setting~2, the off-index quadratic (Online Resource 1, Section 19).

\begin{figure*}[!tbp]
\centering
\fullfig{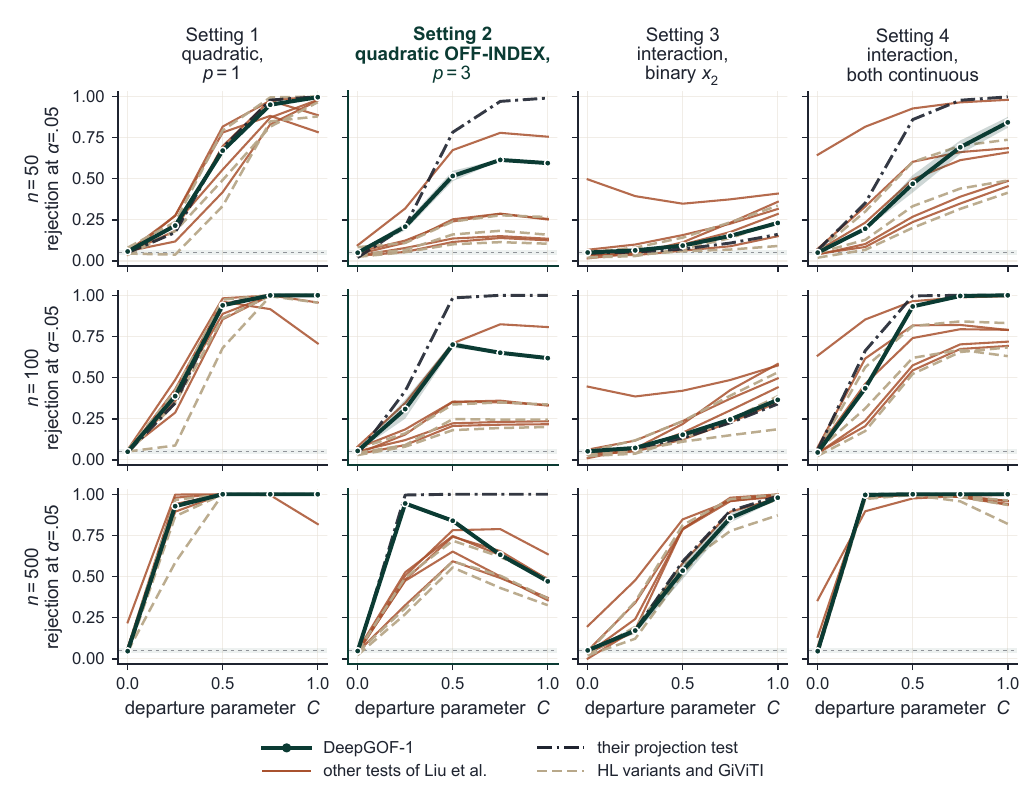}
\caption{The \citet{liu2024} benchmark, extended. Rejection rate at nominal $.05$ against departure
strength, for their four settings by three sample sizes. Their other tests (rust: Hosmer--Lemeshow,
mHL, Osius--Rojek, Stukel, Stute--Zhu), their projection test (dash-dot), the two Hosmer--Lemeshow
variants and GiViTI (tan dashes), and DeepGOF-1 (deep green; ribbon = range over two training
corpora); the band marks $\alpha=.05$. Raw rates are shown, so a test with inflated level starts
above the band. Setting 2 is outlined}
\label{fig:liu}
\end{figure*}

\section{Simulation III: two active covariates among ten}\label{sec:niche}

Section~\ref{sec:irrelevantmain} predicts where DeepGOF-1 should have higher power than the
projection test: when the misfit lives on two covariates and several others carry little signal and
no misfit. This simulation study tests that prediction at finite $n$.

\emph{Design.} Ten independent standard normal covariates, $n\in\{200,500\}$, and
$\eta=-0.5+x_{1}+0.8x_{2}+0.15\sum_{j=3}^{10}x_{j}+C\,f(x_{1},x_{2})$, with an omitted interaction
$f=x_{1}x_{2}$ or an omitted quadratic $f=x_{1}^{2}-1$ and $C\in\{0.3,0.5,0.8\}$: twelve
alternative cells of $1{,}000$ replicates each, and $2{,}000$ null replicates ($C=0$) at each $n$.
The fitted model is linear in all ten covariates. The tests are DeepGOF-1, the projection test of
\citet{liu2024} on the raw covariates, GiViTI, Stukel's test, Hosmer--Lemeshow and mHL, all
computed on the same datasets, and BAGofT at $n=200$ on the first hundred replicates of each cell,
because one BAGofT test takes about three minutes here. Pigeon--Heyse and the equal-width variant
were left out because they track Hosmer--Lemeshow closely (Section~\ref{sec:liubench}), Osius--Rojek
because it does not hold its level, and Stute--Zhu was not part of the study's design. Every test
returned a p-value on every dataset, so the convention for undefined p-values does not arise.

\emph{Analysis.} Each test is placed at exactly level $.05$ on its own null replicates at the same $n$
(Section~\ref{sec:power}), and in each cell the paired difference in matched rejection, DeepGOF-1
minus each other test, is taken over the same datasets. DeepGOF-1 is counted as having higher power in
a cell when it exceeds every other test there by more than two paired standard errors, and the
finding holds only if no test exceeds DeepGOF-1 by that margin in any cell. BAGofT, with a tenth of
the replicates, is outside this rule.

\begin{table*}[!tbp]
\centering
\caption{Two active covariates among ten: matched-level power at $\alpha=.05$ (each test
randomized to exactly $.05$ on its own $2{,}000$ nulls at the same $n$, BAGofT on its $100$; $1{,}000$ replicates per
cell, Monte Carlo standard error at most $.016$), and in the last two rows the raw null size
(binomial standard error about \nichesizese). An asterisk marks a cell in which DeepGOF-1 has
higher power than every other test except BAGofT by more than two paired standard errors.
Projection: the test of \citet{liu2024}; Stukel: Stukel's test in the SstBoth form, as in
Table~\ref{tab:liu}. BAGofT: $n=200$
only, $100$ replicates per cell and $100$ nulls, so its standard error is up to \nichebagsemax;
it is not part of the decision rule.}
\label{tab:niche}
\small\setlength{\tabcolsep}{4.5pt}
\begin{tabular}{@{}lccccccccc@{}}
\toprule
departure & $C$ & $n$ & \textbf{DeepGOF-1} & projection & GiViTI & Stukel & HL & mHL & BAGofT \\
\midrule
\tableniche
\bottomrule
\end{tabular}
\end{table*}

\emph{Results.} DeepGOF-1 has the highest matched power of all \nichentests\ tests in \nichenhighest\
of the twelve settings (Table~\ref{tab:niche}), including all \nichenquadhighest\ quadratic settings,
and exceeds every other test by more than two paired standard errors in \nichenlead\ of them
(asterisks). The exceptions are the two weakest interaction settings at $n=500$: GiViTI is level with
DeepGOF-1 at $C=0.5$ ($.375$ each) and ahead at $C=0.3$ by $.034\pm.016$ ($2.2$ paired standard
errors), as is Stukel's test by $.031\pm.016$, so by the stated rule DeepGOF-1 has higher power in most
of this design but not in all of it. Against the projection test, as Section~\ref{sec:irrelevantmain}
predicts, DeepGOF-1 is ahead in every cell, by $+\nichedminproj$ to $+\nichedmaxproj$ (more than two
paired standard errors in \nicheleadproj\ of twelve), because the projection statistic spreads its
power over all ten directions. Stukel's test (\nichesizestukel) and mHL (\nichesizemhl) run above the
nominal size, and all tests are compared at matched level. BAGofT, at $n=200$ only, is below
DeepGOF-1 in every cell, by more than two paired standard errors in \nichebagnlead\ of the six.

The axis rule is what places DeepGOF-1 here: it selected the two active covariates in $93$ to $100\%$
of the datasets of every cell. Reading every pair of the ten covariates instead, and taking the
largest of the $45$ scores, lands on the active pair in only $2$ to $62\%$ of them, and its matched
power is lower in all twelve cells ($.286$ against $.453$ on average), so the axis rule is the default.

Here the learned score adds power. On a separate set of datasets of the same design ($150$ per
cell), the network on the axis-rule map had matched power $\abNicheNet$ on average, against
$\abNicheSS$ for the sum of squared cells of the same map and $\abNicheMax$ for its largest cell; it
was ahead of the sum of squares by more than two paired standard errors in $\abNicheSSAhead$ of the
twelve cells and behind in none, and ahead of the largest cell in $\abNicheMaxAhead$ (Online Resource 1, Section~21). The misfit fills a few cells of the map; a sum of squares spreads its power
over all $36$, and the largest cell reads only one, while the network was trained on the shapes such
misfit takes.

The opposite case is a covariate whose effect is a pure U-shape, with no linear slope, as in
Section~\ref{sec:support}. A simulation drew five independent standard normal covariates with
$\eta=-0.5+0.8x_{2}+0.6x_{3}+0.4x_{4}+0.3x_{5}+C(x_{1}^{2}-1)$, $C\in\{0.3,0.6\}$ and $n\in\{200,500\}$
($1{,}000$ nulls and $500$ alternatives per cell), and fitted the model linear in all five. The axis
rule selected $x_{1}$ in at most $1.6\%$ of the datasets, and its matched power stayed at the level
($.03$ to $.07$). The all-pairs reading placed its largest score on a pair containing $x_{1}$ in $62$
to $100\%$ of the alternative datasets and had matched power $.13$, $.59$, $.50$ and $.98$, ahead of the
axis rule by $.06$ to $.95$ (paired standard errors at most $.023$). The Hosmer--Lemeshow and Stukel
tests had almost none ($.04$ to $.09$), and the projection test had the most ($.23$, $.72$, $.62$ and
$1.00$); every test held its level ($.045$ to $.060$). A U-shape in a covariate with no linear effect is
therefore the case for the all-pairs reading, and the projection test, which reads every direction, is
stronger still.

\emph{Real covariate designs.} On the covariate designs of the myopia data of \citet{hosmer2013} and
the Pima data distributed with \citet{venables2002}, with the same two departures, the advantage did
not carry over. DeepGOF-1 held
its level ($.048$ and $.057$), but another test had higher power in every cell: the projection test on
the myopia design ($.94$ to $1.00$ against $.16$ to $.91$) and GiViTI on the Pima design ($.28$ to
$.82$ against $.07$ to $.39$). The active effects were weak, an active covariate was strongly skewed,
binary covariates outranked the active pair in the axis rule (the right pair in only $2$ to $38\%$ of
the myopia datasets), or the departure acted mainly as a curve on the risk scale. The advantage of this
section therefore needs continuous, fairly symmetric active covariates with sizeable effects.

\section{Simulation IV: locating the misfit}\label{sec:location}

A test that says where a model fails is useful only if it is right about where. Five standard normal
covariates, $n=500$, and a model fitted linear in all five were used with three omitted terms: a
U-shape in $x_{1}$, a threshold in $x_{2}$, and an interaction of $x_{1}$ and $x_{3}$ ($300$ datasets
per cell, $500$ nulls). DeepGOF-1 reports the pair it displays and, when it rejects, the cells beyond
a simultaneous $95\%$ threshold of that map from the same bootstrap replicates. It was compared with
two tools that examine one covariate at a time: the cumulative sums of residuals of
\citet{linweiying2002}, calibrated by the same refits, and the spline screen of
Section~\ref{sec:support}, both flagging covariates at the Bonferroni level (Online Resource 1,
Section~22).

For the interaction the all-pairs map rejects in $\locInterRej$ of the datasets at the stronger
departure, displays the right pair in $\locInterRight$ of them, and places every flagged cell in the
corners of the true pair in $\locInterLocated$; the two one-covariate tools reject in $\locInterOtherRej$
of them and name the right covariates in none, because an interaction
leaves each covariate's marginal residuals flat. For curvature or a threshold in one covariate the
spline screen locates best ($\locUSplineLocated$ and $\locThreshSpline$ of the datasets, against
$\locUAllLocated$ and $\locThreshAll$ for the map), as a test built for that one-covariate question
should. Under the null the map flags a cell in $\locNullFalseAll$ of the datasets (all-pairs reading)
and $\locNullFalseAxes$ (axis rule). The axis rule, which ranks covariates by their linear effect,
locates neither the U-shape nor the interaction; location belongs to the all-pairs reading.

\section{Simulation V: a published risk model on new patients}\label{sec:external}

A risk model is often published with its coefficients and then checked on patients it was not fitted
to. The null is then simple, $y_{i}\sim\mathrm{Bernoulli}(p_{i})$ with $p_{i}$ given, and nothing is
refitted: DeepGOF-1 draws $y^{*}\sim\mathrm{Bernoulli}(p)$, recomputes the map, standardized by the
given $p$, and scores it, so the law of the score is the same under the whole null and
Lemma~\ref{lem:exact} makes the test exactly valid at every sample size, with no regularity
condition. The same holds for any model that outputs probabilities, logistic or not; the test is
\texttt{deepgof1.external()} in \texttt{ebrahim.gof}, which needs the outcomes, the probabilities and
the covariates, and, without coefficients, ranks the covariates for the axis rule by a least-squares
fit of the logit of the given probabilities. The map lays the
residuals out over the covariates, so the test checks calibration within patient subgroups, the
strong calibration of Van Calster et al.~\citep{vancalster2016}, while the tests that group patients
by predicted risk check calibration along it.

\emph{Design.} A published model in five standard normal covariates (the design of
Section~\ref{sec:location}) was checked at $n=500$ against five kinds of miscalibration, each at two
strengths: a wrong overall rate, an overfitted model (calibration slope $.8$ and $.6$), a U-shape in
$x_{1}$, a threshold in $x_{2}$ and an interaction of $x_{1}$ and $x_{3}$ ($300$ datasets per cell;
$500$ null datasets at $n=250$ and $500$ and $300$ at $n=1{,}000$). The comparison tests are those of
Section~\ref{sec:power} in their external forms: the projection and Pearson (Osius--Rojek) statistics,
calibrated exactly in the same way, the Hosmer--Lemeshow test with ten groups on $10$ degrees of
freedom, Stukel's two terms added to the published linear predictor, and the GiViTI calibration belt
for external validation \citep{nattino2014}, which reduces to the test of calibration intercept and
slope when the calibration curve it selects is linear (Online Resource 1, Section~23).

\emph{Results.} DeepGOF-1 held its level ($\extLevelDGlo$ to $\extLevelDGhi$ over its three readings
and three sample sizes; the other tests up to $\extLevelOthHi$). On miscalibration that differs
between patients with the same predicted risk, DeepGOF-1 has the most power in every cell but one,
where every test is weak. With the combined
reading, a single test fixed in advance, DeepGOF-1 has matched power $\extThreshDG$ for the stronger
threshold, $\extUDG$ and $\extUweakDG$ for the two U-shapes and $\extInterDG$ for the stronger
interaction, against at most $\extThreshBest$, $\extUBest$, $\extUweakBest$ and $\extInterBest$ for
any other test, and GiViTI reaches $\extThreshGiv$, $\extUGiv$, $\extUweakGiv$ and $\extInterGiv$. Only at the
weaker interaction is every test weak ($\extInterWeakDG$ against at most $\extInterWeakBest$). A model
can therefore pass the calibration belt and still be wrong for a recognizable group of patients, and
DeepGOF-1 finds that group and shows it on the map. Miscalibration that is the same for every patient,
a wrong overall rate or slope, is the domain of the tests along the predicted risk (Online Resource 1, Section~23).

\section{Robustness to corrupted records}\label{sec:corrupt}

A logistic model is sometimes fitted to thirty or fifty patients, and one record may carry a
data-entry error: a laboratory value with a slipped decimal, a sign error, a mis-keyed outcome. A
goodness-of-fit test that rejects \emph{because of that record} answers the wrong question
\citep{ylvisaker1977}. For DeepGOF-1 the risk is concrete: at $n=50$ a $6\times6$ map holds about
$1.4$ patients per cell, so one record can \emph{be} a cell, and the recession slope is positive for a
spike in every cell and both signs, so the property that certifies consistency
(Section~\ref{sec:consistency}) also lets a single extreme record push the score up.

\subsection{One corrupted record at $n\le50$}

A first study drew $x_{1},x_{2}\sim N(0,1)$ and $\eta=-0.5+x_{1}+0.6x_{2}+C(x_{1}^{2}-1)$, fitted the
main-effects model, and corrupted one uniformly chosen record after its outcome was generated, by a
decimal slip ($x_{1}\leftarrow10x_{1}$), a sign error or a flipped outcome ($C=0$ for size and
$C\in\{0.5,1.0,1.5\}$ for power; $1{,}000$ null replicates per cell, so the standard error of a $5\%$
rate is about $.007$). A test was taken to misbehave if its size exceeded $.075$, about $3.6$ standard
errors above $.05$. The tests were Hosmer--Lemeshow, Stukel's joint score test, DeepGOF-1 ($B=99$) and
the projection test of \citet{liu2024} ($B=199$; Online Resource 1, Section 15).

\begin{table}[!tbp]
\centering
\caption{Size at $\alpha=.05$ with one corrupted record ($C=0$; 1,000 replicates per cell; standard
error about .007). HL: Hosmer--Lemeshow; Stukel: Stukel's joint score test;
DG-1: DeepGOF-1; projection: the test of Liu et al. An asterisk marks a size
above $.075$. The decimal slip, a leverage point, breaks three of the four tests; sign errors and
flipped outcomes break none, except the projection test marginally at $n=30$.}
\label{tab:corrupt}
\small\setlength{\tabcolsep}{4pt}
\begin{tabular}{@{}llcccc@{}}
\toprule
$n$ & corruption & HL & Stukel (joint) & DG-1 & projection \\
\midrule
30 & none & .046 & .048 & .056 & .046 \\
30 & decimal slip & .049 & .079* & .086* & .143* \\
30 & sign error & .033 & .063 & .057 & .079* \\
30 & flipped outcome & .034 & .071 & .038 & .064 \\
50 & none & .049 & .061 & .055 & .054 \\
50 & decimal slip & .041 & .142* & .118* & .180* \\
50 & sign error & .045 & .051 & .046 & .056 \\
50 & flipped outcome & .041 & .054 & .056 & .039 \\
\bottomrule
\end{tabular}
\end{table}

A slipped decimal pushes Stukel's test, DeepGOF-1 and the projection test above $.075$ at both sample
sizes, the projection test most ($.180$ at $n=50$; Table~\ref{tab:corrupt}). The Hosmer--Lemeshow test
is immune, because the bad record falls into an extreme risk decile and is diluted there. The problem
is leverage, not outcome error.

\subsection{Two repair rules}

A rule that rejects only if the test rejects after deleting each record in turn has size at most the
base test's clean size for a record chosen independently of the data, by the intersection--union
argument of \citet{berger1982}, but at these sample sizes it is far too conservative (size $.000$ to
$.012$) and halves the power of DeepGOF-1 (Online Resource 1, Theorem~S.14 and Section~15). The
repair we recommend deletes at most one record, chosen from the covariates alone: if the largest robust
outlyingness $\max_{j}|x_{ij}-\mathrm{med}_{j}|/\mathrm{MAD}_{j}$ exceeds $4.5$, that record is removed
and the base test is run on the rest. Because the rule never looks at the outcomes, the base test's
validity on clean data is unchanged, and under a covariate corruption the size is at most the base
test's clean size plus the probability that the rule misses the corrupted record (Online Resource 1, Theorem~S.17), about $.04$ to $.05$ for a gross slip. A second study, on fresh seeds, tested the rule (Online Resource 1, Section 15).

\begin{table}[!tbp]
\centering
\caption{The covariate-only deletion rule, confirmation study on fresh seeds (1,000 null replicates
and 300 alternatives per cell). Size under a decimal slip, before and after the rule, and the base
test's clean size. Stukel: Stukel's joint score test; projection: the test of Liu et al.}
\label{tab:trim}
\begin{tabular}{@{}llccc@{}}
\toprule
$n$ & test & slip, no rule & slip, with rule & clean \\
\midrule
30 & Stukel (joint) & .091 & .066 & .053 \\
30 & DeepGOF-1 & .088 & .047 & .055 \\
30 & projection & .139 & .056 & .059 \\
50 & Stukel (joint) & .145 & .073 & .064 \\
50 & DeepGOF-1 & .111 & .055 & .055 \\
50 & projection & .164 & .058 & .045 \\
\bottomrule
\end{tabular}
\end{table}

The rule passed its check in every cell: each broken test returned to within two standard errors of
its own clean size (Table~\ref{tab:trim}), which puts DeepGOF-1 and the projection test back in the
band ($.047$ to $.058$), while Stukel's test improves but stays above it ($.066$ and $.073$). On clean
data, under the null and three strengths of misfit, the rule left every test's rejection rate
unchanged within two paired standard errors in all $24$ cells, so it has no measurable cost in power.
It catches a gross slip in about $95\%$ of cases; the mild slips it misses left DeepGOF-1 at its
nominal size but not Stukel's test ($.124$ at $n=50$). The rule is meant for continuous, roughly normal
or bounded covariates: with ten covariates at $n\le50$ it deletes a good record in up to $15\%$ of
clean datasets, and a heavy-tailed or skewed covariate should be transformed first (Online Resource 1, Section~16); a false fire costs at most one record of power and never the level.

\subsection{A few corrupted records at $n=1{,}000$}

In a large dataset one bad record is diluted, but a data-entry error is rarely alone. A third study
used the design of an earlier study of corrupted records for calibration tests, so that the other
tests' p-values are on the same datasets: $x\sim U(-3,3)$, a binary $d$, the correct logistic model
in $x$ and $d$, and after the outcomes were drawn, $x$ multiplied by $4$ in $1$, $2$, $5$ or $10$ of the
$1{,}000$ records. The outcome is right and the covariate is wrong, so every rejection is a false
alarm (Online Resource 1, Section~25).

\begin{table}[!tbp]
\centering
\caption{False-alarm rate at $.05$ at $n=1{,}000$ when $x$ is multiplied by $4$ in a few records. Standard
error about $.007$ at $1{,}000$ datasets; the projection test and BAGofT were run on the first datasets
of each cell (number in brackets). HL: Hosmer--Lemeshow; Stukel: joint score test.}
\label{tab:corrupt1000}
\small\setlength{\tabcolsep}{3pt}
\begin{tabular}{@{}llrcccccc@{}}
\toprule
$n$ & corruption & datasets & DeepGOF-1 & HL & Stukel & GiViTI & projection & BAGofT \\
\midrule
\tablecorruptmain
\bottomrule
\end{tabular}
\end{table}

DeepGOF-1 kept its level with up to five corrupted records ($\coDGclean$ clean, $\coDGfive$ with five)
and rose to $\coDGten$ with ten, while Stukel's test reached $\coStukfive$ with five and $\coStukten$
with ten, and GiViTI $\coGivfive$ and $\coGivten$ (Table~\ref{tab:corrupt1000}). The Hosmer--Lemeshow
test is again among the least affected ($\coHLten$), for the reason given above. The projection test,
on the first $\coProjN$ datasets of each cell, read $\coProjlo$ to $\coProjhi$, and BAGofT, on the first
$\coBagN$ of the clean and ten-record cells, $\coBagclean$ in both; on the same datasets neither differed
from DeepGOF-1 by more than chance (at most $\coDiscMax$ datasets on which only one of the two
rejected). The map is built on the
ranks of the covariates, so a corrupted value moves its record to an extreme row or column but does
not enter the statistic through its size; Stukel's test and GiViTI look for curvature along the fitted
linear predictor, which a few extreme covariate values bend directly. At $n=500$ with three records multiplied by $8$,
DeepGOF-1 read $\coDGxeight$, the projection test $\coProjxeight$ and Stukel's test $\coStukxeight$ ($100$
datasets; Online Resource 1, Section~25). The deletion rule above removes at most one record and
was not studied at this size; with a percent of the records suspect, the covariates should be screened
for impossible values before any goodness-of-fit test.

\section{Application: in-hospital mortality in the SUPPORT study}\label{sec:support}

The Study to Understand Prognoses and Preferences for Outcomes and Risks of Treatments (SUPPORT)
enrolled seriously ill adults in five US teaching hospitals \citep{knaus1995support}; its public
release \citep{support2data} has $9{,}105$ patients. We model in-hospital death with the logistic
regression a modeller would write first, linear in each covariate: age, mean arterial pressure,
heart rate, respiratory rate, temperature, serum creatinine, serum sodium, white blood cell count,
the number of comorbidities and sex. Complete cases give $n=8{,}873$ patients and $2{,}328$ deaths;
the $232$ patients with a missing value are excluded, which could bias the fitted risks slightly but
not the comparison of the tests, which all see the same data.
The repairs below were tried in a fixed order. The rows were randomly permuted once, and DeepGOF-1 was run with $B=199$.

\emph{Detection.} The axis rule selects mean arterial pressure and the number of comorbidities, and
DeepGOF-1 rejects M0 at the smallest attainable p-value, $.005$ (score $8.87$ against a bootstrap
maximum of $2.98$). The second axis is a heavily tied count, which condition A1 excludes and
Recommendation~3 warns about, so this reading lies outside the theory of Section~\ref{sec:theory};
over $20$ further tie-breaks the p-value is still $.005$ every time (scores $8.6$ to $10.4$). Nor does
the verdict depend on this pair. Because a U-shaped effect has almost no linear slope, mean arterial
pressure led the axis rule only narrowly (the rule picked it in about $40\%$ of subsamples of $250$ to
$1{,}000$ patients), and every pair of the four covariates the rule ranked highest, with the axes
fixed in advance, rejects at $p=.005$ (scores $6.7$ to $13.6$). The all-pairs reading, the largest
score over the $45$ pairs in the data and in every bootstrap replicate, also rejects M0 at $p=.005$;
its largest score is on mean arterial pressure and respiratory rate, and over $20$ further
tie-breaks the p-value stays $.005$ and the top pair always contains mean arterial pressure.

\begin{figure*}[!tbp]
\centering
\fullfig{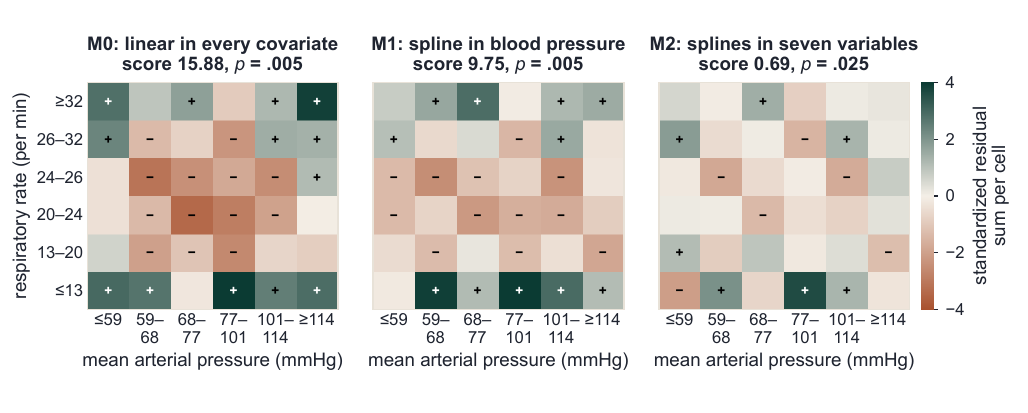}
\caption{The DeepGOF-1 residual map of the SUPPORT in-hospital mortality model along the modelling
ladder ($n=8{,}873$, $2{,}328$ deaths), over the sixths of mean arterial pressure (columns) and
respiratory rate (rows), the pair on which the all-pairs reading of M0 reaches its largest score. Cells
are standardized residual sums (observed minus predicted deaths, divided by the square root of their
predicted variance); a sign marks each cell beyond $\pm1$, a descriptive display (the simultaneous
$95\%$ threshold for M0 is $3.10$). Left: M0, linear in every covariate; deaths exceed the prediction
at both extremes of respiratory rate and of blood pressure. Middle: M1, with a natural spline in mean
arterial pressure; the blood-pressure pattern is gone and the respiratory-rate pattern remains. Right:
M2, with natural splines in seven physiological variables; on this pair M2 still rejects (score
$0.69$, $p=.025$). Every panel reads the same pair with the same tie-break}
\label{fig:support}
\end{figure*}

\emph{Location.} The map of the pair with the largest score shows where the misfit lies
(Figure~\ref{fig:support}, left). Both axes are integer-valued. Respiratory rate is heavily heaped:
$15.0\%$ of the patients have exactly $20$ breaths per minute and $13.5\%$ exactly $24$, and both
values are sextile cut points, so the random tie-break decides which of two adjacent rows those
patients fall in; mean arterial pressure is only mildly tied (at most $2.5\%$ of the patients at any
cut point). This reading, too, lies outside condition A1 and rests on the measured level; over the
$20$ tie-breaks above, its p-value and its leading covariate did not change. In the lowest sixth of respiratory rate ($\le13$ per
minute) and the highest ($\ge32$) more patients died than M0 predicts in most blood-pressure columns,
and in the middle sixths fewer died; the same holds at both extremes of mean arterial pressure
($\le59$ and $\ge114$~mmHg). These are the U-shaped risks that APACHE III builds in for both variables
and that a linear term cannot represent. The test is on this pattern as a whole; single cells are
descriptive. A simultaneous $95\%$ threshold for the largest cell, from $999$ parametric-bootstrap maps
of M0 on this pair, is $|m|=3.10$, and four cells exceed it, the largest $3.98$. On the axes of the
axis rule the U-shape of blood pressure is the dominant pattern: more patients died than predicted at
both extremes of pressure in most comorbidity sixths.

\emph{Repair.} The first repair on the list, a product term between the two covariates of the axis
rule, leaves the rejection in place ($p=.005$). The second, a natural spline with three degrees of
freedom in mean arterial pressure \citep[Ch.~2]{harrell2015rms}, removes the pattern the map showed.
Read on the axes of M0, fixed before the repaired model (M1) was fitted, the score falls from $8.87$
to $-1.97$ ($p=.90$), no cell exceeds $\pm2$ (six did), and the U-shape of blood pressure is gone. The
axis rule scores each covariate by the spread of its terms' joint contribution to the linear
predictor, so a covariate entered as a spline remains one axis; on M1 it selects mean arterial
pressure and heart rate and still rejects ($p=.020$), and the all-pairs reading rejects with its
largest score on respiratory rate and sodium ($p=.005$): the misfit that remains lies elsewhere. On the pair of
Figure~\ref{fig:support} (middle) the blood-pressure extremes are gone as well, but the U-shape of
respiratory rate remains, and the map still rejects (score $9.75$, $p=.005$): the next step of the
ladder below. A likelihood-ratio test, which uses no goodness-of-fit statistic,
agrees: M1 improves on M0 by $115$ on $2$ degrees of freedom ($p<10^{-25}$), and the Akaike criterion
falls by $111$. Because the repair was chosen after looking at these data, neither the $p=.90$ nor
the likelihood-ratio test is a confirmatory test of M1; they show that the pattern the map found is
removed by the term it points to.

\emph{What the usual diagnostics show.} A calibration curve, the tool most modellers look at first,
largely hides this misfit. On the risk scale M0 looks well calibrated: its integrated calibration
index \citep{austin2019ici} is $.019$, and observed and predicted mortality agree within $.03$ in every
sixth of predicted risk. Across mean arterial pressure the same model is off by $.047$ in the lowest
sixth and $.045$ in the highest, in opposite directions to the middle. The excess deaths at both
extremes of blood pressure occur at ordinary predicted risks, so they average out when the residuals
are grouped by risk, and they appear only when the residuals are laid out over the covariates, which
is what the map does. In the calibration hierarchy of Van Calster et al.~\citep{vancalster2016}, M0 is close to calibrated along
predicted risk but not calibrated within covariate patterns, the strong calibration that the map
checks over two covariates at a time. A screen that tries a spline for each continuous covariate in
turn \citep{harrell2015rms} also finds the misfit: at this sample size it flags seven of the eight
covariates, with mean arterial pressure, respiratory rate and heart rate strongest. Spline modelling of
these variables is standard, and the misfit here is two separate U-shapes, one in each covariate, not
a joint pattern of two covariates; the example illustrates detection, location and repair of a known
misfit. What the map adds is that one test shows where the misfit lies and which term to add.

\emph{The modelling ladder.} The classical tests reject M0 as well, so in these data DeepGOF-1 adds
the location of the misfit, not its detection. Their verdict on M1 shows why location matters: all of
them except the equal-width variant and the Osius--Rojek test still reject (Table~\ref{tab:support}),
and none says where the remaining misfit lies. Read on all $28$ pairs of the eight continuous
covariates, the map of M1 flags every pair, with respiratory rate, creatinine, sodium and white cell
count in the strongest; each is a physiological variable whose risk rises at both extremes. Splines in
all seven physiological variables (M2) improve on M1 by $194$ on $12$ degrees of freedom; the map of M2
flags $15$ of $28$ pairs, the six strongest all involving age, the one continuous covariate still
linear, but a spline in age (M3) gains little ($p=.077$). Each reading turns this search into one test
at each step, and the three agree: the axis rule rejects M1 and M2 ($p=.020$ and $.010$) and not M3
($p=.195$); the all-pairs reading rejects M1 ($p=.005$, largest score on respiratory rate and sodium)
and M2 ($p=.020$, on age and the number of comorbidities) and not M3 ($p=.21$); and the combined
reading gives $.010$, $.020$ and $.35$. On the pair of
Figure~\ref{fig:support} the score falls from $15.9$ (M0) to $9.8$ (M1) and $0.7$ (M2), but M2 still
rejects on that pair ($p=.025$): a small misfit remains there at this sample size. A score is
calibrated only against its own model's bootstrap, so the p-values, not the raw scores, are the scale
on which the ladder should be read. At $n=8{,}873$ small misfit may remain in M3, and several omnibus
tests still reject it, as omnibus tests tend to at large $n$ \citep{paul2013,nattino2020}; we stop
there. The optimism-corrected area under the ROC curve rises from $.627$ (M0) to $.671$ (M2), and the
calibration slope stays between $.96$ and $.98$.

The repair changes the risks clinicians act on. For a patient at the median of every other covariate,
M0 predicts in-hospital death with probability $.29$ ($95\%$ interval $.27$--$.31$) at a mean arterial
pressure of $45$~mmHg, and M1 predicts $.36$ ($.33$--$.39$); at $130$~mmHg the two predictions are
$.18$ ($.17$--$.20$) and $.23$ ($.21$--$.25$). The linear model under-states the risk at both ends of
the blood-pressure range, where treatment decisions are made.

\emph{Cost, level and power on this design.} The projection test, the strongest omnibus test on the
benchmark of Section~\ref{sec:power}, needs $O(n^{3})$ operations and an $n\times n$ weight. Run on M0
with all $8{,}873$ patients and $B=199$, it rejected ($p=\tmProjP$) after $\tmProjHours$ on one core,
with a peak memory of $\tmProjGB$~GB; on the same machine and load DeepGOF-1 took $\tmDGaxes$~seconds
and its all-pairs reading $\tmDGall$, about $\tmRatio$ and $\tmRatioAll$ times faster. To check the
level and power on the study's own covariates, outcomes were simulated from M0 (null) and from M1 (a
U-shaped risk of blood pressure) for random subsets of the patients and for all of them (Online Resource 1, Section~24). DeepGOF-1 held its level at every size ($\plDGlo$ to $\plDGhi$ for
$n=250$ to $1{,}000$, and $\plFullDGlo$ to $\plFullDGhi$ on $\plFullReps$ redrawn outcome sets of all
$8{,}873$ patients). Against the U-shape the projection test had the most power at $n=1{,}000$
($\plProjPowK$, against $\plDGPowK$ for DeepGOF-1), and the Hosmer--Lemeshow, Stukel and GiViTI tests
had almost none (at most $\plClassPowK$). Where the projection test can be run, it is the stronger
omnibus check; at the size of a clinical registry DeepGOF-1 is the test that can be run routinely.

\emph{External validation.} The public release has no hospital or date identifier, so a random split
stands in for a new cohort: half of the patients develop the model, which is then frozen and checked
on the other half ($\xsNval$ patients), over $\xsSplits$ random splits (Online Resource 1,
Section~24). Checked in this way, M0 looks calibrated by the usual summaries (median calibration
intercept $\xsMzeroCitl$ and slope $\xsMzeroSlope$), and the GiViTI calibration test passes it in all
but $\xsMzeroGiv$ of the $\xsSplits$ splits (median $p=\xsMzeroGivmed$). DeepGOF-1 for frozen
predictions (Section~\ref{sec:external}) rejects it in all $\xsMzeroDG$ (median $p=\xsMzeroDGmed$), the
Hosmer--Lemeshow test in $\xsMzeroHL$ and Stukel's test in $\xsMzeroStuk$, and the pair its map shows
contains respiratory rate in $\xsMzeroResp$ of the splits. The linear model is calibrated on average and
wrong for patients at the extremes of physiology, which is the kind of miscalibration the map is built to
find. The spline model M2 is in Online Resource 1.

\begin{table*}[!tbp]
\centering
\caption{The SUPPORT modelling ladder: goodness-of-fit p-values and likelihood-ratio tests. M0: linear
in every covariate. M1: natural spline (3 df) in mean arterial pressure. M2: natural splines in all
seven physiological variables. M3: M2 with a spline in age. DeepGOF-1: the three readings of
\texttt{deepgof1()} from one call per model, $B=199$; axis rule: the two covariates whose terms
contribute most to the linear predictor; all pairs: one test over the $45$ pairs of the ten
covariates; combined: the exact minimum of the two. Axes of M0 fixed: the map on mean arterial
pressure $\times$ number of comorbidities, the axes chosen on M0, given for M0 and M1. Pairs flagged:
the number of the $28$ pairs of continuous covariates whose map rejects at $.05$ ($B=99$; a search,
not a single test), and the largest score. Stute--Zhu uses $200$ bootstrap draws, so $0$ is reported
as $<.005$. The projection test was run on M0 only; it took about ten hours (see text). AUC: area under the ROC curve, optimism-corrected by
$200$ bootstrap refits.}
\label{tab:support}
\small\setlength{\tabcolsep}{5pt}
\begin{tabular}{@{}lcccc@{}}
\toprule
 & M0 & M1 & M2 & M3 \\
\midrule
Hosmer--Lemeshow & $<.001$ & $.004$ & $.064$ & $.34$ \\
mHL & $<.001$ & $<.001$ & $<.001$ & $<.001$ \\
Pigeon--Heyse & $<.001$ & $.009$ & $.097$ & $.41$ \\
Hosmer--Lemeshow, equal width & $.007$ & $.15$ & $<.001$ & $<.001$ \\
Osius--Rojek & $<.001$ & $.28$ & $.53$ & $.51$ \\
Stukel & $.005$ & $.009$ & $<.001$ & $<.001$ \\
GiViTI & $<.001$ & $.037$ & $.013$ & $.030$ \\
Stute--Zhu & $<.005$ & $<.005$ & $<.005$ & $.015$ \\
Projection & $.005$ & & & \\
\textbf{DeepGOF-1, axis rule} & $\mathbf{.005}$ & $\mathbf{.020}$ & $\mathbf{.010}$ & $\mathbf{.195}$ \\
\textbf{DeepGOF-1, all pairs} & $\mathbf{.005}$ & $\mathbf{.005}$ & $\mathbf{.020}$ & $\mathbf{.21}$ \\
\textbf{DeepGOF-1, combined} & $\mathbf{.005}$ & $\mathbf{.010}$ & $\mathbf{.020}$ & $\mathbf{.35}$ \\
DeepGOF-1, axes of M0 fixed & $.005$ & $.90$ & & \\
DeepGOF-1, pairs flagged (largest score) & $28$ ($15.9$) & $28$ ($11.1$) & $15$ ($3.1$) & $10$ ($1.6$) \\
\midrule
Likelihood-ratio test against the previous model & & $115$ on $2$ df & $194$ on $12$ df & $5.1$ on $2$ df \\
 & & $p<10^{-25}$ & $p<10^{-34}$ & $p=.077$ \\
AUC, optimism-corrected & $.627$ & $.643$ & $.671$ & $.670$ \\
\bottomrule
\end{tabular}
\end{table*}

\section{Discussion}\label{sec:discussion}

The audit leaves four findings. First, the level of a learned test can be made a property of the
\emph{calibration}, not of the learning: the rank p-value \eqref{eq:pval} inherits its validity from
the bootstrap, so the conditions of Proposition~\ref{prop:valid} constrain the fitted logistic model
rather than the $18{,}273$ frozen numbers, except for the continuity of the score's null limit law,
which is checked numerically. The measurements agree, from the sixty-cell grid and the shipped
network's twenty cells to the external benchmark, the real designs, samples as small as $n=20$ and
all $8{,}873$ SUPPORT patients; where DeepGOF-1 alone stayed inside the band, several other tests were within one or two standard
errors of it, so the stronger evidence is the small spread of its size across blocks and designs.
Second, consistency is not an article of faith about the training distribution: for any alternative a
practitioner can name, one forward pass on the frozen weights decides whether
Theorem~\ref{thm:consistency} applies, and the blind cone, which no binned statistic avoids, is
measured and thin. Third, power depends on where the signal lies. With every covariate carrying signal
the projection test needs fewer observations; with several added covariates that carry no misfit the
order reverses. The shipped network gives up about $.04$ of the power of networks trained on each
setting's design, and its all-pairs reading wins it back on the benchmark, while with many covariates
the axis rule is the stronger reading. The learned score earns its place where misfit is sparse: on
the benchmark a $\chi^{2}$ of the same map does as well, and with two active covariates among ten the
network is ahead of it. Against the most powerful map-based test for its own training
prior, the shipped network reaches about seventy per cent of the attainable power in a Gaussian
surrogate (Online Resource 1, Section 17), which leaves room for a successor trained on a wider
range of designs; an equivalence version, in which adequate fit is the alternative
\citep{baillo2026}, is another direction. Fourth, the map answers questions that tests along the
predicted risk cannot ask: it names the pair of covariates in a missed interaction, and, for a
published model checked on new patients, it is an exact test of calibration within patient subgroups
that has the most power where the calibration belt is nearly blind.

The power comparisons are at $n$ from $20$ to $1{,}000$. The rankings may change as $n$ grows, and
the benchmark shows the direction: DeepGOF-1 ranks \rankdeepnfiftyword\ at $n\le100$ but
\rankdeepnfivehundredword\ at $n=500$, where the tests that use the full covariate information catch
up; on the SUPPORT design the projection test has more power up to $n=1{,}000$, but at the full
$8{,}873$ patients it took about ten hours where DeepGOF-1 took seconds.

We recommend the following.
\begin{enumerate}\itemsep1pt
\item Use the projection test of \citet{liu2024} as the default omnibus check of a logistic model when
$n$ is small enough for it to be computed. Its cost grows as $n^{3}$ in time and $n^{2}$ in memory: at
$n=1{,}000$ it takes about a minute, but on the $8{,}873$ SUPPORT patients it took $\tmProjHours$ and
$\tmProjGB$~GB, which rules it out as a routine check of a registry-sized model.
\item Use DeepGOF-1 when the modeller needs to know where the model fails (the test is on the map's
pattern as a whole, so single cells are descriptive unless a simultaneous threshold is used), when $n$
makes the projection test impractical, or when the misfit is expected on two continuous, fairly symmetric covariates
among several that carry little signal; when a published risk model is checked on new patients and
calibration within patient subgroups matters, alongside the calibration belt, which remains the tool
for the overall rate and the calibration slope; and when a certificate of consistency, or a power
calculation, against a named alternative is wanted before data are collected. At $n\le50$ with only
two covariates it keeps its level where other tests drift, but the projection test, and in most cells
Stukel's test, have more power. It does not detect a wrong link; for that, use Stukel's test or GiViTI.
\item Read the selected axes. Use another test when they are discrete or heavily tied, and, since the
axis rule ranks covariates by their linear effect, fix a covariate whose risk is suspected to be
U-shaped as an axis in advance, as in Section~\ref{sec:support}, or read every pair of covariates
(the all-pairs reading of \texttt{deepgof1()}), which does not depend on the linear effects but, with
many covariates, has less power when one pair carries the misfit (Section~\ref{sec:niche}); the
combined reading guards against both at a small cost. To locate the misfit, read the all-pairs map;
for curvature in a single covariate a spline screen locates it better (Section~\ref{sec:location}).
\item Treat the level as first-order valid and measured rather than exact, and the power certificates
as statements about a Gaussian model of the map.
\item At $n\le50$, screen the covariates for a gross outlier before running a goodness-of-fit test:
one slipped decimal broke the projection test, Stukel's test and DeepGOF-1 in
Section~\ref{sec:corrupt}, and the covariate-only deletion rule brought DeepGOF-1 and the projection
test back into the nominal band; Stukel's test improved but stayed above it. At larger $n$ a few
corrupted values leave DeepGOF-1 near its level but not Stukel's test or GiViTI, and with about $1\%$
of the records corrupted DeepGOF-1 drifts as well ($\coDGten$).
\end{enumerate}

What generalizes beyond logistic regression is the audit itself: render misfit as a statistic-valued
image, train against the bootstrap null you will calibrate against, calibrate by rank, certify
consistency through the recession function, compute local power through the continuous mapping
theorem, and certify power under a Gaussian model of the input by smoothing. Each step applies to any
network built from affine maps, rectifiers and maxima whose output is the test statistic.

\backmatter

\section*{Statements and Declarations}

\noindent\textbf{Funding.} No funding was received for conducting this study.

\smallskip\noindent\textbf{Competing interests.} The authors have no relevant financial or
non-financial interests to disclose.

\smallskip\noindent\textbf{Ethics approval.} Not applicable. This is a methodological and simulation
study. The applications use only public, de-identified data: the SUPPORT study release
\citep{support2data}, institution-level financial data from the FDIC, and the covariate values of two
data sets distributed with R packages (Section~\ref{sec:niche}). No data on human or animal subjects
were collected.

\smallskip\noindent\textbf{Consent.} Not applicable.

\smallskip\noindent\textbf{Author contributions.} E.K.E.\ formulated the method, derived and checked
every proof, directed and verified the simulation, benchmark and application code, produced every
figure and table, and wrote the manuscript. O.A.E.H.\ and A.E.\ supervised the work and reviewed and
edited the manuscript. All authors read and approved the final manuscript.

\smallskip\noindent\textbf{Data availability.} All data are public. The SUPPORT data are available from
the UCI Machine Learning Repository \citep{support2data}; the FDIC data rebuild from two keyless API
calls by the deposited script. Every per-replicate p-value behind every table and figure, the training
corpus and its generator, the frozen weights, and the complete training, benchmark and application
code, including the scripts, seeds and outputs of the SUPPORT application, are archived at Zenodo
\citep{deepgof1archive} (repository \url{https://github.com/ebrahimkhaled/deepgof1}). All computation
used R 4.4.1.

\smallskip\noindent\textbf{Code availability.} The deployable test is \texttt{deepgof1()} in the R
package \texttt{ebrahim.gof} \citep{ebrahimgof2026}, version 2.9.0 or later, which also implements the
other tests compared here (the GiViTI test through its authors' package \texttt{givitiR}) and the
projection test of \citet{liu2024} as \texttt{projection.gof()}; these implementations are benchmarked
by \citet{ebrahim2026benchmark}. BAGofT is available as the R package of \citet{zhang2023}.

\smallskip\noindent\textbf{Acknowledgements.} All praise is due to Allah, who gave us the capacity to
understand, to reason, and to seek knowledge. The authors thank the SUPPORT investigators for making
their data public, and the maintainers of the public FDIC BankFind API.

\smallskip\noindent\textbf{Supplementary information.} Online Resource 1 contains the full proofs of
Lemmas~\ref{lem:exact} and~\ref{lem:recession}, Propositions~\ref{prop:valid} and~\ref{prop:fsbound},
Theorems~\ref{thm:consistency} and~\ref{thm:local}, and Corollaries~\ref{cor:cone}
and~\ref{cor:representation}, the conditions A1--A7 and the auxiliary lemmas A.1--A.6, together with the
supporting studies: covariates that carry no misfit, the calibration gap and the rate of the level
error, the power certificates, the Neyman--Pearson optimality gap, the theory and both studies of
Section~\ref{sec:corrupt}, the study of very small samples and wrong links, the training protocol of the
shipped network and the deposit index.

\bibliography{references}

\end{document}


\maketitle

\noindent This Online Resource proves the results that the paper states, proves the results that
it names without a numbered statement, and reports the supporting studies. Its results carry the
prefix S, its auxiliary lemmas the prefix A, and its conditions are A1--A7. The S-numbers follow the
paper's citations rather than the order of sections, so they do not always rise with the section
number.
It has three parts: Part A proves the results of the paper, Part B reports the power certificates,
the corrupted-record theory and studies, the Neyman--Pearson gap, the study of very small samples
and wrong links, the paired comparison with BAGofT and the FDIC level-transfer study, and Part C records the benchmark
bookkeeping and the deposit folder behind each study. Table~\ref{tab:corresp}
shows where each result of the paper is proved.

\begin{table}[!htbp]
\centering
\caption{Where the results of the paper are proved}
\label{tab:corresp}
\small
\begin{tabular}{@{}llc@{}}
\toprule
result of the paper & results here & section \\
\midrule
Lemma 2 (recession function) & S.1 & \ref{sec:recession} \\
Lemma 1 (exactness), Proposition 1 (first-order validity) & S.2, S.3 & \ref{sec:validity} \\
Proposition 2 (finite-sample size bound) & S.7, S.8; A.3--A.5 & \ref{sec:fsbound} \\
data-computable calibration-gap bound & S.11 & \ref{sec:levelgap} \\
rate of the level error & S.11a, Table S.11a & \ref{sec:rate} \\
Theorem 1 (conditional consistency) & S.4; A.1--A.2 & \ref{sec:consistency}, \ref{sec:aux} \\
Corollaries 1--2 (blind cone, representation limit) & S.5, S.6 & \ref{sec:cone} \\
Theorem 2 (local asymptotic power) & S.9, S.10 & \ref{sec:localpower} \\
covariates that carry no misfit & S.10a, S.10b & \ref{sec:irrelevant} \\
power certificates under a Gaussian model & S.12, S.12a, S.12b; A.6 & \ref{sec:certpowerS} \\
one corrupted record & S.14--S.19; Studies 1--2 & \ref{sec:corrupt}--\ref{sec:touch} \\
Neyman--Pearson gap & S.13 & \ref{sec:np} \\
very small samples and wrong links & -- & \ref{sec:smallnlink} \\
paired comparison with BAGofT & -- & \ref{sec:bagoftS} \\
level transfer to a real design (FDIC) & -- & \ref{sec:fdicS} \\
benchmark: missing p-values, failed refits & -- & \ref{sec:bookkeeping} \\
\bottomrule
\end{tabular}
\end{table}

\tableofcontents

\clearpage
\noindent{\small\itshape Online Resource 1 for ``Where Does a Logistic Risk Model Fail? An Audited Neural Goodness-of-Fit Test for Model Development and External Validation'', by E.~K.~Ebrahim, O.~A.~E.~Hussein and A.~El-Kotory.}\par\medskip
\noindent{\Large\bfseries Part A. Proofs of the results of the paper}\par\medskip
\section{Setting and the deployed algorithm}\label{sec:setting}

Observations $(x_i,y_i)$, $i=1,\dots,n$, are i.i.d.\ from a law $P$ on
$\R^{d}\times\{0,1\}$ with $d\ge 2$, and $y\mid x\sim\mathrm{Bern}(\pi^{*}(x))$ for an
unknown $\pi^{*}$ with $0<\pi^{*}(x)<1$ for $P$-a.e.\ $x$. Write $\tox=(1,x^{\top})^{\top}$.
The null hypothesis is the logistic model
$H_{0}:\ \pi^{*}=\pi_{\beta}\ \text{for some }\beta,\qquad
\pi_{\beta}(x)=\Lambda(\tox^{\top}\beta),\quad \Lambda(u)=(1+e^{-u})^{-1}.$
The quasi--maximum-likelihood estimator (QMLE) $\hat\beta_{n}$ maximizes the Bernoulli
log-likelihood whether or not $H_{0}$ holds.

The deployed test reproduces the released function \texttt{deepgof1()}; the
correspondence between the description below and the released code is line by line except
where a convention is stated explicitly. It computes:

\begin{enumerate}
\item \emph{Axis selection.} For each covariate $j$ let $\hat c_{j}=|\hat\beta_{n,j}|\,
\hat\sigma_{j}$, with $\hat\sigma_{j}$ the sample standard deviation of covariate $j$.
The two largest $\hat c_{j}$ select the covariate pair, returned in model order; call the
selected indices $(a,b)$.
\item \emph{Cells.} With $K=6$: let $r^{a}_{i}$ be the rank of $x_{ia}$ among
$x_{1a},\dots,x_{na}$ (ties broken by one random permutation of the observations, drawn once per
call and held fixed for the observed and every bootstrap map). Observation $i$ lies in marginal
bin $k$ on axis $a$ iff $\lfloor K(r^{a}_{i}-1)/n\rfloor=k-1$ (capped at $K$),
equivalently iff $r^{a}_{i}\in\bigl(\lceil (k-1)n/K\rceil,\ \lceil kn/K\rceil\bigr]$.
Setting $\hat q^{a}_{k}:=x_{a,(\lceil kn/K\rceil)}$ for $k=1,\dots,K-1$ (with
$\hat q^{a}_{0}=-\infty$, $\hat q^{a}_{K}=+\infty$), this says
$x_{ia}\in(\hat q^{a}_{k-1},\hat q^{a}_{k}]$ whenever the relevant boundary order
statistics are distinct; likewise on axis $b$, and $C_{k\ell}$ is the product of the
$k$-th bin on $a$ with the $\ell$-th on $b$. Lemma~\ref{lem:cells} shows the rank cells
coincide a.s.\ eventually with these threshold cells, so no tie-breaking convention
enters the asymptotics; another choice of boundary order statistic changes the map by
$O(n^{-1/2})$.
\item \emph{Map.} $m_{k\ell}=\sum_{i\in C_{k\ell}}(y_{i}-\hat\pi_{i})\big/
\sqrt{\max\bigl(\sum_{i\in C_{k\ell}}\hat\pi_{i}(1-\hat\pi_{i}),\,\varepsilon\bigr)}$
with $\hat\pi_{i}=\pi_{\hat\beta_{n}}(x_{i})$ and $\varepsilon=10^{-8}$; empty cells are
set to $0$. Collect $m_{n}\in\R^{K^{2}}$.
\item \emph{Score.} $T_{n}=S\bigl((m_{n}-a)/s\bigr)$, where $(a,s)\in\R^{K^{2}}\times
\R_{>0}^{K^{2}}$ is the shipped scaler (division componentwise) and $S:\R^{K^{2}}\to\R$
is the shipped network: affine (convolution) layers, componentwise ReLU, max-pooling
over fixed finite windows, a global max/mean readout (serialized as a chain in
Remark~\ref{rem:deployed-recession}), and a linear head. All weights are frozen; the
analyst never trains.
\item \emph{Calibration.} For $b=1,\dots,B$ ($B=199$): draw $y^{*b}_{i}\sim
\mathrm{Bern}(\hat\pi_{i})$ independently \emph{at the observed covariates}, refit the
logistic model, and recompute steps 1--4 on $(x_{i},y^{*b}_{i})$ to get $T^{*b}_{n}$.
If the refit throws an error the released code sets $T^{*b}_{n}=+\infty$, so that
replicate counts \emph{against} rejection; this preserves the bound of
Lemma~\ref{lem:exact} (a $-\infty$ convention would count for rejection and deflate the
p-value). No separation policy is applied to the analyst's observed fit. The reported
p-value is
\begin{equation}\label{eq:pval}
p_{n}\;=\;\frac{1+\#\{b:\ T^{*b}_{n}\ge T_{n}\}}{B+1}.
\end{equation}
\end{enumerate}

Notation: $\hat H_{n}$ denotes the \emph{conditional} distribution function of a single
bootstrap score $T^{*}_{n}$ given the data (the infinite-$B$ object), and $\hat H^{B}_{n}$
the empirical distribution function of the $B$ drawn scores. Equation \eqref{eq:pval} is
the simple Monte Carlo test of \citet[\S 1]{besagclifford1989}; up to its discreteness it
is the Monte Carlo evaluation, through $\hat H^{B}_{n}$, of the first-stage prepivoted
root $1-\hat H_{n}(T_{n}^{-})$ of \citet{beran1988}.

\section{Assumptions}\label{sec:assumptions}

\begin{assumption}\label{a:sampling}
$(x_{i},y_{i})$ are i.i.d.; $0<\pi^{*}(x)<1$ $P$-a.s.; $\E\norm{\tox}^{2}<\infty$. The
marginal distribution functions $F_{a},F_{b}$ of the selected covariates are continuous
and strictly increasing in neighbourhoods of their $K-1$ population quantiles
$q^{a}_{k}=F_{a}^{-1}(k/K)$, $q^{b}_{\ell}=F_{b}^{-1}(\ell/K)$ (so those quantiles are
unique). With $q^{\cdot}_{0}=-\infty$ and $q^{\cdot}_{K}=+\infty$, the population cells
are $C^{\mathrm{pop}}_{k\ell}=\{x:\ x_{a}\in(q^{a}_{k-1},q^{a}_{k}],\
x_{b}\in(q^{b}_{\ell-1},q^{b}_{\ell}]\}$, and each has probability
$p_{k\ell}=P(x\in C^{\mathrm{pop}}_{k\ell})>0$. This is where the theory needs continuous
covariates: a selected covariate with an atom at one of these quantiles, a binary one for example,
is not covered.
\end{assumption}

\begin{assumption}\label{a:qmle}
(White regularity.) There is a unique $\beta^{*}$, interior to a compact parameter set
$\mathcal{K}$, maximizing
$\E\bigl[\pi^{*}(x)\log\pi_{\beta}(x)+(1-\pi^{*}(x))\log(1-\pi_{\beta}(x))\bigr]$, and
the QMLE is asymptotically linear:
\begin{gather*}
\sqrt n(\hat\beta_{n}-\beta^{*})
= A(\beta^{*})^{-1}\,n^{-1/2}\sum_{i}\tox_{i}\bigl(y_{i}-\Lambda(\tox_{i}^{\top}
\beta^{*})\bigr)+o_{p}(1),\\
A(\beta)=\E\bigl[\Lambda'(\tox^{\top}\beta)\,\tox\tox^{\top}\bigr]
\ \text{nonsingular}.
\end{gather*}
Under $H_{0}$, $\beta^{*}$ is the true parameter $\theta_{0}$. (Sufficient primitive
conditions: \citealp[Theorems~2.2 and~3.2]{white1982}.)
\end{assumption}

\begin{assumption}\label{a:axes}
The population criterion $c^{*}_{j}=|\beta^{*}_{j}|\,\sigma_{j}$ has a unique
top-two set $\{a,b\}$.
\end{assumption}

\begin{assumption}\label{a:visible}
(Map-visible misfit; used only for consistency.) With $\delta=\pi^{*}-\pi_{\beta^{*}}$,
the vector $\mu\in\R^{K^{2}}$ with entries
\[
\mu_{k\ell}
=\frac{\E\bigl[\delta(x)\,\ind{x\in C^{\mathrm{pop}}_{k\ell}}\bigr]}
{\sqrt{\E\bigl[\pi_{\beta^{*}}(x)\bigl(1-\pi_{\beta^{*}}(x)\bigr)\,
\ind{x\in C^{\mathrm{pop}}_{k\ell}}\bigr]}}
\]
is nonzero.
\end{assumption}

\begin{assumption}\label{a:certificate}
(Certificate; used only for consistency.) $\Sinf(v)>0$ for $v=\mathrm{diag}(1/s)\,\mu$,
where $\Sinf$ is the recession function of $S$ (Lemma~\ref{lem:recession}). By positive
homogeneity only the direction of $v$ matters.
\end{assumption}

\begin{assumption}\label{a:atomless}
(Used only for Proposition~\ref{prop:asyvalid} and Theorems~\ref{thm:localS},
\ref{thm:G} and \ref{thm:mild}.) Under $H_{0}$ the null limit law
$H(\cdot,\theta_{0})$ of the score (display \eqref{eq:nulllimit} below) has a continuous
distribution function.
\end{assumption}

\begin{assumption}\label{a:compact}
(Used only for Theorems~\ref{thm:looconsistent} and~\ref{thm:mild}.) $x$ has compact support $\mathcal K$, and the limiting risks satisfy
$\pi_{\beta^{*}}(x)\in[\epsilon,1-\epsilon]$ on $\mathcal K$ for some $\epsilon>0$.
\end{assumption}

\section{Two auxiliary lemmas}\label{sec:aux}

\begin{sres}{Lemma}{A.1}[Rank cells are threshold cells, eventually]\settag{A.1}\label{lem:cells}
Let $F$ be the marginal d.f.\ of $x_{a}$, $q_{k}=F^{-1}(k/K)$, $r_{k}=\lceil kn/K\rceil$,
and $\hat q^{a}_{k}=x_{a,(r_{k})}$. Under \ref{a:sampling} the events
$\{x_{a,(r_{k})}<x_{a,(r_{k}+1)},\ k=1,\dots,K-1\}$ and their $x_{b}$ analogues hold
a.s.\ for all large $n$, and on their intersection
\[
C_{k\ell}=\bigl\{i:\ \hat q^{a}_{k-1}<x_{ia}\le\hat q^{a}_{k},\
\hat q^{b}_{\ell-1}<x_{ib}\le\hat q^{b}_{\ell}\bigr\}
\]
irrespective of the tie-breaking rule, and $\hat q^{a}_{k}\to q^{a}_{k}$,
$\hat q^{b}_{\ell}\to q^{b}_{\ell}$ a.s.
\end{sres}

\begin{proof}
By \ref{a:sampling} pick an open $N_{k}\ni q_{k}$ on which $F$ is continuous and
strictly increasing. (a) $\Prob(\exists i\ne j:\ x_{ia}=x_{ja}\in N_{k})=0$: for a pair,
$\Prob(x_{2a}=x_{1a}\in N_{k})=\E[\{F(x_{1a})-F(x_{1a}-)\}\ind{x_{1a}\in N_{k}}]=0$
since $F$ has no atom in $N_{k}$; a union over the countably many pairs disposes of all
ties in $N_{k}$ at once. (b) For small $\varepsilon$ with
$[q_{k}-\varepsilon,q_{k}+\varepsilon]\subset N_{k}$, strict monotonicity gives
$F(q_{k}-\varepsilon)<k/K<F(q_{k}+\varepsilon)$, so by Glivenko--Cantelli and
$r_{k}/n\to k/K$ both $x_{a,(r_{k})}$ and $x_{a,(r_{k}+1)}$ lie in
$(q_{k}-\varepsilon,q_{k}+\varepsilon]$ a.s.\ eventually; with (a) they are then
distinct, and $\hat q^{a}_{k}\to q_{k}$ a.s.\ follows by letting $\varepsilon\downarrow0$.
Given $x_{a,(r_{k})}<x_{a,(r_{k}+1)}$ for every $k$, an observation has rank at most
$r_{k}$ iff its value is at most $\hat q^{a}_{k}$ --- whatever the tie-break, since a tied
block cannot straddle the boundary --- so the marginal bin of \S\ref{sec:setting},
step~2, is exactly $\{i:\hat q^{a}_{k-1}<x_{ia}\le\hat q^{a}_{k}\}$, and the product of
the two coordinatewise identities is the display.
\end{proof}

\begin{sres}{Lemma}{A.2}[Bootstrap QMLE: existence and consistency]\settag{A.2}\label{lem:bootqmle}
Let $\hat\beta_{n}\to_{p}\beta^{*}$ (as under \ref{a:qmle}), let
$y^{*}_{i}\sim\mathrm{Bern}(\hat\pi_{i})$ be conditionally independent given the data at
the fixed observed covariates, and let $\hat\beta^{*}_{n}$ maximize the bootstrap
log-likelihood $Q^{*}_{n}(\beta)=n^{-1}\sum_{i}\{y^{*}_{i}\log\pi_{\beta}(x_{i})
+(1-y^{*}_{i})\log(1-\pi_{\beta}(x_{i}))\}$. Under \ref{a:sampling},
$\hat\beta^{*}_{n}$ exists with probability tending to one and
$\hat\beta^{*}_{n}\to_{p}\beta^{*}$.
\end{sres}

\begin{proof}
Write $\bar Q_{n}(\beta)=\E[Q^{*}_{n}(\beta)\mid\text{data}]
=n^{-1}\sum_{i}\{\hat\pi_{i}\log\pi_{\beta}(x_{i})
+(1-\hat\pi_{i})\log(1-\pi_{\beta}(x_{i}))\}$. Because the logit of the logistic model
is linear, $\log\{\pi_{\beta}(x)/(1-\pi_{\beta}(x))\}=\tox^{\top}\beta$ identically, so
\[
Q^{*}_{n}(\beta)-\bar Q_{n}(\beta)=V_{n}^{\top}\beta,
\qquad V_{n}:=\frac1n\sum_{i}\bigl(y^{*}_{i}-\hat\pi_{i}\bigr)\tox_{i},
\]
an \emph{exact identity, linear in $\beta$}. Conditionally, $\E[V_{n}\mid\text{data}]=0$
and
$\E[\norm{V_{n}}^{2}\mid\text{data}]
=n^{-2}\sum_{i}\hat\pi_{i}(1-\hat\pi_{i})\norm{\tox_{i}}^{2}
\le(4n)^{-1}\,n^{-1}\sum_{i}\norm{\tox_{i}}^{2}=O_{p}(n^{-1})$
by \ref{a:sampling}, so $\norm{V_{n}}=O_{p}(n^{-1/2})$ and, on any compact $\mathcal{K}$,
$\sup_{\mathcal{K}}|Q^{*}_{n}-\bar Q_{n}|\le\norm{V_{n}}\sup_{\mathcal{K}}\norm{\beta}
=O_{p}(n^{-1/2})$: uniformity is inherited from linearity, no equicontinuity argument is
needed. Next, $\bar Q_{n}(\beta)$ differs from
$n^{-1}\sum_{i}\{\pi_{\beta^{*}}(x_{i})\log\pi_{\beta}(x_{i})
+(1-\pi_{\beta^{*}}(x_{i}))\log(1-\pi_{\beta}(x_{i}))\}$ by
$n^{-1}\sum_{i}(\hat\pi_{i}-\pi_{\beta^{*}}(x_{i}))\,\tox_{i}^{\top}\beta$ (the same
logit identity), bounded by
$\tfrac14\norm{\hat\beta_{n}-\beta^{*}}\,\sup_{\mathcal{K}}\norm{\beta}\cdot
n^{-1}\sum_{i}\norm{\tox_{i}}^{2}=o_{p}(1)$ uniformly on $\mathcal{K}$; and the class
$\{\pi_{\beta^{*}}\log\pi_{\beta}+(1-\pi_{\beta^{*}})\log(1-\pi_{\beta}):
\beta\in\mathcal{K}\}$ has integrable envelope
$\log2+\sup_{\mathcal{K}}\norm{\beta}\,\norm{\tox}$ and is Lipschitz in $\beta$, hence
Glivenko--Cantelli, so $\bar Q_{n}\to\bar Q$ uniformly on compacts, where
$\bar Q(\beta)=\E[\pi_{\beta^{*}}\log\pi_{\beta}+(1-\pi_{\beta^{*}})\log(1-\pi_{\beta})]$.

\emph{Uniqueness of the limit maximizer.} For each $x$ the map
$u\mapsto\pi_{\beta^{*}}(x)\log\Lambda(u)+(1-\pi_{\beta^{*}}(x))\log(1-\Lambda(u))$ is
strictly concave with unique maximum at $u=\tox^{\top}\beta^{*}$, so the maximizer set of
$\bar Q$ is $\{\beta:\tox^{\top}\beta=\tox^{\top}\beta^{*}\ P\text{-a.s.}\}$; if it
contained $\beta\ne\beta^{*}$, then $\pi_{\beta}=\pi_{\beta^{*}}$ a.s.\ and the
population criterion of \ref{a:qmle} would have two maximizers, contradicting its
uniqueness. Hence $\beta^{*}$ is the unique maximizer.

\emph{Conclusion by concavity.} $Q^{*}_{n}$ is concave in $\beta$ (logistic
log-likelihood), and by the two previous paragraphs it converges in (conditional)
probability, pointwise on a dense countable set, to the strictly-at-$\beta^{*}$-maximized
concave $\bar Q$. The convexity lemma of \citet{pollard1991}, applied to $-Q^{*}_{n}$,
upgrades pointwise convergence in probability of concave random functions on a dense set to
uniform convergence in probability on compact sets. Then for any $\epsilon>0$,
$\bar Q<\bar Q(\beta^{*})$ on the sphere $\{\norm{\beta-\beta^{*}}=\epsilon\}$ by the
uniqueness above and compactness, so with probability tending to one
$Q^{*}_{n}(\beta^{*})>\max_{\norm{\beta-\beta^{*}}=\epsilon}Q^{*}_{n}(\beta)$, and by
concavity $Q^{*}_{n}$ then attains its maximum inside the ball. This yields existence
w.p.\ $\to1$ and $\hat\beta^{*}_{n}\to_{p}\beta^{*}$ (the deterministic-limit case of the
argmax continuous-mapping theorem, \citealp[Theorem~2.7]{kimpollard1990}); no
compactness of the parameter space or well-separation condition is needed. (When a refit nonetheless fails at
finite $n$, the deployed convention of \S\ref{sec:setting}, step~5 applies.)
\end{proof}

\section{The recession function of the network}\label{sec:recession}

\begin{sres}{Lemma}{S.1}[Recession function; closed form and uniform bias bound]\settag{S.1}\label{lem:recession}
Let $\mathcal{N}$ be the class of functions $\R^{d_{0}}\to\R^{d_{L}}$ expressible as
finite compositions of (i) affine maps $z\mapsto Wz+c$, (ii) componentwise ReLU, and
(iii) componentwise maxima over fixed finite index sets. For $F\in\mathcal{N}$ define
$F_{0}$ as the same composition with every offset $c$ replaced by $0$. Then:
\begin{enumerate}
\item[(a)] $F_{0}$ is continuous, piecewise linear, and positively homogeneous:
$F_{0}(tu)=tF_{0}(u)$ for all $t>0$. ($F$ itself is piecewise \emph{affine}; only
continuity and homogeneity of $F_{0}$ are used below.)
\item[(b)] $\sup_{z}\norm{F(z)-F_{0}(z)}_{\infty}\le C_{F}<\infty$, where $C_{F}$ may be
taken as $\sum_{\text{affine layers }r}\norm{c_{r}}_{\infty}\prod_{r'>r}L_{r'}$ with
$L_{r'}$ the Lipschitz constant of layer $r'$ for $\norm{\cdot}_{\infty}$ --- the
induced $\ell_{\infty}$ operator norm $\norm{W_{r'}}_{\infty}$ of the linear part for an
affine layer, and $1$ for ReLU and max layers.
\item[(c)] For every $z_{0}$ and $u$,
$\lim_{t\to\infty}F(z_{0}+tu)/t=F_{0}(u)$; in particular the limit exists, is
independent of $z_{0}$, and equals the zero-offset network. We write
$F_{\infty}:=F_{0}$.
\end{enumerate}
\end{sres}

\begin{proof}
(a) Affine maps with zero offset are linear; ReLU and componentwise maxima are
continuous, piecewise linear, and positively homogeneous
($\max_{j}(t u_{j})=t\max_{j}u_{j}$ and $\mathrm{relu}(tu)=t\,\mathrm{relu}(u)$ for
$t>0$); all three properties are preserved under composition (for piecewise linearity,
by the standard common-refinement of the polyhedral pieces).

(b) Induction over the layers. Let $G$ denote the composition of the first $r$ layers
and $G_{0}$ its zero-offset counterpart, and suppose
$\sup_{z}\norm{G(z)-G_{0}(z)}_{\infty}\le C_{r}$; the base case (empty composition,
identity) gives $C_{0}=0$. If layer $r+1$ is affine with linear part $W$ and offset $c$,
\[
\norm{WG(z)+c-WG_{0}(z)}_{\infty}
\le \norm{W}_{\infty}\,\norm{G(z)-G_{0}(z)}_{\infty}+\norm{c}_{\infty}
\le L_{r+1}C_{r}+\norm{c_{r+1}}_{\infty}.
\]
If layer $r+1$ is ReLU or a componentwise maximum, it is $1$-Lipschitz for
$\norm{\cdot}_{\infty}$ (as $|\max_{j}g_{j}-\max_{j}h_{j}|\le\max_{j}|g_{j}-h_{j}|$),
so the bound propagates unchanged. Unrolling the recursion gives the stated $C_{F}$.

(c) By (b), $\bigl|F(z_{0}+tu)/t-F_{0}(z_{0}+tu)/t\bigr|\le C_{F}/t\to0$. By positive
homogeneity, $F_{0}(z_{0}+tu)/t=F_{0}(z_{0}/t+u)\to F_{0}(u)$ by continuity of $F_{0}$.
\end{proof}

\begin{remark}[The deployed statistic is in $\mathcal{N}$]\label{rem:deployed-recession}
The deployed statistic is $T=G(m)$ with $G(m)=S((m-a)/s)$; the scaler is an affine layer
and the recession function is $G_{\infty}(u)=\Sinf(u/s)$. Membership of the shipped $S$
in $\mathcal{N}$ requires one observation, because the architecture is not a chain at
the readout: it concatenates two branches read off the \emph{same} activation,
$e=[\mathrm{gmax}(32),\allowbreak\mathrm{gmean}(32)]$ (global max and mean over the
$32\times3\times3$ activation). The index sets in (iii) may be
arbitrary finite subsets --- in particular singletons, for which the maximum is the
identity coordinate --- so $\mathcal{N}$ is closed under such parallel branching:
write $e=P\circ A$ with $A(z)=(z,Mz)$ affine of type (i) (zero offset, $M$ the
per-channel averaging matrix) and $P$ a single type-(iii) layer whose first $32$ outputs
are maxima over each channel's nine spatial indices and whose last $32$ outputs are
maxima over singletons. Since $\norm{[I;M]}_{\infty}=1$ and $A$ has zero offset, the
constant $C_{F}$ of (b) is unchanged. Any directed acyclic arrangement of layers of
types (i)--(iii) serializes the same way, so the chain induction of
Lemma~\ref{lem:recession} covers the deployed network.

\emph{Check on the shipped weights} (\texttt{benchmark/theory/} in the deposit): on five
canonical directions the zero-offset forward pass matches the slope measured between
$t=2000$ and $4000$ to $3.5\times10^{-6}$, and the slope between $t=10^{6}$ and
$2\times10^{6}$, past every kink, to $1.3\times10^{-15}$.
\end{remark}

\section{Exactness and first-order validity}\label{sec:validity}

\begin{sres}{Lemma}{S.2}[Exactness in the pivotal idealization]\settag{S.2}\label{lem:exact}
Fix a covariate configuration $X=(x_{1},\dots,x_{n})$ and suppose there is a
distribution $L_{X}$, not depending on $\theta$, such that for every parameter $\theta$
of the null family the score $T$ computed from $(X,y)$ with
$y_{i}\sim\mathrm{Bern}(\pi_{\theta}(x_{i}))$ independent has law $L_{X}$. Then, under
$H_{0}$, for every $k\in\{1,\dots,B+1\}$,
\[
\Prob\Bigl(p_{n}\le \tfrac{k}{B+1}\ \Bigm|\ X\Bigr)\;\le\;\frac{k}{B+1},
\]
with equality for all $k$ when $L_{X}$ is atomless. In particular, since $(B+1)p_{n}$ is
integer-valued, the test rejecting when $p_{n}\le\alpha$ has conditional size at most
$\lfloor \alpha(B+1)\rfloor/(B+1)\le\alpha$, with size exactly
$\lfloor \alpha(B+1)\rfloor/(B+1)$ when $L_{X}$ is atomless. If the hypothesis holds for
$P$-a.e.\ configuration, taking expectation over the covariate law of \ref{a:sampling}
gives the same unconditional bounds (and the same equality when $L_{X}$ is atomless for
a.e.\ $X$).
\end{sres}

\begin{proof}
Work conditionally on $X$. Under $H_{0}$ the observed score satisfies $T_{n}\sim L_{X}$.
Conditionally on the data, the bootstrap scores $T^{*1},\dots,T^{*B}$ are i.i.d.\ with
the law of $T$ under parameter $\hat\theta_{n}$ at configuration $X$, which by
hypothesis is $L_{X}$ \emph{whatever} $\hat\theta_{n}$ is; hence their conditional law
is a fixed distribution not depending on $y$, so they are i.i.d.\ $L_{X}$ and
independent of $T_{n}$. Therefore
$(T^{(0)},T^{(1)},\dots,T^{(B)}):=(T_{n},T^{*1},\dots,T^{*B})$ is an i.i.d.\ ---
in particular exchangeable --- vector.

Let $U_{0},\dots,U_{B}$ be i.i.d.\ uniform on $(0,1)$, independent of everything; say
that $(t,u)$ \emph{outranks} $(t',u')$ iff $t>t'$, or $t=t'$ and $u>u'$. Let $\tilde R$
be one plus the number of indices $j\ge1$ whose pair outranks the observed pair, so that
\[
\tilde R\;=\;1+\#\{b:\ T^{*b}>T_{n}\}+\#\{b:\ T^{*b}=T_{n},\ U_{b}>U_{0}\}
\;\le\;1+\#\{b:\ T^{*b}\ge T_{n}\}\;=\;(B+1)\,p_{n}
\]
pointwise, the middle tie-count being over a subset of the full tie set. By
exchangeability of the $B+1$ i.i.d.\ pairs, $\tilde R$ is uniform on $\{1,\dots,B+1\}$
\citep[\S1]{besagclifford1989}, whence
$\Prob(p_{n}\le k/(B+1)\mid X)\le\Prob(\tilde R\le k)=k/(B+1)$. When $L_{X}$ is
atomless, ties have probability zero, $p_{n}=\tilde R/(B+1)$ a.s., and equality holds.
\end{proof}

\begin{remark}[What deployment-matched training is for]\label{rem:pivotality}
The hypothesis of Lemma~\ref{lem:exact} is pivotality. A fixed network cannot satisfy it
exactly, but it is what training against bootstrap-replicate nulls aims at, and its failure
is measured by the calibration gap $\Delta_{n}$ of Proposition~\ref{prop:fsbound}, of which
Lemma~\ref{lem:exact} is the $\Delta_{n}=0$ corner. The AUC diagnostic of the paper
($.513$ on average over the sixty cells of the level study) is a necessary check, not a
certificate (Remark~\ref{rem:auc}).
\end{remark}

\begin{sres}{Proposition}{S.3}[First-order validity of the deployed test]\settag{S.3}\label{prop:asyvalid}
Let $H_{0}$ hold with true parameter $\theta_{0}$ satisfying \ref{a:sampling},
\ref{a:qmle} (with $\beta^{*}=\theta_{0}$), \ref{a:axes} and \ref{a:atomless}. Then
$T_{n}\rightsquigarrow H(\cdot,\theta_{0})$ (display \eqref{eq:nulllimit}), and
\[
\sup_{t\in\R}\bigl|\hat H_{n}(t)-H(t,\theta_{0})\bigr|\;\xrightarrow{\ p\ }\;0 .
\]
Consequently:
\begin{enumerate}
\item[(i)] \emph{(the deployed test: $B$ fixed, here $B=199$.)} For every
$k\in\{1,\dots,B+1\}$, $\Prob\bigl(p_{n}\le k/(B+1)\bigr)\to k/(B+1)$; equivalently,
for every $\alpha\in(0,1)$,
$\Prob(p_{n}\le\alpha)\to\lfloor\alpha(B+1)\rfloor/(B+1)\le\alpha$. The test is
asymptotically exact at the $B+1$ attainable levels and asymptotically conservative
elsewhere, matching the finite-sample statement of Lemma~\ref{lem:exact}. The nominal
level $\alpha=.05$ is attainable ($k=10$ of $B+1=200$).
\item[(ii)] \emph{(idealized regime.)} If instead $B=B_{n}\to\infty$ jointly with $n$,
then $p_{n}\rightsquigarrow\mathrm{Unif}(0,1)$ and $\Prob(p_{n}\le\alpha)\to\alpha$ for
every $\alpha\in(0,1)$.
\end{enumerate}
Clause (i) is the one that covers the shipped test.
\end{sres}

\begin{proof}
Throughout, all statements hold on an event of probability tending to one and we
intersect finitely many such events without further comment; here and below, ``on an
event of probability tending to one'' always means the quantities in question coincide
with the stated surrogates on that event --- we never condition on such an event, so no
conditional law is altered.

\emph{Step 1: a joint limit for the map under the null.}
By \ref{a:qmle} and \ref{a:axes} the selected pair equals $(a,b)$ on such an event, and
by Lemma~\ref{lem:cells} the rank cells are the threshold cells of the empirical
quantiles. Consider on $\R^{d}\times\{0,1\}$ the classes
\begin{align*}
\mathcal{G}&=\bigl\{(x,y)\mapsto \ind{x_{a}\le s}\ind{x_{b}\le t}\,
\bigl(y-\Lambda(\tox^{\top}\beta)\bigr):\ s,t\in\bar\R,\ \beta\in
\mathcal{K}\bigr\},\\
\mathcal{W}&=\bigl\{(x,y)\mapsto \ind{x_{a}\le s}\ind{x_{b}\le t}\,w_{\beta}(x):\
s,t\in\bar\R,\ \beta\in\mathcal{K}\bigr\},
\qquad w_{\beta}=\Lambda(\tox^{\top}\beta)\{1-\Lambda(\tox^{\top}\beta)\},
\end{align*}
with $\mathcal{K}$ the compact set of \ref{a:qmle}, and the finite score class
$\mathcal{S}=\{(x,y)\mapsto\tox_{j}(y-\Lambda(\tox^{\top}\theta_{0})):j=0,\dots,d\}$
(envelope $\norm{\tox}$, square-integrable by \ref{a:sampling}). Both
$\mathcal{G}$ and $\mathcal{W}$ are uniformly bounded and we bound their
$L_{2}$-bracketing numbers factor by factor. \emph{(Orthant indicators, any law.)} For
every law $Q$ and $\epsilon\in(0,1)$,
$N_{[\,]}(\epsilon,\{\ind{u\le s\,}:s\in\bar\R\},L_{2}(Q))\le C\epsilon^{-2}$: use
degenerate brackets at the atoms of mass exceeding $\epsilon^{2}$ (at most
$\epsilon^{-2}$ of them) and, between them, brackets
$[\ind{u\le s_{j-1}},\ind{u\le s_{j}}]$ over a $Q$-quantile grid whose slabs carry mass
at most $\epsilon^{2}$. For $[0,1]$-valued factors with brackets in $[0,1]$, products
bracket with widths adding, so
$N_{[\,]}(\epsilon,\{\ind{x_{a}\le s}\ind{x_{b}\le t}\},L_{2}(Q))=O(\epsilon^{-4})$
\emph{for every} $Q$. \emph{(Parametric factor.)} On $\mathcal{K}$,
$\beta\mapsto\Lambda(\tox^{\top}\beta)$ and $\beta\mapsto w_{\beta}$ are Lipschitz with
constant $\le\tfrac14\norm{\tox}\in L_{2}(P)$, so
$N_{[\,]}(\epsilon,\cdot,L_{2}(P))=O(\epsilon^{-(d+1)})$
\citep[Example~19.7]{vaart1998}. \emph{(Assembly.)} $\mathcal{W}$ is a product of
nonnegative bounded factors; $\mathcal{G}$ is a difference of two such products
($\ind{\cdot}y$ and $\ind{\cdot}\Lambda$), and brackets of a difference have adding
widths, so no signed factor is ever multiplied. Hence
$\log N_{[\,]}(\epsilon,\mathcal{G}\cup\mathcal{W},L_{2}(P))\lesssim\log(1/\epsilon)$,
the bracketing integral converges, and
$\mathcal{F}=\mathcal{G}\cup\mathcal{W}\cup\mathcal{S}$ (a finite union of Donsker
classes) is $P$-Donsker, a fortiori Glivenko--Cantelli
\citep[Theorem~19.5]{vaart1998}; $\mathbb{G}_{n}=\sqrt n(P_{n}-P)$ converges weakly in
$\ell^{\infty}(\mathcal{F})$ to a tight Gaussian process $\mathbb{G}$.

Write $\psi(s,t,\beta)=P\,g_{s,t,\beta}$ for $g\in\mathcal{G}$. Under $H_{0}$ the
integrand is a conditionally centred residual, so
\[
\psi(s,t,\theta_{0})
=\E\bigl[\ind{x_{a}\le s}\ind{x_{b}\le t}\bigl(\E[y\mid x]
-\Lambda(\tox^{\top}\theta_{0})\bigr)\bigr]=0
\qquad\text{for every }(s,t):
\]
\emph{the population functional does not move with the cell boundaries at the null
parameter.} No expansion in $(s,t)$ is required, hence no Bahadur representation and no
marginal density; all we use of the quantiles is $\hat q\to q$ a.s.\
(Lemma~\ref{lem:cells}). Every cell sum is a $\pm$-combination of at most four corner
functions by inclusion--exclusion (legitimate by Lemma~\ref{lem:cells}), and
\[
n^{-1/2}\!\!\sum_{i\in C_{k\ell}}\!(y_{i}-\hat\pi_{i})
=\mathbb{G}_{n}\,g^{(4\text{ corners})}_{\hat q,\hat\beta_{n}}
+\sqrt n\,\bigl\{\psi(\hat q,\hat\beta_{n})-\psi(\hat q,\theta_{0})\bigr\},
\]
using $\psi(\hat q,\theta_{0})=0$. For the first term,
$(s,t,\beta)\mapsto g_{s,t,\beta}$ is $L_{2}(P)$-continuous at $(q,q,\theta_{0})$
(indicator increments carry mass
$\le|F_{a}(s)-F_{a}(q)|+|F_{b}(t)-F_{b}(q)|\to0$ by continuity of the marginals near
the quantiles, and $\Lambda$ is Lipschitz), so asymptotic equicontinuity of the Donsker
class \citep[Lemma~19.24]{vaart1998} gives
$\mathbb{G}_{n}g_{\hat q,\hat\beta_{n}}=\mathbb{G}_{n}g_{q,\theta_{0}}+o_{p}(1)$. For
the second, the integral mean-value form gives
\[
\sqrt n\,\bigl\{\psi(\hat q,\hat\beta_{n})-\psi(\hat q,\theta_{0})\bigr\}
=\Bigl\{\int_{0}^{1}\partial_{\beta}\psi\bigl(\hat q,\theta_{0}
+u(\hat\beta_{n}-\theta_{0})\bigr)\,du\Bigr\}^{\!\top}\sqrt n(\hat\beta_{n}-\theta_{0}),
\]
where $\partial_{\beta}\psi(s,t,\beta)
=-\E[\ind{x_{a}\le s}\ind{x_{b}\le t}\,\Lambda'(\tox^{\top}\beta)\,\tox]$ is jointly
continuous in $(s,t,\beta)$ near $(q,q,\theta_{0})$ by dominated convergence
($|\Lambda'|\le\tfrac14$, $\E\norm{\tox}<\infty$, marginals atomless near $q$). Hence
the term equals $-A_{k\ell}^{\top}\sqrt n(\hat\beta_{n}-\theta_{0})+o_{p}(1)$ with
$A_{k\ell}=\E[\ind{C^{\mathrm{pop}}_{k\ell}}\Lambda'(\tox^{\top}\theta_{0})\tox]$, and
by the linearity in \ref{a:qmle},
$\sqrt n(\hat\beta_{n}-\theta_{0})=A(\theta_{0})^{-1}\mathbb{G}_{n}[\mathcal{S}]
+o_{p}(1)$. Each scaled numerator is therefore a fixed continuous linear functional of
the single Gaussian limit $\mathbb{G}$ on $\mathcal{F}$; the denominators divided by
$n$ converge in probability to
$w_{k\ell}=P\,\ind{C^{\mathrm{pop}}_{k\ell}}w_{\theta_{0}}>0$ (Glivenko--Cantelli on
$\mathcal{W}$ plus continuity, as in Theorem~\ref{thm:consistency}, Step D). By the
continuous mapping theorem the map converges jointly,
\begin{equation}\label{eq:nulllimit}
m_{n}\ \rightsquigarrow\ Z(\theta_{0})\sim N\bigl(0,\Sigma(\theta_{0})\bigr),
\qquad
T_{n}\ \rightsquigarrow\ H(\cdot,\theta_{0})
:=\mathcal{L}\Bigl(S\bigl((Z(\theta_{0})-a)/s\bigr)\Bigr),
\end{equation}
where $\Sigma(\theta)$ is the covariance of the influence functionals
$w_{k\ell}^{-1/2}\{\ind{C^{\mathrm{pop}}_{k\ell}}
-A_{k\ell}^{\top}A(\theta)^{-1}\tox\}\,(y-\pi_{\theta})$, continuous in $\theta$ by
dominated convergence. (The estimation of $\beta$ \emph{reduces and correlates} the
cell variances --- the Chernoff--Lehmann effect; the estimated boundaries do not affect
the first-order limit, since the null drift $\psi(\cdot,\cdot,\theta_{0})$ vanishes
identically --- the classical random-cells phenomenon, cf.\
\citealp[Theorem~4.1]{moorespruill1975}. None of what follows requires $\Sigma$ to be
the identity.)

\emph{Step 2: the bootstrap estimates the limit.}
Conditionally on the data, the bootstrap responses are \emph{independent but not
identically distributed}: $y^{*}_{i}\sim\mathrm{Bern}(\hat\pi_{i})$ at the fixed
observed covariates --- a fixed-design parametric bootstrap; no covariates are
resampled. Write $\Probs=\Prob(\cdot\mid\text{data})$. (i) The bootstrap axis selection
uses the unchanged $\hat\sigma_{j}$ and the bootstrap QMLE, which is consistent for
$\theta_{0}$ by Lemma~\ref{lem:bootqmle} (its proof used only \ref{a:sampling} and
$\hat\beta_{n}\to_{p}\theta_{0}$); with \ref{a:axes} the bootstrap pair equals $(a,b)$
with $\Probs$-probability $\to1$ in probability. (ii) On that event the bootstrap rank
cells \emph{are} the original index sets $C_{k\ell}$ --- functions of the covariates
alone --- so there is no cell-boundary randomness. (iii) The logistic Hessian is free
of $y$, so a Taylor expansion of the bootstrap score about $\hat\beta_{n}$ gives the
conditional linearization
$\sqrt n(\hat\beta^{*}_{n}-\hat\beta_{n})
=\hat I_{n}^{-1}n^{-1/2}\sum_{i}\tox_{i}(y^{*}_{i}-\hat\pi_{i})+o_{p^{*}}(1)$ with
$\hat I_{n}=n^{-1}\sum_{i}\hat w_{i}\tox_{i}\tox_{i}^{\top}\to_{p}A(\theta_{0})$; the
remainder is controlled by
$|w_{\beta}(x)-w_{\beta'}(x)|\le\min\{c\,\norm{\tox}\norm{\beta-\beta'},\tfrac14\}$ and
truncation under $\E\norm{\tox}^{2}<\infty$. (iv) Apply the Lindeberg--Feller central
limit theorem (Cram\'er--Wold over the $K^{2}$ cells) to
$n^{-1/2}\sum_{i}c_{n,i}(y^{*}_{i}-\hat\pi_{i})$ with
$c_{n,i}=\ind{i\in C_{k\ell}}-A_{k\ell,n}^{\top}\hat I_{n}^{-1}\tox_{i}$: the summands
are bounded by $C(1+\norm{\tox_{i}})$ and Lindeberg holds since
$n^{-1}\sum_{i}\norm{\tox_{i}}^{2}\ind{\norm{\tox_{i}}>\epsilon\sqrt n}\to0$ a.s.\
under $\E\norm{\tox}^{2}<\infty$; the conditional covariance is the sample analogue of
the Step-1 influence covariance and converges to $\Sigma(\theta_{0})$ in probability
(law of large numbers with uniform integrability from the same second moment, plus
$\hat\beta_{n}\to_{p}\theta_{0}$). (v) Hence
$\sup_{f\in\mathrm{BL}_{1}}|\Es f(m^{*}_{n})-\E f(Z(\theta_{0}))|\to_{p}0$; since
$G=S((\cdot-a)/s)$ is globally Lipschitz (Lemma~\ref{lem:recession}(b) plus piecewise
affineness, or directly: each layer is Lipschitz), $f\circ G\in c\,\mathrm{BL}_{1}$
whenever $f\in\mathrm{BL}_{1}$, so the conditional law of $T^{*}_{n}$ converges weakly,
in probability, to $H(\cdot,\theta_{0})$. \ref{a:atomless} and Polya's theorem upgrade
this to $\sup_{t}|\hat H_{n}(t)-H(t,\theta_{0})|\to_{p}0$.

\emph{Step 3: the p-value.}
Write $H:=H(\cdot,\theta_{0})$ and $V_{n}:=1-\hat H_{n}(T_{n}^{-})$. Conditionally on
the data, $T^{*1}_{n},\dots,T^{*B}_{n}$ are i.i.d.\ with distribution function
$\hat H_{n}$ while $T_{n}$ is data-measurable, so
\[
\#\{b:\ T^{*b}_{n}\ge T_{n}\}\ \bigm|\ \text{data}\ \sim\ \mathrm{Bin}(B,V_{n}).
\]
By Step~2, $\sup_{t}|\hat H_{n}(t)-H(t)|\to_{p}0$, so
$V_{n}=1-H(T_{n}^{-})+o_{p}(1)=1-H(T_{n})+o_{p}(1)$ (\ref{a:atomless} makes $H$
continuous); and $T_{n}\rightsquigarrow H$ with $H$ continuous gives, by the
probability integral transform, $V_{n}\rightsquigarrow U\sim\mathrm{Unif}(0,1)$.

\emph{Fixed $B$ (clause (i)).} For $k\in\{1,\dots,B+1\}$,
$\{p_{n}\le k/(B+1)\}=\{\#\{b:T^{*b}_{n}\ge T_{n}\}\le k-1\}$ by \eqref{eq:pval}, so
\[
\Prob\bigl(p_{n}\le k/(B+1)\bigr)=\E\bigl[\Psi_{k}(V_{n})\bigr],
\qquad
\Psi_{k}(v)=\sum_{j=0}^{k-1}\binom{B}{j}v^{j}(1-v)^{B-j},
\]
and $\Psi_{k}$ is a polynomial, hence bounded and continuous, so
\[
\E[\Psi_{k}(V_{n})]\ \to\ \E[\Psi_{k}(U)]
=\sum_{j=0}^{k-1}\int_{0}^{1}\binom{B}{j}v^{j}(1-v)^{B-j}\,dv
=\sum_{j=0}^{k-1}\frac{1}{B+1}=\frac{k}{B+1}.
\]
The $\alpha$-form follows because $p_{n}$ is
supported on $\{1/(B+1),\dots,1\}$. This is the same attainable-level grid as in
Lemma~\ref{lem:exact}, but obtained \emph{without} pivotality; Lemma~\ref{lem:exact}'s
hypothesis is not assumed here and its finite-$n$ conclusion is not used.

\emph{$B=B_{n}\to\infty$ (clause (ii)).} Applied conditionally, the
Dvoretzky--Kiefer--Wolfowitz inequality gives
$\sup_{t}|\hat H^{B}_{n}(t)-\hat H_{n}(t)|=O_{p}(B^{-1/2})$, so
\[
p_{n}=V_{n}+O_{p}(B^{-1/2})+O(B^{-1})\rightsquigarrow\mathrm{Unif}(0,1),
\]
whence
$\Prob(p_{n}\le\alpha)\to\alpha$ for every $\alpha\in(0,1)$.
\end{proof}

\begin{remark}[The all-pairs reading]\label{rem:allpairs}
The software also offers the statistic
$T^{\max}_{n}=\max_{j<\ell}S\bigl((m^{(j\ell)}_{n}-a)/s\bigr)$, the largest score over the maps
$m^{(j\ell)}_{n}$ of every pair of covariates, with the same maximum taken in every bootstrap
replicate. Proposition~\ref{prop:asyvalid} holds for it without \ref{a:axes}, when \ref{a:sampling}
holds for every covariate and \ref{a:atomless} for the null limit law of every pair. \ref{a:axes}
enters the proof only through the selection of the pair, in Step~1 and Step~2(i), and the maximum
selects nothing. In Step~1 the classes $\mathcal{G}$ and $\mathcal{W}$ indexed by all pairs form a
finite union of Donsker classes, so the maps of all pairs converge jointly to a Gaussian vector
$(Z^{(j\ell)})$, each component as in \eqref{eq:nulllimit}; the maximum of finitely many continuous
functions of that vector is continuous, so
$T^{\max}_{n}\rightsquigarrow H^{\max}(\cdot,\theta_{0})
:=\mathcal{L}\bigl(\max_{j<\ell}S((Z^{(j\ell)}-a)/s)\bigr)$. In Step~2 the bootstrap cells of every
pair are the original index sets, and the conditional central limit theorem applies to the stacked
vector unchanged. $H^{\max}$ is continuous whenever each pair's null limit law is, because
$\Prob(\max_{q}T_{q}=t)\le\sum_{q}\Prob(T_{q}=t)=0$ for every $t$, so Step~3 goes through as
written. With a single covariate the map is $K^{2}$ quantile slabs of that covariate, and the
proof applies with the orthant classes replaced by half-lines.
\end{remark}

\section{A finite-sample size bound}\label{sec:fsbound}

Lemma~\ref{lem:exact} is exact but idealized; Proposition~\ref{prop:asyvalid} covers the
deployed test but is an $o(1)$ statement. The result of this section lies between
them. It holds at every $n$, $B$ and $k$, assumes nothing beyond $H_{0}$ (in particular
nothing about the network), and collects everything unknown into one scalar, a one-sided
Kolmogorov distance between two laws of the \emph{score}.

Throughout the section fix the covariate configuration $X=(x_{1},\dots,x_{n})$ and
work conditionally on it. For a parameter $\theta$ of the null family let $L_{X,\theta}$
be the law of the score produced by steps 1--4 of \S\ref{sec:setting} from $(X,y)$ with
$y_{i}\sim\mathrm{Bern}(\pi_{\theta}(x_{i}))$ independent, and let $F_{X,\theta}$ be its
distribution function. Two features of the deployed algorithm are used, and nothing
else. First, given the data the replicates $T^{*1}_{n},\dots,T^{*B}_{n}$ are i.i.d.\
$L_{X,\hat\beta_{n}}$ \emph{exactly}: the released code draws
$y^{*b}_{i}\sim\mathrm{Bern}(\pi_{\hat\beta_{n}}(x_{i}))$ at the observed covariates,
refits, and applies the same function $T(X,\cdot)$, re-selecting the axes from
$y^{*b}$, so that $\hat H_{n}=F_{X,\hat\beta_{n}}$. It is this self-consistency of the
pipeline that makes $L_{X,\hat\beta_{n}}$ the right reference law. Second, under $H_{0}$
the observed responses are $y_{i}\sim\mathrm{Bern}(\pi_{\theta_{0}}(x_{i}))$
independent, so $T_{n}\sim L_{X,\theta_{0}}$.

Write, as in Step 3 of the proof above,
\[
V_{n}\;:=\;1-\hat H_{n}(T_{n}^{-})\;=\;\Prob\bigl(T^{*}_{n}\ge T_{n}\bigm|\text{data}
\bigr),
\qquad
V^{0}_{n}\;:=\;1-F_{X,\theta_{0}}(T_{n}^{-}) .
\]
$V_{n}$ is the infinite-$B$ bootstrap p-value, the first-stage prepivoted root of
\citet{beran1988}; $V^{0}_{n}$ is the tail p-value the analyst would report if
$\theta_{0}$ were known. The \emph{calibration gap} and its one-sided and tail-localized
versions are
\begin{align}
\Delta_{n}&:=\sup_{t\in\R}\bigl|F_{X,\hat\beta_{n}}(t)-F_{X,\theta_{0}}(t)\bigr|,
\qquad
\Delta^{+}_{n}:=\sup_{t\in\R}\bigl(F_{X,\hat\beta_{n}}(t)-F_{X,\theta_{0}}(t)\bigr)_{+}
\ \le\ \Delta_{n},\label{eq:gap}\\
\Delta^{+}_{n,\tau}&:=\sup\bigl\{\bigl(F_{X,\hat\beta_{n}}(t)-F_{X,\theta_{0}}(t)
\bigr)_{+}:\ 1-F_{X,\theta_{0}}(t^{-})\le\tau\bigr\}\ \le\ \Delta^{+}_{n},
\qquad \tau\in(0,1).\label{eq:tailgap}
\end{align}
All three are random variables, through $\hat\beta_{n}$; all three are distances between
two laws of the scalar score, not between two laws of the map; and $\Delta^{+}_{n}$
retains only the direction in which the bootstrap reference is stochastically
\emph{smaller} than the truth, which is the only direction that can inflate the size.
The quantity $\Delta^{+}_{n,\tau}$ restricts the comparison further, to the upper
$\tau$-fraction of the true null law.

Three elementary facts do all the work. In all of them
\begin{equation}\label{eq:psik}
\Psi_{k}(v)\;:=\;\Prob\bigl(\mathrm{Bin}(B,v)\le k-1\bigr)
\;=\;\sum_{j=0}^{k-1}\binom{B}{j}v^{j}(1-v)^{B-j},
\qquad k\in\{1,\dots,B+1\},
\end{equation}
the polynomial already used in Step 3 of the proof of Proposition~\ref{prop:asyvalid}.

\begin{sres}{Lemma}{A.3}[Exact conditional binomial representation]\settag{A.3}\label{lem:binom}
For every $n$, every $B\ge1$ and every $k\in\{1,\dots,B+1\}$: conditionally on the data,
$\#\{b:T^{*b}_{n}\ge T_{n}\}\sim\mathrm{Bin}(B,V_{n})$; the events
$\{p_{n}\le k/(B+1)\}$ and $\{\#\{b:T^{*b}_{n}\ge T_{n}\}\le k-1\}$ coincide; and
\[
\Prob\Bigl(p_{n}\le\tfrac{k}{B+1}\Bigm|X\Bigr)\;=\;\E\bigl[\Psi_{k}(V_{n})\bigm|X\bigr].
\]
Moreover $\Psi_{k}$ is continuous and nonincreasing on $[0,1]$ with values in $[0,1]$,
$\Psi_{k}(0)=1$, and $\int_{0}^{1}\Psi_{k}(v)\,dv=k/(B+1)$.
\end{sres}

\begin{proof}
$\hat\beta_{n}$ and $T_{n}$ are data-measurable and the $T^{*b}_{n}$ are conditionally
i.i.d.\ $L_{X,\hat\beta_{n}}$, so each $\ind{T^{*b}_{n}\ge T_{n}}$ is conditionally
Bernoulli with success probability
$\Prob(T^{*}_{n}\ge T_{n}\mid\text{data})=1-\hat H_{n}(T_{n}^{-})=V_{n}$, and the
$B$ indicators are conditionally independent. The set identity is
$(1+N)/(B+1)\le k/(B+1)\iff N\le k-1$ with $N=\#\{b:T^{*b}_{n}\ge T_{n}\}$; taking
conditional expectations and then the tower property over the data given $X$ gives the
display. $\Psi_{k}$ is a polynomial, hence continuous; it is nonincreasing because
$\mathrm{Bin}(B,v)$ is stochastically nondecreasing in $v$ (couple through
$\sum_{b}\ind{U_{b}\le v}$ with $U_{b}$ i.i.d.\ uniform); $\Psi_{k}(0)=1$ since
$k\ge1$; and
$\int_{0}^{1}\binom{B}{j}v^{j}(1-v)^{B-j}\,dv
=\binom{B}{j}\,j!\,(B-j)!/(B+1)!=1/(B+1)$ for each $j$, so the integral is $k/(B+1)$.
\end{proof}

\begin{sres}{Lemma}{A.4}[Super-uniformity of the ideal tail p-value; atoms allowed]
\settag{A.4}\label{lem:superunif}
Under $H_{0}$, $\Prob(V^{0}_{n}\le u\mid X)\le u$ for every $u\in[0,1]$, with equality
for every $u$ when $L_{X,\theta_{0}}$ is atomless.
\end{sres}

\begin{proof}
Let $T\sim L_{X,\theta_{0}}$ and $G(t):=\Prob(T\ge t)=1-F_{X,\theta_{0}}(t^{-})$, which
is nonincreasing, so $V^{0}_{n}=G(T_{n})$ and, under $H_{0}$, $T_{n}\sim L_{X,\theta_{0}}$.
Fix $u\in[0,1]$ and put $A=\{t:G(t)\le u\}$. Since $G$ is nonincreasing, $A$ is an
up-set, hence $A=[t_{u},\infty)$ or $A=(t_{u},\infty)$ with $t_{u}=\inf A$ (the cases
$A=\emptyset$ and $A=\R$ are trivial). If $A=[t_{u},\infty)$ then
$\Prob(T\in A)=\Prob(T\ge t_{u})=G(t_{u})\le u$, because $t_{u}\in A$. If
$A=(t_{u},\infty)$ then
$\Prob(T\in A)=\Prob(T>t_{u})=\lim_{t\downarrow t_{u}}G(t)\le u$, because $G(t)\le u$
for every $t>t_{u}$. As $\{V^{0}_{n}\le u\}=\{T_{n}\in A\}$, the inequality follows. If
$L_{X,\theta_{0}}$ is atomless then $G(T_{n})=1-F_{X,\theta_{0}}(T_{n})$ and the
probability integral transform gives $V^{0}_{n}\sim\mathrm{Unif}(0,1)$.
\end{proof}

Lemma~\ref{lem:superunif} is where, and the only place where, $H_{0}$ is used.

\begin{sres}{Lemma}{A.5}[Shift lemma]\settag{A.5}\label{lem:shift}
Let $W$ be a $[0,1]$-valued random variable with $\Prob(W\le u)\le u$ for every
$u\in[0,1]$, let $\varepsilon\ge0$, and let $\psi:[0,1]\to[0,1]$ be nonincreasing. Then
\[
\E\bigl[\psi\bigl((W-\varepsilon)_{+}\bigr)\bigr]\ \le\
\varepsilon\,\psi(0)+\int_{0}^{1}\psi(v)\,dv .
\]
If moreover $W\sim\mathrm{Unif}(0,1)$, then
$\E\bigl[\psi\bigl(\min(W+\varepsilon,1)\bigr)\bigr]\ \ge\
\int_{0}^{1}\psi(v)\,dv-\varepsilon\,\psi(0)$.
\end{sres}

\begin{proof}
Put $Y=(W-\varepsilon)_{+}$ and $Y'=(U-\varepsilon)_{+}$ with $U\sim\mathrm{Unif}(0,1)$.
For $w\ge0$, $\Prob(Y\le w)=\Prob(W\le w+\varepsilon)\le\min(w+\varepsilon,1)
=\Prob(Y'\le w)$, and both are $0$ for $w<0$; hence $Y$ stochastically dominates $Y'$.
A nonincreasing $\psi$ is Borel and here bounded, so $\E[\psi(Y)]\le\E[\psi(Y')]$. If
$\varepsilon\ge1$ then $Y'\equiv0$ and
$\E[\psi(Y')]=\psi(0)\le\varepsilon\psi(0)$. If $\varepsilon<1$ then
\[
\E[\psi(Y')]=\int_{0}^{\varepsilon}\psi(0)\,du+\int_{\varepsilon}^{1}
\psi(u-\varepsilon)\,du
=\varepsilon\,\psi(0)+\int_{0}^{1-\varepsilon}\psi(v)\,dv
\ \le\ \varepsilon\,\psi(0)+\int_{0}^{1}\psi(v)\,dv,
\]
using $\psi\ge0$. For the lower bound with $U\sim\mathrm{Unif}(0,1)$ and
$\varepsilon<1$,
$\E[\psi(\min(U+\varepsilon,1))]=\int_{\varepsilon}^{1}\psi(v)\,dv
+\varepsilon\,\psi(1)\ge\int_{0}^{1}\psi-\int_{0}^{\varepsilon}\psi
\ge\int_{0}^{1}\psi-\varepsilon\,\psi(0)$, since $\psi\le\psi(0)$ on $[0,\varepsilon]$;
for $\varepsilon\ge1$ the left side is $\psi(1)\ge0$ while the right side is at most
$\int_{0}^{1}\psi-\psi(0)\le0$.
\end{proof}

Only monotonicity of $\psi$ is used. A Lipschitz argument would carry the maximum of the
$\mathrm{Beta}(k,B+1-k)$ density $|\Psi_{k}'|$, which grows like $\sqrt{B}$ ($26.8$ at $B=199$,
$k=10$), and a bound in the mean with an absolute constant is false: with
$V^{0}_{n}\sim\mathrm{Unif}(0,1)$ and gap $\varepsilon'\ind{V^{0}_{n}\in[a,a+\varepsilon']}$,
$\E[\Delta_{n}]=\varepsilon'^{2}$ while the size rises by order $\sqrt{B}\,\varepsilon'^{2}$. The
bound below therefore uses an upper \emph{quantile} of the gap, not its mean.

\begin{sres}{Proposition}{S.7}[Finite-sample size bound with a measurable calibration gap]
\settag{S.7}\label{prop:fsbound}
Assume only that $H_{0}$ holds, that is, $\pi^{*}=\pi_{\theta_{0}}$ for some
$\theta_{0}$. Fix the covariate configuration $X$, let $\varepsilon\ge0$ be any
constant (unrelated to the numerical floor of \S\ref{sec:setting}, step~3) and put
$\eta:=\Prob(\Delta^{+}_{n}>\varepsilon\mid X)$ with $\Delta^{+}_{n}$ as in
\eqref{eq:gap}. Then for every $n$, every $B\ge1$ and every $k\in\{1,\dots,B+1\}$,
\begin{equation}\label{eq:fsbound}
\Prob\Bigl(p_{n}\le\frac{k}{B+1}\Bigm|X\Bigr)\ \le\ \frac{k}{B+1}+\varepsilon+\eta ,
\end{equation}
and consequently, for every $\alpha\in(0,1)$,
\begin{equation}\label{eq:fsboundalpha}
\Prob\bigl(p_{n}\le\alpha\bigm|X\bigr)\ \le\
\frac{\lfloor\alpha(B+1)\rfloor}{B+1}+\varepsilon+\eta\ \le\ \alpha+\varepsilon+\eta .
\end{equation}
If $\varepsilon(X)$ is any measurable function of the configuration and
$\eta(X)=\Prob(\Delta^{+}_{n}>\varepsilon(X)\mid X)$, then averaging over the covariate
law gives $\Prob(p_{n}\le\alpha)\le\alpha+\E_{X}[\varepsilon(X)]+\E_{X}[\eta(X)]$.
\end{sres}

\begin{proof}
Work conditionally on $X$ and let $A=\{\Delta^{+}_{n}\le\varepsilon\}$, so
$\Prob(A^{c}\mid X)=\eta$. Since
$V^{0}_{n}-V_{n}=F_{X,\hat\beta_{n}}(T_{n}^{-})-F_{X,\theta_{0}}(T_{n}^{-})$ and each
left limit of $F_{X,\hat\beta_{n}}-F_{X,\theta_{0}}$ is a limit of values of that
difference, we have $V^{0}_{n}-V_{n}\le\Delta^{+}_{n}$ pointwise, hence on $A$
\[
V_{n}\ \ge\ V^{0}_{n}-\varepsilon,
\qquad\text{and therefore}\qquad
V_{n}\ \ge\ \bigl(V^{0}_{n}-\varepsilon\bigr)_{+}
\]
because $V_{n}\ge0$. As $\Psi_{k}$ is nonincreasing,
$\Psi_{k}(V_{n})\le\Psi_{k}((V^{0}_{n}-\varepsilon)_{+})$ on $A$, while
$0\le\Psi_{k}\le1$ everywhere. Lemma~\ref{lem:binom} then gives
\[
\Prob\Bigl(p_{n}\le\tfrac{k}{B+1}\Bigm|X\Bigr)=\E\bigl[\Psi_{k}(V_{n})\bigm|X\bigr]
\ \le\ \E\bigl[\Psi_{k}\bigl((V^{0}_{n}-\varepsilon)_{+}\bigr)\bigm|X\bigr]
+\Prob(A^{c}\mid X).
\]
By Lemma~\ref{lem:superunif} the variable $V^{0}_{n}$ satisfies the hypothesis of
Lemma~\ref{lem:shift}; applying that lemma with $\psi=\Psi_{k}$, and using
$\Psi_{k}(0)=1$ together with $\int_{0}^{1}\Psi_{k}=k/(B+1)$ from
Lemma~\ref{lem:binom}, bounds the first term by $k/(B+1)+\varepsilon$. This is
\eqref{eq:fsbound}. For \eqref{eq:fsboundalpha}, $p_{n}$ takes values in
$\{1/(B+1),\dots,1\}$, so $\{p_{n}\le\alpha\}
=\{p_{n}\le\lfloor\alpha(B+1)\rfloor/(B+1)\}$, which is empty when
$\lfloor\alpha(B+1)\rfloor=0$; and $\lfloor\alpha(B+1)\rfloor/(B+1)\le\alpha$. The final
sentence holds because, given $X$, the quantity $\varepsilon(X)$ is a constant, so
\eqref{eq:fsboundalpha} applies configuration by configuration and integrates linearly
over the covariate law.
\end{proof}

\begin{remark}[What the bound assumes]\label{rem:fsassumes}
Only $H_{0}$: none of A1--A6 and no condition on the network, $n$, $d$, $K$ or the event
rate. The estimator may be any measurable function of the data, since $\Delta^{+}_{n}$ is
defined from the realized $\hat\beta_{n}$; a poor estimator inflates the gap rather than
breaking the bound. The $+\infty$ convention of \S\ref{sec:setting}, step~5 leaves
\eqref{eq:fsbound} unchanged, because a stochastically larger reference law can only raise
$V_{n}$ and $\Psi_{k}$ is nonincreasing.
\end{remark}

\begin{sres}{Proposition}{S.8}[Tail-localized refinement]\settag{S.8}\label{prop:fstail}
Assume $H_{0}$. Fix $\tau\in(0,1)$, fix $\bar\Delta\in[0,\tau)$, let
$\varepsilon_{\tau}\ge0$ be any constant, and put
$\eta_{\tau}:=\Prob(\Delta^{+}_{n,\tau}>\varepsilon_{\tau}\mid X)$, with
$\Delta^{+}_{n,\tau}$ as in \eqref{eq:tailgap}. Then for every $n$, every $B\ge1$ and
every $k\in\{1,\dots,B+1\}$,
\begin{equation}\label{eq:fstail}
\Prob\Bigl(p_{n}\le\frac{k}{B+1}\Bigm|X\Bigr)\ \le\ \frac{k}{B+1}+\varepsilon_{\tau}
+\eta_{\tau}+\Psi_{k}(\tau-\bar\Delta)+\Prob\bigl(\Delta^{+}_{n}>\bar\Delta\bigm|X\bigr).
\end{equation}
If $\Delta^{+}_{n,\tau}\le\varepsilon_{\tau}$ holds on the whole event
$\{\Delta^{+}_{n}\le\bar\Delta\}$, the term $\eta_{\tau}$ may be dropped. If
$\varepsilon_{\tau}$ and $\bar\Delta$ do not depend on $X$, averaging over the covariate law
gives the same bound unconditionally, with $\eta_{\tau}$ and
$\Prob(\Delta^{+}_{n}>\bar\Delta\mid X)$ replaced by their averages.
\end{sres}

\begin{proof}
Let $E=\{\Delta^{+}_{n}\le\bar\Delta\}$ and $G=\{\Delta^{+}_{n,\tau}\le\varepsilon_{\tau}\}$,
so $\Prob(G^{c}\mid X)=\eta_{\tau}$, and split $\E[\Psi_{k}(V_{n})]$ over
$E\cap G\cap\{V^{0}_{n}\le\tau\}$, $E\cap G^{c}\cap\{V^{0}_{n}\le\tau\}$,
$E\cap\{V^{0}_{n}>\tau\}$ and $E^{c}$, using $0\le\Psi_{k}\le1$ on the second and the
last, which contribute at most $\eta_{\tau}$ and $\Prob(E^{c}\mid X)$. On
$\{V^{0}_{n}\le\tau\}$ the observed score lies in
$A_{\tau}=\{t:1-F_{X,\theta_{0}}(t^{-})\le\tau\}$, so the definition \eqref{eq:tailgap}
gives $V^{0}_{n}-V_{n}\le\Delta^{+}_{n,\tau}$, which is at most $\varepsilon_{\tau}$ on $G$.
On $E\cap G\cap\{V^{0}_{n}\le\tau\}$, therefore,
$\Psi_{k}(V_{n})\le\Psi_{k}((V^{0}_{n}-\varepsilon_{\tau})_{+})$ and, by
Lemmas~\ref{lem:superunif} and~\ref{lem:shift} exactly as in
Proposition~\ref{prop:fsbound},
$\E[\Psi_{k}(V_{n})\ind{E\cap G\cap\{V^{0}_{n}\le\tau\}}]\le k/(B+1)+\varepsilon_{\tau}$.
On $E\cap\{V^{0}_{n}>\tau\}$ we have $V_{n}\ge V^{0}_{n}-\bar\Delta>\tau-\bar\Delta$, so
$\Psi_{k}(V_{n})\le\Psi_{k}(\tau-\bar\Delta)$ by monotonicity. Adding the four
contributions gives \eqref{eq:fstail}. If $G^{c}\cap E$ is empty, the second piece is
empty and $\eta_{\tau}$ drops out. The unconditional form follows because the right-hand
side is linear in the two conditional probabilities and the other terms are constants.
\end{proof}

The term $\Psi_{k}(\tau-\bar\Delta)$ is negligible at the deployed configuration: with $B=199$,
$k=10$, $\tau=0.30$ and $\bar\Delta=0.05$, $\Psi_{10}(0.25)=9.1\times10^{-15}$. Apart from the
probability that the whole gap exceeds $\bar\Delta$, what matters in \eqref{eq:fstail} is the
discrepancy over the upper $30\%$ of the null law and the probability $\eta_{\tau}$ that it
exceeds $\varepsilon_{\tau}$; a reference law may be badly wrong in the body of the
distribution without harm.

\begin{remark}[The constant is one, and it cannot be improved]\label{rem:sharp}
No constant smaller than $1$ multiplies $\varepsilon$ in \eqref{eq:fsbound}, and none
larger is needed. Take $L_{X,\theta_{0}}$ uniform on $(0,1)$ and a reference law that is
the same law shifted left by $\varepsilon\in(0,1)$, so that
$F_{X,\hat\beta_{n}}(t)=\min(t+\varepsilon,1)$ on $[0,1]$ and hence
$\Delta_{n}=\Delta^{+}_{n}=\varepsilon$ with $\eta=0$. Then $V^{0}_{n}$ is uniform,
$V_{n}=(V^{0}_{n}-\varepsilon)_{+}$ identically, and by the computation in the proof of
Lemma~\ref{lem:shift},
\[
\Prob\Bigl(p_{n}\le\tfrac{k}{B+1}\Bigm|X\Bigr)=\E\bigl[\Psi_{k}((V^{0}_{n}
-\varepsilon)_{+})\bigr]
=\frac{k}{B+1}+\varepsilon-\int_{1-\varepsilon}^{1}\Psi_{k}(v)\,dv ,
\]
and the subtracted integral is at most $\varepsilon\,\Psi_{k}(1-\varepsilon)$, exponentially small
in $B$ when $k/(B+1)<1-\varepsilon$. So \eqref{eq:fsbound} is attained up to an exponentially small
deficit, with a constant free of $B$ and $k$. The example concerns the class of perturbations the
proposition covers; no fitted logistic model is claimed to realize it.
\end{remark}

\begin{remark}[The exact test that the gap is the price of]\label{rem:mmc}
The maximized Monte Carlo p-value of \citet{dufour2006},
$p^{\mathrm{MMC}}_{n}=\sup_{\theta\in\Theta_{0}}p_{n}(\theta)$ with $p_{n}(\theta)$ computed
from replicates drawn at $\theta$, has size at most $\lfloor\alpha(B+1)\rfloor/(B+1)$ at every
$n$, because the supremum dominates the exactly valid $p_{n}(\theta_{0})$. It needs a supremum
of a Monte Carlo objective over a $(d+1)$-dimensional surface, a fresh bootstrap at every
evaluated $\theta$ and a compact $\Theta_{0}$, and a search that misses the supremum loses the
exactness. The deployed test evaluates $p_{n}$ at $\hat\beta_{n}$ only, and $\varepsilon$ is
the price: restricting the supremum to a ball of radius $\delta$ about $\hat\beta_{n}$,
Proposition~\ref{prop:fsbound} applies with $\eta=\Prob(\norm{\hat\beta_{n}-\theta_{0}}>\delta\mid X)$
and $\varepsilon$ equal to the modulus of continuity
\begin{equation}\label{eq:modulus}
\omega_{n}(X,\delta)\ :=\ \sup_{\norm{\theta-\theta'}\le\delta}\ \sup_{t\in\R}
\bigl|F_{X,\theta}(t)-F_{X,\theta'}(t)\bigr| .
\end{equation}

\emph{Feasibility.} We built the exact variant with $\Theta_{0}$ replaced by a confidence set
$C_{\gamma}$ from Bernstein's inequality applied coordinatewise to the score, with a union
bound, so that its level is at most $\alpha+\gamma$ at every $n$. On design D1 of the level
study ($p=2$, independent covariates, $50\%$ events) at $n=100$, with $\gamma=.01$, $B=199$ and
$117$ null and $117$ alternative datasets paired with the deployed test on common random
numbers, it rejected $0$ of $117$ nulls at $.05$ and at $.04$. Its power at $.04$, the
threshold that guarantees level $.05$, was $.171$, against $.350$ for the deployed test
(paired loss $.18$, confidence interval $[.11,.25]$), at a median of $87$ times the deployed
test's refits (maximum $442$). The Bernstein set was numerically unbounded on $35\%$ of the
datasets, so it had to be capped, and the search reached its time budget on $15\%$, where the
reported value is a lower bound and exactness is void. This is one design at one sample size.
Scripts, per-dataset rows and report: \texttt{theory/exact\_mmc/}.
\end{remark}

\begin{remark}[What the AUC diagnostic measures, and what it does not]\label{rem:auc}
Let $P$ and $Q$ be laws on $\R$ --- here $P=L_{X,\theta_{0}}$, scores on true-null data,
and $Q=L_{X,\hat\beta_{n}}$, scores on bootstrap replicates. Write $A_{\mathrm{MW}}$ for
the Mann--Whitney area under the curve traced by threshold rules on the score itself,
$A_{\mathrm{MW}}=\Prob(T_{P}>T_{Q})+\tfrac12\Prob(T_{P}=T_{Q})$ with $T_{P}\sim P$ and
$T_{Q}\sim Q$ independent; write $A^{\star}$ for the area attained by the
likelihood-ratio rule, the largest attainable by any measurable rule; and write
$d_{\mathrm{K}}$ and $\mathrm{TV}$ for the Kolmogorov and total-variation distances
between $P$ and $Q$. Then
\begin{equation}\label{eq:aucchain}
\bigl|A_{\mathrm{MW}}-\tfrac12\bigr|\ \le\ d_{\mathrm{K}}(P,Q)\ \le\ \mathrm{TV}(P,Q)
\ \le\ 2\bigl(A^{\star}-\tfrac12\bigr),
\end{equation}
and any fixed rule $g$ has population area $A_{g}\le A^{\star}$.

\emph{Proof sketch.} With mid-distribution functions $\tilde F$, $A_{\mathrm{MW}}-\tfrac12
=\int(\tilde F_{Q}-\tilde F_{P})\,dP$, of modulus at most $d_{\mathrm{K}}$; the second
inequality is immediate; for the third, the concave likelihood-ratio curve lies above the
polygon through its point of largest vertical deviation $\mathrm{TV}$, whose area is
$\tfrac12(1+\mathrm{TV})$; the last claim is the Neyman--Pearson lemma. $\square$

So $A_{\mathrm{MW}}$ bounds $d_{\mathrm{K}}$ only from \emph{below}: $P=N(0,1)$ and
$Q=N(0,\sigma^{2})$ give $A_{\mathrm{MW}}=\tfrac12$ for every $\sigma$, while
$d_{\mathrm{K}}(P,Q)\to\tfrac12$ as $\sigma\to\infty$. And a fitted classifier's area is a
lower bound on $A^{\star}$, which bounds neither $\mathrm{TV}$ nor $d_{\mathrm{K}}$. The
diagnostic is therefore a necessary check; the quantity Proposition~\ref{prop:fsbound} needs
is $\Delta^{+}_{n}$ or $\Delta^{+}_{n,\tau}$.
\end{remark}

\subsection*{Estimating the gap, and a pilot measurement}\label{rem:estgap}

Unlike $A^{\star}$, the gap has a finite-sample upper confidence bound, because it is a
two-sample Kolmogorov--Smirnov distance between two sets of scalars. In a design where
$\theta_{0}$ is known, as in every cell of the level study, the protocol for one replicate
dataset is:
\begin{enumerate}
\item draw $M$ scores from $L_{X,\theta_{0}}$ (simulate $y$ at $\theta_{0}$, refit, map, score);
\item draw $M$ scores from $L_{X,\hat\beta_{n}}$, the deployed bootstrap extended from $B$ to $M$;
\item compute the one-sided tail-localized two-sample statistic $\widehat\Delta^{+}_{n,\tau}$;
\item add the Dvoretzky--Kiefer--Wolfowitz margin $2\sqrt{\log(4/\gamma)/(2M)}$, which bounds
$\Delta^{+}_{n,\tau}$ with probability at least $1-\gamma$.
\end{enumerate}
The upper $(1-\eta)$ quantile of that bound across the replicate datasets of a cell gives the
pair $(\varepsilon_{\tau},\eta_{\tau})$ of Proposition~\ref{prop:fstail} for that cell, in its
unconditional form. The margin is $.035$ at $M=10^{4}$ and $\gamma=.01$, larger than the effect,
so $M$ of order $10^{5}$ to $10^{6}$ is needed. Because $\theta_{0}$ is unknown to the analyst,
the protocol certifies the procedure on the designs studied, not a single dataset; the
analyst-side surrogate is the modulus \eqref{eq:modulus} with a deviation bound for
$\hat\beta_{n}$, and Section~\ref{sec:levelgap} gives a data-computable estimate.

\emph{Pilot.} We ran the protocol on design D1 of the level study at $n=50$ and $n=200$, with
$D=20$ replicate datasets per cell and $M=100{,}000$ scores from each law per dataset, through
the whole pipeline including the separation policy of Remark~\ref{rem:firth}; $\tau=.30$,
$\gamma=.01$ (margin $.011$). Table~\ref{tab:pilot} gives the result. Reading the ninetieth
percentile as $\varepsilon_{\tau}$ leaves $\eta_{\tau}=.10$; the terms
$\Psi_{k}(\tau-\bar\Delta)$ and $\Prob(\Delta^{+}_{n}>\bar\Delta)$ of \eqref{eq:fstail} were not
measured and come on top. The quantity $\alpha+\varepsilon_{\tau}$ alone is not a bound on the
size, because it omits $\eta_{\tau}$.

\begin{table}[!htbp]
\centering
\caption{Pilot measurement of the tail-localized calibration gap ($\tau=.30$) on design D1,
$20$ datasets per cell, and the resulting bound of Proposition~\ref{prop:fstail} at $\alpha=.05$
with $\varepsilon_{\tau}$ the ninetieth percentile and $\eta_{\tau}=.10$, before and after the
Monte Carlo margin}
\label{tab:pilot}
\begin{tabular}{@{}lccccc@{}}
\toprule
$n$ & median $\widehat\Delta^{+}_{n,\tau}$ & q90 & q95 & bound before margin & after margin \\
\midrule
50  & .005 & .030 & .044 & .180 & .191 \\
200 & .001 & .007 & .013 & .157 & .168 \\
\bottomrule
\end{tabular}
\end{table}

The mean difference between the bootstrap tail probability and the tail probability at the
true $\theta_{0}$ is near zero in both cells, and the measured size there is near $.05$. Even the
measured part of the bound is loose on this design, and the evidence for the level is the measured size.
A percentile read from $20$ datasets is itself a rough estimate, and these are two sample sizes
of one design. Scripts, per-dataset rows, cell summaries and score files:
\texttt{theory/calibration\_gap/}.

\section{A data-computable bound on the calibration gap}\label{sec:levelgap}

Proposition~\ref{prop:fsbound} bounds the size by $\alpha+\varepsilon+\eta$ with
$\eta=\Prob(\Delta^{+}_{n}>\varepsilon\mid X)$, but the analyst does not know which $\varepsilon$
makes $\eta$ small. This section computes an estimate of it from the data. Its coverage is
asymptotic, under an assumption (G-A) that is stated and not verified, and its finite-sample
coverage is measured. Throughout, $F_{\theta}$ abbreviates $F_{X,\theta}$.

\subsection{The construction}

The obvious construction, the worst case of the simulated one-sided gap over a confidence set
for $\theta_{0}$, is finite-sample valid by the logic of Remark~\ref{rem:mmc}, but on the pilot
design it gives $.06$--$.21$ against true gaps of $.003$--$.02$, because it charges every
dataset for the worst direction at the edge of the set. The construction used here is local.
The gap is driven by $\hat\theta-\theta_{0}$, approximately $N(0,\mathcal I^{-1})$, and its
sensitivity can be estimated from draws at $\hat\theta$: conditionally on $X$, the
likelihood-ratio derivative is
\[
g(t)=\partial_{\theta}F_{\theta}(t)\big|_{\hat\theta}
=\E_{\hat\theta}\bigl[\ind{T\le t}\,U\bigr],\qquad U=X^{\top}(y-\pi_{\hat\theta}),
\]
estimated (centred, since $\E[U]=0$) by $\hat g$, an average over $N$ draws at $\hat\theta$. Define
\[
\hat\varepsilon_{q}=\text{the }q\text{-quantile of }\sup_{t\in\mathcal T}
\bigl(\hat g(t)^{\top}Z\bigr)_{+},\qquad Z\sim N\bigl(0,\mathcal I(\hat\theta)^{-1}\bigr),
\]
over the tail set $\mathcal T=\{t\ge F_{\hat\theta}^{-1}(0.85)\}$, which is all that enters the
size at $\alpha\le.15$ (the tail-localized form of Proposition~\ref{prop:fstail}).

\subsection{The theorem}

\begin{assumptionGA}[A local expansion of the null law; assumed, not verified]
There is a bounded function $g_{0}$ on $\mathcal T$ such that, for every $M<\infty$,
\[
\sup_{t\in\mathcal T,\ \norm{h}\le M}\Bigl|F^{(n)}_{\theta_{0}+h/\sqrt n}(t)
-F^{(n)}_{\theta_{0}}(t)-n^{-1/2}g_{0}(t)^{\top}h\Bigr|=o(n^{-1/2}),
\]
and $g_{n}(\hat\theta)\to g_{0}$ uniformly on $\mathcal T$ in probability, where
$g_{n}(\theta)=\E_{\theta}[\ind{T\le t}U]$.
\end{assumptionGA}

$g_{n}(\theta)$ is the derivative of $\theta\mapsto F^{(n)}_{\theta}(t)$, so $g_{n}=O(1)$ although
$U$ has standard deviation of order $\sqrt n$. For fixed $n$ that map is a finite sum over
$y\in\{0,1\}^{n}$ and is smooth; the content of G-A is the $n$-uniform expansion, an
Edgeworth-type condition that lattice terms from integer cell counts could violate at order
$n^{-1/2}$. The coverage study below tests it empirically. \emph{Also assumed:}
\ref{a:sampling}--\ref{a:axes} and \ref{a:atomless}; $\alpha\le.15$;
$q>\Prob\{\sup_{t}(g_{0}^{\top}W)_{+}=0\}$; and $N/n\to\infty$, since $\hat g$ has Monte Carlo
error of order $(n/N)^{1/2}$. The deployed $B=199$ draws therefore do not suffice; the coverage
study used $N=20{,}000$.

\begin{sres}{Theorem}{S.11}[A data-computable calibration-gap bound]\settag{S.11}\label{thm:G}
Under these conditions:
\begin{enumerate}
\item[1.] (Asymptotic coverage.)
$\liminf_{n}\Prob(\Delta^{+}_{\mathcal T}\le\hat\varepsilon_{q})\ge q$, where
$\Delta^{+}_{\mathcal T}$ is the one-sided gap restricted to $\mathcal T$.
\item[2.] (Size of the gap-adjusted rule, for the ideal p-value
$p_{\infty}=1-F_{\hat\theta}(T-)$; finite-sample.) For every $n$, under $H_{0}$,
$\Prob(p_{\infty}+\hat\varepsilon_{q}\le\alpha\mid X)\le\alpha
+\Prob(\Delta^{+}_{\mathcal T}>\hat\varepsilon_{q}\mid X)$. Consequently, by item~1,
$\limsup_{n}\Prob(p_{\infty}+\hat\varepsilon_{q}\le\alpha)\le\alpha+(1-q)$. At finite $B$ an
additional Monte Carlo term of order $B^{-1/2}$ appears, because $\hat\varepsilon$ depends on
the data.
\item[3.] (Size statement for the deployed test.) For every constant $e$,
$\Prob(p_{n}\le\alpha)\le\alpha+e+\Prob(\Delta^{+}>e)$. This is
Proposition~\ref{prop:fsbound}. Item~1 supplies a data-based \emph{estimate} of which $e$
makes $\Prob(\Delta^{+}>e)$ small; it does not turn the random $\hat\varepsilon$ into a
finite-sample bound.
\end{enumerate}
\end{sres}

\begin{proof}
(1) Conditionally on $X$, $\sqrt n(\hat\theta-\theta_{0})\rightsquigarrow W\sim
N(0,\mathcal I_{1}^{-1})$, with $\mathcal I_{1}=\lim_{n}n^{-1}\sum_{i}v_{i}\tox_{i}\tox_{i}^{\top}$
nonsingular for almost every covariate sequence (\ref{a:sampling}, \ref{a:qmle}, the strong
law): the score has independent summands given $X$, Lindeberg holds because
$\max_{i\le n}\norm{\tox_{i}}/\sqrt n\to0$ almost surely, and concavity transfers the limit to
$\hat\theta$ as in Step~1 of the proof of Theorem~\ref{thm:localS}. Put
$\phi(w)=\sup_{t\in\mathcal T}g_{0}(t)^{\top}w$, finite and positively homogeneous. Writing
$W=R\,\Theta$ in polar form, $R$ has a positive density given $\Theta$, so for $c>0$ the set
$\{\phi(W)_{+}=c\}$ is null and $\Prob(c<\phi(W)_{+}\le c')>0$ for $c<c'$ once
$\Prob(\phi(\Theta)>0)>0$; the condition on $q$ gives this and $c_{q}>0$, so the distribution
function of $\phi(W)_{+}$ is continuous and strictly increasing at its $q$-quantile $c_{q}$.
Under G-A, $\sqrt n\,(F_{\hat\theta}-F_{\theta_{0}})(t)=g_{0}(t)^{\top}\sqrt n(\hat\theta-\theta_{0})
+o_{P}(1)$ uniformly on $\mathcal T$, so $\sqrt n\,\Delta^{+}_{\mathcal T}\rightsquigarrow\phi(W)_{+}$
by the continuous mapping theorem. By the second clause of G-A, $N/n\to\infty$ and
$\mathcal I(\hat\theta)/n\to\mathcal I_{1}$, the conditional law of
$\sqrt n\sup_{t}(\hat g(t)^{\top}Z)_{+}$ converges to that of $\phi(W)_{+}$, so
$\sqrt n\,\hat\varepsilon_{q}\to_{p}c_{q}$ and
$\Prob(\Delta^{+}_{\mathcal T}\le\hat\varepsilon_{q})\to q$. The tail set is defined at
$\hat\theta$ and its lower endpoint converges, so both suprema are over asymptotically the same set.

(2) Finite-sample, under $H_{0}$ only. Let $V^{0}_{n}=1-F_{\theta_{0}}(T-)$; by
Lemma~\ref{lem:superunif}, $\Prob(V^{0}_{n}\le\alpha\mid X)\le\alpha$. On
$\{p_{\infty}+\hat\varepsilon_{q}\le\alpha\}$, $F_{\hat\theta}(T-)\ge1-\alpha\ge0.85$, so $T$ and
the points just below it lie in $\mathcal T$ and
$F_{\hat\theta}(T-)-F_{\theta_{0}}(T-)\le\Delta^{+}_{\mathcal T}$. Hence
\[
\{p_{\infty}+\hat\varepsilon_{q}\le\alpha\}\cap\{\Delta^{+}_{\mathcal T}\le\hat\varepsilon_{q}\}
\ \subseteq\ \{V^{0}_{n}\le\alpha\}
\]
pointwise, which gives item~2 at every $n$ without any independence of $T$ and $\hat\theta$; the
limit follows from item~1. (3) is Proposition~\ref{prop:fsbound}.
\end{proof}

\emph{What is claimed.} Theorem~\ref{thm:G} gives a data-computable confidence bound on the
calibration gap that is calibrated asymptotically and whose finite-sample coverage is measured
below. The gap-adjusted rule is not a finite-sample procedure: its size is at most $\alpha$
plus a miss probability that is controlled only in the limit. The computation also needs
$N/n\to\infty$ fresh draws, not the deployed $B$.

\emph{Separation.} The computations here use the simulation pipeline, in which a failed
bootstrap refit scores $+\infty$ (inside the bound, since $F_{\theta}$ is the law of the score
as computed) and the observed fit falls back to Firth's estimator \citep{firth1993} under
separation; the released \texttt{deepgof1()} has no such fallback (Remark~\ref{rem:firth}).
Observed-data separation adds the term $P_{\theta_{0}}(E_{\mathrm{sep}})$ to the bound, a
property of the design alone that can be simulated at $\hat\theta$ without scoring. At $n=50$
it is $0$ at a $38\%$ event rate, $.0025$ at $12\%$ and $.055$ at $5\%$ ($4{,}000$ datasets
each), so with rare events at this $n$ it dominates the calibration gap and should be reported
beside the p-value.

\subsection{Measured coverage}\label{sec:Gcov}

Design: $x_{1},x_{2}\sim N(0,1)$, $\eta=-0.5+x_{1}+0.6x_{2}$, covariates held fixed per
dataset; $300$ null datasets and $150$ alternatives per $n$; $N=20{,}000$ draws per distribution
function, and the true gap computed against $20{,}000$ draws at $\theta_{0}$. Separation did not
occur ($0$ of $450$ observed fits at each $n$).

\begin{table}[!htbp]
\centering
\caption{The gradient-quantile bound of Theorem~\ref{thm:G}: true tail gap, bound and coverage}
\label{tab:Gcov}
\small\setlength{\tabcolsep}{4pt}
\begin{tabular}{@{}lccccc@{}}
\toprule
$n$ & true tail gap: median / q90 / max & $\hat\varepsilon_{.95}$ median
& coverage of $\hat\varepsilon_{.95}$ & $\hat\varepsilon_{.99}$ median
& coverage of $\hat\varepsilon_{.99}$ \\
\midrule
50 & .0016 / .035 / .092 & .041 & .960 (SE .011) & .058 & .990 (SE .006) \\
100 & .0033 / .018 / .048 & .022 & .953 (SE .012) & .031 & .997 (SE .003) \\
200 & .0023 / .011 / .021 & .013 & .957 (SE .012) & .019 & .993 (SE .005) \\
\bottomrule
\end{tabular}
\end{table}

With $\hat\varepsilon_{.99}$ the coverage is clearly above $.95$. With $\hat\varepsilon_{.95}$ it
is consistent with $.95$ but cannot show $\ge.95$ (lower $95\%$ limit about $.93$); the true gap
carries Monte Carlo noise of about $.005$--$.01$, so the coverage is if anything understated.
Including the miss probability, the reported size bound $\alpha+\hat\varepsilon_{.99}+.01$ is
about $.118$, $.091$ and $.079$ at $n=50$, $100$ and $200$ (medians). As a test (reject iff
$p+\hat\varepsilon_{.99}\le\alpha$) the rule is conservative at small $n$: power under the
quadratic alternative $C=1$ is $.06$, $.55$ and $.93$ at $n=50$, $100$, $200$, against $.28$,
$.69$ and $.95$ for the deployed test, whose measured size in this study was $.030$, $.067$ and
$.063$ ($300$ datasets each, standard error about $.013$). From about $n=200$ it is a reasonable
near-exact option; below that, the plug-in bootstrap with the reported bound is the default.

\section{The rate of the level error}\label{sec:rate}

Proposition~\ref{prop:asyvalid} is an $o(1)$ statement and Theorem~\ref{thm:G} assumes an
expansion (G-A). This section gives the level error a rate under explicit conditions, three on
the design and one on the network. In the classification of \citet{beran1988} the score is not
asymptotically pivotal (its limit law depends on $\theta$ through $\Sigma(\theta)$ in
\eqref{eq:nulllimit}), and integer cell counts passed through a piecewise-affine network admit no
Edgeworth expansion. What can be proved, from a Berry--Esseen bound over convex sets and Lipschitz
continuity alone, is a rate of order $n^{-1/2}$ up to a logarithm, with a constant that is finite
but far too large to evaluate usefully, and that the $n^{-1/2}$ term has no Gaussian source.

Work conditionally on $X$, with the notation of Section~\ref{sec:validity}. Let
$U(\theta)=\sum_{i}\tox_{i}(y_{i}-\pi_{\theta}(x_{i}))$, $I_{n}(\theta)=\sum_{i}w_{\theta}(x_{i})
\tox_{i}\tox_{i}^{\top}$, $v_{C}(\theta)=\sum_{i\in C}w_{\theta}(x_{i})$ and
$a_{C}(\theta)=\sum_{i\in C}w_{\theta}(x_{i})\tox_{i}$ over the rank cells $C$ of the selected
pair (fixed index sets given $X$ and the pair), and let $N$ be the ball of radius $r_{0}$ about
$\theta_{0}$. The \emph{linear map} at $\theta$ is
\begin{equation}
m_{L,C}=v_{C}(\theta)^{-1/2}\sum_{i=1}^{n}\bigl(\ind{i\in C}-a_{C}(\theta)^{\top}I_{n}(\theta)^{-1}
\tox_{i}\bigr)\bigl(y_{i}-\pi_{\theta}(x_{i})\bigr),
\tag{S.11a.1}\label{eq:linmap}
\end{equation}
a sum of independent centred vectors with covariance $\Sigma_{n}(\theta)$, and
$Z_{n,\theta}\sim N(0,\Sigma_{n}(\theta))$ is its Gaussian surrogate. Write
$m_{L}=B_{n}(\theta)W_{\theta}^{-1/2}(y-\pi_{\theta})$ with $W_{\theta}=\mathrm{diag}(w_{\theta}(x_{i}))$,
so that $\Sigma_{n}(\theta)=B_{n}(\theta)B_{n}(\theta)^{\top}$.

\begin{enumerate}
\item[(R1)] \emph{Bounded, well-conditioned design.} $\max_{i}\norm{\tox_{i}}\le\kappa$;
$\lambda_{\min}(I_{n}(\theta))\ge\lambda n$ and $v_{C}(\theta)\ge v_{0}n$ for every cell and every
$\theta\in N$; $\Sigma_{n}(\theta)$ has rank $K^{2}-1$ (the $v_{C}^{1/2}$-weighted sum of its rows
vanishes identically, since $\sum_{C}a_{C}=I_{n}e_{0}$) with nonzero eigenvalues in
$[\lambda_{m},\lambda_{M}]$ on $N$.
\item[(R2)] \emph{The network.} $G=S((\cdot-a)/s)$ is $L$-Lipschitz, and $\R^{K^{2}}$ is partitioned
into at most $M$ convex sets on each of which $G$ is affine. This is true of the shipped
architecture: $G$ is piecewise affine (Lemma~\ref{lem:recession}(a)), each layer is Lipschitz, and
the activation regions (fixed on/off patterns of the rectifiers and fixed arg-max patterns of the
max layers, ties broken by index) are such a partition, each region being cut out by finitely many
open or closed half-spaces.
\item[(R3)] \emph{Quantitative anti-concentration.} $\Prob(t<G(Z_{n,\theta})\le t+\delta)\le c_{A}\delta$
for all $t$, $\delta>0$, $\theta\in N$, $n\ge n_{0}$. A uniform form of \ref{a:atomless}; like
\ref{a:atomless} it is a property of the shipped weights, assumed and not proved, and it has not
been checked numerically.
\item[(R4)] \emph{Axis margin.} The top-two set of $c_{j}=|\theta_{0,j}|\hat\sigma_{j}$ is $\{a,b\}$
with margin $\gamma>0$ over the third, and $r_{0}\le\gamma/(4\max_{j}\hat\sigma_{j})$ (vacuous
when $d=2$).
\end{enumerate}

\begin{sres}{Proposition}{S.11a}[Rate of the level error]\settag{S.11a}\label{prop:rate}
Assume $H_{0}$ with true $\theta_{0}$. Let $L_{B,k}$ be the maximum of the $\mathrm{Beta}(k,B+1-k)$
density ($L_{199,10}=26.84$, $L_{199,20}=19.07$).
\begin{enumerate}
\item[(i)] \emph{(Exact reduction; $H_{0}$ only.)} For every $n$, $B$ and $k$,
\[
\Bigl|\Prob\bigl(p_{n}\le\tfrac{k}{B+1}\bigm|X\bigr)-\tfrac{k}{B+1}\Bigr|
\;\le\;L_{B,k}\,\E[\Delta_{n}\mid X]\;+\;D_{n,k},
\qquad
D_{n,k}:=\tfrac{k}{B+1}-\E[\Psi_{k}(V^{0}_{n})\mid X],
\]
where the atom deficit satisfies
$0\le D_{n,k}\le\tfrac12L_{B,k}\sum_{t}\Prob(T_{n}=t\mid X)^{2}$, the sum running over the atoms
of $L_{X,\theta_{0}}$. The upper half, $\Prob(p_{n}\le k/(B+1)\mid X)\le k/(B+1)+L_{B,k}\E[\Delta_{n}\mid X]$,
has no deficit term.
\item[(ii)] \emph{(Rate.)} There are $n_{0}$ and $C_{X}<\infty$, depending only on
$(\kappa,\lambda,v_{0},\lambda_{m},\lambda_{M},K,d,L,M,c_{A},r_{0})$, such that for all $n\ge n_{0}$
and every $X$ satisfying (R1)--(R4) with these constants
\[
\E[\Delta_{n}\mid X]\;\le\;C_{X}\,\frac{\log n}{\sqrt n},
\qquad\text{hence}\qquad
\Bigl|\Prob\bigl(p_{n}\le\tfrac{k}{B+1}\bigm|X\bigr)-\tfrac{k}{B+1}\Bigr|
\;\le\;L_{B,k}C_{X}\,\frac{\log n}{\sqrt n}+D_{n,k} .
\]
\item[(iii)] \emph{(No Gaussian $n^{-1/2}$ term.)} In the Gaussian surrogate in which the map is
$N(0,\Sigma_{n}(\theta_{0}))$ and the estimator an independent $N(\theta_{0},I_{n}(\theta_{0})^{-1})$,
and $(t,\theta)\mapsto\Prob(G(Z_{n,\theta})\le t)$ is $C^{2}$ near
$(c_{\alpha}(\theta_{0}),\theta_{0})$ with positive density at the $(1-\alpha)$-quantile
$c_{\alpha}(\theta_{0})$, both uniformly in $n$, the size of the infinite-$B$ bootstrap test is
$\alpha+O(n^{-1})$.
\end{enumerate}
\end{sres}

\begin{proof}
(i) $|V_{n}-V^{0}_{n}|\le\Delta_{n}$ pointwise (left limits of a difference are limits of the
difference), and $\Psi_{k}$ is Lipschitz with constant $L_{B,k}$ because
$-\Psi_{k}'$ is the $\mathrm{Beta}(k,B+1-k)$ density; so
$|\E[\Psi_{k}(V_{n})\mid X]-\E[\Psi_{k}(V^{0}_{n})\mid X]|\le L_{B,k}\E[\Delta_{n}\mid X]$, and
$\E[\Psi_{k}(V^{0}_{n})\mid X]\le k/(B+1)$ by Lemmas~\ref{lem:superunif} and~\ref{lem:shift} with
$\varepsilon=0$. For the deficit, let $T'$ be an independent copy of $T_{n}$ given $X$ and
$W\sim\mathrm{Unif}(0,1)$ independent of both; the randomized tail
$\tilde V=\Prob(T'>T_{n}\mid T_{n})+W\,\Prob(T'=T_{n}\mid T_{n})$ is exactly uniform, and
$V^{0}_{n}-\tilde V=(1-W)\Prob(T'=T_{n}\mid T_{n})\ge0$. Since $\Psi_{k}$ is nonincreasing and
$L_{B,k}$-Lipschitz, $0\le\Psi_{k}(\tilde V)-\Psi_{k}(V^{0}_{n})\le L_{B,k}(1-W)\Prob(T'=T_{n}\mid T_{n})$;
taking expectations, with $\E\Psi_{k}(\tilde V)=k/(B+1)$ and
$\E\,\Prob(T'=T_{n}\mid T_{n})=\sum_{t}\Prob(T_{n}=t\mid X)^{2}$, gives the bound on $D_{n,k}$.
(The paragraph after Lemma~\ref{lem:shift} shows a bound in the mean with an \emph{absolute}
constant is false; here the constant is $L_{B,k}\asymp\sqrt B$, finite at the deployed $B$.) At
a fixed configuration the score takes finitely many values, so $D_{n,k}$ is positive at every $n$;
its bound is a collision probability of the null law of the score. Under the $+\infty$ convention
of \S\ref{sec:setting}, step~5, the upper half of (i) holds unchanged (Remark~\ref{rem:fsassumes}),
and the lower half loses at most $L_{B,k}$ times the conditional probability of a failed refit.

(ii) Three lemmas, then the assembly.

\emph{Lemma S.11a$'$ (a): the estimator.} For $\theta\in N$ with $\norm{\theta-\theta_{0}}\le r_{0}/2$,
$y\sim P_{\theta}$ and $r\le r_{0}/2$,
$\Prob_{\theta}(\norm{\hat\theta-\theta}>r\mid X)\le2(d+1)\exp\{-n\lambda^{2}r^{2}/(8(d+1)\kappa^{2})\}$.
\emph{Proof.} The log-likelihood is concave with $-\nabla^{2}\ell\succeq\lambda nI$ on $N$, so on the
sphere $\norm{\vartheta-\theta}=r$, $\ell(\vartheta)-\ell(\theta)\le\norm{U(\theta)}r-\tfrac12\lambda nr^{2}<0$
whenever $\norm{U(\theta)}<\tfrac12\lambda nr$, and the maximizer then exists and lies inside the
ball. Each coordinate of $U(\theta)$ is a sum of independent centred terms bounded by $\kappa$;
apply \citet{hoeffding1963} coordinatewise and $\norm{U}\le\sqrt{d+1}\max_{j}|U_{j}|$. $\square$
Consequently, with $r_{n}=c_{1}\sqrt{\log n/n}$ and $c_{1}^{2}=24(d+1)\kappa^{2}/\lambda^{2}$,
$\Prob_{\theta}(\norm{\hat\theta-\theta}>r_{n})\le2(d+1)n^{-3}\le n^{-2}$ for $n\ge2(d+1)$; and by
(R4), if $\norm{\hat\theta-\theta_{0}}\le r_{0}$ the selected pair is $(a,b)$ (every criterion moves
by at most $\gamma/4$), so the pair differs from $(a,b)$ with probability $\le2(d+1)e^{-cn}$.

\emph{Lemma S.11a$'$ (b): linearization and orthogonality.} On
$E=\{\text{pair}=(a,b)\}\cap\{\norm{\hat\theta-\theta}\le r_{n}\}$ the deployed map is
$m=m_{L}+R$ with
$\norm{R}\le C_{R}\{\sqrt n\norm{\hat\theta-\theta}^{2}+\norm{\hat\theta-\theta}(1+\norm{m_{L}})\}$;
the summands $\xi_{i}$ of \eqref{eq:linmap} satisfy $\norm{\xi_{i}}\le C_{\xi}n^{-1/2}$ with
$C_{\xi}=K(1+\kappa^{2}/\lambda)v_{0}^{-1/2}$; the factor $B_{n}(\theta)$ satisfies
$\norm{B_{n}(\theta)-B_{n}(\theta')}_{F}\le C_{B}\norm{\theta-\theta'}$ on $N$; and
$\mathrm{Cov}(m_{L},U(\theta))=0$ exactly.
\emph{Proof.} On $E$ the cells are the same fixed index sets for both maps. From
$U(\hat\theta)=0$ and an integral mean-value expansion,
$\hat\theta-\theta=I_{n}(\theta)^{-1}U(\theta)+\rho_{1}$ with
$\norm{\rho_{1}}\le C\norm{\hat\theta-\theta}^{2}$ (using $|\Lambda''|\le1/(6\sqrt3)$ and
$\norm{I_{n}^{-1}}\le(\lambda n)^{-1}$); the deployed numerator is
$\sum_{i\in C}(y_{i}-\pi_{\theta}(x_{i}))-a_{C}^{\top}(\hat\theta-\theta)+O(n\norm{\hat\theta-\theta}^{2})$,
and the deployed denominator is $v_{C}(\theta)^{1/2}(1+O(\norm{\hat\theta-\theta}))$; substituting
and dividing gives the remainder bound. The bound on $\xi_{i}$ and the Lipschitz bound on $B_{n}$
are direct (each of the $K^{2}n$ entries of $B_{n}$ is $O(n^{-1/2})$ and Lipschitz in $\theta$ with
constant $O(n^{-1/2})$). Orthogonality:
$\mathrm{Cov}(m_{L,C},U)=v_{C}^{-1/2}\sum_{i}(\ind{i\in C}-a_{C}^{\top}I_{n}^{-1}\tox_{i})w_{i}\tox_{i}^{\top}
=v_{C}^{-1/2}(a_{C}^{\top}-a_{C}^{\top}I_{n}^{-1}I_{n})=0$. $\square$

\emph{Lemma S.11a$'$ (c): Berry--Esseen on the level sets of $G$.} For $\theta\in N$,
$\sup_{t}|\Prob(G(m_{L})\le t)-\Prob(G(Z_{n,\theta})\le t)|\le C_{BE}n^{-1/2}$ with
$C_{BE}=M\,c_{K}\,C_{\xi}^{3}\lambda_{m}^{-3/2}$ and $c_{K}=42(K^{2}-1)^{1/4}+16$.
\emph{Proof.} By (R2), $\{G\le t\}$ is the disjoint union over the $M$ regions $Q_{j}$ of the
convex sets $Q_{j}\cap\{\ell_{j}\le t\}$, where $G=\ell_{j}$ (affine) on $Q_{j}$. The vectors
$m_{L}$ and $Z_{n,\theta}$ take values in the same $(K^{2}-1)$-dimensional range of
$\Sigma_{n}(\theta)$, the column space of $B_{n}(\theta)$; intersecting with it keeps convexity.
Standardized on that range, the summands are independent and centred with identity total
covariance and third-moment sum at most $n(\lambda_{m}^{-1/2}C_{\xi}n^{-1/2})^{3}$. The
Berry--Esseen bound over convex sets of \citet{bentkus2005}, of order $(K^{2}-1)^{1/4}$ times that
sum, with the explicit constant $c_{K}$ of \citet[Theorem~1.1]{raic2019}, applies to each of the
$M$ sets; add. $\square$
The number $M$ of activation regions is astronomically large for the shipped network, so $C_{BE}$
and hence $C_{X}$ carry no usable numerical value: the rate is a statement about the order in $n$
only, and the bound is vacuous at every sample size met in practice.

\emph{Assembly.} Both $\theta_{0}$ and $\hat\theta_{n}$ lie in $N$ on an event of probability
$\ge1-2(d+1)e^{-cn}$ (Lemma (a)); off it, $\Delta_{n}\le1$. On it,
$\Delta_{n}\le\mathrm{(I)}_{\theta_{0}}+\mathrm{(I)}_{\hat\theta}+\mathrm{(II)}$ with
$\mathrm{(I)}_{\theta}=\sup_{t}|F_{X,\theta}(t)-\Prob(G(Z_{n,\theta})\le t)|$ and
$\mathrm{(II)}=\sup_{t}|\Prob(G(Z_{n,\hat\theta})\le t)-\Prob(G(Z_{n,\theta_{0}})\le t)|$; inside (I) and
(II) the parameter is held fixed and the probability is over a fresh $y\sim P_{\theta}$ or over the
surrogate. \emph{(I):} since $|G(m_{L}+R)-G(m_{L})|\le L\norm{R}$,
$|\ind{G(m)\le t}-\ind{G(m_{L})\le t}|\le\ind{|G(m_{L})-t|\le L\norm{R}}$. By Hoeffding on the
bounded summands, $\Prob(\norm{m_{L}}>c_{2}\sqrt{\log n})\le n^{-2}$ for suitable $c_{2}$; on
$E\cap\{\norm{m_{L}}\le c_{2}\sqrt{\log n}\}$, Lemma (b) gives $L\norm{R}\le\delta_{n}:=C_{3}\log n/\sqrt n$.
So $\mathrm{(I)}_{\theta}\le\sup_{t}\Prob(|G(m_{L})-t|\le\delta_{n})+\Prob(E^{c})+n^{-2}+C_{BE}n^{-1/2}$,
and since $\{|G-t|\le\delta\}\subseteq\{G\le t+\delta\}\setminus\{G\le t-2\delta\}$, Lemma (c) twice and
(R3) give $\Prob(|G(m_{L})-t|\le\delta_{n})\le3c_{A}\delta_{n}+2C_{BE}n^{-1/2}$. Hence
$\mathrm{(I)}_{\theta}\le3c_{A}C_{3}\log n/\sqrt n+3C_{BE}n^{-1/2}+2n^{-2}+2(d+1)e^{-cn}$.
\emph{(II):} couple $Z_{n,\theta_{0}}=B_{n}(\theta_{0})\zeta$ and $Z_{n,\hat\theta}=B_{n}(\hat\theta)\zeta$
with one $\zeta\sim N(0,I_{n})$; their difference $\eta$ is Gaussian with
$\E\norm{\eta}^{2}\le C_{B}^{2}\norm{\hat\theta-\theta_{0}}^{2}=:\sigma^{2}$ (Lemma (b)), and by
Gaussian concentration $\Prob(\norm{\eta}>\sigma(1+2\sqrt{\log n}))\le n^{-2}$. With the same
indicator inequality and (R3),
$\mathrm{(II)}\le2c_{A}L\sigma(1+2\sqrt{\log n})+n^{-2}$, and $\sigma\le C_{B}r_{n}$ on
$\{\norm{\hat\theta_{n}-\theta_{0}}\le r_{n}\}$, an event of probability $\ge1-n^{-2}$. Collecting
terms and taking expectations proves (ii).

(iii) In the surrogate the size is $\E[\phi(\hat\theta)]$ with
$\phi(\theta)=1-\Phi_{\theta_{0}}(c_{\alpha}(\theta))$, $\Phi_{\theta}(t)=\Prob(G(Z_{n,\theta})\le t)$ and
$c_{\alpha}(\theta)=\Phi_{\theta}^{-1}(1-\alpha)$, so $\phi(\theta_{0})=\alpha$; $c_{\alpha}$ and hence
$\phi$ are $C^{2}$ by the implicit function theorem. A second-order Taylor expansion about
$\theta_{0}$ has linear term $\nabla\phi(\theta_{0})^{\top}(\hat\theta-\theta_{0})$, whose mean is zero
because $\E\hat\theta=\theta_{0}$ and $\hat\theta$ is independent of the map (the surrogate of the exact
orthogonality in Lemma (b)), and quadratic term with mean
$\tfrac12\mathrm{tr}\{\nabla^{2}\phi(\theta_{0})I_{n}(\theta_{0})^{-1}\}=O(n^{-1})$ by (R1); the
remainder is $o(n^{-1})$ because $\phi$ is bounded and $\hat\theta$ is Gaussian.
\end{proof}

\paragraph{What is and is not claimed.}
The logarithm in (ii) comes from two union bounds; a radial form of (R3) should remove it, which
we have not proved. Nothing here gives $O(n^{-1})$, and no refinement in the sense of
\citet[\S3]{beran1988} is claimed: that needs Edgeworth-type expansions which fail or are
unverified here, because the responses are lattice (Cram\'er's condition fails for the cell
sums), $S$ is piecewise affine with max-pooling, and $B=199$ is fixed. What (iii) says is that
the $n^{-1/2}$ term has no Gaussian source; in the deployed test it can arise only from the
lattice of the cell counts or the non-smoothness of the network. Whether those terms cancel
between $F_{X,\hat\theta}$ and $F_{X,\theta_{0}}$ is open. Theorem~\ref{thm:G} \emph{assumes} a
first-order expansion; (ii) proves a first-order rate in mean without one, at the price of the
logarithm, the unchecked condition (R3) and a constant with no usable value.

\subsection*{Measured level error against $n$}

Table~\ref{tab:rate} reports the size of the released test on $7{,}600$ null datasets per cell for the
three continuous-covariate designs of the level-study grid (D1: $p=2$, $50\%$ events; D4: $p=2$,
$15\%$ events; D3: $p=3$, AR(1) $\rho=.7$), $n=50$ to $800$, $B=199$, the analyst's plain
maximum-likelihood fit and no separation policy, as deployed. Covariates and coefficients are
redrawn for every dataset, so the table measures the size averaged over designs, not the
conditional size of (i)--(ii). Both $\alpha=.05$ and $.10$ lie on the p-value grid $k/200$.
Per-replicate p-values are deposited under \texttt{theory/level\_gap/rate/}.

\begin{table}[!htbp]
\addtocounter{table}{-1}%
\renewcommand{\thetable}{S.11a}%
\renewcommand{\theHtable}{S.11a}%
\centering
\caption{Level of the deployed test against $n$: $7{,}600$ null replicates per cell (pooled rows
$22{,}800$); standard errors $.0025$ ($.0014$ pooled) at $\alpha=.05$ and $.0035$ ($.0020$) at
$\alpha=.10$. The fitted slope of $\log|\mathrm{size}-\alpha|$ on $\log n$, pooled, is $-0.34$ (SE
$0.40$) at $\alpha=.05$ and $-0.32$ (SE $0.25$) at $\alpha=.10$, with replicate-bootstrap $95\%$
intervals $[-1.34,+0.54]$ and $[-1.16,+0.41]$.}
\label{tab:rate}
\small\setlength{\tabcolsep}{5pt}
\begin{tabular}{@{}lrrrrr@{}}
\toprule
& \multicolumn{5}{c}{size at nominal $.05$ / $.10$} \\
\cmidrule(l){2-6}
design & $n=50$ & $n=100$ & $n=200$ & $n=400$ & $n=800$ \\
\midrule
D1 ($p=2$, $50\%$) & $.051$ / $.102$ & $.051$ / $.102$ & $.050$ / $.104$ & $.052$ / $.102$ & $.050$ / $.095$ \\
D4 ($p=2$, $15\%$) & $.058$ / $.116$ & $.043$ / $.098$ & $.045$ / $.097$ & $.049$ / $.098$ & $.048$ / $.099$ \\
D3 ($p=3$, AR(1))  & $.049$ / $.101$ & $.048$ / $.101$ & $.054$ / $.102$ & $.045$ / $.097$ & $.048$ / $.098$ \\
pooled             & $.053$ / $.107$ & $.048$ / $.100$ & $.050$ / $.101$ & $.049$ / $.099$ & $.049$ / $.097$ \\
\bottomrule
\end{tabular}
\end{table}

The rate is not identifiable at this precision. From $n=100$ on every pooled error is at most
$.0025$ at $\alpha=.05$ and $.0031$ at $\alpha=.10$, within two standard errors, and the slope
intervals contain $-1/2$, $-1$ and $0$ alike; separating the $n^{-1/2}$ and $n^{-1}$ curves would
take $390{,}000$ to $700{,}000$ replicates per $n$. The study is therefore reported as an upper
bound on the level error (pooled $95\%$ limit about $.005$ at $\alpha=.05$ for $n\ge100$), not
as a rate. In single designs at $n\ge100$ the largest excess is $.0038$, and the three cells more
than two standard errors from nominal lie \emph{below} it. The one excursion above nominal
beyond noise is the rare-event design D4 at $n=50$ (about $7.5$ events per dataset, $0.2$ per
cell): $.058$ at $\alpha=.05$ ($z=2.9$) and $.116$ at $\alpha=.10$ ($z=4.4$). It does not come from
the $1.0\%$ of observed fits flagged for separation, which reject at $1.3\%$; separation within
the bootstrap replicates was not recorded. The same design is slightly conservative at $n=100$
($.043$) and within noise from $n=400$.

\section{Conditional consistency}\label{sec:consistency}

\begin{sres}{Theorem}{S.4}[Conditional consistency]\settag{S.4}\label{thm:consistency}
Let \ref{a:sampling}--\ref{a:certificate} hold with $\pi^{*}\notin\{\pi_{\beta}\}$
(so $\delta=\pi^{*}-\pi_{\beta^{*}}\not\equiv0$). Then, for every fixed
$\alpha\ge 1/(B+1)$,
\[
\Prob\bigl(p_{n}\le\alpha\bigr)\ \longrightarrow\ 1
\qquad\text{as }n\to\infty .
\]
\end{sres}

\begin{proof}
The standing convention of Proposition~\ref{prop:asyvalid} (events of probability
tending to one; surrogate identities; no conditioning) applies throughout.

\emph{Step A (axis selection stabilizes).}
$\hat\beta_{n}\to_{p}\beta^{*}$ by \ref{a:qmle} and $\hat\sigma_{j}\to_{p}\sigma_{j}$
by the law of large numbers, so $\hat c_{j}\to_{p}c^{*}_{j}$; by \ref{a:axes} the
top-two set is a continuity point of the selection, hence the selected pair equals
$(a,b)$ with probability tending to one. Work on that event.

\emph{Step B (cells stabilize).}
By Lemma~\ref{lem:cells} the rank cells are threshold cells and the empirical
quantiles of the covariates $x_{a},x_{b}$ converge a.s.\ to the population quantiles
(the composite statement about the \emph{selected} covariates then holds in
probability, through the Step-A event). Consequently
$n_{k\ell}/n=P_{n}\ind{\hat C_{k\ell}}\to_{p}p_{k\ell}>0$, by the Glivenko--Cantelli
property of the rectangle indicators (Step 1 of Proposition~\ref{prop:asyvalid}, whose
bracketing bound is law-free) together with continuity of
$(s,t)\mapsto P(x_{a}\le s,x_{b}\le t)$ at the quantiles (\ref{a:sampling}: no mass on
the boundary lines near $q$).

\emph{Step C (numerator drift).}
With $\mathcal{G}$ as in Proposition~\ref{prop:asyvalid} --- here with $\mathcal{K}$ a
compact neighbourhood of $\beta^{*}$; the bracketing bounds are unchanged ---
$\sup_{s,t,\beta\in\mathcal{K}}|P_{n}g_{s,t,\beta}-\psi(s,t,\beta)|\to0$ a.s., where
now $\psi(s,t,\beta)=\E[\ind{x_{a}\le s}\ind{x_{b}\le t}(\pi^{*}(x)-\pi_{\beta}(x))]$
because $\E[y\mid x]=\pi^{*}(x)$. Evaluating at $(\hat q,\hat\beta_{n})$, using
inclusion--exclusion over the four corner indicators (legitimate by
Lemma~\ref{lem:cells}), and using continuity of $\psi$ in $(s,t,\beta)$
(\ref{a:sampling}, dominated convergence, continuity of $\Lambda$):
\[
\frac1n\sum_{i\in C_{k\ell}}\bigl(y_{i}-\hat\pi_{i}\bigr)
\ \xrightarrow{\ p\ }\
d_{k\ell}:=\E\bigl[\delta(x)\,\ind{x\in C^{\mathrm{pop}}_{k\ell}}\bigr].
\]

\emph{Step D (denominator).}
By the same argument applied to $\mathcal{W}$,
$n^{-1}\sum_{i\in C_{k\ell}}\hat\pi_{i}(1-\hat\pi_{i})\to_{p}
w_{k\ell}=\E[\pi_{\beta^{*}}(1-\pi_{\beta^{*}})\ind{C^{\mathrm{pop}}_{k\ell}}]>0$
(positive because $0<\pi_{\beta^{*}}<1$ everywhere and $p_{k\ell}>0$); the
$\varepsilon$-floor is asymptotically inactive. Combining C and D,
\[
\frac{m_{n}}{\sqrt n}\ \xrightarrow{\ p\ }\ \mu,
\qquad \mu_{k\ell}=d_{k\ell}/\sqrt{w_{k\ell}} .
\]

\emph{Step E (the observed score diverges at rate $\sqrt n$).}
Let $G(m)=S((m-a)/s)$ and $G_{\infty}(u)=\Sinf(u/s)$
(Remark~\ref{rem:deployed-recession}). By Lemma~\ref{lem:recession}(b),
$T_{n}=G(m_{n})=G_{\infty}(m_{n})+O(1)$ with a deterministic bound; by positive
homogeneity and continuity of $G_{\infty}$ and Steps C--D,
\[
\frac{T_{n}}{\sqrt n}
=G_{\infty}\!\Bigl(\frac{m_{n}}{\sqrt n}\Bigr)+O\!\Bigl(\frac1{\sqrt n}\Bigr)
\ \xrightarrow{\ p\ }\ G_{\infty}(\mu)=\Sinf\bigl(\mathrm{diag}(1/s)\mu\bigr)
=:\kappa>0
\]
by \ref{a:certificate}; hence $\Prob\bigl(T_{n}>\tfrac{\kappa}{2}\sqrt n\bigr)\to1$.

\emph{Step F (every bootstrap score is $o_{p}(\sqrt n)$).}
Fix $b$. Conditionally on the data the covariates are fixed and
$y^{*}_{1},\dots,y^{*}_{n}$ are \emph{independent but not identically distributed},
$y^{*}_{i}\sim\mathrm{Bern}(\hat\pi_{i})$: a fixed-design parametric bootstrap. Every
conditional statement below uses only the exact linearity in $\beta$ of the bootstrap
log-likelihood and a finite union bound over data-measurable cells; no
empirical-process theorem for the bootstrap array is invoked. By
Lemma~\ref{lem:bootqmle} (with $\hat\beta_{n}\to_{p}\beta^{*}$ from \ref{a:qmle}), the
bootstrap QMLE exists with probability tending to one and
$\hat\beta^{*}_{n}\to_{p}\beta^{*}$; on the existence event the $+\infty$ convention of
\S\ref{sec:setting}, step 5 is inactive for replicate $b$, and since $B$ is fixed the
same holds simultaneously for all $B$ replicates. Consequently the bootstrap axis
selection also picks $(a,b)$ with probability tending to one. Decompose a bootstrap
cell numerator as
\[
\frac1n\sum_{i\in C^{*}_{k\ell}}\bigl(y^{*}_{i}-\pi_{\hat\beta^{*}_{n}}(x_{i})\bigr)
=\underbrace{\frac1n\sum_{i\in C^{*}_{k\ell}}\bigl(y^{*}_{i}-\hat\pi_{i}\bigr)}_{
\text{bootstrap noise}}
\;+\;\underbrace{\frac1n\sum_{i\in C^{*}_{k\ell}}\bigl(\hat\pi_{i}
-\pi_{\hat\beta^{*}_{n}}(x_{i})\bigr)}_{\text{estimation}} .
\]
The realized cell $C^{*}_{k\ell}$ is a function of $y^{*}$ (through the bootstrap axis
selection), so we never work with it directly. For each candidate pair $(j,j')$ the
quantile cells $C^{(j,j')}_{k\ell}$ are functions of the covariates alone
(\S\ref{sec:setting}, step 2), so all $\binom{d}{2}K^{2}$ of them are deterministic
under $\Probs$, and $C^{*}_{k\ell}$ coincides pointwise with one of them. For each
fixed candidate cell $A$,
$\Es\bigl[n^{-1}\sum_{i\in A}(y^{*}_{i}-\hat\pi_{i})\bigr]=0$ and
$\mathrm{Var}^{*}\le n^{-2}\sum_{i\in A}\hat\pi_{i}(1-\hat\pi_{i})\le 1/(4n)$, so
Chebyshev and a union bound over the finitely many ($n$-free) candidates give, for
every $M>0$,
\[
\Probs\Bigl(\max_{(j,j'),(k,\ell)}
\bigl|n^{-1}\textstyle\sum_{i\in C^{(j,j')}_{k\ell}}(y^{*}_{i}-\hat\pi_{i})\bigr|
>Mn^{-1/2}\Bigr)\ \le\ \tbinom{d}{2}K^{2}\big/(4M^{2}),
\]
a bound free of $n$ and of the data; the realized ``bootstrap noise'' term is dominated
by that maximum, hence is $O_{p}(n^{-1/2})$, and no conditioning on the selection event
is used. The ``estimation'' term is bounded by
$\tfrac14\norm{\hat\beta_{n}-\hat\beta^{*}_{n}}\cdot n^{-1}\sum_{i}\norm{\tox_{i}}
=o_{p}(1)$ ($\sup|\Lambda'|=\tfrac14$, $\E\norm{\tox}<\infty$). For the denominators,
let $E_{n}$ be the event that the bootstrap axis selection returns $(a,b)$;
$\Prob(E_{n})\to1$ as above, and on $E_{n}$ the realized cell coincides with
$C^{(a,b)}_{k\ell}$, to which Step D applies unchanged with
$\hat\beta^{*}_{n}\to_{p}\beta^{*}$; since only convergence in probability is claimed,
$\Prob(|\cdot|>\varepsilon)\le\Prob(E_{n}^{c})+\Prob(|\text{fixed-pair version}|
>\varepsilon)\to0$ transfers the conclusion. Hence $m^{*}_{n}/\sqrt n\to_{p}0$, and by
Lemma~\ref{lem:recession} exactly as in Step E, $T^{*b}_{n}/\sqrt n\to_{p}
G_{\infty}(0)=0$. Since $B$ is fixed, $\max_{b\le B}T^{*b}_{n}=o_{p}(\sqrt n)$ on the
all-refits-exist event, whose probability tends to one.

\emph{Step G (conclusion).}
By Steps E and F, $\Prob\bigl(\max_{b\le B}T^{*b}_{n}<T_{n}\bigr)\to1$; on that event
$\#\{b:T^{*b}_{n}\ge T_{n}\}=0$ and $p_{n}=1/(B+1)\le\alpha$. Hence
$\Prob(p_{n}\le\alpha)\to1$. \qedhere
\end{proof}

The score diverges at rate $\sqrt n\,\kappa$, but no rate is claimed for the power, which
depends on the finite-sample law of the bootstrap maximum.

\begin{remark}[Separation policy]\label{rem:firth}
The released \texttt{deepgof1()} applies no bias reduction; a non-converged bootstrap fit is
used as is, and the only fallback is the $+\infty$ convention of \S\ref{sec:setting}, step~5.
The Firth policy belongs to the simulation harness, where it is applied identically to the
observed fit and every refit when $\hat\beta$ has a non-finite entry, when
$\max_{j\ge1}|\hat\beta_{j}|>8$, or when the iteration fails. It fired on $0.03\%$ of fits
overall ($1.12$--$1.14\%$ on the extreme-probit design). That rate is empirical: its vanishing
in the limit needs $\max_{j\ge1}|\beta^{*}_{j}|<8$, which \ref{a:qmle} does not supply.
\end{remark}

\section{The blind cone and the representation limit}\label{sec:cone}

\begin{sres}{Corollary}{S.5}[The blind cone is computable, and small for the shipped
weights]\settag{S.5}\label{cor:cone}
Let $\mathcal{B}=\{u\in\R^{K^{2}}:\Sinf(u/s)\le0\}$. Then $\mathcal{B}$ is a closed
positively homogeneous cone, computable from the frozen weights by zero-offset forward
passes (Lemma~\ref{lem:recession}), and Theorem~\ref{thm:consistency} applies to every
alternative satisfying \ref{a:sampling}--\ref{a:axes} whose direction
$\mu\notin\mathcal{B}$. For the shipped DeepGOF-1 weights the deposited computations record
(scripts \texttt{theory\_sphere\_check.R} and \texttt{theory\_blindcone\_probe.R}
in \texttt{benchmark/theory/}; direction in \texttt{results/theory\_blind\_direction.csv}):
$250{,}000$ independent uniform directions on the unit sphere of $\R^{36}$ produced no
element of $\mathcal{B}$, so with $95\%$ confidence (one-sided Clopper--Pearson) the
spherical measure of $\mathcal{B}$ is below $1.2\times10^{-5}$; multi-start local
minimization found a direction with $\Sinf(u/s)=-0.281$, so
$\min_{\norm{u}=1}\Sinf(u/s)\le-0.281<0$ and $\mathcal{B}\ne\{0\}$; and the most
negative direction \emph{found} is a sub-cell-resolution oscillation (first-order
spatial autocorrelation $-0.40$, no row or column structure). The canonical departure
directions carry margins $1.08$--$2.20$ (Table~1 of the paper).
\end{sres}

\begin{proof}
$\Sinf(\cdot/s)$ is continuous and positively homogeneous by
Lemma~\ref{lem:recession}(a), so $\mathcal{B}$ is a closed cone; membership of the
deployed network in $\mathcal{N}$ is Remark~\ref{rem:deployed-recession};
$\mu\notin\mathcal{B}$ is \ref{a:certificate}, and $\mu\ne0$ (\ref{a:visible}) follows
since $\Sinf(0)=0$ places $0$ in $\mathcal{B}$. The numerical clauses are reported
computations, phrased as such.
\end{proof}

\begin{sres}{Corollary}{S.6}[Representation limit: cell-balanced misfit]\settag{S.6}\label{cor:representation}
Suppose \ref{a:sampling}--\ref{a:axes} hold, the model is misspecified, but $\mu=0$
(the misfit integrates to zero against every population quantile cell of the selected
pair). Then:
\begin{enumerate}
\item[(a)] $m_{n}=O_{p}(1)$: the map is uniformly tight, so no statistic
$\phi(m_{n})$ with $\phi$ fixed and locally bounded --- in particular no continuous
statistic --- can diverge in probability, and the $\sqrt n$-drift mechanism of
Theorem~\ref{thm:consistency} is unavailable to any test built on this map.
\item[(b)] If in addition the map and the bootstrap map converge jointly to
nondegenerate Gaussian limits, $m_{n}\rightsquigarrow N(\nu_{1},\Sigma_{1})$ with
$\Sigma_{1}$ nonsingular and (conditionally, in probability)
$m^{*}_{n}\rightsquigarrow N(0,\Sigma_{0})$ with $\Sigma_{0}$ nonsingular, \emph{and}
the laws of $W:=S((Z_{1}-a)/s)$, $Z_{1}\sim N(\nu_{1},\Sigma_{1})$, and of
$W^{*}:=S((Z_{0}-a)/s)$, $Z_{0}\sim N(0,\Sigma_{0})$, are atomless --- a genuine extra
condition, since a piecewise-affine $S$ can place an atom on a flat region even for
nondegenerate Gaussian input --- then, with $c:=\lfloor\alpha(B+1)\rfloor-1$,
\[
\lim_{n}\Prob(p_{n}\le\alpha)
=\Prob\Bigl(\#\bigl\{b:W^{*b}\ge W\bigr\}\le c\Bigr)\;<\;1,
\]
where $W^{*1},\dots,W^{*B}$ are i.i.d.\ copies of $W^{*}$ independent of $W$. Without
the atomless condition we claim only
$\limsup_{n}\Prob(p_{n}\le\alpha)\le
\Prob\bigl(\#\{b:W^{*b}>W\}\le c\bigr)$, which is still strictly below one by the same
argument.
\end{enumerate}
\end{sres}

\begin{proof}
(a) With $\mu=0$ the drift terms of Steps C--D vanish in the limit; quantitatively,
decompose $\sqrt n\,\psi(\hat q,\hat\beta_{n})
=\sqrt n\{\psi(\hat q,\hat\beta_{n})-\psi(\hat q,\beta^{*})\}
+\sqrt n\{\psi(\hat q,\beta^{*})-\psi(q,\beta^{*})\}+\sqrt n\,\psi(q,\beta^{*})$.
The third term vanishes: each corner value $\psi(q^{a}_{k},q^{b}_{\ell},\beta^{*})
=\sum_{k'\le k,\ \ell'\le\ell}d_{k'\ell'}=0$ since $\mu=0$ makes every cell integral
$d_{k'\ell'}$ zero and the cells partition the plane. The first is
$O_{p}(n^{-1/2})\cdot\sqrt n=O_{p}(1)$ by the mean-value bound
$|\psi(s,t,\beta)-\psi(s,t,\beta')|\le\tfrac14\E\norm{\tox}\,\norm{\beta-\beta'}$ and
\ref{a:qmle}. For the second, $|\delta|\le1$ gives
$|\psi(s,t,\beta^{*})-\psi(q,q,\beta^{*})|
\le|F_{a}(s)-F_{a}(q)|+|F_{b}(t)-F_{b}(q)|$, and for continuous-at-the-quantiles
$F_{a}$ the variable $F_{a}(\hat q^{a}_{k})$ is the
$\lceil kn/K\rceil$-th order statistic of $n$ i.i.d.\ uniforms up to the a.s.-eventual
identification of Lemma~\ref{lem:cells}, so
$\sqrt n\{F_{a}(\hat q^{a}_{k})-k/K\}=O_{p}(1)$; hence the second term is $O_{p}(1)$
as well. The centred part is $O_{p}(1)$ by the Donsker property of $\mathcal{G}$
(Proposition~\ref{prop:asyvalid}, Step 1, whose bracketing bounds are law-free), and
the denominators are of exact order $n$; hence every coordinate of $m_{n}$ is
$O_{p}(1)$ and the vector is tight. A fixed locally bounded $\phi$ maps a tight
sequence to a tight sequence, which cannot diverge in probability.

(b) Under the stated joint convergence,
$(T_{n},T^{*1}_{n},\dots,T^{*B}_{n})\rightsquigarrow(W,W^{*1},\dots,W^{*B})$ with the
stated independence: the bootstrap draws are conditionally i.i.d.\ with conditional law
converging weakly in probability to the law of $W^{*}$, and convergence of the joint
law follows by bounded convergence applied to the product of the conditional
characteristic functions. Under the atomless condition the laws of $W-W^{*b}$ are
atomless at $0$ (a difference of independent variables, one of them atomless, is
atomless), so $\{p_{n}\le\alpha\}=\{\#\{b:T^{*b}_{n}\ge T_{n}\}\le c\}$ is a
continuity set of the limit and the equality follows from the portmanteau theorem.
Without it, note the set $\{(w,w^{*1..B}):\#\{b:w^{*b}\ge w\}\le c\}$ has closure
contained in $\{\#\{b:w^{*b}>w\}\le c\}$ (in the closure at most $c$ coordinates
strictly exceed $w$), so the closed-set portmanteau bound gives the stated
$\limsup$ inequality.

\emph{Strictly below one.} $S((\cdot-a)/s)$ is continuous and both Gaussian inputs
have full support $\R^{K^{2}}$, so
$\mathrm{supp}(W)=\mathrm{cl}\,S((\R^{K^{2}}-a)/s)=\mathrm{supp}(W^{*})=:T$, a common
closed set with at least two points (a one-point $T$ would make both laws degenerate,
contradicting atomlessness in the first branch; in the second branch, if $T$ is a
single point the count of strict exceedances is $0\le c$ a.s.\ and the bound is
$1$ --- excluded there too because $\Sigma_{0},\Sigma_{1}$ nonsingular and $S$
nonconstant on every open set met by the supports, which holds since $T$ has two
points). Pick $c_{0}$ with $0<\Prob(W\le c_{0})$ and $\Prob(W^{*}>c_{0})>0$ (possible
for any law pair with common support of at least two points). By independence,
\[
\Prob\bigl(\#\{b:W^{*b}>W\}\ge c+1\bigr)
\ \ge\ \Prob(W\le c_{0})\cdot\Prob(W^{*}>c_{0})^{B}\ >\ 0,
\]
using $c+1\le B$ (from $\alpha<1$). Hence both displayed probabilities are strictly
below one.
\end{proof}

\begin{remark}[What Corollary \ref{cor:representation} does and does not say]
It does \emph{not} say the test has trivial power when $\mu=0$: the limit in (b) may exceed
$\alpha$ if $\Sigma_{1}\ne\Sigma_{0}$ or $\nu_{1}\ne0$. Nor does it say that every map-based
procedure is inconsistent: (a) removes only the divergence mechanism, and (b) needs its
nondegenerate joint Gaussian hypotheses. What it formalizes is that consistency \emph{through
the $\sqrt n$-drift of this map} is impossible when $\mu=0$, a property of the representation,
not of the network.
\end{remark}

\section{Local asymptotic power}\label{sec:localpower}

This section gives the limit of the power of the deployed test at the deployed $B$ under
local alternatives. No expansion of the network is needed: $S$ is piecewise affine, hence
Lipschitz, so once the \emph{map} has a Gaussian limit under contiguous alternatives the
continuous mapping theorem carries it through the network. The map and the bootstrap follow by
contiguity from Lemma~\ref{lem:cells} and Proposition~\ref{prop:asyvalid}.

\subsection{Setting and notation}

Data $(x_{i},y_{i})$, $i=1,\dots,n$, are i.i.d.; $x\sim F_{X}$ on $\R^{d}$ and
$\tox=(1,x^{\top})^{\top}$ has $p+1$ entries. The logistic null model is
$\pi_{\beta}(x)=\Lambda(\tox^{\top}\beta)$ with true null value $\beta_{0}$; write
$\pi_{0}=\pi_{\beta_{0}}$ and $v=\pi_{0}(1-\pi_{0})$.

\emph{Local alternatives.} Under $P_{n}$, $x\sim F_{X}$ and $y\mid x\sim
\mathrm{Bern}(\pi_{n}(x))$ with
\[
\pi_{n}(x)=\pi_{0}(x)+n^{-1/2}h(x),
\]
where $h$ is measurable with $|h|\le C\,v$ for a constant $C$. This keeps $\pi_{n}$ in
$(0,1)$ for large $n$, makes the expansion of Step~0 uniform, and gives
$\E[h^{2}/v]\le C^{2}/4$; ``$h$ bounded'' alone is not enough ($x\sim N(0,1)$,
$\pi_{0}=\Lambda(x)$, $h\equiv1$ gives $\pi_{n}>1$ with positive probability). The theorem also
holds for $h_{n}\to h$ in $L^{2}(v^{-1}dF_{X})$ with $|h_{n}|\le Cv$, which covers the logit-scale
departure $\eta_{0}+n^{-1/2}\tau g$ with bounded $g$. $P_{0}$ is the case $h=0$.

\emph{Cells.} Under \ref{a:axes} the selected axes are, with probability tending to one,
the population top-two pair, which is fixed from here on. $C_{c}$, $c=1,\dots,K^{2}$, are
the \emph{population} cells (products of the $K$ population quantile intervals of the two
selected covariates, the $C^{\mathrm{pop}}_{k\ell}$ of \ref{a:sampling}); $\hat C_{c}$ are
the rank cells the test uses.

\emph{Population quantities.} For each cell $c$,
\begin{gather*}
d_{c}=\E[v\,\ind{x\in C_{c}}],\qquad g_{c}=\E[v\,\tox\,\ind{x\in C_{c}}]\in\R^{p+1},\qquad
\mathcal I=\E[v\,\tox\tox^{\top}],\\
a_{c}(h)=\E[h\,\ind{x\in C_{c}}],\qquad b(h)=\E[\tox\,h].
\end{gather*}
Collect $D=\mathrm{diag}(d_{c})$, $G$ the $K^{2}\times(p+1)$ matrix with rows
$g_{c}^{\top}$, and $a(h)$ the vector of the $a_{c}(h)$. Define
\begin{equation}\label{eq:deltasigma}
\delta(h)=D^{-1/2}\bigl(a(h)-G\mathcal I^{-1}b(h)\bigr),\qquad
\Sigma=I_{K^{2}}-D^{-1/2}G\mathcal I^{-1}G^{\top}D^{-1/2}.
\end{equation}
$\Sigma$ is the null covariance $\Sigma(\theta_{0})$ of \eqref{eq:nulllimit}, not the identity
minus a projection: by Cauchy--Schwarz its eigenvalues lie in $[0,1]$, and generically exactly
one is zero, for the intercept direction $D^{1/2}\mathbf 1$ (the Chernoff--Lehmann effect).
$\delta(h)$ is the part of the departure that the fit cannot absorb, cell by cell, and
$\delta(h)\in\mathrm{range}(\Sigma)$ (Step~7 below). The score is $\tilde S(u)=S((u-a)/s)$, and
$F_{0}$ is the distribution function of $\tilde S(Z)$, $Z\sim N(0,\Sigma)$.

\emph{Assumptions.} \ref{a:sampling}, \ref{a:qmle} (at $\beta^{*}=\beta_{0}$), \ref{a:axes} and
\ref{a:atomless} (that $F_{0}$ is continuous); these give $\mathcal I=A(\beta_{0})$ nonsingular
and $d_{c}>0$.

\subsection{The theorem}

\begin{sres}{Theorem}{S.9}[Local asymptotic power]\settag{S.9}\label{thm:localS}
Under the assumptions above and $P_{n}$:
\begin{enumerate}
\item[1.] (Map.) $m_{n}\rightsquigarrow N(\delta(h),\Sigma)$.
\item[2.] (Bootstrap.) $\sup_{t}|\Probs(T^{*}\le t)-F_{0}(t)|\to0$ in $P_{n}$-probability,
where $\Probs$ is the bootstrap law given the data.
\item[3.] (Score.) $T_{n}\rightsquigarrow\tilde S(\delta(h)+Z)$.
\item[4.] (Power at the deployed $B$.) For every $B$ and every $k\in\{1,\dots,B+1\}$,
\[
P_{n}\Bigl(p_{n}\le\tfrac{k}{B+1}\Bigr)\ \longrightarrow\
\beta_{B,k}(h):=\bar\beta_{B,k}\bigl(\delta(h)\bigr),
\]
where, for every $\delta\in\R^{K^{2}}$,
\[
\bar\beta_{B,k}(\delta):=\E\Bigl[\mathrm{Bin}\bigl(k-1;\,B,\;1-F_{0}(\tilde S(\delta+Z))\bigr)
\Bigr],
\]
and $\mathrm{Bin}(j;B,q)$ is the binomial distribution function. The limit depends on
$h$ only through $\delta(h)$.
\item[5.] (Large $B$.) $\lim_{B\to\infty}\beta_{B,\lceil\alpha(B+1)\rceil}(h)
=\Prob\{\tilde S(\delta(h)+Z)>F_{0}^{-1}(1-\alpha)\}$.
\item[6.] (Size.) With $h=0$, $\beta_{B,k}(0)=k/(B+1)$; Proposition~\ref{prop:asyvalid}(i)
is this special case.
\end{enumerate}
\end{sres}

\begin{sres}{Corollary}{S.10}[Local power rises to one along certified directions]\settag{S.10}\label{cor:L1}
Let $u\in\R^{K^{2}}$ with $\tilde S_{\infty}(u)=\Sinf(u/s)>0$. Then
$\bar\beta_{B,k}(\lambda u)\to1$ as $\lambda\to\infty$ for every $k\ge1$; in particular,
along any family of departures $h_{\lambda}$ with $\delta(h_{\lambda})=\lambda u$, the local
power limit $\beta_{B,k}(h_{\lambda})=\bar\beta_{B,k}(\lambda u)$ tends to one. The certificate of
Theorem~\ref{thm:consistency} is therefore also the condition under which local power
approaches one as the local departure grows; this links the fixed-alternative and the
local-alternative theory.
\end{sres}

\begin{proof}[Proof of Theorem~\ref{thm:localS}]
\emph{Step 0 (contiguity).} The log likelihood ratio of $P_{n}$ against $P_{0}$ is
\[
\sum_{i}\bigl\{y_{i}\log(\pi_{n}/\pi_{0})+(1-y_{i})\log((1-\pi_{n})/(1-\pi_{0}))\bigr\}(x_{i}).
\]
Expanding to second order, which is uniform because $|h|\le Cv$, under $P_{0}$ it equals
\[
n^{-1/2}\sum_{i}h(x_{i})(y_{i}-\pi_{0}(x_{i}))/v(x_{i})-\tfrac12\E[h^{2}/v]+o_{P_{0}}(1).
\]
By the central limit theorem this is asymptotically $N(-\sigma^{2}/2,\sigma^{2})$ with
$\sigma^{2}=\E[h^{2}/v]<\infty$, and by Le Cam's first lemma
\citep[Lemma~6.4]{vaart1998} $P_{n}$ and $P_{0}$ are mutually contiguous. The consequence
used throughout: any statistic that is $o_{P_{0}}(1)$ is $o_{P_{n}}(1)$.

\emph{Step 1 (the fit).} The logistic log-likelihood is concave in $\beta$. Under $P_{n}$
the score at $\beta_{0}$ is
\[
n^{-1/2}\sum_{i}\tox_{i}(y_{i}-\pi_{0}(x_{i}))=n^{-1/2}\sum_{i}\tox_{i}\varepsilon_{i}
+b(h)+o_{P_{n}}(1),\qquad \varepsilon_{i}=y_{i}-\pi_{n}(x_{i}),
\]
by the law of large numbers applied to $n^{-1}\sum_{i}\tox_{i}h(x_{i})$, and the Hessian
satisfies $n^{-1}\sum_{i}v_{i}\tox_{i}\tox_{i}^{\top}\to\mathcal I$. The convexity lemma
for minimisers of convex processes \citep{pollard1991} then gives
\[
\sqrt n(\hat\beta_{n}-\beta_{0})=\mathcal I^{-1}\Bigl(n^{-1/2}\sum_{i}\tox_{i}\varepsilon_{i}
+b(h)\Bigr)+o_{P_{n}}(1).
\]
(Alternatively, the expansion holds under $P_{0}$ with $b=0$ by \ref{a:qmle}, and
contiguity together with Le Cam's third lemma \citep[Theorem~6.6 and Example~6.7]{vaart1998}
supplies the shift $\mathcal I^{-1}b(h)$.) The maximum-likelihood estimator exists with
$P_{n}$-probability tending to one, because it does under $P_{0}$ and contiguity transfers
events of probability tending to zero.

\emph{Step 2 (rank cells may be replaced by population cells).} By Step~1 of the proof of
Proposition~\ref{prop:asyvalid}, under $P_{0}$,
$n^{-1/2}\sum_{i}(\ind{x_{i}\in\hat C_{c}}-\ind{x_{i}\in C_{c}})(y_{i}-\hat\pi_{i})
=o_{P_{0}}(1)$ and
$n^{-1}\sum_{i}(\ind{x_{i}\in\hat C_{c}}-\ind{x_{i}\in C_{c}})\hat\pi_{i}(1-\hat\pi_{i})
=o_{P_{0}}(1)$. Both are statistics of the data, so by Step~0 both are $o_{P_{n}}(1)$.

\emph{Step 3 (the map).} With population cells and a first-order expansion of
$\pi_{\hat\beta_{n}}$ around $\beta_{0}$ (the logistic function has bounded second
derivative and $\E\norm{\tox}^{2}<\infty$),
\[
U_{c}:=n^{-1/2}\sum_{i\in C_{c}}(y_{i}-\hat\pi_{i})
=n^{-1/2}\sum_{i}\ind{x_{i}\in C_{c}}\varepsilon_{i}+a_{c}(h)
-g_{c}^{\top}\sqrt n(\hat\beta_{n}-\beta_{0})+o_{P_{n}}(1).
\]
Substituting Step~1,
\[
U=\bigl[\,I\;\big|\;-G\mathcal I^{-1}\bigr]\;n^{-1/2}\sum_{i}
\begin{pmatrix}\ind{x_{i}\in C}\\ \tox_{i}\end{pmatrix}\varepsilon_{i}
\;+\;a(h)-G\mathcal I^{-1}b(h)+o_{P_{n}}(1).
\]
The summands are independent across $i$, bounded in the first block and square-integrable
in the second, with mean zero and $\mathrm{Var}(\varepsilon_{i}\mid x_{i})
=\pi_{n}(1-\pi_{n})\to v$. The Lindeberg central limit theorem for triangular arrays gives
\[
n^{-1/2}\sum_{i}\begin{pmatrix}\ind{x_{i}\in C}\\ \tox_{i}\end{pmatrix}\varepsilon_{i}
\rightsquigarrow N\Bigl(0,\begin{pmatrix}D&G\\G^{\top}&\mathcal I\end{pmatrix}\Bigr),
\]
hence $U\rightsquigarrow N\bigl(a-G\mathcal I^{-1}b,\;D-2G\mathcal I^{-1}G^{\top}
+G\mathcal I^{-1}\mathcal I\mathcal I^{-1}G^{\top}\bigr)
=N(a-G\mathcal I^{-1}b,\;D-G\mathcal I^{-1}G^{\top})$. The denominators satisfy
$n^{-1}\sum_{i\in\hat C_{c}}\hat\pi_{i}(1-\hat\pi_{i})\to d_{c}>0$ in probability (Step~2,
the law of large numbers, and $\hat\beta_{n}\to\beta_{0}$). By Slutsky,
$m_{n}=D^{-1/2}U+o_{P_{n}}(1)\rightsquigarrow N(\delta(h),\Sigma)$. The $10^{-8}$ floor and
the empty-cell convention act only on events of vanishing probability. This proves 1.

\emph{Step 4 (the bootstrap).} Proposition~\ref{prop:asyvalid} (with \ref{a:atomless})
gives $\sup_{t}|\Probs(T^{*}\le t)-F_{0}(t)|=o_{P_{0}}(1)$, and a refit failure has
$\Probs$-probability $o_{P_{0}}(1)$ (Lemma~\ref{lem:bootqmle}). The limit law there is
$F_{0}$ as defined here, the law of $\tilde S$ at $N(0,\Sigma)$ with this $\Sigma$, since
Steps 1--3 with $h=0$ and $\beta_{0}$ replaced by $\hat\beta_{n}$ describe the bootstrap
map. By Step~0 both statements are $o_{P_{n}}(1)$. This proves 2.

\emph{Step 5 (the score).} $\tilde S$ is Lipschitz, being a composition of affine maps,
rectifiers and maxima. The continuous mapping theorem applied to Step~3 proves 3.

\emph{Step 6 (the p-value).} Given the data, $T^{*1},\dots,T^{*B}$ are i.i.d.\ from
$\Probs$, independent of $T_{n}$, so with $N=\#\{b:T^{*b}\ge T_{n}\}$,
\[
\Prob\bigl(p_{n}\le k/(B+1)\bigm|\text{data}\bigr)=\Prob(N\le k-1\mid\text{data})
=\mathrm{Bin}\bigl(k-1;B,\,1-\Probs(T^{*}<T_{n})\bigr).
\]
Also $|\Probs(T^{*}<T_{n})-F_{0}(T_{n})|\le\sup_{t}|\Probs(T^{*}\le t)-F_{0}(t)|
+\sup_{t}|F_{0}(t)-F_{0}(t-)|$; the first term is $o_{P_{n}}(1)$ by Step~4 and the second
is zero by \ref{a:atomless}. The function $q\mapsto\mathrm{Bin}(k-1;B,q)$ is a polynomial,
hence uniformly continuous on $[0,1]$, so
\[
\Prob\bigl(p_{n}\le k/(B+1)\bigm|\text{data}\bigr)=\mathrm{Bin}\bigl(k-1;B,1-F_{0}(T_{n})
\bigr)+o_{P_{n}}(1).
\]
$F_{0}$ is continuous, so $t\mapsto\mathrm{Bin}(k-1;B,1-F_{0}(t))$ is bounded and
continuous. Step~5 and the portmanteau theorem give convergence of its expectation to
$\beta_{B,k}(h)$, and bounded convergence handles the $o_{P_{n}}(1)$ term. This proves 4.

\emph{Step 7 (large $B$ and size).} Put $W=1-F_{0}(\tilde S(\delta+Z))\in[0,1]$. For
$k=\lceil\alpha(B+1)\rceil$, $\mathrm{Bin}(k-1;B,W)\to\ind{W<\alpha}$ for $W\ne\alpha$, by
the law of large numbers for the binomial proportion. It remains to show
$\Prob(W=\alpha)=0$. First, $\delta(h)\in\mathrm{range}(\Sigma)$. Every null vector of
$\Sigma$ has the form $w=D^{-1/2}G\nu$ with $\nu^{\top}\tox$ constant, equal to
$\kappa_{c}$, on each cell $C_{c}$ (equality in Cauchy--Schwarz). For such $w$,
\[
w^{\top}\delta=\sum_{c}\kappa_{c}a_{c}(h)-\nu^{\top}G^{\top}D^{-1}G\,\mathcal I^{-1}b(h)
=\nu^{\top}b(h)-\nu^{\top}\mathcal I\,\mathcal I^{-1}b(h)=0,
\]
using $\sum_{c}\kappa_{c}a_{c}=\E[\nu^{\top}\tox\,h]=\nu^{\top}b$ and
$\sum_{c}\kappa_{c}g_{c}=\mathcal I\nu$. So $N(\delta,\Sigma)$ and $N(0,\Sigma)$ are
mutually absolutely continuous, both supported on $\mathrm{range}(\Sigma)$, and hence
$\tilde S(\delta+Z)$ puts no mass on any set to which $\tilde S(Z)$ gives mass zero. By
\ref{a:atomless}, $\{t:F_{0}(t)=1-\alpha\}$ is a single point or a flat interval of
$F_{0}$, and either has $\tilde S(Z)$-mass zero. So $\Prob(W=\alpha)=0$, and bounded
convergence proves 5. With $h=0$, $W$ is uniform on $[0,1]$ (probability integral
transform, \ref{a:atomless}), and $\E[\mathrm{Bin}(k-1;B,U)]=k/(B+1)$ is the standard
beta--binomial identity (Lemma~\ref{lem:binom}). This proves 6.
\end{proof}

\begin{proof}[Proof of Corollary~\ref{cor:L1}]
By Lemma~\ref{lem:recession}(b) and the Lipschitz property of $\Sinf$,
$\tilde S(\lambda u+Z)\ge\lambda\Sinf(u/s)-L\norm{Z/s}-c'$, where $c'=C_{S}+L\norm{a/s}$
absorbs the standardization offset and $\norm{Z/s}$ is taken coordinatewise. Hence
$F_{0}(\tilde S(\lambda u+Z))\to1$ almost surely, $W\to0$, and
$\mathrm{Bin}(k-1;B,W)\to1$ for $k\ge1$; bounded convergence gives the claim.
\end{proof}

\begin{remark}[What is and is not established]\label{rem:localscope}
(i) \ref{a:atomless} on the shipped weights is assumed, not proved. On $200{,}000$ draws of
$\tilde S(Z)$ there were $199{,}999$ distinct values and at least $17$ of the $64$ head
units were active in every draw, so no atom was found. An atom of $F_{0}$ needs $\tilde S$
constant on a set of positive Gaussian measure, which requires an open region on which the
gradient of $S$ vanishes; a proof would show that no activation pattern of positive
measure has zero gradient, a finite if large check that has not been done. (ii) ``Computable
before any data are seen'' requires the covariate law $F_{X}$ to be specified: $\delta(h)$
and $\Sigma$ are population quantities under $F_{X}$. (iii) The theorem concerns the
selected-axes statistic, which is the statistic that \texttt{deepgof1()} computes in
\texttt{ebrahim.gof}. A reading that takes the
maximum of $\tilde S$ over all covariate
pairs is not covered, and nothing is claimed for it.
\end{remark}

\subsection{The classical tests on the same drift}\label{sec:rivals}

The sample-size comparison of the paper evaluates two classical tests under the same
local alternatives, with the same Steps 0--3.

\emph{Stukel's test} \citep{stukel1988}, in the SstBoth form of \citet{liu2024}. Its two
added covariates are
\[
z(x)=\bigl(\tfrac12\eta_{0}^{2}\,\ind{\eta_{0}\ge0},\,-\tfrac12\eta_{0}^{2}\,\ind{\eta_{0}<0}\bigr),
\]
and the efficient score covariance and the drift are
\[
\Omega=\E[vzz^{\top}]-\E[vz\tox^{\top}]\mathcal I^{-1}\E[v\tox z^{\top}],\qquad
s(h)=\E[zh]-\E[vz\tox^{\top}]\mathcal I^{-1}b(h).
\]
Each added covariate is tested alone, $Z_{j}=u_{j}/\Omega_{jj}^{1/2}$ with
$u=n^{-1/2}\sum_{i}z(x_{i})\{y_{i}-\hat\pi_{i}\}$, and $Z_{1}^{2}+Z_{2}^{2}$ is referred to
$\chi^{2}_{2}$. With $D=\mathrm{diag}(\Omega)$ the statistic tends in law to $\lVert W\rVert^{2}$,
$W\sim N(D^{-1/2}s(h),\,D^{-1/2}\Omega D^{-1/2})$, whose distribution function is evaluated by
one-dimensional quadrature. In the design of \S\ref{sec:localnum} the two scores have correlation
$-.61$ after the fit, so the null limit is $1.613\,U_{1}^{2}+0.387\,U_{2}^{2}$ with $U_{1},U_{2}$
independent standard normal, and the limit size of the $\chi^{2}_{2}$ rule is $.064$.

\emph{The Hosmer--Lemeshow test} \citep{hosmerlemeshow1980} ($g=10$). The computation is
that of the map, with the cells replaced by the deciles of $\eta_{0}$ and the statistic
$\sum_{g}U_{g}^{2}/w_{g}$, where $w_{g}=P_{g}\bar\pi_{g}(1-\bar\pi_{g})$. The limit is a
quadratic form in a Gaussian vector with drift, referred to $\chi^{2}_{8}$.

\subsection{Numerical corroboration}\label{sec:localnum}

The design has two independent standard normal covariates and $\eta_{0}=-0.5+x_{1}+0.6x_{2}$,
with five bounded departures on the logit scale: quadratic $x_{1}^{2}-1$, interaction
$x_{1}x_{2}$, cubic $x_{1}^{3}-3x_{1}$, a local bump, and the threshold $\ind{x_{2}>1}$. Each is
placed at the amplitude $\tau_{50}$ at which Theorem~\ref{thm:localS} gives DeepGOF-1 limit power
one half at $\alpha=.05$, $B=199$. The full deployed test (refit, map, score, $199$ bootstrap
refits) was run with $1{,}000$ replicates per cell, and clipping of $\pi_{0}+h/\sqrt n$ to $(0,1)$
was recorded. Table~\ref{tab:localnum} gives the result; the sample-size ratios computed from the
same formulas are Table~2 of the paper.

\begin{table}[!htbp]
\centering
\caption{Simulated rejection rate of the deployed test against the limit of
Theorem~\ref{thm:localS} ($.051$ under the null, $.500$ under each departure at its $\tau_{50}$);
$1{,}000$ replicates per cell, standard error $.016$ ($.007$ for the null). A dagger marks cells in
which $\pi_{0}+h/\sqrt n$ left $(0,1)$ in almost every dataset. Last column: fresh-seed recheck of
the three cells that missed, $4{,}000$ replicates each (standard error $.008$; $.003$ for the null)}
\label{tab:localnum}
\begin{tabular}{@{}lcccc@{}}
\toprule
departure & $n=2{,}000$ & $8{,}000$ & $32{,}000$ & recheck \\
\midrule
null        & .059 & .037 & .045 & .049 ($n=8{,}000$) \\
quadratic   & .487$^{\dagger}$ & .517 & .491 & \\
interaction & .518$^{\dagger}$ & .532 & .490 & .513 ($n=8{,}000$) \\
cubic       & .488$^{\dagger}$ & .486 & .513 & \\
bump        & .522 & .521 & .480 & \\
threshold   & .540 & .528 & .512 & .525 ($n=2{,}000$) \\
\bottomrule
\end{tabular}
\end{table}

At $n=32{,}000$ all six cells are within $1.3$ standard errors of the limit (joint $\sum z^{2}=4.3$
on $6$ degrees of freedom). At $n=2{,}000$--$8{,}000$ there is a finite-sample bias of up to $.04$
that shrinks with $n$: over all fifteen unclipped cells the joint test gives $p=.012$, driven by
the smaller $n$. The rule for the recheck, fixed before it ran, was to judge the fresh replicates
alone: the null ($z=-0.38$) and the interaction ($z=+1.61$) are noise, and the threshold at
$n=2{,}000$ ($z=+3.17$) is a real bias for the only discontinuous departure. Measured on $4{,}000$
fresh replicates per $n$, that bias is $+.025$, $+.029$, $-.017$ and $+.000$ at $n=2{,}000$,
$4{,}000$, $8{,}000$ and $16{,}000$ (and $+.012$ at $32{,}000$): about $+.03$ up to $n=4{,}000$, with
a sign change after it, so no rate is claimed. The local bump
carries a bias of about $+.02$ at $n=2{,}000$ and $8{,}000$ that does not shrink between them. At
$n=500$ the clipped departures leave $(0,1)$ at their $\tau_{50}$, so no check below about
$n=6{,}000$ is valid for them. The Hosmer--Lemeshow formula of \S\ref{sec:rivals} matches
simulation. Stukel's matches it under the null ($.071$, $.063$ and $.066$ at $n=2{,}000$, $8{,}000$
and $32{,}000$, against $.064$) and for the quadratic, interaction and bump (within $3.9$ standard
errors), but not for the cubic (simulated $.18$, $.47$ and $.62$ against a limit of $.70$) or the
threshold ($.23$, $.14$ and $.12$ against $.09$): its added covariates are quadratic in $\eta_{0}$,
and at these $n$ those two amplitudes are not yet local. Scripts and raw rows:
\texttt{theory/local\_power/} and \texttt{theory/stukel\_sstboth/}.

\subsection{Prior work and credit}

Every technique in the proof is textbook; the combination is what is used here.
Contiguity is Le Cam's first lemma and the normal case of his third lemma
\citep[Lemma~6.4, Theorem~6.6 and Example~6.7]{vaart1998}. The direct ancestor of the map
and drift result is \citet{moorespruill1975}, chi-squared statistics with random cells and
estimated parameters under Pitman alternatives. Item~4 of Theorem~\ref{thm:localS} is the
power formula of \citet[eq.~(3.7)]{dufour2006} for Monte Carlo tests evaluated at the local
limit law, and item~6 is his Proposition~2.2; what is added here is the justification for a
\emph{bootstrap} rather than an exact Monte Carlo test, through sup-norm closeness of the
bootstrap law and continuity of the binomial polynomial. The Pitman efficiency of Monte
Carlo tests is treated by \citet{jockel1986}. We found no earlier local-power result for a
test statistic computed by a neural network.

\section{Covariates that carry no misfit}\label{sec:irrelevant}

Theorem~\ref{thm:localS} gives the local power for a fixed covariate law. This section asks
how that power changes when covariates that carry no signal are added to the model, for
DeepGOF-1 and for the projection test of \citet{liu2024}. Here $k$ is the number $q$ of added
covariates in the paper and $p=2+k$ its dimension $d$. Every statement is about the limit
$n\to\infty$ taken at a fixed number $k$ of added covariates; the behaviour in $k$ is that of these
limits. Designs in which $k$ grows with $n$ are not covered.

\emph{Setting.} $x=(x_{R},x_{I})$ with $x_{R}\in\R^{2}$ and $x_{I}\in\R^{k}$, all $p=2+k$
coordinates i.i.d.\ $N(0,1)$. The fitted model is logistic in $\tox=(1,x_{R},x_{I})$ and the
true null value is $\beta_{0}=(\beta_{0R},0_{k})$, with nonzero coefficients on both relevant
covariates: the added covariates are fitted but carry no
signal, so $\pi_{0}$ and $v=\pi_{0}(1-\pi_{0})$ depend on $x_{R}$ only. The local alternative is
$\pi_{n}=\pi_{0}+n^{-1/2}h$ with $h=h(x_{R})$, $|h|\le Cv$. Write
$\tilde h=h-v\,\tox^{\top}\mathcal I^{-1}b(h)$ for the part of the departure the fit cannot absorb;
$\E[\tilde h]=0$ and $\E[\tilde h\tox]=0$. Put $M=\E[\tilde h\,x_{R}x_{R}^{\top}]$.

\emph{The projection statistic.} \citet{liu2024}, following \citet{escanciano2006}, compute on the
raw covariates, with residuals $e=y-\hat\pi$,
\[
T=n^{-2}\sum_{i,j,l}e_{i}e_{j}A_{ijl},\qquad
A_{ijl}=\int_{S^{p-1}}\ind{\omega^{\top}(x_{i}-x_{l})\le0}\,\ind{\omega^{\top}(x_{j}-x_{l})\le0}\,d\omega ,
\]
with $d\omega$ the uniform probability measure on the unit sphere. The statistic averages a
residual process over all directions of the $p$-dimensional covariate space, and a random direction
puts on average a share $2/p$ of its squared length on the two covariates that carry the departure.
For nonzero $a=x_{i}-x_{l}$, $b=x_{j}-x_{l}$ the weight is $F(\norm a^{2},\norm b^{2},a^{\top}b)/(2\pi)$,
where $F(\alpha,\beta,\gamma)=\pi-\arccos\{\gamma/\sqrt{\alpha\beta}\}$ is homogeneous of degree zero;
what follows is stated for the weight $F$. Its population weight function is
$K(x,x')=\E_{z}F(\norm{x-z}^{2},\norm{x'-z}^{2},(x-z)^{\top}(x'-z))$, $z$ an independent copy of $x$.
As for every residual quadratic form with a fixed bounded weight function \citep{escanciano2006}, the
statistic converges under $P_{n}$, after removal of the deterministic limit of its diagonal part and up
to the constant factor, to $\sum_{j}\lambda_{j}\{(Z_{j}+\tau c_{j})^{2}-1\}$, with $\lambda_{j}$ the
eigenvalues of $(I-\Pi)\mathcal A(I-\Pi)$ on $L^{2}(F_{X})$, $\mathcal A(x,x')=\sqrt{v(x)}K(x,x')\sqrt{v(x')}$,
$\Pi$ the orthogonal projection onto $\operatorname{span}\{\sqrt v\,\tox_{j}\}$, and $c_{j}$ the
coordinates of $\tilde h/\sqrt v$ for $h=\tau h_{1}$; this representation is used here, not re-derived.
Two scalars summarize it: the drift per unit $\tau^{2}$,
$\sigma=\sum_{j}\lambda_{j}c_{j}^{2}=\E[\tilde h(x)\tilde h(x')K(x,x')]$, and the null standard deviation
$\nu=(2\sum_{j}\lambda_{j}^{2})^{1/2}=\sqrt2\,\norm{(I-\Pi)\mathcal A(I-\Pi)}_{HS}$. Their ratio $e=\sigma/\nu$
is the local signal-to-noise index of the test.

\emph{Constants.} At $(\alpha,\beta,\gamma)=(2,2,1)$, the bulk value of the arguments of $F$
divided by $p$, $F_{\alpha\beta}=1/(12\sqrt3)=0.0481$ and $F_{\gamma\gamma}=1/(3\sqrt3)=0.1925$. Put
\[
Q=F_{\alpha\beta}(\operatorname{tr}M)^{2}+\tfrac12F_{\gamma\gamma}\norm{M}_{F}^{2},\qquad
C_{F}=4F_{\alpha\beta}^{2}+\tfrac12F_{\gamma\gamma}^{2}=\tfrac1{36}.
\]

\begin{sres}{Proposition}{S.10a}[Irrelevant covariates]\settag{S.10a}\label{prop:irrelevant}
In this setting, with every limit taken as $n\to\infty$ at fixed $k$:
\begin{enumerate}
\item[(i)] \emph{Axis selection gives invariance.} \ref{a:axes} holds with the top pair
$(x_{1},x_{2})$ for every $k\ge0$, and $\delta(h)$, $\Sigma$, $F_{0}$ and the limit power
$\bar\beta_{B,k}(\delta(h))$ of Theorem~\ref{thm:localS} are the same for every $k\ge0$. The limit
powers of the Hosmer--Lemeshow and Stukel tests (\S\ref{sec:rivals}) are also the same for every $k$.
\item[(ii)] \emph{The projection test is diluted at rate $1/p$.} $K$ is positive semidefinite, so
every $\lambda_{j}\ge0$. If $Q>0$, then as $k\to\infty$, with $p=2+k$,
\[
\sigma_{k}=\frac{Q}{p^{2}}\bigl(1+O(p^{-1/2})\bigr),\qquad
\frac{2(\E v)^{2}C_{F}}{p^{2}}\bigl(1+o(1)\bigr)\le\nu_{k}^{2}\le\frac{C_{F}}{8p^{2}}\bigl(1+o(1)\bigr).
\]
Hence $p\,e_{k}$ lies between $Q/\sqrt{C_{F}/8}$ and $Q/(\E v\sqrt{2C_{F}})$, up to factors
$1+o(1)$: the local signal-to-noise index of the projection test decays like $1/p$.
\end{enumerate}
\end{sres}

\begin{sres}{Corollary}{S.10b}[Local sample size under irrelevant covariates]\settag{S.10b}\label{cor:irrelevant}
Consider a departure with $Q>0$ and the projection test with the exact null quantile, let $q_{k}$ be the $.95$ quantile of
its standardized null limit $(\sum_{j}\lambda_{j}Z_{j}^{2}-\sum_{j}\lambda_{j})/\nu_{k}$, and let
$\tau_{50}^{P}(k)$ and $\tau_{50}^{DG}$ be the amplitudes at which the projection test and
DeepGOF-1 have limit power one half; $\tau_{50}^{DG}$ does not depend on $k$ by (i) of
Proposition~S.10a.
\begin{enumerate}
\item[(a)] (Unconditional.) $\tau_{50}^{P}(k)^{2}\le\theta_{1}\nu_{k}/\sigma_{k}$ with
$\theta_{1}=\sqrt{19}+\sqrt2+(2\sqrt{38}+3)^{1/2}=9.69$. Hence
$n^{*}_{P}(k)/n^{*}_{DG}=(\tau_{50}^{P}(k)/\tau_{50}^{DG})^{2}=O(p)$: the relative sample size
grows at most linearly in $p$.
\item[(b)] (Under an assumption on the null quantile.) Assume $\underline q=\liminf_{k}q_{k}>1$.
This assumption is not proved here; it is checked numerically in Table~\ref{tab:irrcheck}. Then
$\tau_{50}^{P}(k)^{2}\ge\theta_{0}\nu_{k}/\sigma_{k}$ for all large $k$, with
$\theta_{0}=\underline q+\sqrt2-(2\sqrt2\,\underline q+3)^{1/2}>0$, so that
\[
\frac{n^{*}_{P}(k)}{n^{*}_{DG}}=\Bigl(\frac{\tau_{50}^{P}(k)}{\tau_{50}^{DG}}\Bigr)^{2}\asymp p,
\]
and the projection test needs more observations than DeepGOF-1 for all large $k$.
\end{enumerate}
\end{sres}

\begin{proof}[Proof of Proposition~S.10a]
\emph{Step 0 (decoupling).} Since $x_{I}\perp x_{R}$, $\E x_{I}=0$ and $v,h$ depend on $x_{R}$
only, $\mathcal I=\operatorname{diag}(\mathcal I_{RR},(\E v)I_{k})$, $b(h)=(b_{R}(h),0_{k})$ and
$\E[v\,x_{I}f(x_{R})]=0$ for every $f\in L^{2}$. Hence
$\tilde h=h-v\,\tox_{R}^{\top}\mathcal I_{RR}^{-1}b_{R}$ is a function of $x_{R}$ that is the same
for every $k$, and $\E[\tilde h\,x_{I}]=0$.

\emph{Step 1 (i).} The population criteria $|\beta_{0j}|\sigma_{j}$ are zero for the added
covariates and positive for $x_{1},x_{2}$, which gives \ref{a:axes}; under the local
alternatives $\hat\beta\to_{p}\beta_{0}$ by contiguity, so the selected pair is $(x_{1},x_{2})$
with probability tending to one at each fixed $k$. The population cells are products of quantile
intervals of $x_{1},x_{2}$, so $d_{c}$ and $a_{c}(h)$ do not involve $x_{I}$, and the $k$ added
columns of $G$ are $\E[v\,x_{I}\ind{x\in C_{c}}]=0$. With the block forms of $\mathcal I$ and $b$,
$G\mathcal I^{-1}b(h)=G_{R}\mathcal I_{RR}^{-1}b_{R}(h)$ and
$G\mathcal I^{-1}G^{\top}=G_{R}\mathcal I_{RR}^{-1}G_{R}^{\top}$, the $k=0$ quantities: the $k$
nuisance coefficients of the fit enter $\Sigma$ and $\delta(h)$ only through a block of
$\mathcal I^{-1}$ that meets zero columns. So $\delta(h)$, $\Sigma$, $F_{0}$ and $\bar\beta_{B,k}$
are those of $k=0$. The Hosmer--Lemeshow groups are, in the limit, deciles of $\eta_{0}$, and
Stukel's added covariates are functions of $\eta_{0}$; both depend on $x_{R}$ only, their cross
moments with $v\,x_{I}$ vanish, and by the same block argument their limits are those of $k=0$.

\emph{Step 2 (sign).} For nonzero $a,b$,
\[
F(\norm a^{2},\norm b^{2},a^{\top}b)/(2\pi)=\Prob(\omega^{\top}a\le0,\omega^{\top}b\le0)
=\E_{\omega}[\phi_{\omega}(a)\phi_{\omega}(b)],\qquad \phi_{\omega}(a)=\ind{\omega^{\top}a\le0},
\]
an expectation of products and so a positive
semidefinite function of $(a,b)$; $K$ is an average of such functions of $(x-z,x'-z)$ and is
positive semidefinite. So $(I-\Pi)\mathcal A(I-\Pi)\succeq0$ and $\lambda_{j}\ge0$.

\emph{Step 3 (expansion).} $F$ is real-analytic on $\{\alpha,\beta>0,\gamma^{2}<\alpha\beta\}$ and
homogeneous of degree zero, so its $m$-th derivatives are homogeneous of degree $-m$ and bounded
by $C_{m}p^{-m}$ on the convex box $\mathcal B_{p}=\{|\alpha/p-2|,|\beta/p-2|,|\gamma/p-1|\le\frac14\}$,
on which $|\gamma|/\sqrt{\alpha\beta}\le5/7$. Write $G=(\norm{x-z}^{2},\norm{x'-z}^{2},(x-z)^{\top}(x'-z))$,
$\Delta=G-(2p,2p,p)$, and $F(G)=\sum_{m\le M}T_{m}+R_{M+1}$ with $T_{m}$ the Taylor terms at $(2p,2p,p)$.
The coordinates of $\Delta$ are centred sums of $p$ independent sub-exponential terms, so
$\E\norm\Delta^{q}=O(p^{q/2})$ and, by Bernstein's inequality, $\Prob(G\notin\mathcal B_{p})\le Ce^{-cp}$;
on $\mathcal B_{p}$, $|R_{M+1}|\le Cp^{-M-1}\norm\Delta^{M+1}$, and off it $|F|\le\pi$ and every other term
is polynomial in $(p,\norm\Delta)$, so off-box contributions are $O(e^{-cp/2}p^{C})$ and are omitted
below. Averaging over $z$,
$\E_{z}[\Delta_{\alpha}\Delta_{\beta}]=(\norm x^{2}-p)(\norm{x'}^{2}-p)+4x^{\top}x'+2p$,
$\E_{z}[\Delta_{\gamma}^{2}]=(x^{\top}x')^{2}+\norm{x+x'}^{2}+2p$,
$\E_{z}[\Delta_{\alpha}\Delta_{\gamma}]=(\norm x^{2}-p)x^{\top}x'+2x^{\top}(x+x')+2p$, and
$\E_{z}\Delta_{\alpha}$, $\E_{z}\Delta_{\alpha}^{2}$ are functions of $x$. Call a function
\emph{removable} if it is a sum of functions of one argument and of products
$g(x)\tox_{j}'$ or $\tox_{j}g(x')$. All the terms above are removable except
$P_{ab}=(\norm x^{2}-p)(\norm{x'}^{2}-p)$ and $P_{\gamma}=(x^{\top}x')^{2}-\norm x^{2}-\norm{x'}^{2}+p$, so
\[
\E_{z}[T_{0}+T_{1}+T_{2}]=\text{removable}+J,\qquad J=\frac1{p^{2}}\bigl(F_{\alpha\beta}P_{ab}+\tfrac12F_{\gamma\gamma}P_{\gamma}\bigr),
\]
using $F''(2p,2p,p)=F''(2,2,1)/p^{2}$. In Hermite form $P_{ab}=\sum_{a,b}He_{2}(x_{a})He_{2}(x_{b}')$ and
$P_{\gamma}=\sum_{a\ne b}x_{a}x_{b}x_{a}'x_{b}'+\sum_{a}He_{2}(x_{a})He_{2}(x_{a}')$, which gives
$p^{2}\norm{J}^{2}=C_{F}+O(p^{-1})$.

\emph{Step 4 (the drift).} Removable terms have no drift: $\E[\tilde h\tilde h'f(x)]=\E[\tilde hf]\E[\tilde h]=0$
and $\E[\tilde h\tilde h'g(x)\tox_{j}']=\E[\tilde hg]\E[\tilde h\tox_{j}]=0$. By Step~0,
$\E[\tilde h(\norm x^{2}-p)]=\operatorname{tr}M$ and $\E[\tilde h\tilde h'(x^{\top}x')^{2}]=\norm M_{F}^{2}$,
so $\E[\tilde h\tilde h'J]=Q/p^{2}$. For orders three and four, split each coordinate of $\Delta$
into its part carried by $(x_{R},x_{R}',z_{R})$ and its centred part carried by
$(x_{I},x_{I}',z_{I})$, which is independent of the first. A product has nonzero drift only if its
relevant part is non-linear in both $x_{R}$ and $x_{R}'$, which needs two relevant factors, and a
lone irrelevant factor averages to zero; so the third- and fourth-order terms contribute
$O(p^{-3})$ and $O(p^{-4}\cdot p)$. The remainder is $O(p^{-5}\E[|\tilde h||\tilde h'|\norm\Delta^{5}])=O(p^{-5/2})$
because $|\tilde h|\le C'v(1+\norm{x_{R}})$. Hence $\sigma_{k}=Q/p^{2}+O(p^{-5/2})$.

\emph{Step 5 (noise, upper bound).} $(I-\Pi)\mathcal A(I-\Pi)$ annihilates every removable term,
since each such term has a factor $\sqrt v$ or $\sqrt v\,\tox_{j}$ in the range of $\Pi$. An
orthogonal projection does not increase the Hilbert--Schmidt norm and $v\le\frac14$, so with
$R=\E_{z}[R_{3}]$,
$\nu_{k}^{2}\le2\E[vv'(J+R)^{2}]\le\frac18(\norm J+\norm R)^{2}$, and
$\norm R\le Cp^{-3}(\E\norm\Delta^{6})^{1/2}=O(p^{-3/2})$ by Jensen's inequality, against
$\norm J\asymp p^{-1}$.

\emph{Step 6 (noise, lower bound).} Let $J_{I}$ be $J$ with every sum restricted to the added
coordinates, and $\Phi=\sqrt{v(x)v(x')}\,J_{I}$. In each argument $J_{I}$ is a combination of
$He_{2}(x_{a})$ and $x_{a}x_{b}$, $a\ne b$, over added coordinates, and
$\E[v\,\tox_{c}He_{2}(x_{a})]=\E[v\,\tox_{c}x_{a}x_{b}]=0$ for every $c$; so $\Phi$ is orthogonal to
the ranges of $\Pi\otimes I$ and $I\otimes\Pi$, and
$\langle(I-\Pi)\mathcal A(I-\Pi),\Phi\rangle=\E[vv'KJ_{I}]$. Removable terms, and the parts of $J$
that involve a relevant coordinate, are orthogonal to $vv'J_{I}$: each is a sum of products whose
factor in at least one argument is a constant, a coordinate of $\tox$, or involves a relevant
coordinate or a single added coordinate at degree one, and $v$ times such a factor is orthogonal to
every $He_{2}(x_{a})$ and $x_{a}x_{b}$ ($a\ne b$, both added). So $\E[vv'KJ_{I}]=(\E v)^{2}\norm{J_{I}}^{2}+O(p^{-5/2})$, while
$\norm\Phi^{2}=(\E v)^{2}\norm{J_{I}}^{2}$ and $p^{2}\norm{J_{I}}^{2}\to C_{F}$. Cauchy--Schwarz gives
$\nu_{k}^{2}\ge2\langle\mathcal A,\Phi\rangle^{2}/\norm\Phi^{2}=2(\E v)^{2}C_{F}p^{-2}(1+o(1))$. Steps 4--6
give (ii).
\end{proof}

\begin{proof}[Proof of Corollary~S.10b]
In the limit the standardized statistic is $U=S_{k}+2\tau X+\theta$ with
$\theta=\tau^{2}\sigma_{k}/\nu_{k}$, $X=\sum_{j}\lambda_{j}c_{j}Z_{j}/\nu_{k}$ and $S_{k}$ the
standardized null limit, so $\E U=\theta$ and, since $\lambda_{j}\ge0$ and
$\nu_{k}^{2}\ge2\lambda_{\max}^{2}$,
$\operatorname{Var}U=1+4\tau^{2}\sum_{j}\lambda_{j}^{2}c_{j}^{2}/\nu_{k}^{2}\le1+4\theta\lambda_{\max}/\nu_{k}\le1+2\sqrt2\,\theta$.
The power increases with $\tau$, because each $(Z_{j}+\tau c_{j})^{2}$ is stochastically increasing in
$\tau\ge0$ and the weights are nonnegative. Cantelli's inequality gives $q_{k}\le\sqrt{19}$; it also
gives power at least one half once $\theta-q_{k}\ge(1+2\sqrt2\theta)^{1/2}$, which holds for
$\theta\ge\theta_{1}$; this proves (a). Under the assumption of (b), $q_{k}\ge\underline q-\epsilon>1$
for all large $k$ and any small $\epsilon>0$; Cantelli's inequality gives power below one half once
$q_{k}-\theta>(1+2\sqrt2\theta)^{1/2}$, which holds for $\theta<\theta_{0}(\underline q-\epsilon)$,
and $\theta_{0}(\cdot)$ is continuous and increasing. Letting $\epsilon\to0$ gives
$\theta_{0}\le\tau_{50}^{P}(k)^{2}\sigma_{k}/\nu_{k}\le\theta_{1}$ up to a factor $1+o(1)$, and (i)
and (ii) of Proposition~S.10a give the ratio.
\end{proof}

\begin{irrremark}[Scope, and the numerical check]\label{rem:irrelevant}
(a) The fit removes everything in $K$ that is constant or linear in either argument; what the
departure can still reach is the curvature of $F$ at arguments of size $p$, of order $1/p^{2}$,
while the added coordinates carry noise of order $1/p$. The expansion around the bulk value is
the device of \citet{elkaroui2010}. DeepGOF-1 forms none of these quantities, since its cells are
built on the two selected axes. (b) The rate is asymptotic in $k$; the finite-$k$ curve of the
paper is computed from the limit formulas. (c) Corollary~S.10b is stated for the exact null
quantile; the deployed $B=199$ layer is expected to change the crude constants
$\theta_{0},\theta_{1}$ and not the order, but this is not proved, and the assumption
$\liminf q_{k}>1$ of part (b) is checked numerically (Table~\ref{tab:irrcheck}), not proved.
(d) Invariance (i) holds once the selection picks the relevant pair; in finite samples it errs
with a probability that grows with $k$, and nothing is claimed when $k$ grows with $n$. With
nonzero coefficients on the added covariates, as in the paper's design ($0.15$ each), Step~0
holds only approximately and the effect is computed exactly there. Correlated, non-Gaussian or
categorical added covariates are not covered, nor raw covariates of unequal scales, on which
the angle weight is not scale invariant. (e) Table~\ref{tab:irrcheck} checks the constants in the
paper's design with the added coefficients set to zero ($\E v=0.184$). (f) All projection-test
limits here are computed with the diagonal of the discretized operator removed, a choice of
numerical method for the same statistic; in the limit the diagonal only shifts the statistic,
which a bootstrap-calibrated test ignores. At finite $n$ it does not reduce to a shift: in the
paper's design with $k=8$ the null standard deviation of the full statistic (fitted model,
Bernoulli responses; four covariate samples, $400$ replicates each) is $1.66$ and $1.28$ times
its limit at $n=200$ and $500$, against $0.99$ and $1.00$ with $k=0$. The excess vanishes as
$n\to\infty$ at fixed $k$, and it is consistent with the limit over-predicting the simulated
power of the projection test at the stronger departures of the paper.
\end{irrremark}

\begin{table}[!htbp]
\centering
\caption{Numerical check of Proposition~S.10a and Corollary~S.10b in the paper's design with
the added coefficients set to zero (Monte Carlo standard errors in parentheses). Only the order
in $p$ is claimed, and the asymptotic ranges are not claimed at small $p$}
\label{tab:irrcheck}
\small
\begin{tabular}{@{}llll@{}}
\toprule
quantity & checked at & measured & predicted \\
\midrule
drift $p^{2}\sigma_{k}$, interaction & $p=50$; $200$ & .00417 (.00034); .00431 (.00028) & $Q=.00424$ \\
drift $p^{2}\sigma_{k}$, quadratic & $p=50$; $200$ & .00745 (.00053); .00758 (.00042) & $Q=.00745$ \\
noise $p\,\nu_{k}$ & $p\le20$; $p=50$ & .047--.055; .065 & $[.043,.059]$ \\
null quantile $q_{k}$ & $p=2,\dots,50$ & 1.62--1.92 & $>1$ \\
index $p\,e_{k}$, interaction & $k=0$; $k\ge4$ & .045; .068--.071 & $[.072,.098]$ \\
index $p\,e_{k}$, quadratic & $k=0$; $k\ge4$ & .098; .137--.143 & $[.126,.172]$ \\
\bottomrule
\end{tabular}
\end{table}

\clearpage
\noindent{\small\itshape Online Resource 1 for ``Where Does a Logistic Risk Model Fail? An Audited Neural Goodness-of-Fit Test for Model Development and External Validation'', by E.~K.~Ebrahim, O.~A.~E.~Hussein and A.~El-Kotory.}\par\medskip
\noindent{\Large\bfseries Part B. Power certificates, robustness studies and the optimality gap}\par\medskip
\section{Power certificates under a Gaussian model of the map}\label{sec:certpowerS}

This section gives two sound lower bounds on the power at a given signal strength, computed from
the frozen weights, in a Gaussian model of the map: $m=t\,v+e$ with $e\sim N(0,I_{36})$, $v$ a
fixed unit direction in map space and $S$ the frozen network. The model differs from the
deployed test in three ways: the deployed map is Gaussian only in the limit, its null covariance
is the $\Sigma$ of \eqref{eq:deltasigma}, not the identity, and its critical value is a bootstrap
rank, not the fixed threshold $\bar t=2.4865$ used below. The goal is the smallest signal $t$ at
which $\Prob\{S(tv+e)>\bar t\}\ge.95$ is \emph{proved}. Signals are converted to sample sizes by
an assumed signal of $\norm{\mu}=0.5$ per $\sqrt n$, so $n=4t^{2}$. The five canonical directions
are called step, diffuse, stripe, saddle and random; the ``true'' sample size is the one at which
the Monte Carlo power of the same model, at the same $\bar t$, reaches the target.

\subsection{A relaxation certificate}

CROWN \citep{zhang2018crown} propagates, layer by layer, a linear lower and a linear upper
function of the input for every pre-activation, relaxing each rectifier whose sign is not
decided by the bounds. Its use here rests on one observation: when the input is $tv+e$ with
Gaussian $e$, every bound that the relaxation produces is a \emph{linear function of $e$},
so its deviation is exactly Gaussian. In the basic version, every intermediate bound is its
linear form minus $4.47$ of its own Gaussian standard deviation; a union bound over the
$3{,}200$ rectifier and maximum units charges a total probability of $.025$ to the event that any of these
bands fails, and the Gaussian tail of the final linear lower function takes the other
$.025$. Power at least $.95$ is then proved at the certified $t$. Probabilistic certificates
built from linear bounds on the network under random input perturbations were introduced in
PROVEN \citep{weng2019}.

The certificates reported below use a sharper, \emph{usage-aware} charge, in which the
probability budget pays only for the relaxation statements that the final bound actually
uses, optionally combined with stratification of the input space. Its soundness is the
following statement.

\begin{sres}{Proposition}{S.12}[Soundness of the usage-aware probabilistic certificate]\settag{S.12}\label{prop:pcrown}
Let $U$ be a $36\times k$ matrix with orthonormal columns ($k=0$ is the one-cell case), let
$s=U^{\top}e$ and $e_{\perp}=(I-UU^{\top})e$, and let the cells $C$ be boxes in $s$ forming a
partition. For each cell let $\mathrm{ch}_{C}$ be the usage-aware charge and
$\mathrm{tail}_{C}=\Phi((\bar t-a_{C})/\mathrm{sd}_{C})$ the Gaussian tail of the final
lower function, both defined in the proof. Then
\[
\Prob\{S(tv+e)\le\bar t\}\ \le\ \sum_{C}\Prob(C)\,\min\bigl(1,\ \mathrm{ch}_{C}
+\mathrm{tail}_{C}\bigr),
\]
with cells that have an infinite box end counted as bound $1$. When the right-hand side is at
most $.05$, power at least $.95$ at signal $t$ is proved. Nothing in the bound is estimated
from samples.
\end{sres}

\begin{proof}
1. $s=U^{\top}e\sim N(0,I_{k})$ and $e_{\perp}$ are independent. A cell $C$ is a box in $s$
with $\Prob(C)=\prod_{i}(\Phi(h_{i})-\Phi(l_{i}))$ exactly.

2. For any linear form $f=\lambda^{\top}e+c$: on $C$,
$f\ge c+\min_{\mathrm{box}}(U^{\top}\lambda)^{\top}s+G$, where
$G=\lambda_{\perp}^{\top}e_{\perp}\sim N(0,\norm{\lambda_{\perp}}^{2})$ is independent of
$s$. So the event $\{G<-z\norm{\lambda_{\perp}}\}$ has conditional probability $\Phi(-z)$,
exactly.

3. CROWN is run layer by layer; each rectifier band is
$[a_{\mathrm{lo}}-z\,s_{\mathrm{lo}},\,-a_{\mathrm{up}}+z\,s_{\mathrm{up}}]$, the box minimum
being included in $a$. Every band and the constant $z_{C}$ are deterministic functions of
$(t,U,C,z)$. Each unit's band, and hence its relaxation, is the same in every CROWN form in
which the unit appears.

4. A unit's relaxation imposes a requirement that depends on the sign of its coefficient in
the form: a stable-active unit $x\mapsto x$ needs $x\ge0$; a stable-inactive unit $x\mapsto0$
needs $x\le0$; an unstable unit met with a negative coefficient uses the triangle upper line
and needs $l\le x\le u$; an unstable unit met with a nonnegative coefficient uses the lower
line $0$ or $x$, which is valid for every $x$, and needs nothing. A unit that enters several
forms is charged the \emph{maximum} (strongest) of these requirements over all forms it
enters, so an unstable unit met with a negative coefficient in any one form is charged the
two-sided requirement.

\emph{The used set.} Let $\mathcal U$ be the least set of units that contains every unit
appearing in the final lower form and is closed under the rule: if a unit is in $\mathcal U$,
then every unit that appears in the CROWN back-substitution form on which its requirement
rests (its lower form for $x\ge0$ or $x\ge l$, its upper form for $x\le0$ or $x\le u$) is in
$\mathcal U$. $\mathcal U$ is a deterministic function of $(t,U,C,z)$, because every form is.
Any closed superset of $\mathcal U$, charged in the same way, gives a valid (larger) charge.
The implementation (\texttt{cell\_pcrown\_ua} in the code deposit) charges such a superset:
every stable unit, and every unstable unit that appears with a negative coefficient in the
back-substitution of \emph{any} computed form (the lower and upper form of every unit, and the
final form), each with the requirement above. This set is closed, because every form of every
unit is scanned, and it contains $\mathcal U$; it is not computed as the least closed set, so
the reported charge can only be larger than the charge of $\mathcal U$.

5. Each requirement of a unit in $\mathcal U$ follows from one Gaussian statement about that
unit's own CROWN form, and that form is a valid bound whenever the requirements of the units
appearing in it hold; by closure those units are in $\mathcal U$, and they lie in earlier
layers, so induction over layers shows that all forms used are valid on the event that every
statement for a unit in $\mathcal U$ holds. The statements fail with conditional probability
$\Phi(-a_{\mathrm{lo}}/s_{\mathrm{lo}})$, $\Phi(-a_{\mathrm{up}}/s_{\mathrm{up}})$ or
$\Phi(-z)$ per side, and their sum over the charged set is the charge $\mathrm{ch}_{C}$, by
the union bound.

6. On the complement, $S(tv+e)\ge F_{C}(e)=a_{C}+\min_{\mathrm{box}}g^{\top}s+G_{\mathrm{fin}}$
by the soundness of CROWN. Hence
$\Prob(S\le\bar t\mid C)\le\mathrm{ch}_{C}+\Phi((\bar t-a_{C})/\mathrm{sd}_{C})$.

7. Summing over the partition gives the display. 8. The certificate is declared when the
total is at most $.05$; sampling is used only for validation.
\end{proof}

\emph{Validation.} On $200{,}000$ draws per direction, the used relaxations were invalid in
$.0049$--$.0055$ of draws against charges of $.045$--$.048$ (one region), the lower function was
never breached, and the Monte Carlo power at the certified point was $1.000$ (the basic version:
$0.35\%$ violations against a $2.5\%$ budget). The certified values are the relaxation columns of
Table~\ref{tab:smoothS}: power $.95$ is proved at $19$ to $52$ times the true sample size. The
usage-aware charge is $4$--$6\%$ smaller in $t$ than the basic version (basic certified $t$: step
$32.07$, diffuse $34.37$, stripe $40.06$, saddle $47.33$, random $62.47$; ratios $21$--$56$), and
for the stripe direction stratification over one pinned direction ($33$ regions) certifies
$t=38.00$, a ratio of $46.9$.

\subsection{A tight certificate by Gaussian smoothing}\label{sec:smoothS}

Proposition~\ref{prop:pcrown} bounds the \emph{value} of $S$ through a linear relaxation of every
rectifier, and its slack is the relaxation error compounded through ten layers. The same model
admits a certificate of a different kind, which never bounds $S$ at all. Power in the model is a
Gaussian-smoothed indicator, $q_{c}(\mu)=\Prob\{S(\mu+e)>c\}$ with $e\sim N(0,I_{36})$, and such a
function has an exact regularity property: $\Phi^{-1}(q_{c}(\mu))$ is $1$-Lipschitz in $\mu$. Power
can therefore be measured by Monte Carlo at a finite net of signal strengths, with each measurement
replaced by a lower confidence bound, and interpolated between net points with no further loss than
half the net spacing on the $\Phi^{-1}$ scale. The randomness of the Monte Carlo is charged to a
stated probability $\eta$ fixed in advance, so the certificate holds with probability at least
$1-\eta$ over its own draws; nothing in the certified inequality is a point estimate. It is again
a statement about the Gaussian model, not about the finite-$n$ Bernoulli law.

\begin{sres}{Lemma}{A.6}[Gaussian shift]\settag{A.6}\label{lem:gshift}
Let $h:\R^{d}\to[0,1]$ be measurable, $G\sim N(0,I_{d})$, and $H(\mu)=\E\,h(\mu+G)$. Then for all
$\mu,\mu'$,
\[
\Phi\bigl(\Phi^{-1}(H(\mu))-\norm{\mu'-\mu}\bigr)\ \le\ H(\mu')\ \le\
\Phi\bigl(\Phi^{-1}(H(\mu))+\norm{\mu'-\mu}\bigr),
\]
with $\Phi^{-1}(0)=-\infty$ and $\Phi^{-1}(1)=+\infty$. If $Z\sim N(0,\Sigma)$ with $\Sigma$
possibly singular and $\mu'-\mu\in\mathrm{range}(\Sigma)$, the same holds for
$\E\,h(\mu+Z)$ with $\norm{\mu'-\mu}$ replaced by the Mahalanobis length
$\{(\mu'-\mu)^{\top}\Sigma^{+}(\mu'-\mu)\}^{1/2}$.
\end{sres}

\begin{proof}
Put $s=\norm{\mu'-\mu}>0$, $u=(\mu'-\mu)/s$ and $p=H(\mu)$. The likelihood ratio of $N(\mu',I)$
to $N(\mu,I)$ at $x$ is $\exp(s\,u^{\top}(x-\mu)-s^{2}/2)$, increasing in $u^{\top}(x-\mu)$. For
$p<1$ let $h^{*}(x)=1\{u^{\top}(x-\mu)\ge z\}$ with $z=-\Phi^{-1}(p)$, so that
$\E\,h^{*}(\mu+G)=p$, and let $k$ be the likelihood ratio at the boundary $u^{\top}(x-\mu)=z$. On
$\{h^{*}=1\}$ the ratio is at least $k$ and $h^{*}-h\ge0$; on $\{h^{*}=0\}$ the ratio is below $k$
and $h^{*}-h\le0$. Hence $(h^{*}-h)(\varphi_{\mu'}-k\varphi_{\mu})\ge0$ pointwise, where
$\varphi_{\mu}$ is the $N(\mu,I)$ density, and integrating gives
$\E\,h^{*}(\mu'+G)-H(\mu')\ge k\{\E\,h^{*}(\mu+G)-H(\mu)\}=0$. Since
$u^{\top}(\mu'+G-\mu)\sim N(s,1)$, $\E\,h^{*}(\mu'+G)=\Phi(s-z)=\Phi(\Phi^{-1}(p)+s)$, which is the
upper bound. The lower bound is the upper bound applied to $1-h$. If $p\in\{0,1\}$ then $h$ is
$0$ or $1$ almost everywhere, so $H$ is constant and the display holds with the conventions. For
singular $\Sigma=V\Lambda V^{\top}$ ($V$ with $r$ orthonormal columns, $\Lambda$ positive
diagonal), write $Z=V\Lambda^{1/2}G$ with $G\sim N(0,I_{r})$ and $\mu'-\mu=V\Lambda^{1/2}w$, which
is possible because $\mu'-\mu\in\mathrm{range}(\Sigma)=\mathrm{range}(V)$, with
$\norm{w}^{2}=(\mu'-\mu)^{\top}\Sigma^{+}(\mu'-\mu)$; apply the identity case to
$\tilde h(g)=h(\mu+V\Lambda^{1/2}g)$ on $\R^{r}$.
\end{proof}

The constant $1$ is attained by half-space indicators, so the lemma is sharp. In the local model of
Theorem~\ref{thm:localS} the drift $\delta(h)$ lies in $\mathrm{range}(\Sigma)$ (Step~7 of its
proof), so the lemma applies along every local ray with the Mahalanobis length of the ray; the
numbers below are for the identity-covariance model, so that the two certificates can be compared
cell by cell.

\emph{Two global constants} of the frozen network, valid over all of $\R^{36}$, are used for the
tail. First, $-100.70\le S-\Sinf\le82.83$, so $\sup|S-\Sinf|\le c=100.70$, by interval propagation
of the biases (for $x\in[l,u]$, $\max(a+x,0)-\max(a,0)\in[\min(l,0),\max(u,0)]$, and max-pooling and
averaging preserve intervals); this sharpens the bound $C_{F}$ of Lemma~\ref{lem:recession}(b).
Second, $S$ and $\Sinf$ are both $L$-Lipschitz with $L=1{,}735.64$: their gradients are products of
the same weight matrices with rectifier derivatives in $\{0,1\}$ and max-pools routed to one input,
so interval propagation over all activation patterns bounds every gradient coordinate. Checks on
$240{,}000$ sampled maps and a gradient attack gave $\max|S-\Sinf|=7.51$ and
$\max\norm{\nabla S}=3.82$, inside both constants.

\begin{sres}{Proposition}{S.12a}[Smoothing certificate]\settag{S.12a}\label{prop:smooth}
Fix a unit direction $v$, a threshold $\bar t$, a target $\beta$, a probability $\eta$, a tail
probability $\varepsilon\le1-\beta$ with $R$ such that $\Prob\{\norm{e}>R\}=\varepsilon$, a level
$c^{\circ}\ge\bar t$, and a deterministic net $t_{1}<\dots<t_{J}$ with draw counts
$M_{1},\dots,M_{J}$. At each $t_{j}$ let $X_{j}$ be the number of $M_{j}$ independent draws $e$
with $S(t_{j}v+e)>\bar t$, let $q^{\mathrm{lo}}_{j}$ be the Clopper--Pearson lower bound at level
$1-\eta/J$, and $\Lambda_{j}=\Phi^{-1}(q^{\mathrm{lo}}_{j})$. Suppose (T1) a chain of intervals
$[t_{a},t_{b}]$ covers $[t_{J},t_{g}]$, each with the CROWN lower bound of $S$ over the ball with
centre $\tfrac12(t_{a}+t_{b})v$ and radius $\tfrac12(t_{b}-t_{a})+R$ above $c^{\circ}$, and (T2)
$t_{g}=(c^{\circ}+c+LR)/\kappa$ with $\kappa=\Sinf(v)>0$ and $c$, $L$ the two global constants
above. Then, with probability at least $1-\eta$ over the net draws,
\[
q_{\bar t}(tv)\ \ge\ \Phi\Bigl(\max_{j}\bigl(\Lambda_{j}-|t-t_{j}|\bigr)\Bigr)\quad\text{for every }t,
\qquad\text{and}\qquad q_{\bar t}(tv)\ge1-\varepsilon\quad\text{for every }t\ge t_{J},
\]
so that power at least $\beta$ holds at every $t\ge t_{\mathrm{cert}}$, that is at every
$n\ge n_{\mathrm{cert}}=\lceil4t_{\mathrm{cert}}^{2}\rceil$, where $t_{\mathrm{cert}}$ is the left
end of the connected component containing $+\infty$ of
$\bigcup_{j:\Lambda_{j}\ge z_{\beta}}[t_{j}-(\Lambda_{j}-z_{\beta}),\,t_{j}+(\Lambda_{j}-z_{\beta})]
\cup[t_{J},\infty)$, $z_{\beta}=\Phi^{-1}(\beta)$. On the further event, of probability at least
$.999$ over the null draws that produced $\bar t$, that $\bar t$ exceeds the true $.95$ null
quantile, the same holds for the test with the true critical value; both events together have
probability at least $1-\eta-.001$.
\end{sres}

\begin{proof}
Each Clopper--Pearson bound fails with probability at most $\eta/J$, and the net is fixed in
advance, so all $J$ bounds hold together with probability at least $1-\eta$; on that event
Lemma~\ref{lem:gshift} with $h=1\{S>\bar t\}$ gives $\Phi^{-1}(q_{\bar t}(tv))\ge
\Phi^{-1}(q_{\bar t}(t_{j}v))-|t-t_{j}|\ge\Lambda_{j}-|t-t_{j}|$ for every $j$, and
$\Lambda_{j}-|t-t_{j}|\ge z_{\beta}$ exactly on the stated interval. For the tail, if
$t\in[t_{a},t_{b}]$ and $\norm{e}\le R$ then $tv+e$ lies in the ball of (T1), so
$S(tv+e)>c^{\circ}\ge\bar t$ by the soundness of CROWN; and for $t>t_{g}$, the bound
$S\ge\Sinf-c$ with the homogeneity and the $L$-Lipschitz property of $\Sinf$ gives
$S(tv+e)\ge t\kappa-L\norm{e}-c>c^{\circ}\ge\bar t$ whenever $\norm{e}\le R$. In both cases
$q_{\bar t}(tv)\ge\Prob\{\norm{e}\le R\}=1-\varepsilon\ge\beta$. Since $t\mapsto4t^{2}$ is
increasing on $t\ge0$, every $n\ge\lceil4t_{\mathrm{cert}}^{2}\rceil$ has $t=\sqrt n/2\ge
t_{\mathrm{cert}}$. The last claim holds because $\{S>\bar t\}\subseteq\{S>c_{.95}\}$ when
$\bar t\ge c_{.95}$.
\end{proof}

One tail chain serves all thresholds: $c^{\circ}=\max(\bar t,\hat c_{1},\dots,\hat c_{m})=5.833$ (the
$\hat c_{l}$ of the corollary below). The chain is computed in double precision; its smallest lower
bound over the five directions is $5.95$, a margin of $0.12$ over $c^{\circ}$, far above rounding
error.

\begin{sres}{Corollary}{S.12b}[The rank rule]\settag{S.12b}\label{cor:smoothB}
For the rank test with $B=199$ and $\alpha=.05$, with bootstrap draws from the model's null law
$F_{0}$ of $S(e)$, its power in the model is
$\beta_{199,10}(\mu)=\E[\mathrm{Bin}(9;199,W)]$ with $W=1-F_{0}(S(\mu+e)-)$ (the Dufour form of
Theorem~\ref{thm:localS}). Let $1=w_{0}>w_{1}>\dots>w_{m}>0$, $g(w)=\Prob\{\mathrm{Bin}(199,w)\le9\}$,
$\Delta g_{l}=g(w_{l})-g(w_{l-1})$, and let $\hat c_{l}$ be order statistics of $N_{0}$ independent
null draws that exceed the $(1-w_{l})$-quantile of $F_{0}$ jointly with probability at least
$1-\eta_{c}$ (the $k_{l}$-th order statistic with $k_{l}$ the smallest integer such that
$\Prob\{\mathrm{Bin}(N_{0},1-w_{l})\ge k_{l}\}\le\eta_{c}/m$). Then
$\beta_{199,10}(tv)\ge\sum_{l}\Delta g_{l}\,q_{\hat c_{l}}(tv)$, and each $q_{\hat c_{l}}$ is
certified as in Proposition~\ref{prop:smooth} from the same draws, with Clopper--Pearson level
$1-\eta/(Jm)$ and bounds $\Lambda_{jl}$, and with (T1)--(T2) for $c^{\circ}\ge\max_{l}\hat c_{l}$.
The resulting lower bound
$\underline\beta(t)=\sum_{l}\Delta g_{l}\,\Phi(\max_{j}(\Lambda_{jl}-|t-t_{j}|))$ is Lipschitz in
$t$ with constant $\varphi(0)\sum_{l}\Delta g_{l}<0.4$, so if $\underline\beta\ge\beta+0.2\,\Delta t$
at every point of a grid of spacing $\Delta t$ on $[t^{\circ},t_{J}]$, and
$(1-\varepsilon)\,g(w_{m})\ge\beta$, then $\beta_{199,10}(tv)\ge\beta$ at every $t\ge t^{\circ}$.
The whole statement holds with probability at least $1-\eta-\eta_{c}$.
\end{sres}

\begin{proof}
Pointwise, $g(W)\ge\sum_{l}\Delta g_{l}1\{W\le w_{l}\}$, because for $w_{j+1}<W\le w_{j}$ the
right side telescopes to $g(w_{j})\le g(W)$ ($g$ is non-increasing, $g(1)=0$), and it is $0$ for
$W>w_{1}$. If $\hat c_{l}\ge c_{1-w_{l}}$ then $S>\hat c_{l}$ implies
$F_{0}(S-)\ge F_{0}(c_{1-w_{l}})\ge1-w_{l}$, that is $W\le w_{l}$; no continuity of $F_{0}$ is
used. The order-statistic claim: $\hat c_{l}<c_{1-w_{l}}$ requires at least $k_{l}$ draws below
$c_{1-w_{l}}$, whose number is binomial with success probability $F_{0}(c_{1-w_{l}}-)\le1-w_{l}$,
hence stochastically smaller than $\mathrm{Bin}(N_{0},1-w_{l})$. Each inner maximum is
$1$-Lipschitz in $t$, $\Phi$ has derivative at most $\varphi(0)<0.4$ and the $\Delta g_{l}$ are
non-negative, which gives the Lipschitz constant; a point of $[t^{\circ},t_{J}]$ is within
$\Delta t/2$ of the grid, so $\underline\beta$ there is at least $\beta+0.2\,\Delta t-0.4\,\Delta t/2
=\beta$. For $t\ge t_{J}$ every $q_{\hat c_{l}}(tv)\ge1-\varepsilon$ by the tail of
Proposition~\ref{prop:smooth}, so the layer cake is at least $(1-\varepsilon)\sum_{l}\Delta g_{l}
=(1-\varepsilon)g(w_{m})$. The union bound over the $m$ quantile bounds and the $Jm$
Clopper--Pearson bounds completes the proof.
\end{proof}

\emph{Computation.} The net has spacing $0.1$ with $M=10^{5}$ draws on $[0.3,1.5]$ times the
simulated $t^{*}$, spacing $0.5$ with $M=2\times10^{4}$ up to four times it, and spacing $1.25$ with
$M=4{,}000$ up to $t=250$; beyond it the CROWN chain (T1) runs with $R=8.74$ ($\varepsilon=10^{-4}$)
to $t_{g}$ ($6{,}944$ to $20{,}908$), where (T2) takes over. Net, draw counts and seeds were fixed
before any draw. The budget is $\eta=.001$ per direction, and the critical values carry $.001$
($\bar t$ from $400{,}000$ null draws; the fourteen levels $w\in\{.05,.045,\dots,.01,.007,.005,
.003,.002,.001\}$ of the corollary from $2{,}000{,}000$ fresh null draws); $\underline\beta$ is
evaluated on a grid of spacing $.001$, and certified values are rounded up. The true $t^{*}$ is the
signal at which the Monte Carlo power of the same model reaches the target ($200{,}000$ draws per
bisection step, delta-method standard error).

\emph{What the probability covers.} Each fixed-rule row of Table~\ref{tab:smoothS} holds with
probability at least $1-\eta-.001=.998$, and each bootstrap-rule result with probability at least
$1-\eta-\eta_{c}=.998$; the two targets $.95$ and $1/2$ of one rule and direction share the same
event. By the union bound all twenty certified cells (two rules, two targets, five directions)
hold together with probability at least $.988$.

\begin{table}[!htbp]
\centering
\caption{Certified against true sample size at power $.95$ and fixed threshold $\bar t$, for the
relaxation certificate (Proposition~S.12, usage-aware charge, one region) and the smoothing
certificate (Proposition~S.12a). Certified $t$ is rounded up and certified
$n^{*}=\lceil4t_{\mathrm{cert}}^{2}\rceil$; true $n^{*}=4t^{*2}$ to the nearest integer. Ratios are
certified over true, from unrounded values. Each smoothing row holds with probability at least
$.998$, the five jointly with probability at least $.994$}
\label{tab:smoothS}
\small
\setlength{\tabcolsep}{4pt}
\begin{tabular}{@{}lcccccccc@{}}
\toprule
& \multicolumn{2}{c}{true} & \multicolumn{3}{c}{relaxation} & \multicolumn{3}{c}{smoothing} \\
\cmidrule(lr){2-3}\cmidrule(lr){4-6}\cmidrule(l){7-9}
direction & $t^{*}$ (SE) & $n^{*}$ & $t_{\mathrm{cert}}$ & $n^{*}$ & ratio & $t_{\mathrm{cert}}$ & $n^{*}$ & ratio \\
\midrule
step & 4.27 (0.006) & 73 & 30.78 & 3{,}790 & 51.9 & 4.33 & 75 & 1.02 \\
diffuse & 5.10 (0.006) & 104 & 32.83 & 4{,}311 & 41.4 & 5.16 & 107 & 1.02 \\
stripe & 5.55 (0.006) & 123 & 38.34 & 5{,}881 & 47.8 & 5.61 & 126 & 1.02 \\
saddle & 7.16 (0.007) & 205 & 44.84 & 8{,}044 & 39.3 & 7.22 & 209 & 1.02 \\
random & 13.53 (0.013) & 732 & 58.99 & 13{,}919 & 19.0 & 13.71 & 751 & 1.03 \\
\bottomrule
\end{tabular}
\end{table}

The smoothing certificate proves power $.95$ at $1.02$--$1.03$ times the true sample size, against
$19$--$52$ for the relaxation certificate; moving the true value by two standard errors leaves every
ratio in $[1.01,1.03]$. For power one half the certified sample sizes are step $22$ (true $21$),
diffuse $39$ ($37$), stripe $46$ ($45$), saddle $92$ ($88$) and random $325$ ($313$), ratios
$1.03$--$1.04$. For the rank rule (Corollary~\ref{cor:smoothB}) they are, at power $.95$, step $79$
(true $74$), diffuse $112$ ($106$), stripe $132$ ($125$), saddle $221$ ($209$) and random $794$
($746$), ratios $1.05$--$1.06$, and at power one half $24$ ($22$), $42$ ($38$), $50$ ($45$), $97$
($89$) and $347$ ($316$), ratios $1.08$--$1.10$; the true values are from the Dufour form with $W$
evaluated against the $2{,}000{,}000$ null draws. Nothing was retrained, and no bound on $S$ was
needed at the signal strengths that matter.

\emph{The relaxation certificate cannot be tightened with current tools:} more than $2{,}000$ of
the $3{,}200$ rectifier and maximum units are undecided at the signal strengths that matter, and every route we
measured (tighter relaxations, oracle intermediate bounds, Lipschitz constants) stays at about
ten times the true sample size or worse; the runs are in the deposit. The smoothing certificate
shows that a tight certificate can nevertheless be computed from frozen weights, within a
Gaussian model of the map.

\section{One corrupted record at small $n$: theory}\label{sec:corrupt}

A logistic model fitted to $30$--$50$ patients, one of whose records carries a data-entry
error, raises a question the rest of this Online Resource 1 does not address: does a
goodness-of-fit test reject \emph{because of the one bad record}? This section gives the
theory; Section~\ref{sec:studies} gives the two studies.

\subsection{Setting and the contamination model}

Data $D=\{(x_{i},y_{i})\}_{i=1}^{n}$ have binary $y_{i}$. A \emph{base goodness-of-fit test}
$\varphi$ maps any dataset $D'$ to a p-value $p(D',U)$, where $U$ is the test's own
randomness (the parametric bootstrap draws; for an asymptotic test $U$ is absent). The base
test rejects at level $\alpha$ when $p(D',U)\le\alpha$.

\begin{assumptionA0}[Permutation invariance]
The law of $p(D',U)$ depends on $D'$ only through the multiset of its records.
\end{assumptionA0}
A0 holds for maximum-likelihood fits and for rank-based maps with continuous covariates,
where ties occur with probability zero.

\begin{definition}[Clean size]
$\mathrm{size}_{m}(\varphi)=\Prob\{p(D_{c},U)\le\alpha\}$, where $D_{c}$ consists of $m$
records drawn from a law under which the logistic model is correctly specified and $U$ is
independent of $D_{c}$.
\end{definition}

\begin{definition}[Contamination model $\mathcal C_{k}$, insertion]
$D$ consists of $n-k$ i.i.d.\ clean records and $k$ \emph{arbitrary} records, with any
values at any positions, possibly chosen as a function of the clean records. No
distributional assumption is made about the arbitrary records.
\end{definition}

Two conditions matter.
\begin{itemize}
\item \emph{The adversary does not see the test's randomness $U$.} A fixed seed that the
adversary can anticipate breaks the argument.
\item \emph{Insertion, not value-dependent replacement.} If \emph{which} record gets
corrupted depends on its own values (slipped decimals on large values; an adversary
overwriting the record that best supports the model), the remaining $n-k$ records are a
selected sample, not i.i.d., and the bound can fail. The precise condition is that the index
$r$ of the corrupted record be independent of \emph{all} $n$ records' original values;
uniformly-at-random corruption qualifies, while independence of the target's own values alone
does not. The corrupted \emph{values} may depend on anything except $U$.
\end{itemize}

\emph{The leave-one-out unanimity rule.} With independent draws $U_{1},\dots,U_{n}$,
\[
\varphi_{\mathrm{LOO}}(D)=\ind{\max_{1\le i\le n}p(D\setminus i,\,U_{i})\le\alpha}:
\]
reject the model only if the base test rejects after deleting each record in turn. The rule
never needs to know which record, if any, is corrupted.

\subsection{Robustness of the level}

\begin{sres}{Theorem}{S.14}[Single-record robustness; an intersection--union argument]\settag{S.14}\label{thm:iut}
Under $\mathcal C_{1}$ and A0,
$\Prob\{\varphi_{\mathrm{LOO}}(D)=1\}\le\mathrm{size}_{n-1}(\varphi)$, whatever the corrupted
record contains and wherever it is inserted. When the corruption instead overwrites an existing
record, the same bound holds provided the index of the overwritten record is independent of all
$n$ records' original values, as with a uniformly chosen record (the second condition above).
\end{sres}

\begin{proof}
Let $r$ be the index of the arbitrary record, so $D\setminus r=D_{c}$ as a multiset. If
$\varphi_{\mathrm{LOO}}(D)=1$ then $\max_{i}p(D\setminus i,U_{i})\le\alpha$, and in
particular $p(D\setminus r,U_{r})\le\alpha$. So
$\{\varphi_{\mathrm{LOO}}(D)=1\}\subseteq\{p(D_{c},U_{r})\le\alpha\}$. The index $r$ is a
function of the data alone and the $U_{i}$ are i.i.d.\ and independent of the data, so
$U_{r}$ has the law of $U$ and is independent of $D_{c}$. Under insertion, $D_{c}$ is an
i.i.d.\ clean sample of size $n-1$ by definition of $\mathcal C_{1}$, whatever $r$ is. Under
overwriting, $D_{c}$ is the original sample less record $r$, and it is an i.i.d.\ clean sample
of size $n-1$ when $r$ is independent of all $n$ original records. In either case, by A0 the
probability of the right-hand event is $\mathrm{size}_{n-1}(\varphi)$.
\end{proof}

The inclusion is pointwise, so nothing about the corrupted record enters: a slipped decimal,
a sign error, a mis-keyed outcome, or a record chosen adversarially to make the test reject
are all covered. Theorem~\ref{thm:iut} is the intersection--union argument of
\citet{berger1982}: the null ``some record is corrupted'' is the union over $r$ of
``$D\setminus r$ is clean''. The proof extends it slightly, to an inserted record at a position
chosen from the clean data. It also holds with one common $U$ in place of independent $U_{i}$;
only blindness to $U$ is needed.

\begin{remark}[The unavoidable price]\label{rem:indist}
Let $\mathcal M_{1}$ be any alternative in which $n-1$ records follow the model and one
record does not, a genuine misfit carried by a single patient. For every test $\psi$,
\[
\sup_{\text{law in }\mathcal M_{1}}\Prob\{\psi=1\}\ \le\
\sup_{\text{law in }\mathcal C_{1}}\Prob\{\psi=1\}.
\]
Indeed every data law in $\mathcal M_{1}$ is also a data law in $\mathcal C_{1}$, which allows
the single record any law. So a test whose level is at most $\alpha$ under one arbitrary record
has power at most $\alpha$ against every misfit carried by one record: a genuine outlying patient
and a data-entry error cannot be told apart from the data. This is the finite-sample, one-record
form of the identifiability breakdown function of \citet[eq.~(2.5)]{he1990}.
\end{remark}

\begin{sres}{Theorem}{S.15}[Consistency is preserved]\settag{S.15}\label{thm:looconsistent}
Let \ref{a:sampling}--\ref{a:certificate} hold for DeepGOF-1 at a fixed alternative with
certificate $\kappa=\Sinf(\mu/s)>0$; for $p>2$ covariates let the population axis criterion
have a strict gap between its second and third values; let \ref{a:compact} hold; and assume
positive population cell masses $d_{c}>0$ and a nonsingular $\E[\tox\tox^{\top}]$. Fix $B$.
Then $\Prob\{\max_{i}p(D\setminus i,U_{i})=1/(B+1)\}\to1$, so
$\Prob\{\varphi_{\mathrm{LOO}}=1\}\to1$ at every attainable $\alpha\ge1/(B+1)$.
\end{sres}

(Every p-value is at least $1/(B+1)$, so ``$\max_{i}p\to0$'' would be false.)

\emph{A perturbation step, used here and for Theorem~\ref{thm:mild}.} Let $D'$ differ from $D$
by deleting, or replacing, one record, and let \ref{a:compact} hold. (P1) With probability tending
to one the log-likelihood has Hessian eigenvalues at least $c_{0}n$ on a fixed ball around
$\beta^{*}$, and the one-record change moves its gradient at $\hat\beta$ by at most
$\sup_{\mathcal K}\norm{\tox}$, so $\norm{\hat\beta(D')-\hat\beta(D)}\le C/n$, uniformly over the
changed record. (P2) The rank bins move each of the $K-1$ boundaries per axis by at most two
ranks, so at most $4(K-1)+1$ records change cell or residual, by at most $1$ each; the fitted risks
move by $C/n$ by (P1); the denominators are at least $c_{1}n$ and change by $O(1)$. Hence
$\norm{m(D')-m(D)}\le Cn^{-1/2}$, and $|\tilde S(m(D'))-\tilde S(m(D))|\le LCn^{-1/2}$. (P3) With
two covariates the axes are fixed (the code orders them by model position); for $p>2$, (P1) moves
every $|\hat\beta_{j}|\hat\sigma_{j}$ by $O(1/n)$, and a strict gap between the second and third
population criteria keeps the selected pair with probability tending to one.

\begin{proof}[Proof sketch]
Write $T_{i}=\tilde S(m(D\setminus i))$ and $T^{*}_{i,b}$ for its bootstrap scores; it suffices
that, with probability tending to one, $\min_{i}T_{i}>\max_{i,b}T^{*}_{i,b}$ and no refit fails.
By (P1)--(P3), uniformly in $i$, $\min_{i}T_{i}\ge\tilde S(m(D))-LCn^{-1/2}$. By Step~D of the proof
of Theorem~\ref{thm:consistency}, $m(D)=\sqrt n\,\mu+o_{P}(\sqrt n)$, and by
Lemma~\ref{lem:recession}(b) and the homogeneity and Lipschitz property of $\Sinf$,
$\tilde S(m(D))\ge\sqrt n\,\kappa-o_{P}(\sqrt n)$; so $\min_{i}T_{i}\ge\tfrac12\kappa\sqrt n$. For
the bootstrap, fix $(i,b)$: by Hoeffding, each raw cell sum and each coordinate of the bootstrap
score is at most $n^{3/4}$ except with probability $Ce^{-c\sqrt n}$; on that event strong concavity
gives a refit within $Cn^{-1/4}$ of $\hat\beta_{-i}$, so $\norm{m^{*}}\le Cn^{1/4}$ and
$T^{*}_{i,b}<\tfrac14\kappa\sqrt n$, and no refit fails. The union bound over the $nB$ pairs costs
$nB\,Ce^{-c\sqrt n}\to0$, and a refit moved by $O(n^{-1/4})$ keeps the axes by the gap of (P3).
\end{proof}

\subsection{Why the base network needs protection: the certificate cuts both ways}

\begin{sres}{Proposition}{S.16}[Single-record breakdown of the unmodified network]\settag{S.16}\label{prop:spike}
Let $e_{j}$ be the $j$-th cell direction and $\tilde S(z)=S((z-a)/s)$ the deployed score. If
$\Sinf(e_{j}/s)>0$, then for any fixed remainder $z$ of the map,
$\tilde S(z+t\,e_{j})\to\infty$ as $t\to\infty$.
\end{sres}

\begin{proof}
By Lemma~\ref{lem:recession}(b), $S\ge\Sinf-c$. $\Sinf$ is positively homogeneous and
$L$-Lipschitz, so
$\tilde S(z+te_{j})\ge\Sinf\bigl((z-a)/s+te_{j}/s\bigr)-c
\ge t\,\Sinf(e_{j}/s)-L\norm{(z-a)/s}-c\to\infty$.
\end{proof}

\emph{Scope.} Proposition~\ref{prop:spike} concerns the map. Applied to the whole test it
leaves out two effects: the refit pulls $\hat\pi$ toward the corrupted record, and the
bootstrap reproduces a spike with probability $\hat p$. The first study below measures the
net effect, and the spike does break the test under a covariate slip.

\emph{Measured on the shipped weights.} $\Sinf(+e_{j}/s)\in[0.12,1.07]$ and
$\Sinf(-e_{j}/s)\in[0.18,0.84]$: positive for all $36$ cells and both signs. A record alone in
its cell with fitted risk $p$ contributes $\sqrt{(1-p)/p}$ standardized units ($10$ at $p=.01$),
and on null maps a spike of $10$ is rejected $91\%$ of the time (median over cells), a spike of
$20$ always. At $n\le50$, with about $1.4$ patients per cell, one record routinely is a cell. The
recession function that certifies consistency (Theorem~\ref{thm:consistency}) therefore also
certifies this fragility.

\subsection{A covariate-only deletion rule}

The first study (\S\ref{sec:studyone}) showed that only \emph{covariate} corruption, a
leverage point, breaks the tests, and that unanimity over all deletions costs most of the
power. Both point to a rule that deletes at most one record, chosen from the covariates
alone.

Work conditionally on the observed covariate matrix $X$, as
Proposition~\ref{prop:fsbound} does. $\mathrm{size}_{m}(\varphi\mid X')$ is the conditional
probability that the base test rejects a clean sample whose covariates are $X'$, with $m$
rows.

\begin{definition}[Covariate contamination $\mathcal C^{x}_{1}$]
For one record $r$, the observed covariate $\tilde x_{r}$ is arbitrary, and so is $y_{r}$.
Precisely: $r$, $\tilde x_{r}$ and $y_{r}$ are measurable functions of the original
covariates $x_{1},\dots,x_{n}$ and of randomness independent of all outcomes and of $U$. In
particular the mechanism may depend on all the covariates but not on the clean outcomes
$y_{i}$, $i\ne r$. (A data-entry error does not look at the other patients' outcomes; an
adversary who does is excluded here, and Theorem~\ref{thm:iut} covers that case.)
\end{definition}

\emph{The rule.} $\rho:\R^{n\times d}\to\{0,1,\dots,n\}$ is any covariate-only deletion map,
where $0$ means ``delete nothing''. The trimmed test is
$\varphi_{\rho}(D)=\varphi\bigl(D\setminus\rho(X)\bigr)$, calibrated by the base test's own
bootstrap on the trimmed data.

\begin{sres}{Theorem}{S.17}[Exact protection when the bad record is caught]\settag{S.17}\label{thm:trim}
Under $\mathcal C^{x}_{1}$,
\[
\Prob\{\varphi_{\rho}=1\}\ \le\ \E\bigl[\mathrm{size}_{n-1}(\varphi\mid X\setminus r)\bigr]
+\Prob\{\rho(X)\ne r\}
\ \le\ \alpha+\E[\varepsilon+\eta]+\Prob\{\rho(X)\ne r\},
\]
where the last step applies Proposition~\ref{prop:fsbound} conditionally on the clean design.
Here $X\setminus r$ is a \emph{selected} design: the index $r$ may depend on all the
covariates, so $X\setminus r$ need not have the law of $n-1$ i.i.d.\ clean covariate rows,
and both expectations are taken over the law of $X\setminus r$ induced by the corruption
mechanism, not over i.i.d.\ designs. (The conditional terms $\varepsilon$ and $\eta$ are
evaluated at the realized $X\setminus r$.) With no contamination, $\Prob\{\varphi_{\rho}=1\mid X\}=\mathrm{size}(\varphi\mid
X\setminus\rho(X))$: the rule leaves the conditional validity of the base test untouched.
\end{sres}

\begin{proof}
Condition on the original covariates, the observed $X$ and $r$ (not on $y_{r}$). By the
mechanism, the clean outcomes are still independent with
$y_{i}\sim\mathrm{Bern}(\pi_{\beta_{0}}(x_{i}))$, $i\ne r$. The event $\{\rho(X)=r\}$ is
determined by what is conditioned on, and on it $D\setminus\rho(X)$ is exactly a clean sample
with covariates $X\setminus r$, so the conditional rejection probability is
$\mathrm{size}_{n-1}(\varphi\mid X\setminus r)$. Off it, bound the rejection probability by
one. Taking expectations gives the first claim. For the second, with no contamination
$\rho(X)$ is a function of $X$ and the outcomes are clean given $X$.
\end{proof}

A $\sup_{X'}$ form of the first term can be vacuous, because near-separated designs have
conditional size close to one; the bound is therefore stated in expectation.

\begin{sres}{Corollary}{S.18}[The miss probability is a covariate-only quantity]\settag{S.18}\label{cor:miss}
For any covariate-only rule $\rho$ and any corruption mechanism that acts on the covariates,
the term $\Prob\{\rho(X)\ne r\}$ of Theorem~\ref{thm:trim} depends only on the law of the
clean covariates and on the mechanism. It does not depend on the outcomes, the model or the
test, and it is computed exactly, up to Monte Carlo error, by simulating covariates alone.
\end{sres}

\begin{proof}
$\{\rho(X)\ne r\}$ is an event of $X$ alone, and under covariate corruption $X$ is generated
by the clean covariate law and the mechanism.
\end{proof}

Take for $\rho$ the rule that deletes the record with the largest robust outlyingness
\[
o_{i}=\max_{j}|x_{ij}-\mathrm{med}_{j}|/(1.4826\,\mathrm{MAD}_{j})
\]
(sample median and MAD) if it exceeds $c=4.5$. With standard normal covariates, $d=2$, and a decimal slip
$x_{r1}\leftarrow10x_{r1}$ on a uniformly chosen record, Table~\ref{tab:miss} gives the
miss probabilities ($400{,}000$ replicates).

\begin{table}[!htbp]
\centering
\caption{Miss probability of the covariate-only rule (Corollary~\ref{cor:miss}), overall and
split by the slipped covariate $x$: gross, $|x|>.45$; mild, $|x|\le.45$.}
\label{tab:miss}
\begin{tabular}{@{}lcccc@{}}
\toprule
$n$ & clean data touched & slip missed, overall & missed, gross & missed, mild \\
\midrule
30 & .032 & .355 & .049 & .928 \\
50 & .019 & .353 & .037 & .943 \\
\bottomrule
\end{tabular}
\end{table}

So Theorem~\ref{thm:trim} gives a \emph{finite-sample} bound for gross slips, size at most
$\alpha+\E[\varepsilon+\eta]+.04$ to $.05$. Mild slips are rarely deleted: a value inside the
bulk is not detectable from the covariates, and Theorem~\ref{thm:mild} is what covers it.

\begin{sres}{Theorem}{S.19}[Mild errors are asymptotically harmless for the base test]\settag{S.19}\label{thm:mild}
Assume \ref{a:sampling}, \ref{a:qmle}, \ref{a:axes}, \ref{a:atomless} and \ref{a:compact}; a
nonsingular $\mathcal I$; $d_{c}>0$ for every cell; and an adversary blind to $U$. Let one
record be replaced by an arbitrary record whose covariate lies in $\mathcal K$ and whose
outcome is arbitrary. Then the size of the deployed (base) test at fixed $B$ converges to the
same limit as on clean data, $\lfloor\alpha(B+1)\rfloor/(B+1)$.
\end{sres}

\begin{proof}
Couple the corrupted dataset $D'$ with the clean dataset $D$ that differs from it in that one
record. The perturbation step before the proof of Theorem~\ref{thm:looconsistent} applies to
replacement (the strict gap of (P3) is supplied by \ref{a:axes}), so the axes agree with
probability tending to one, $\norm{\hat\beta(D')-\hat\beta(D)}\le C/n$,
$\norm{m(D')-m(D)}\le Cn^{-1/2}$ and $T(D')-T(D)=o_{P}(1)$. \emph{The bootstrap, by direct coupling:} generate both bootstrap
samples from common uniforms $u^{*}_{j}$ for $j\ne r$. Their success probabilities differ by
at most $C/n$ for $j\ne r$, so the number of records whose $y^{*}$ differ is $O_{P}(1)$, plus
the one replaced row; the two bootstrap datasets therefore differ in $O_{P}(1)$ records. By
the same concavity argument their refits differ by $O_{P}(1/n)$, their maps by
$O_{P}(n^{-1/2})$ and their scores by $o_{P}(1)$, given the data. Hence
$|P^{*\prime}(T^{*}\le t)-\Probs(T^{*}\le t)|\le\Probs(|\Delta T^{*}|>\delta)
+\sup_{t}[F^{*}(t+\delta)-F^{*}(t-\delta)]\to0$, using Proposition~\ref{prop:asyvalid} for
the clean $D$ and the continuity of $F_{0}$ (\ref{a:atomless}). Every bound is uniform over
the replaced record $(x',y')\in\mathcal K\times\{0,1\}$, so the argument also covers
value-dependent or adversarial replacement (blind to $U$). \emph{The p-value:} Step~6 of the
proof of Theorem~\ref{thm:localS} (with $h=0$) then gives the same limit for
$\Prob(p\le k/(B+1))$ on $D'$ as on $D$, namely $k/(B+1)$. An $o_{P}(1)$ perturbation of the
observed score and a vanishing perturbation of the bootstrap law do not change a limit whose
law is continuous.
\end{proof}

No rate is claimed for Theorem~\ref{thm:mild}: conditional on $X$ the score has no density and
the critical value moves by $O(n^{-1/2})$. The finite-sample size under mild slips is measured
instead, in Study~2.

\emph{Together.} Gross covariate errors are deleted with probability about $95$--$96\%$ (normal
covariates, the slip mechanism above), and Theorem~\ref{thm:trim}, which does not need
\ref{a:compact}, bounds the size in finite samples; over all decimal slips the miss term is
$.355$. Mild errors stay in, and Theorem~\ref{thm:mild} makes them asymptotically harmless for the
base test; the trimmed-test version needs a two-record extension, not written out, and Study~2
measures its size directly (DeepGOF-1 $.047$--$.055$). Clean data are touched in about $2$--$3\%$
of datasets at $n\le50$ (Table~\ref{tab:miss}), which never affects validity and costs one
record of sample size.

\emph{Prior work.} The problem is stated by \citet[\S3]{ylvisaker1977} for $\chi^{2}$ goodness of
fit: with low resistance to rejection, a few unfortunate errors can cause a rejection. Ylvisaker
uses resistance only to help choose the critical value and builds no robust test. His test
resistance and the sample breakdown points of \citet{zhang1996} use replacement rather than
deletion.

\section{One corrupted record: the two studies}\label{sec:studies}
\subsection{Study 1: is the problem real, and does unanimity repair it?}\label{sec:studyone}

\emph{Design}:
\begin{enumerate}
\item design: $x_{1},x_{2}\sim N(0,1)$ independent, true $\eta=-0.5+x_{1}+0.6x_{2}+C(x_{1}^{2}-1)$,
fitted model $y\sim x_{1}+x_{2}$;
\item corruption of one uniformly chosen record after its outcome is generated: decimal
($x_{1}\leftarrow10x_{1}$), sign ($x_{1}\leftarrow-x_{1}$) or outcome ($y\leftarrow1-y$);
\item cells: $n\in\{30,50\}$; null ($C=0$) with no, decimal, sign and outcome corruption,
$1{,}000$ replicates each; $C\in\{0.5,1.0,1.5\}$ with no or decimal corruption, $300$ replicates
each; $11{,}600$ datasets in all;
\item tests: Hosmer--Lemeshow, Stukel's joint score test and DeepGOF-1 with the shipped weights, map axes fixed on
the observed fit and $B=99$ (the leave-one-out rule needs $n$ further bootstraps per dataset);
Firth fallback as in the benchmark, and a deletion whose fit fails gives $p=1$;
\item rules: a test \emph{misbehaves} if its size under contamination exceeds $.075$ (about $3.6$
standard errors above $.05$); the leave-one-out rule passes if, under each contamination, its size
does not exceed the base test's clean size by more than $2$ standard errors.
\end{enumerate}

\emph{Results.} The sizes are Table~5 of the paper. A slipped decimal, a leverage point, breaks
Stukel's test and DeepGOF-1 (up to $13$ standard errors above $.05$ at $n=50$) and the projection
test, computed on the same datasets (up to $18$ standard errors); Hosmer--Lemeshow is immune, because
the bad record lands in an extreme risk decile and is diluted there; sign errors and flipped outcomes
break none of the three original tests, and the projection test only marginally, under a sign error
at $n=30$ ($.079$). Theorem~\ref{thm:iut} holds in all $24$ cells: the size of the
leave-one-out rule lies between $.000$ and $.012$ and never exceeds the clean base size. The rule
is correct but far too conservative: at $n=50$, $C=1.5$, on clean data, it lowers power from $.55$
to $.26$ for DeepGOF-1, $.47$ to $.24$ for Stukel's test and $.29$ to $.08$ for Hosmer--Lemeshow.

\subsection{Study 2: the covariate-only deletion rule, confirmed on fresh seeds}

\emph{Design}. The rule, suggested by the data of Study~1, is tested here on fresh seeds. It
is that of Section~\ref{sec:corrupt}: if $\max_{i}o_{i}>4.5$, delete the one record with the
largest $o_{i}$, otherwise nothing, and run the base test on the rest with its own calibration;
it never looks at $y$. The design is that of Study~1, with DeepGOF-1 at the deployed $B=199$
($11{,}600$ datasets). The checks: (1) protection, the trimmed size under a decimal slip within $2$
standard errors of the base test's clean size for every test that failed in Study~1; (2) no harm,
trimmed and base rejection rates within $2$ standard errors on clean data, under the null and at
$C>0$; (3) the mechanism; (4) sign and outcome corruption, reported with no claim.

\emph{Results.} (1) Every cell passes; the sizes are Table~6 of the paper. Hosmer--Lemeshow was
not broken and is unchanged ($.035\to.037$ at $n=30$, $.044\to.042$ at $n=50$). (2) All $24$ cells
agree within $2$ paired standard errors; for example DeepGOF-1 at $n=50$ gives $.055/.055$,
$.073/.077$, $.307/.307$ and $.537/.540$ (trimmed/base) under the null and at $C=.5,1,1.5$, so
the rule costs no power. (3) The corrupted record is deleted in $65\%$ of decimal-corrupted
datasets and nothing is deleted in $34\%$, the mild slips of Corollary~\ref{cor:miss}. Among null
datasets whose slip was not deleted, the size is $.048/.053$ for DeepGOF-1, $.026/.047$ for
Hosmer--Lemeshow and $.077/.124$ for Stukel's test ($n=30/50$). These sizes are conditional on a
covariate-selected event, at $n\le50$ and with normal covariates outside \ref{a:compact}, so they
are consistent with Theorem~\ref{thm:mild} without testing it; Stukel's test is sensitive even to
mild slips, because its added covariates are squares of the linear predictor. (4) Sign and outcome
corruption are harmless before and after trimming (outcome, $n=50$: DeepGOF-1 $.051\to.050$).

\emph{Verdict.} One corrupted record at $n\le50$ is a leverage problem. The covariate-only rule
removes it for DeepGOF-1 at no measurable power cost; Stukel's test passes the check but stays
above the nominal band $[.035,.065]$ of the level study. Leave-one-out unanimity is correct but not
usable at these $n$, and remains the guarantee against adversarial contamination.

\section{The deletion rule with more, or non-normal, covariates}\label{sec:touch}

The rule was designed and tested with two standard normal covariates. Because it takes the
maximum of $nd$ robust $z$-scores, it fires on clean data more often as $d$ grows, and because
the scores are standardised by a sample MAD, it fires far more often than the normal tail
suggests. By Corollary~\ref{cor:miss} both the probability that it fires on clean data (the touch
rate) and the probability that it misses a slipped record are functions of the covariate law
alone, so they were computed from covariates only, with $20{,}000$ replicates per cell (binomial
standard errors at most $.004$). The $d=2$ cells reproduce Table~\ref{tab:miss}.

\begin{table}[!htbp]
\centering
\caption{Touch rate of the rule ($c=4.5$) on clean data, by covariate law, number of
covariates $d$ and sample size $n$ ($20{,}000$ replicates per cell; standard errors at most $.004$,
and below $.001$ for the entries at or near $1$). ``liu2'' is the three-covariate design of Liu et
al.'s Setting~2 (uniform, normal and $t_{4}$).}
\label{tab:touch}
\begin{tabular}{@{}llccccc@{}}
\toprule
covariate law & $d$ & $n=30$ & $50$ & $100$ & $200$ & $500$ \\
\midrule
normal & 2  & .032 & .019 & .012 & .009 & .011 \\
normal & 5  & .079 & .047 & .030 & .022 & .030 \\
normal & 10 & .152 & .094 & .053 & .045 & .056 \\
uniform $(-3,3)$ & 2, 5, 10 & $\le.001$ & .000 & .000 & .000 & .000 \\
$t_{4}$ & 2  & .44 & .58 & .80 & .96 & 1.00 \\
$t_{4}$ & 10 & .95 & .99 & 1.00 & 1.00 & 1.00 \\
liu2 & 3 & .27 & .36 & .56 & .80 & .98 \\
lognormal & 2 & .97 & 1.00 & 1.00 & 1.00 & 1.00 \\
Bernoulli$(.5)$ & 2 & .98 & .99 & .99 & 1.00 & 1.00 \\
\bottomrule
\end{tabular}
\end{table}

\emph{Normal covariates.} The touch rate grows roughly in proportion to $d$ at small $n$ ($.15$
at $d=10$, $n=30$; $.09$ at $n=50$) and from $n=100$ on stays at or below about $.05$ at every $d$
(Table~\ref{tab:touch}). The naive union approximation $1-(1-2\bar\Phi(c))^{nd}$ under-predicts
these rates by a factor of $40$ to $75$ at $n=30$, because the driver is the sampling variability of
the MAD, not the multiplicity of the tail; for $n\le50$ the table should be used. An approximation
that integrates over the MAD, and its errors, are in the deposit (\texttt{theory/small\_n/touch\_rate/}).

\emph{Bounded and binary covariates.} A uniform covariate cannot produce a robust score above
about $1.4$, so the rule is inert on it. A binary covariate has MAD zero whenever its sample is
unbalanced, which makes every minority record infinitely outlying. The rule is therefore for
continuous covariates only.

\emph{Heavy-tailed and skewed covariates.} With a $t_{4}$ or lognormal covariate the rule fires on
most or all clean datasets (Table~\ref{tab:touch}); the level is still protected, since
Theorem~\ref{thm:trim} needs only that the deletion be a function of $X$. What erodes is the
protection: the slip competes with genuine tail records, so its miss probability rises with $n$
($t_{4}$, $d=2$: $.41$ at $n=30$, $.62$ at $n=500$; lognormal, $d=10$: $.71$ to $.90$; liu2: $.36$
to $.58$). Such a covariate should be transformed first, or the threshold set from its own law as
the $95$th percentile of $\max_{i}o_{i}$ on clean designs, which Corollary~\ref{cor:miss} lets one
simulate from covariates alone.

\emph{The cost of a false fire.} With no contamination, Theorem~\ref{thm:trim} gives
$\Prob\{\varphi_{\rho}=1\mid X\}=\mathrm{size}(\varphi\mid X\setminus\rho(X))$ at every $d$, so
the only cost is power at $n-1$, and of a high-leverage record. A paired check at $d=10$
($n=100$, ten standard normal covariates,
$\eta=-0.5+x_{1}+0.6x_{2}+0.3x_{3}-0.3x_{4}+(x_{1}^{2}-1)$, fitted model with all ten; $400$
datasets; on each the base test run on all $n$ rows and on the $n-1$ rows left after deleting
the most outlying record, deleted unconditionally) gave DeepGOF-1 ($B=199$) $.443\to.460$,
difference $+.018$ with paired standard error $.018$ ($22{:}29$ discordant pairs, exact McNemar
$p=.40$). The check resolves a loss of about $.035$, not one of $.01$; it bounds the power cost of
one deletion for DeepGOF-1 at $.017$ at the $95\%$ level, so the expected loss of a false fire
under normal covariates is below $.004$ everywhere in Table~\ref{tab:touch}.

\emph{A threshold for high $d$ (a recommendation; the rule itself is unchanged).} The empirical
$95$th percentile of $\max_{i}o_{i}$ on clean normal designs holds the touch rate at $.04$ to
$.05$ by construction (Table~\ref{tab:thresh}); a Bonferroni threshold does not, for the same
reason as the union approximation. At $d=10$, $n=30$, raising the threshold from $4.5$ to $5.2$
lowers the touch rate from $.15$ to $.05$ and raises the miss probability on gross slips
($|x|>.45$) from $.05$ to $.10$; at $n=50$ the threshold $4.8$ moves the touch rate from $.09$ to
$.05$ and the gross miss from $.04$ to $.06$. For $d\le5$, and for every $d$ from $n=100$ on, the
threshold $4.5$ is already within $.03$ of the target.

\begin{table}[!htbp]
\caption{Threshold holding the clean touch rate at $5\%$ (the empirical $95$th percentile of
$\max_{i}o_{i}$ under standard normal covariates, rounded up to one decimal), and the miss
probability on gross decimal slips at that threshold against $4.5$.}
\label{tab:thresh}
\centering
\begin{tabular}{@{}lccccc@{}}
\toprule
$d$ & $n=30$ & $50$ & $100$ & $200$ & $500$ \\
\midrule
2  & 4.3 & 4.1 & 4.0 & 4.1 & 4.2 \\
5  & 4.8 & 4.5 & 4.4 & 4.3 & 4.4 \\
10 & 5.2 & 4.8 & 4.6 & 4.5 & 4.6 \\
\midrule
gross miss at $d=10$, adjusted & .102 & .060 & .033 & .019 & .019 \\
gross miss at $d=10$, $c=4.5$   & .054 & .040 & .026 & .019 & .013 \\
\bottomrule
\end{tabular}
\end{table}

\section{The Neyman--Pearson optimality gap}\label{sec:np}

The network was trained to separate null maps from alternative maps drawn from a stated prior,
so the best it can do is the most powerful test of the null against the prior-averaged
alternative. This section measures how close it gets. The result is a diagnostic, not an
optimality property of the test.

\subsection{Setting}

The map is $m\in\R^{36}$ and $P_{0}$ is its law under the null; the null class is bootstrap
replicates, as in training. $\Pi$ is the shipped departure prior of the training corpus and
$P_{1}=\int P_{h}\,d\Pi(h)$ the prior-averaged alternative law. Suppose $P_{1}\ll P_{0}$ with
density ratio $f=dP_{1}/dP_{0}$. A \emph{map-based test} is any measurable
$\psi(m)\in[0,1]$; its level is $\E_{0}\psi$ and its prior-averaged power $\E_{1}\psi$.

$\varphi^{*}=\ind{f>k}$ (randomized on $\{f=k\}$) with $\E_{0}\varphi^{*}=\alpha$ is the
Neyman--Pearson test, and $\varphi_{S}=\ind{S>c}$, randomized on ties so that
$\E_{0}\varphi_{S}=\alpha$ \emph{exactly}, is the score test at the same level; with only
$\E_{0}\varphi_{S}\le\alpha$ the bound below fails. $\varphi_{S}$ is the \emph{oracle} score
test, thresholded at the exact null quantile, not the deployed bootstrap test;
Theorem~\ref{thm:localS} and Proposition~\ref{prop:asyvalid} connect the two. The upper bound
$\E_{1}\varphi^{*}$ holds for every test of level at most $\alpha$ under each null law in the
mixture.

\begin{sres}{Theorem}{S.13}[Neyman--Pearson gap]\settag{S.13}\label{thm:np}
Suppose $P_{1}\ll P_{0}$, with $f=dP_{1}/dP_{0}$, and let $\varphi_{S}=\ind{S>c}$ be
randomized on $\{S=c\}$ so that $\E_{0}\varphi_{S}=\alpha$ \emph{exactly}.
For every map-based test $\psi$ with $\E_{0}\psi\le\alpha$, $\E_{1}\psi\le\E_{1}\varphi^{*}$.
For every constant $b\in\R$,
\[
0\ \le\ \E_{1}\varphi^{*}-\E_{1}\varphi_{S}\ \le\ \E_{0}\bigl|f-e^{S-b}\bigr|.
\]
More sharply, with $\hat f=e^{S-b}$ and $\hat k=e^{c-b}$ the gap equals
$\E_{0}[(\varphi^{*}-\varphi_{S})(f-\hat f)]+\E_{0}[(\varphi^{*}-\varphi_{S})(\hat f-\hat k)]$,
and the second term is at most $0$.
\end{sres}

\begin{proof}
The first claim is the Neyman--Pearson lemma. For the second, $\E_{1}\psi=\E_{0}[\psi f]$ for
any test. Since $\E_{0}\varphi^{*}=\E_{0}\varphi_{S}=\alpha$, for any constant $\hat k$,
\[
\E_{1}\varphi^{*}-\E_{1}\varphi_{S}=\E_{0}[(\varphi^{*}-\varphi_{S})f]
=\E_{0}[(\varphi^{*}-\varphi_{S})(f-\hat f)]+\E_{0}[(\varphi^{*}-\varphi_{S})(\hat f-\hat k)].
\]
Take $\hat k=e^{c-b}$. Then $\varphi_{S}=\ind{\hat f>\hat k}$ up to ties, and pointwise
$(\varphi^{*}-\varphi_{S})(\hat f-\hat k)\le0$: where $\varphi_{S}=1$, $\hat f\ge\hat k$ and
$\varphi^{*}-\varphi_{S}\le0$; where $\varphi_{S}=0$, $\hat f\le\hat k$ and
$\varphi^{*}-\varphi_{S}\ge0$; and where $\varphi_{S}\in(0,1)$ (the tie $S=c$),
$\hat f=\hat k$ and the product is $0$ whatever the randomization. So the second term is at
most $0$. The first is at most $\E_{0}|f-\hat f|$ because $|\varphi^{*}-\varphi_{S}|\le1$.
Nonnegativity of the gap is the first claim.
\end{proof}

\begin{remark}[Reading the bound]\label{rem:npread}
(i) The bound involves only the null law and the two density ratios, and it holds for every
$b$, so it can be minimized over $b$. A network trained by cross-entropy on balanced classes
estimates $\log f$ up to the constant $b$ (the log prior odds), which is why this is the
natural quantity; no claim links the bound to the training loss. (ii) When $f$ is computable
the gap itself can be computed, which is what the measurement below does under a Gaussian
surrogate for the map law. (iii) Optimality here is \emph{prior-averaged} and
\emph{map-based}. It says nothing about a single named alternative, where a test aimed at
that alternative is more powerful than any omnibus test, and nothing about tests that use more than the map;
Corollary~\ref{cor:representation} is that boundary.
\end{remark}

\subsection{Measurement}

Held-out maps were drawn from the shipped prior: $300$ per family, for the $100$ training
families and for $100$ fresh families, with $30{,}000$ bootstrap-null maps, all on fresh
dataset seeds. In the \emph{Gaussian surrogate}, $P_{0}=N(\mu_{0},\Sigma_{0})$ and $P_{1}$ is
the equal-weight mixture of the $N(m_{f},\Sigma_{0})$ over the families, so the exact
Neyman--Pearson test is available ($200{,}000$ draws; standard error $.001$).

\begin{table}[!htbp]
\centering
\caption{Neyman--Pearson gap in the Gaussian surrogate. ``Share'' is DeepGOF-1 power as a
share of the Neyman--Pearson power; the last column is the bound of
Theorem~\ref{thm:np}, $\min_{b}\E_{0}|f-e^{S-b}|$.}
\label{tab:np}
\begin{tabular}{@{}lccccc@{}}
\toprule
families & NP power & DeepGOF-1 power & gap (SE) & share & bound \\
\midrule
training & .273 & .202 & .071 (.001) & 73.9\% & .702 \\
fresh & .289 & .204 & .085 (.001) & 70.5\% & .711 \\
\bottomrule
\end{tabular}
\end{table}

\emph{Reading.} Within the Gaussian surrogate, DeepGOF-1 attains $73.9\%$ (training
families) and $70.5\%$ (fresh families) of the power of the most powerful map-based test for
its prior. The two shares differ by far more than their standard errors (about $.001$), so
there is a small generalization loss to new families. The generic bound is vacuous here
($.70$), because the network's score is not calibrated as a log likelihood ratio; the
measured gap is the usable quantity, which is why Remark~\ref{rem:npread}(ii) matters. On
real (non-surrogate) held-out maps the comparator available is a surrogate
likelihood-ratio statistic, not the Neyman--Pearson test, so no share of the optimum is
claimed there. In that fair out-of-sample comparison, with both tests built from the training
families and scored on the fresh families' maps, the closed-form Gaussian-mixture
likelihood-ratio statistic attains power $.246$ (SE $.004$) and DeepGOF-1 $.225$ (SE
$.003$), a difference of about four standard errors. The level of that closed-form statistic
under the bootstrap has not been checked, and nothing is claimed for it.

\emph{Credit.} The inequality of Theorem~\ref{thm:np} follows in one line from the identity for
the excess type-II error of plug-in Neyman--Pearson classifiers in \citet[proofs of
Propositions~1 and~2, eq.~(5)]{tong2013} and \citet[proof of Proposition~2.4]{zhao2016}, whose level-difference
term vanishes at exact level. What is new is only its use: a measured optimality gap for a frozen
goodness-of-fit statistic against the Neyman--Pearson test for its own published training prior.

\section{Very small samples and wrong links}\label{sec:smallnlink}

The design has two standard normal covariates and linear predictor $\eta=0.25+1.5x_{1}+x_{2}$, about half
events; the fitted model is the logistic regression in $x_{1}$ and $x_{2}$, fitted by a plain \texttt{glm()}.
Null datasets: $2{,}000$ at $n=20$ and $30$ and $1{,}000$ at $n=50$ and $100$. Wrong links: the outcome is
drawn with probability $F(\eta)$ for $F$ the probit, the complementary log-log, and Stukel's generalized
logistic with heavier tails ($\alpha_{1}=\alpha_{2}=-1$) and with asymmetric tails ($\alpha_{1}=0.5$,
$\alpha_{2}=-1$), at $n=20$, $30$, $50$ and $100$. Misfit in the covariates: $\eta+C\{x_{1}^{2}-1\}$ (omitted
quadratic) or $\eta+Cx_{1}x_{2}$ (omitted interaction), $C\in\{0.5,1\}$, at $n=20$, $30$ and $50$. Every
alternative cell has $1{,}000$ datasets. The tests are DeepGOF-1 (shipped network, $B=199$; with two
covariates both readings coincide), the projection test ($B=199$), Stukel's joint score test and its SstBoth
form, GiViTI through \texttt{givitiR}, and the Hosmer--Lemeshow test. Power is at matched level $.05$: each
test is placed at exactly $.05$ on its own null datasets at the same $n$ by the randomized threshold of the
main text, and is evaluated on the datasets where it returns a p-value.

\begin{table}[h]
\centering
\caption{Size at nominal $.05$ on the null datasets; in brackets, the share of datasets with no p-value
where it is not zero. Separated: share of fits that did not converge or had a coefficient above $8$ in
absolute value. Band: $[.035,.065]$.}
\label{tab:smalln_size}
\small
\begin{tabular}{@{}rrcccccc@{}}
\toprule
$n$ & separated & DG-1 & proj. & Stukel J & SstBoth & GiViTI & HL \\
\midrule
20 & .073 & .036 & .034 & .030 & .031 & .093 [0.06] & .018 \\
30 & .008 & .045 & .045 & .039 & .056 & .061 & .027 \\
50 & .000 & .047 & .042 & .048 & .063 & .045 & .045 \\
100 & .000 & .055 & .070 & .051 & .068 & .044 & .056 \\
\bottomrule
\end{tabular}
\end{table}

\begin{table}[h]
\centering
\caption{Matched-level power against misfit in the covariates. Monte Carlo standard error at most $.016$.}
\label{tab:smalln_power}
\small
\begin{tabular}{@{}lrrcccccc@{}}
\toprule
misfit & $C$ & $n$ & DG-1 & proj. & Stukel J & SstBoth & GiViTI & HL \\
\midrule
quadratic & 0.5 & 20 & .063 & .098 & .096 & .093 & .065 & .061 \\
quadratic & 0.5 & 30 & .070 & .107 & .102 & .091 & .092 & .072 \\
quadratic & 0.5 & 50 & .080 & .142 & .104 & .114 & .105 & .060 \\
quadratic & 1.0 & 20 & .124 & .246 & .184 & .195 & .151 & .079 \\
quadratic & 1.0 & 30 & .153 & .259 & .191 & .203 & .184 & .106 \\
quadratic & 1.0 & 50 & .249 & .450 & .269 & .292 & .278 & .115 \\
interaction & 0.5 & 20 & .058 & .078 & .080 & .082 & .058 & .043 \\
interaction & 0.5 & 30 & .060 & .070 & .077 & .067 & .061 & .064 \\
interaction & 0.5 & 50 & .080 & .086 & .065 & .073 & .068 & .052 \\
interaction & 1.0 & 20 & .097 & .122 & .105 & .112 & .081 & .068 \\
interaction & 1.0 & 30 & .116 & .134 & .105 & .105 & .104 & .085 \\
interaction & 1.0 & 50 & .137 & .182 & .168 & .169 & .165 & .071 \\
\bottomrule
\end{tabular}
\end{table}

\begin{table}[h]
\centering
\caption{Matched-level power when the true link is not the logit. Monte Carlo standard error at most $.015$.}
\label{tab:smalln_link}
\small
\begin{tabular}{@{}lrcccccc@{}}
\toprule
true link & $n$ & DG-1 & proj. & Stukel J & SstBoth & GiViTI & HL \\
\midrule
probit & 20 & .026 & .059 & .019 & .021 & .037 & .009 \\
probit & 30 & .068 & .048 & .018 & .017 & .068 & .042 \\
probit & 50 & .048 & .048 & .025 & .027 & .029 & .044 \\
probit & 100 & .032 & .033 & .018 & .020 & .015 & .041 \\
cloglog & 20 & .017 & .044 & .051 & .049 & .065 & .013 \\
cloglog & 30 & .044 & .048 & .046 & .045 & .095 & .026 \\
cloglog & 50 & .061 & .052 & .065 & .071 & .073 & .026 \\
cloglog & 100 & .032 & .045 & .093 & .105 & .093 & .033 \\
Stukel, heavier tails & 20 & .070 & .095 & .114 & .115 & .071 & .073 \\
Stukel, heavier tails & 30 & .061 & .089 & .092 & .078 & .076 & .072 \\
Stukel, heavier tails & 50 & .061 & .071 & .108 & .097 & .099 & .053 \\
Stukel, heavier tails & 100 & .062 & .048 & .132 & .097 & .104 & .054 \\
Stukel, asymmetric & 20 & .053 & .086 & .118 & .125 & .091 & .032 \\
Stukel, asymmetric & 30 & .067 & .082 & .096 & .103 & .108 & .049 \\
Stukel, asymmetric & 50 & .090 & .120 & .154 & .198 & .209 & .061 \\
Stukel, asymmetric & 100 & .097 & .120 & .290 & .350 & .369 & .096 \\
\bottomrule
\end{tabular}
\end{table}

DeepGOF-1 is the only test whose size lies inside the band at every $n$
(Table~\ref{tab:smalln_size}), although several other sizes are within one or two standard errors of
the band's edges. Against misfit in the covariates it has more power than the Hosmer--Lemeshow
test in ten of the twelve cells, and less than the projection test in all twelve and than Stukel's joint
score test in ten; in the other two its lead over Stukel's test is within Monte Carlo error
(Table~\ref{tab:smalln_power}); with only two covariates, both carrying signal, this is the order the local
power comparison of the main text predicts. Against a wrong link DeepGOF-1 and the projection test stay close
to their level (Table~\ref{tab:smalln_link}): a wrong link bends the risk along the linear predictor and
leaves little pattern over the covariate ranks, while Stukel's test and GiViTI look along the linear
predictor and are the tests to use; even they reach appreciable power only for the asymmetric link at
$n\ge50$. The scripts and every per-dataset p-value are in \texttt{theory/smalln\_link/} of the deposit.

\section{A paired comparison with the adaptive-partition test}\label{sec:bagoftS}

The closest rival to a partition learned once is a partition learned on the analyst's own data:
BAGofT \citep{zhang2023} splits the sample, grows a tree on one part to find where the model fails,
and tests there. In the extended benchmark of \citet{liu2024} (Section~6 and Table~3 of the paper)
it was run, because of its cost (\bagsecslo--\bagsecshi\ seconds per test), at a hundred replicates
per power cell and two hundred per null block, on one training corpus. Its level is nominal on
average ($\bagsize$) and its power, $\bagpow$, places it below DeepGOF-1 and above the partition
tests of its own family. Because the two tests saw identical datasets, they can be compared dataset
by dataset (Table~\ref{tab:bagoftS}). Over all \bagpairs\ alternative datasets they disagree
\bagdiscord\ times (only BAGofT rejects, only DeepGOF-1 rejects), so DeepGOF-1 is ahead at an exact
McNemar p-value of \bagmcnemar; allowing for the clustering of the datasets across sample sizes moves
it to \bagmcnemarcl. DeepGOF-1 is ahead at $n=50$ ($p={}$\bagpnfifty), on Settings~1, 3 and~4 and at
the two weakest departures; BAGofT is ahead on Setting~2 alone ($p={}$\bagptwo), the off-index
quadratic on which the two-axis grid carries least information. Level accounts for little of the
difference: at exactly matched level the gap is $\bagmatchedgap$ against $\bagrawgap$ raw (on the one
corpus and the replicates both tests share, which is why it differs slightly from the two rows of
Table~3 of the paper).

\begin{table}[!htbp]
\centering
\caption{DeepGOF-1 against BAGofT \citep{zhang2023} on identical datasets: the paired half of the
comparison whose averages are two rows of Table~3 of the paper. Both tests reject at $p\le.05$;
\emph{discordant} counts the datasets only BAGofT rejects against those only DeepGOF-1 rejects, and
$p$ is McNemar's exact p-value for that pair. The first row is the primary comparison; the rows below
it are subgroups, with p-values Holm-adjusted across the eleven tested (three sample sizes, four
settings, four departure strengths; the deposit lists them all). Setting~1 required a one-line repair
to the released BAGofT code before it would run on a single-covariate model}
\label{tab:bagoftS}
\small
\begin{tabular}{@{}lccccc@{}}
\toprule
comparison & datasets & BAGofT & DeepGOF-1 & discordant & $p$ \\
\midrule
\tablebagoft
\bottomrule
\end{tabular}
\end{table}

\section{Level transfer to a real design: bank failures after 2008}\label{sec:fdicS}

This study asks whether the level of DeepGOF-1 transfers to real covariate geometry. It is not
medical, but a bank-failure model is the same kind of rare-event logistic risk model as a clinical
score, fitted to a few hundred units with few events, which is where the classical tests drift. All
FDIC-insured banks at 2008-12-31 ($8{,}395$ institutions, $397$ failures in 2009--2011, rebuilt from
two keyless FDIC BankFind API calls by a deposited script) are modelled within each state with at
least fifteen failures by a logistic regression on equity, nonperforming assets, return on assets,
brokered deposits and commercial real estate, each scaled by total assets; the primary cell is
Georgia. For each cell we simulate $2{,}000$ null outcome vectors on the \emph{real} design matrix
(Table~\ref{tab:fdicS}). The classical size distortion transfers: the Hosmer--Lemeshow and
Pigeon--Heyse tests exceed $.065$ in the two cells with the fewest events for their size, up to
$\fdicpartmax$. The level of DeepGOF-1 largely transfers too: its size ranges from $.044$ to $.068$,
five of the six cells lie inside the band $[.035,.065]$, and the exception ($.068$) is the cell with
sixteen failures, where the Hosmer--Lemeshow test sits at $.078$. On the observed data DeepGOF-1
rejects no cell (p-values $.175$ to $.905$).

\begin{table}[!htbp]
\centering
\caption{The FDIC transfer study, by state cell: the primary cell, Georgia, first, then the others
ordered by $n$. Size columns: rejection rate at nominal $.05$ on $2{,}000$ nulls simulated on the real
design, for Hosmer--Lemeshow (HL), Pigeon--Heyse (PH), the equal-width variant (HLw) and DeepGOF-1
(DG-1); a test that fails to compute counts as not rejecting. Observed $p$: the adequacy p-values of
the fitted model, with HL and PH combined by taking the smaller of the two}
\label{tab:fdicS}
\small\setlength{\tabcolsep}{4pt}
\begin{tabular}{@{}lrrccccccc@{}}
\toprule
& & & \multicolumn{4}{c}{size} & \multicolumn{3}{c}{observed $p$} \\
\cmidrule(lr){4-7}\cmidrule(lr){8-10}
state & $n$ & events & HL & PH & HLw & DG-1 & HL/PH & HLw & DG-1 \\
\midrule
GA & 334 & 69 & .055 & .049 & .057 & .044 & .115 & .120 & .350 \\
\midrule
IL & 655 & 46 & .088 & .077 & .048 & .047 & .878 & .048 & .785 \\
MN & 431 & 16 & .078 & .070 & .048 & .068 & .994 & .004 & .905 \\
CA & 311 & 34 & .045 & .040 & .048 & .047 & .743 & .161 & .655 \\
FL & 307 & 56 & .048 & .037 & .051 & .045 & .142 & .782 & .175 \\
WA & 97 & 17 & .048 & .036 & .048 & .061 & .319 & .280 & .330 \\
\bottomrule
\end{tabular}
\end{table}

\section{The network against simple statistics of the same map}\label{sec:ablationS}

Does the trained network add anything to the residual map, or would a plain statistic of the same $36$ cells do as
well? On every dataset one set of $B=199$ parametric-bootstrap refits gives five p-values: the shipped network on the
map of the axis rule (the default test), the largest network score over every pair of covariates (the all-pairs
reading), the smaller of these two p-values calibrated exactly over the $B+1$ draws (the combined reading), the sum of
the squared cells of the axis-rule map (a $\chi^{2}$ statistic over a partition of two covariates) and the largest
absolute cell of the same map. Each statistic is placed at exactly level $.05$ on its own nulls of the same design and
$n$ by the randomized threshold of the main text, and the paired difference in matched rejection, network minus each
simple statistic, is taken over the same datasets. Designs: Liu et al.'s Settings 2--4 at $n=50,100,500$ (nulls $500$,
alternatives $250$ per strength $C\in\{.25,.5,.75,1\}$); the design of Section~7 of the main text, ten standard normal
covariates with an omitted interaction or quadratic on the first two ($C\in\{.3,.5,.8\}$, $n=200,500$; nulls $600$,
alternatives $150$); and a pure U-shape in one of five covariates ($C\in\{.3,.6\}$, $n=200,500$; nulls $1{,}000$,
alternatives $400$).

\begin{table}[h]
\centering
\caption{Mean matched-level power per design (each cell weighted equally), and the number of cells in which the
network (axis rule) is ahead of, behind, or level with the sum of squares and the largest cell by more than two paired
standard errors. Last column: range of raw null size over the five statistics and all sample sizes.}
\label{tab:ablation}
\small\setlength{\tabcolsep}{3pt}
\begin{tabular}{@{}lrccccccccc@{}}
\toprule
design & cells & network & all pairs & combined & $\Sigma$ squares & max cell & vs $\Sigma$ sq. & vs max & size \\
 & & (axis rule) & & & & & a/b/l & a/b/l & \\
\midrule
Liu et al.\ Settings 2--4 & 36 & .535 & .612 & .613 & .510 & .365 & 5/5/26 & 20/4/12 & .012--.078 \\
two active among ten & 12 & .452 & .297 & .409 & .359 & .220 & 6/0/6 & 10/0/2 & .037--.067 \\
U-shape, five covariates & 4 & .054 & .510 & .479 & .037 & .046 & 0/0/4 & 0/0/4 & .041--.063 \\
\bottomrule
\end{tabular}
\end{table}

\begin{table}[h]
\centering
\caption{Per cell, the designs with many covariates: matched-level power of the five statistics, the paired difference
network minus sum of squares (standard error), and the share of datasets in which the axis rule selected a pair
containing the misfit covariate(s).}
\label{tab:ablationcells}
\small\setlength{\tabcolsep}{3pt}
\begin{tabular}{@{}llrrcccccrc@{}}
\toprule
covariates & departure & $C$ & $n$ & network & all pairs & combined & $\Sigma$ sq. & max & net $-$ $\Sigma$ sq. & axes hit \\
\midrule
ten & interaction & 0.3 & 200 & .051 & .027 & .031 & .033 & .035 & $+0.018$ (.022) & 1.00 \\
ten & interaction & 0.5 & 200 & .173 & .080 & .121 & .129 & .087 & $+0.044$ (.037) & 1.00 \\
ten & interaction & 0.8 & 200 & .416 & .073 & .281 & .285 & .197 & $+0.130$ (.046) & 1.00 \\
ten & quadratic & 0.3 & 200 & .122 & .040 & .085 & .072 & .055 & $+0.050$ (.031) & 1.00 \\
ten & quadratic & 0.5 & 200 & .342 & .187 & .301 & .141 & .120 & $+0.201$ (.040) & .993 \\
ten & quadratic & 0.8 & 200 & .796 & .660 & .754 & .453 & .210 & $+0.342$ (.043) & .960 \\
ten & interaction & 0.3 & 500 & .133 & .082 & .130 & .175 & .123 & $-0.042$ (.034) & 1.00 \\
ten & interaction & 0.5 & 500 & .393 & .177 & .313 & .418 & .222 & $-0.025$ (.044) & 1.00 \\
ten & interaction & 0.8 & 500 & .820 & .360 & .787 & .828 & .570 & $-0.008$ (.034) & 1.00 \\
ten & quadratic & 0.3 & 500 & .393 & .202 & .333 & .238 & .120 & $+0.155$ (.040) & 1.00 \\
ten & quadratic & 0.5 & 500 & .787 & .685 & .777 & .570 & .303 & $+0.217$ (.037) & 1.00 \\
ten & quadratic & 0.8 & 500 & 1.00 & .993 & 1.00 & .967 & .597 & $+0.033$ (.015) & 1.00 \\
five & quadratic & 0.3 & 200 & .056 & .131 & .123 & .036 & .044 & $+0.020$ (.014) & .000 \\
five & quadratic & 0.6 & 200 & .044 & .547 & .500 & .035 & .039 & $+0.009$ (.013) & .013 \\
five & quadratic & 0.3 & 500 & .053 & .382 & .330 & .039 & .054 & $+0.014$ (.014) & .000 \\
five & quadratic & 0.6 & 500 & .065 & .978 & .964 & .038 & .048 & $+0.028$ (.014) & .000 \\
\bottomrule
\end{tabular}
\end{table}

On the benchmark of Liu et al.\ the network and the $\chi^{2}$ of the same map are level in most cells: the learned
score adds nothing there, and the power comes from the map and its calibration. With two active covariates among ten,
the setting of Section~7 of the main text, the network is ahead of the sum of squares in half of the cells and behind
in none, because the misfit is concentrated in a few cells and a sum of squares spreads its power over all $36$; the
largest-cell statistic, which reads only one cell, is weaker still. In the U-shape design the axis rule misses the
misfit covariate and every statistic of its map stays near the level; the all-pairs reading recovers it.

\section{Locating the misfit}\label{sec:locS}

Five standard normal covariates, $n=500$, $\eta_{0}=-0.5+0.5x_{1}+0.8x_{2}+0.6x_{3}+0.4x_{4}+0.3x_{5}$, and a model fitted
linear in all five. Truths: $\eta_{0}+C(x_{1}^{2}-1)$ (U-shape), $\eta_{0}+C\max(x_{2}-1,0)$ (threshold),
$\eta_{0}+Cx_{1}x_{3}$ (interaction), and $\eta_{0}$ ($500$ null datasets; $300$ per alternative cell). One set of
$B=199$ refits per dataset serves every tool. DeepGOF-1: its p-value, the pair it displays, and the cells beyond the
simultaneous $95\%$ threshold of that map (the $95\%$ quantile of its largest absolute cell over the bootstrap
replicates), read only when the test rejects. The true region is both extreme sixths of $x_{1}$ (U-shape), the top two
sixths of $x_{2}$ (threshold) and the four corner blocks of the $(x_{1},x_{3})$ map (interaction). Lin--Wei--Ying: for
each covariate the supremum of the cumulative residual over its ranks, calibrated by the same refits, and the spline
screen: for each covariate the likelihood-ratio test of a natural spline with three degrees of freedom in place of the
linear term; both flag covariates at the Bonferroni level $.05/5$ and reject when they flag any.

\begin{table}[h]
\centering
\caption{Locating the misfit ($n=500$). Rejects: rejection rate at $.05$. Right: the displayed pair contains the true
covariate(s) (DeepGOF-1), or the flagged set is exactly the true covariate(s) (the other two). Located: the tool rejects
and every flag lies in the truth (for DeepGOF-1, at least one flagged cell and every flagged cell inside the true
region). False: the tool flags something outside the truth; under the null, any flag. Monte Carlo standard error at most
$.029$.}
\label{tab:loc}
\small\setlength{\tabcolsep}{3pt}
\begin{tabular}{@{}llllcccc@{}}
\toprule
truth & $C$ & tool & datasets & rejects & right & located & false \\
\midrule
none & -- & DeepGOF-1, axis rule & 500 & .044 & -- & -- & .006 \\
none & -- & DeepGOF-1, all pairs & 500 & .056 & -- & -- & .012 \\
none & -- & Lin--Wei--Ying & 500 & .056 & -- & -- & .056 \\
none & -- & spline screen & 500 & .050 & -- & -- & .050 \\
\addlinespace
U-shape in $x_1$ & 0.4 & DeepGOF-1, axis rule & 300 & .197 & 25\% & 5\% & .013 \\
U-shape in $x_1$ & 0.4 & DeepGOF-1, all pairs & 300 & .650 & 93\% & 14\% & .020 \\
U-shape in $x_1$ & 0.4 & Lin--Wei--Ying & 300 & .713 & 65\% & 65\% & .063 \\
U-shape in $x_1$ & 0.4 & spline screen & 300 & .980 & 95\% & 95\% & .033 \\
\addlinespace
U-shape in $x_1$ & 0.8 & DeepGOF-1, axis rule & 300 & .210 & 18\% & 10\% & .020 \\
U-shape in $x_1$ & 0.8 & DeepGOF-1, all pairs & 300 & 1.00 & 100\% & 62\% & .100 \\
U-shape in $x_1$ & 0.8 & Lin--Wei--Ying & 300 & 1.00 & 96\% & 96\% & .037 \\
U-shape in $x_1$ & 0.8 & spline screen & 300 & 1.00 & 95\% & 95\% & .050 \\
\addlinespace
threshold in $x_2$ & 2 & DeepGOF-1, axis rule & 300 & .143 & 100\% & 1\% & .013 \\
threshold in $x_2$ & 2 & DeepGOF-1, all pairs & 300 & .103 & 52\% & 0\% & .030 \\
threshold in $x_2$ & 2 & Lin--Wei--Ying & 300 & .150 & 9\% & 9\% & .063 \\
threshold in $x_2$ & 2 & spline screen & 300 & .463 & 41\% & 41\% & .050 \\
\addlinespace
threshold in $x_2$ & 4 & DeepGOF-1, axis rule & 300 & .403 & 100\% & 2\% & .050 \\
threshold in $x_2$ & 4 & DeepGOF-1, all pairs & 300 & .270 & 77\% & 2\% & .040 \\
threshold in $x_2$ & 4 & Lin--Wei--Ying & 300 & .413 & 39\% & 39\% & .027 \\
threshold in $x_2$ & 4 & spline screen & 300 & .867 & 83\% & 83\% & .037 \\
\addlinespace
interaction $x_1x_3$ & 0.6 & DeepGOF-1, axis rule & 300 & .067 & 2\% & 0\% & .010 \\
interaction $x_1x_3$ & 0.6 & DeepGOF-1, all pairs & 300 & .290 & 73\% & 13\% & .020 \\
interaction $x_1x_3$ & 0.6 & Lin--Wei--Ying & 300 & .067 & 0\% & 0\% & .030 \\
interaction $x_1x_3$ & 0.6 & spline screen & 300 & .077 & 0\% & 0\% & .040 \\
\addlinespace
interaction $x_1x_3$ & 1 & DeepGOF-1, axis rule & 300 & .110 & 1\% & 1\% & .007 \\
interaction $x_1x_3$ & 1 & DeepGOF-1, all pairs & 300 & .863 & 97\% & 73\% & .020 \\
interaction $x_1x_3$ & 1 & Lin--Wei--Ying & 300 & .067 & 0\% & 0\% & .020 \\
interaction $x_1x_3$ & 1 & spline screen & 300 & .127 & 0\% & 0\% & .053 \\
\bottomrule
\end{tabular}
\end{table}

A missed interaction of two covariates is invisible to tools that look at one covariate at a time, and the all-pairs
map both detects it and names the pair. Curvature or a threshold in one covariate is a one-covariate problem, and the
spline screen locates it best; the axis rule, which ranks covariates by their linear effect, locates neither the
U-shape nor the interaction.

\section{External validation of a published risk model}\label{sec:extS}

A published model with fixed coefficients, $\eta_{0}=-0.5+0.5x_{1}+0.8x_{2}+0.6x_{3}+0.4x_{4}+0.3x_{5}$, is checked on new
patients with five independent standard normal covariates. The null is $y_{i}\sim\mathrm{Bernoulli}(p_{i})$ with
$p_{i}=\mathrm{logit}^{-1}(\eta_{0i})$ given; nothing is refitted. Truths in the validation population: $\eta_{0}+C$
(overall rate, $C\in\{.2,.4\}$), $C\eta_{0}$ (an overfitted model with calibration slope $C\in\{.8,.6\}$),
$\eta_{0}+C(x_{1}^{2}-1)$ (U-shape, $C\in\{.4,.8\}$), $\eta_{0}+C\max(x_{2}-1,0)$ (threshold, $C\in\{2,4\}$) and
$\eta_{0}+Cx_{1}x_{3}$ (interaction, $C\in\{.6,1\}$). Null datasets: $500$ at $n=250$ and $500$, $300$ at $n=1{,}000$;
alternatives: $300$ per cell at $n=500$.

With $p$ fixed, a statistic calibrated by drawing $y^{*}\sim\mathrm{Bernoulli}(p)$ and recomputing it is an exact Monte
Carlo test at any $n$ \citep{besagclifford1989}. DeepGOF-1 (its map over the covariates, standardized by the given $p$;
the axis rule uses the published coefficients), the projection statistic and Pearson's statistic (the statistic of the
Osius--Rojek test) are calibrated in this way from $B=199$ shared draws. The others use their usual external
references: Hosmer--Lemeshow with ten groups against $\chi^{2}_{10}$; Stukel's two directions added to the offset
$\eta_{0}$, likelihood ratio against $\chi^{2}_{2}$; and the GiViTI calibration belt in its external form, which reduces
to the test of calibration intercept and slope when the selected calibration curve is linear.

\begin{table}[h]
\centering
\caption{External validation: size at nominal $.05$ (standard error about $.010$ at $500$ datasets and $.013$ at
$300$).}
\label{tab:extlevel}
\small\setlength{\tabcolsep}{3pt}
\begin{tabular}{@{}rcccccccc@{}}
\toprule
$n$ & DG-1 axes & DG-1 all & DG-1 comb. & proj. & Pearson & HL & Stukel & GiViTI \\
\midrule
250 & .040 & .044 & .046 & .068 & .050 & .060 & .060 & .070 \\
500 & .038 & .044 & .042 & .046 & .076 & .056 & .066 & .062 \\
1000 & .063 & .053 & .053 & .050 & .043 & .060 & .053 & .053 \\
\bottomrule
\end{tabular}
\end{table}

\begin{table}[h]
\centering
\caption{External validation, $n=500$: matched-level power (each test placed at exactly $.05$ on its own $500$ nulls;
$300$ datasets per cell, standard error at most $.029$). Bold: the highest in the row.}
\label{tab:extpower}
\small\setlength{\tabcolsep}{3pt}
\begin{tabular}{@{}lrcccccccc@{}}
\toprule
miscalibration & $C$ & DG-1 axes & DG-1 all & DG-1 comb. & proj. & Pearson & HL & Stukel & GiViTI \\
\midrule
overall rate & 0.2 & .24 & .29 & .28 & \textbf{.46} & .07 & .17 & .11 & .32 \\
overall rate & 0.4 & .78 & .83 & .80 & \textbf{.96} & .22 & .73 & .45 & .92 \\
slope & 0.8 & .10 & .07 & .09 & .15 & \textbf{.39} & .29 & .29 & .36 \\
slope & 0.6 & .36 & .38 & .42 & .45 & .97 & .92 & .95 & \textbf{.98} \\
U-shape & 0.4 & .04 & \textbf{.47} & .40 & .05 & .10 & .08 & .08 & .06 \\
U-shape & 0.8 & .06 & \textbf{1.00} & .99 & .20 & .42 & .35 & .31 & .34 \\
threshold & 2 & \textbf{.43} & .26 & .38 & .17 & .02 & .09 & .13 & .17 \\
threshold & 4 & \textbf{.84} & .71 & .80 & .35 & .01 & .14 & .24 & .36 \\
interaction & 0.6 & .08 & \textbf{.16} & .15 & .09 & .16 & .12 & .12 & .11 \\
interaction & 1 & .07 & \textbf{.61} & .51 & .04 & .43 & .27 & .36 & .24 \\
\bottomrule
\end{tabular}
\end{table}

Miscalibration that is the same for every patient, a wrong overall rate or an overfitted slope, is found best by the
tests built along the predicted risk. Miscalibration that differs between patients with the same predicted risk is
invisible to them, and DeepGOF-1, which reads the residuals over the covariates, has the most power on every such cell
but the weaker interaction, where every test is weak.

\section{The SUPPORT design: level, power, computing time and a split-sample validation}\label{sec:plasmodeS}

Each replicate draws $n$ of the $8{,}873$ complete-case patients without replacement, keeps their real covariates, and
simulates in-hospital death from a true model fitted on all of them: the null is M0, linear in every covariate, and the
alternative is M1, with a natural spline in mean arterial pressure (a U-shaped risk). The analyst fits M0. DeepGOF-1's
three readings share one set of $B=199$ refits; the projection test ($B=199$), Stukel's joint and SstBoth tests, the
Hosmer--Lemeshow test and GiViTI were run up to $n=1{,}000$, and at the full $n=8{,}873$ DeepGOF-1 alone (every patient,
outcomes redrawn).

\begin{table}[h]
\centering
\caption{SUPPORT design: size at nominal $.05$ (standard error about $.013$ at $300$ datasets, $.015$ at $200$ and
$.022$ at $100$).}
\label{tab:plasmodelevel}
\small\setlength{\tabcolsep}{3pt}
\begin{tabular}{@{}rrcccccccc@{}}
\toprule
$n$ & datasets & DG-1 axes & DG-1 all & DG-1 comb. & proj. & Stukel J & SstBoth & HL & GiViTI \\
\midrule
250 & 300 & .030 & .040 & .040 & .080 & .053 & .048 & .067 & .037 \\
500 & 300 & .053 & .057 & .057 & .040 & .047 & .068 & .043 & .050 \\
1{,}000 & 200 & .060 & .035 & .040 & .065 & .070 & .070 & .080 & .065 \\
8{,}873 & 100 & .020 & .040 & .030 & -- & -- & -- & -- & -- \\
\bottomrule
\end{tabular}
\end{table}

\begin{table}[h]
\centering
\caption{SUPPORT design, U-shaped risk of blood pressure: matched-level power (raw power in brackets); $200$ datasets
at $n=250$ and $500$, $150$ at $1{,}000$.}
\label{tab:plasmodepower}
\small\setlength{\tabcolsep}{2pt}
\begin{tabular}{@{}rcccccccc@{}}
\toprule
$n$ & DG-1 axes & DG-1 all & DG-1 comb. & proj. & Stukel J & SstBoth & HL & GiViTI \\
\midrule
250 & .10 (.07) & .10 (.07) & .09 (.06) & .12 (.17) & .04 (.04) & .05 (.05) & .06 (.06) & .07 (.03) \\
500 & .10 (.10) & .10 (.12) & .10 (.10) & .35 (.32) & .07 (.07) & .06 (.06) & .06 (.04) & .05 (.06) \\
1{,}000 & .21 (.22) & .25 (.23) & .23 (.22) & .57 (.65) & .04 (.05) & .03 (.03) & .01 (.05) & .03 (.03) \\
\bottomrule
\end{tabular}
\end{table}

\emph{Computing time.} On M0 with all $8{,}873$ patients and $B=199$, on one core of the same machine under the same
load, DeepGOF-1 took 13~s (all-pairs and combined readings 71~s), and
\texttt{projection.gof()} took 35{,}568~s (9~h~53~min), with a peak memory of
4.5~GB; both returned $p=.005$.

\emph{Split-sample external validation.} The public release has no hospital or date identifier, so the split is
random: half of the $8{,}873$ patients develop a model, which is frozen and checked on the other half ($4{,}437$
patients), repeated over 20 random splits. The frozen models are M0 and M2 of the main text.
DeepGOF-1 is \texttt{deepgof1.external()} ($B=199$) on the ten covariates (sex coded $0/1$); GiViTI is the calibration
test in its external form; Hosmer--Lemeshow uses ten groups against $\chi^{2}_{10}$; Stukel's two terms are added to
the frozen linear predictor.

\begin{table}[h]
\centering
\caption{SUPPORT, split-sample external validation: number of the 20 splits in which each test
rejects at $.05$; median p-values of DeepGOF-1 (combined reading) and GiViTI; median calibration intercept (slope fixed
at one) and calibration slope on the validation half.}
\label{tab:extsplit}
\small\setlength{\tabcolsep}{3pt}
\begin{tabular}{@{}lcccccccccc@{}}
\toprule
model & DG-1 comb. & DG-1 axes & DG-1 all & GiViTI & HL & Stukel & med.\ p DG-1 & med.\ p GiViTI & intercept & slope \\
\midrule
M0 & 20 & 17 & 20 & 2 & 11 & 7 & .007 & .336 & .01 & .99 \\
M2 & 8 & 1 & 8 & 8 & 1 & 16 & .110 & .068 & .01 & .91 \\
\bottomrule
\end{tabular}
\end{table}

\section{A few corrupted records at $n=500$ and $1{,}000$}\label{sec:corrupt1000S}

The design of the corrupted-record studies of the directed calibration tests, so that the stored p-values of the other
tests are on the same datasets: $x\sim U(-3,3)$, $d\sim\mathrm{Bernoulli}(.5)$,
$y\sim\mathrm{Bernoulli}\{\mathrm{logit}^{-1}(.6x+.5d)\}$, fitted as $y\sim x+d$, the correct model. After the outcome is
drawn, $k$ records chosen at random have $x$ multiplied by $4$ or $8$; the outcome is right and the covariate is wrong,
so every rejection is a false alarm caused by the corruption. Each dataset was rebuilt from its stored seed and checked
against the stored number of events and fitted slope (no mismatch), and DeepGOF-1 ($B=199$; with one pair of covariates
its readings coincide) was run on it. The projection test ($B=199$) and BAGofT (package defaults) at $n=1{,}000$ were run
on the first $200$ and $100$ datasets of their cells (number in brackets).

\begin{table}[h]
\centering
\caption{False-alarm rate at $.05$ with corrupted covariate values. Standard error about $.007$ at $1{,}000$ datasets,
$.015$ at $200$ and $.022$ at $100$. HL: Hosmer--Lemeshow, ten groups; Stukel: joint score test; GiViTI: internal form.}
\label{tab:corrupt1000}
\small\setlength{\tabcolsep}{3pt}
\begin{tabular}{@{}llrcccccc@{}}
\toprule
$n$ & corruption & datasets & DeepGOF-1 & HL & Stukel & GiViTI & projection & BAGofT \\
\midrule
1{,}000 & none & 1{,}000 & .051 & .040 & .044 & .042 & .055 (200) & .030 (100) \\
1{,}000 & $\times4$, 1 record & 1{,}000 & .047 & .051 & .121 & .089 & .080 (200) & -- \\
1{,}000 & $\times4$, 2 records & 1{,}000 & .050 & .044 & .217 & .146 & .040 (200) & -- \\
1{,}000 & $\times4$, 5 records & 1{,}000 & .055 & .042 & .398 & .251 & .100 (200) & -- \\
1{,}000 & $\times4$, 10 records & 1{,}000 & .088 & .058 & .627 & .397 & .085 (200) & .030 (100) \\
\addlinespace
500 & none & 100 & .060 & .020 & .060 & .070 & .080 (100) & .060 (100) \\
500 & $\times4$, 3 records & 100 & .060 & .000 & .450 & .290 & .060 (100) & .010 (100) \\
500 & $\times8$, 3 records & 100 & .160 & .060 & .490 & .370 & .220 (100) & .050 (100) \\
\bottomrule
\end{tabular}
\end{table}

\clearpage
\noindent{\small\itshape Online Resource 1 for ``Where Does a Logistic Risk Model Fail? An Audited Neural Goodness-of-Fit Test for Model Development and External Validation'', by E.~K.~Ebrahim, O.~A.~E.~Hussein and A.~El-Kotory.}\par\medskip
\noindent{\Large\bfseries Part C. Reproducibility record}\par\medskip
\section{Benchmark bookkeeping: missing p-values and failed refits}\label{sec:bookkeeping}

This section reports two counts behind Table~3 of the paper (the extended benchmark of
\citet{liu2024}, $24$ blocks, $143{,}999$ null and alternative datasets) and behind its
simulation with two active covariates among ten ($16{,}000$ datasets). Both are computed from the
stored per-replicate rows.

\emph{Stukel's test without a p-value.} Stukel's test in the SstBoth form of \citet{liu2024}
returns no p-value on some datasets, for example when every fitted risk lies on one side of one
half, so that one of its two added variables is identically zero. In Table~3 of the paper each
test is evaluated on the datasets where it returns a p-value, for nulls and alternatives alike.
Table~\ref{tab:stukelna} gives the share. Over the benchmark it is $28.6\%$ of the null and $5.7\%$ of the alternative
datasets; it is largest in Setting~3, where the true linear predictor is negative everywhere
under the null, and near zero at $n=500$ in Settings~1 and~4. In the two-active-covariate simulation it
is zero in every cell. Under the original convention, which counts such a dataset as not
rejecting, the missing values, falling mostly on nulls, lower this test's raw size, and at matched
level they let its threshold rise above $.05$ wherever the null share is large: the median matched
threshold is then about $p\le.17$, $.27$ and $.997$ in Setting~3 at $n=50$, $100$ and $500$.

\begin{table}[!htbp]
\caption{Share (\%) of datasets on which Stukel's test (SstBoth) returns no p-value in the benchmark
of Table~3 of the paper, pooled over the two training corpora ($4{,}000$ nulls per cell, $3{,}999$
in Setting~2 at $n=50$; $8{,}000$ alternatives per cell). Over all cells: $28.6\%$ of nulls and
$5.7\%$ of alternatives}
\label{tab:stukelna}
\centering
\begin{tabular}{@{}lcccccc@{}}
\toprule
& \multicolumn{2}{c}{$n=50$} & \multicolumn{2}{c}{$n=100$} & \multicolumn{2}{c}{$n=500$} \\
\cmidrule(lr){2-3}\cmidrule(lr){4-5}\cmidrule(l){6-7}
setting & nulls & alternatives & nulls & alternatives & nulls & alternatives \\
\midrule
1 & 8.1  & 3.6  & 2.5  & 1.1  & 0.0  & 0.0 \\
2 & 37.2 & 4.7  & 36.1 & 2.7  & 11.7 & 0.1 \\
3 & 68.7 & 21.7 & 81.0 & 16.2 & 97.6 & 10.0 \\
4 & 0.7  & 5.6  & 0.1  & 3.2  & 0.0  & 0.2 \\
\bottomrule
\end{tabular}
\end{table}

\emph{The Osius--Rojek test.} Its size is far above nominal in Settings~3 and~4 of the benchmark at
every $n$ ($\osiussizebadthree$ and $\osiussizebadfour$ even at $n=500$), because the variance of its
normal reference collapses when the design absorbs almost all of
$(1-2\hat\pi)/\{\hat\pi(1-\hat\pi)\}$: only $\osiusvarthree$\% and $\osiusvarfour$\% of that variance
survives there.

\emph{Failed bootstrap refits.} The two bootstrap-calibrated tests treat a refit that fails
differently. The projection test scores a failed refit $-\infty$, that is, as a bootstrap statistic
below the observed one, which counts toward rejection. DeepGOF-1 as released scores it $+\infty$
(step~5 of Section~\ref{sec:setting}), which counts toward acceptance, and so does the harness of
the two-active-covariate simulation; the benchmark harness of Table~3 substitutes an all-zero map.
Both tests refit with the same routine (maximum likelihood, with a Firth fallback), which returns
no fit only when the fitting routine stops with an error. The projection rows store the count:
no refit failed in the $28{,}655{,}801$ refits of the benchmark ($143{,}999$ datasets $\times$
$199$) or in the $3{,}184{,}000$ refits of the two-active-covariate simulation. The rows of
DeepGOF-1 do not store the count, but its refits use the same routine on the same datasets under
the same bootstrap law, so the same rate applies. A rate of zero in $31{,}839{,}801$ refits puts
the failure probability per refit below $9.4\times10^{-8}$ at the $95\%$ level (below
$1.05\times10^{-7}$ on the benchmark alone), so the chance that any one dataset has a failed refit
is below $2.1\times10^{-5}$. One failed refit moves a p-value by at most $1/200$. At this rate
neither convention can change a reported number.

\section{Deposit index}\label{sec:hashes}

Table~\ref{tab:deposit} lists, for each study of this Online Resource 1, the
folder of the public archive that holds its scripts and per-replicate output.

\begin{table}[!htbp]
\centering
\caption{Deposit folders of the studies in this Online Resource 1}
\label{tab:deposit}
\small
\begin{tabular}{@{}lll@{}}
\toprule
study & section & folder \\
\midrule
slope check of the recession function; blind-cone search & \ref{sec:recession}, \ref{sec:cone} & \texttt{benchmark/theory/}, \texttt{results/} \\
exact maximized Monte Carlo variant & \ref{sec:fsbound} & \texttt{theory/exact\_mmc/} \\
calibration-gap pilot (Table~\ref{tab:pilot}) & \ref{sec:fsbound} & \texttt{theory/calibration\_gap/} \\
gap-bound coverage (Table~\ref{tab:Gcov}) & \ref{sec:levelgap} & \texttt{theory/level\_gap/} \\
level error against $n$ (Table~S.11a) & \ref{sec:rate} & \texttt{theory/level\_gap/rate/} \\
local power: formulas and convergence (Table~\ref{tab:localnum}) & \ref{sec:localpower} & \texttt{theory/local\_power/} \\
covariates that carry no misfit (Table~\ref{tab:irrcheck}) & \ref{sec:irrelevant} & \texttt{theory/local\_power/irrelevant/} \\
power certificates and their limits & \ref{sec:certpowerS} & \texttt{theory/finite\_power/} \\
touch rate and miss probability of the deletion rule & \ref{sec:touch} & \texttt{theory/small\_n/touch\_rate/} \\
Neyman--Pearson gap (Table~\ref{tab:np}) & \ref{sec:np} & \texttt{theory/optimality/} \\
Studies 1 and 2 & \ref{sec:studies} & \texttt{theory/small\_n/} \\
projection test on the datasets of Studies 1 and 2 & \ref{sec:studies} & \texttt{theory/small\_n/proj\_addon/} \\
shipped network: level study and benchmark, both readings & main text & \texttt{theory/shipped\_network/}, \texttt{theory/allpairs\_study/} \\
a U-shaped covariate among linear ones & main text & \texttt{theory/ushape/} \\
the SUPPORT application & main text & \texttt{application/support/} \\
very small samples and wrong links & \ref{sec:smallnlink} & \texttt{theory/smalln\_link/} \\
paired comparison with BAGofT (Table~\ref{tab:bagoftS}) & \ref{sec:bagoftS} & \texttt{bagoft/} \\
FDIC level transfer (Table~\ref{tab:fdicS}) & \ref{sec:fdicS} & \texttt{application/fdic/}, \texttt{results/fdic/} \\
\bottomrule
\end{tabular}
\end{table}

\emph{Training protocol of the shipped network.} Table~\ref{tab:training} summarizes how the
frozen network was built; the files named are in \texttt{training/} of the archive. Every DeepGOF-1
computation in this Online Resource 1 uses this network, except the classifier diagnostic quoted in
Remark~\ref{rem:pivotality}; that diagnostic and the benchmark simulations of Sections~5 and~6 of
the paper use networks with the same architecture and training loop trained per simulation block
(\texttt{benchmark/grid/s3\_score.py}, \texttt{benchmark/liu/liu\_score.py}).

\begin{table}[!htbp]
\centering
\caption{Training protocol of the shipped network}
\label{tab:training}
\small
\begin{tabular}{@{}p{0.2\textwidth}p{0.74\textwidth}@{}}
\toprule
item & protocol \\
\midrule
training datasets & $n=200$ each; map with $K=6$ ($36$ cells) as in Section~\ref{sec:setting} \\
null class & one parametric-bootstrap replicate of a correctly specified fit (fit, simulate from the
fit, refit, map); covariates $x_{1}\sim U(-3,3)$, $x_{2}\sim N(0,1)$, $\eta_{0}=0.3+0.8x_{1}-0.5x_{2}$ \\
alternative class & $100$ random departure families; each has two covariates as above (probability
$.7$) or four, $x_{1}\sim U(-3,3)$, $x_{2},x_{3}\sim N(0,1)$, $x_{4}\sim U(-2,2)$ (probability $.3$) \\
departure terms & one to three per family, drawn from quadratic, cubic, interaction, sine, bump,
step, absolute value, and a quadratic and a sine of the linear predictor (probabilities $.15$, $.10$,
$.15$, $.12$, $.12$, $.10$, $.06$, $.10$, $.10$), with random signs and shapes; their sum is rescaled to
a standard deviation drawn from $U(.25,.75)$ on the logit scale \\
corpus & $8{,}400$ labelled maps (\texttt{ship\_corpus.csv}); the generator
\texttt{01\_corpus.R} reproduces the recipe with fresh seeds \\
network & three $3\times3$ convolutions ($16$, $32$, $32$ channels), one $2\times2$ max-pooling stage,
global max and mean readout ($64$), a fully connected layer of $64$ and a linear head; $18{,}273$
parameters, trained as a classifier of null against alternative maps \\
export and check & \texttt{03\_export\_to\_R.py}; \texttt{VERIFY.R} compares the base-R forward pass
with the training framework (agreement $5\times10^{-11}$) \\
\bottomrule
\end{tabular}
\end{table}

\bibliographystyle{plainnat}
\bibliography{esm}